%% file: s0_main.tex
\documentclass[twocolumn]{aastex63}

\shorttitle{COMs in HMYSOs with ALMA}
\shortauthors{Baek et al.}

\graphicspath{{./}{Figures/}}
\DeclareUnicodeCharacter{2212}{-}
\begin{document}

\title{Complex organic molecules detected in twelve high-mass star-forming regions with Atacama Large Millimeter/submillimeter Array (ALMA)}

\author{Giseon Baek}
\affiliation{School of Space Research, Kyung Hee University, 1732, Deogyeong-daero, Giheung-gu, Yongin-si, Gyeonggi-do 17104, Republic of Korea}
\email{giseon@khu.ac.kr}

\author{Jeong-Eun Lee}
\affiliation{School of Space Research, Kyung Hee University, 1732, Deogyeong-daero, Giheung-gu, Yongin-si, Gyeonggi-do 17104, Republic of Korea}
\email{jeongeun.lee@khu.ac.kr}

\author{Tomoya Hirota}
\affiliation{Mizusawa VLBI Observatory, National Astronomical Observatory of Japan, Osawa 2-21-1, Mitaka-shi, Tokyo 181-8588, Japan}
\affiliation{Department of Astronomical Sciences, SOKENDAI (The Graduate University for Advanced Studies), Osawa 2-21-1, Mitaka-shi, Tokyo 181-8588, Japan}

\author{Kee-Tae Kim}
\affiliation{Korea Astronomy and Space Science Institute, 776 Daedeok-daero, Yuseong, Daejeon 34055, Republic of Korea}
\affiliation{University of Science and Technology, Korea (UST), 217 Gajeong-ro, Yuseong-gu, Daejeon 34113, Republic of Korea}
 
\author{Mi Kyoung Kim}
\affiliation{Department of Child Studies, Faculty of Home Economics, Otsuma Women's University, 12 Sanban-cho, Chiyoda-ku 102-8357, Tokyo, Japan}

\begin{abstract}
Recent astrochemical models and experiments have explained that complex organic molecules (COMs; molecules composed of six or more atoms) are produced on the dust grain mantles in cold and dense gas in prestellar cores. However, the detailed chemical processes and the roles of physical conditions on chemistry are still far from understood. To address these questions, we investigated twelve high-mass star-forming regions using the ALMA band 6 observations. They are associated with 44/95GHz class I and 6.7 GHz class II CH$_{3}$OH masers, indicative of undergoing active accretion. We found 28 hot cores with COMs emission among 68 continuum peaks at 1.3 mm and specified 10 hot cores associated with 6.7 GHz class II CH$_{3}$OH masers. Up to 19 COMs are identified including oxygen- and nitrogen-bearing molecules and their isotopologues in cores. The derived abundances show a good agreement with those from other low- and high-mass star-forming regions, implying that the COMs chemistry is predominantly set by the ice chemistry in the prestellar core stage. One clear trend is that the COMs detection rate steeply grows with the gas column density, which can be attributed to the efficient formation of COMs in dense cores. In addition, cores associated with a 6.7 GHz class II CH$_{3}$OH maser tend to be enriched with COMs. Finally, our results suggest that the enhanced abundances of several molecules in our hot cores could be originated by the active accretion as well as different physical conditions of cores.

\end{abstract}

\section{Introduction}\label{sec:intro}
\input{s1_introduction}

\section{Observation} \label{sec:obs}
\input{s2_observation}
\input{s2_tab1_obstable}

\section{Results} \label{sec:results}
\input{s3_results}

\section{Discussion} \label{sec:discuss}
\input{s4_discussion}

\section{Summary} \label{sec:summary}
We aim to study the chemical characteristics of COMs in hot cores and the roles of physical conditions on their chemistry.
The results of our study is summarized as follows.
\begin{enumerate}
    \item We investigate twelve high-mass star forming regions associated with class I and class II CH$_3$OH masers, using the ALMA band 6 observations. In twelve regions, total 68 continuum peaks are resolved. Among them, COMs are detected in 28 hot cores, in which 10 cores are associated with 6.7 GHz class II CH$_3$OH maser.
    \item Up to 19 COMs are identified in a core, including rotational transitions in torsionally excited state. $^{13}$C, $^{18}$O, and $^{15}$N isotopologues and deuterated species are also detected.
    \item COMs detection rate increases in cores with high N(H$_2$). We divide COMs into two groups of which the first one shows flat ratio of their abundance relative to CH$_3$OH with N(H$_2$), in contrast to positive slopes found in second group. Similar tendency is found with T$_{ex}$ but less sensitive than N(H$_2$). It indicates that the COMs complexity is efficiently elevated in denser and hotter core. 
    \item The cores associated with the class II CH$_3$OH maser show higher COMs detection rates, suggesting that the 6.7 GHz class II CH$_3$OH maser can be a COMs rich hot core tracer.
    \item By the comparison of our sample with various types of COM-rich sources, we found a general similarity in the relative COMs abundances to CH$_3$OH over six orders of magnitudes in source luminosity. It suggests that the COMs chemistry is not dominantly affected by the current luminosity, but is predominantly set by the ice chemistry in the cold prestellar core stage. On the other hand, different COMs abundances within an order of magnitude may be influenced by the local physical conditions.
    \item The correlations among COMs detected in our observation are mostly explained by the current chemical models, however, the distinct roles of grain-surface and gas-phase chemistry are needed to be examined by models to understand what causes the relative abundances of COMs including CH$_3$CHO and HCOCH$_2$OH.
\end{enumerate}

\acknowledgments
We thank the anonymous referee for detailed and helpful comments. 
We thank J.-H. Kang for comments on the data reduction and manuscript. We Thank J. Kim and KaVA team for sharing the distance information toward a part of the regions. We also thank Y.-L. Yang for sharing the results of PEACHES survey and related discussions.  This work was supported by the National Research Foundation of Korea (NRF) grant funded by the Korea government (MSIT) (grant number 2021R1A2C1011718). TH is financially supported by the MEXT/JSPS KAKENHI Grant Numbers 17K05398, 18H05222, and 20H05845. This paper makes use of the following ALMA data: ADS/JAO.ALMA\#2015.1.01571.S.
ALMA is a partnership of ESO (representing its member states), NSF (USA) and NINS (Japan), together with NRC (Canada), NSC and ASIAA (Taiwan), and KASI (Republic of Korea), in cooperation with the Republic of Chile. The Joint ALMA Observatory is operated by ESO, AUI/NRAO and NAOJ.\\
 
\bibliographystyle{aasjournal}
\bibliography{myso12}{} 

\input{s3_tab1_1.3cont_peak}
\input{s3_tab2_216GHz_info}
\input{s3_tab3_maser_association}
\input{s3_tab4_rotation_diagram_all.tex}
\input{s3_tab5_columndensity_XCLASS}

\input{s3_f1_figure1_continuum_figureset}
\input{s3_f2_spectrum_figureset}

\pagebreak
\input{s5_appendix}

\end{document}

%% file: s1_introduction.tex
Complex organic molecules (COMs; molecules composed of six or more atoms) are the building blocks of prebiotic molecules, such as amino acids. As such, one of the key questions on star formation is how COMs are synthesized and delivered to the planets and comets.

Recent astrochemical models and laboratory experiments have explained that COMs are produced on the icy surface and mantle of dust grain in cold ($\sim$ 10 K) and dense (10$^{4}$--10$^{7}$ cm$^{-3}$) gas in star-forming regions, even in environments where the diffusion of large radicals cannot occur \citep{oberg2021}. As a protostar forms, it warms up the innermost envelope, liberating the molecules from grain mantles into the gas phase when the grain surface temperature is high enough to evaporate the species. The dust grains are also heated by the shock energy generated by protostellar jets and outflows. As the protostar evolves, the temperature of the inner envelope reaches around 100 -- 300 K, forming a so-called hot core, or hot corino for low-mass counterpart, where the molecules are completely desorbed from dust grain mantles \citep{Herbst2009,Garrod2013,Oberg14,Oberg2016,Garrod2022}.

According to current models and laboratory studies, in cold and dense environments such as prestellar stage, radicals are produced on the icy mantle via successive atom-addition reactions or dissociation of present molecules. The production of COMs could occur in non-diffusive manners \citep{Jin2020,Garrod2022}, and if the mantle is warmed up (10 -- 100 K), radicals become mobile in the icy mantle, combine with other species and form COMs \citep{Oberg2016,oberg2021}. 
Once a protostar has formed, the protostellar luminosity determines the physical condition of its natal dense core, which in turn affects the further chemical characteristics.

High-mass young stellar objects (HMYSOs; M $\geq$ 8 M$_{\odot}$) undergo a dense cold phase, and rapidly evolve \citep{Motte2018}. Since the production of the ice mantle on dust grain is sensitive to the density and timescale, the chemical richness and complexity depend on the initial conditions of ice formation in the early stage of each star-forming region \citep{Aikawa2020}.

If an HMYSO is formed via a scaled-up version of its low mass counterpart, its luminosity is mostly an accretion luminosity \citep{Fischer2022}. The degree of accretion has a significant effect on the temperature of a hot core region and the UV radiation strength. The temperature is one of the key factors that control the chemistry. The temperature regulates the synthesis and destruction rates of species, determining the abundances of COMs, even affecting subsequent chemical networks in both ice and gas phases. Strong UV radiation generated by a luminous protostar promotes photochemistry, causing abundant COMs, especially in icy grain mantles \citep{Oberg2016}. 

In spite of the expectations, the chemical characteristics in hot cores of HMYSOs are largely unknown because of their rapid formation timescales ($\sim$ 10$^{5}$ years) and large distances (d $>$ 1kpc), making structural identification very challenging. Since the development of facilities which enable to resolve the hot cores such as Atacama Large Millimeter/submillimeter
Array (ALMA), tens of COMs have been found in hot cores in different physical conditions \citep{Herbst2009,Jorgensen2020,McGuire2022}.

On the other hand, the maser lines from various molecular species, such as OH, H$_{2}$O, CH$_{3}$OH, and SiO, are useful probes for the high-mass star forming regions to trace energetic processes in very specific conditions associated with HMYSOs. In particular, the CH$_{3}$OH maser at 6.6685192~GHz (5$_{1}$-6$_{0}$A$^{+}$) is exclusively associated with high-mass star-forming regions (HMSFRs; \citealt{Minier2003,Breen2013}) since it is pumped by strong far-infrared emission from warm dust heated by HMYSOs \citep[e.g.][]{Cragg2005}. VLBI observations have revealed that the CH$_{3}$OH masers could trace circumstellar disks, torus, or outflows \citep{Bartkiewicz2009, Bartkiewicz2016, Fujisawa2014}. Although the timescales and sequence of the evolutionary phases of HMYSOs associated with various masers are still a matter of debate, the 6.7~GHz CH$_{3}$OH maser tends to be associated with the early phase of high-mass star-formation with active accretion \citep{Walsh1998,Cyganowski2009, Breen2010, deVilliers2015}. 

Since the majority of previous maser studies have focused on kinematics and measuring distances, the physical and chemical characteristics combined with maser properties have not been sufficiently studied. The spatial resolutions to study thermal emission are so low that the information given by maser observation (e.g., \citealt{Green2009,Hu2016}) is diluted when it is compared with thermal line observation, making the comprehensive interpretation difficult. 
In addition, while it has become known that the formation of COMs is a natural consequence of star formation, their forming mechanism in such environment is still unclear. To address these questions, we aim to study the HMYSOs in different physical conditions by taking advantage of their strong emission and larger samples than their low mass counterparts, and with constraints provided by the class II CH$_{3}$OH maser association.

We investigate the twelve HMSFRs observed with ALMA in previously unexplored spatial resolution, to trace how physical properties affect their chemistry, in particular, COMs. The twelve regions are the sources of KVN (Korean VLBI Network) and VERA (VLBI Exploration of Radio Astrometry) array (KaVA) Large program \citep{Kim2018}. They are associated with class I or II CH$_{3}$OH masers (6.7 GHz; e.g., \citealt{Green2010,Breen2015,Hu2016}, and 44 and 95 GHz; \citealt{Kang2016}). 
The class I CH$_{3}$OH maser associations hint {\it indirectly} the active mass accretion process in our targets because the class I CH$_{3}$OH maser emerges in the shocked regions by outflows and jets located along various offsets from the central source. The presence of protostars is also endorsed by the class II CH$_{3}$OH maser emission since it is radiatively pumped by the strong infrared radiation field. Therefore, their association implies that the active accretion process is ongoing in the regions. In particular, the class II CH$_{3}$OH masers are a strong evidence of heated dust material, which can develop hot cores enriched with COMs. 

One of our samples, G25.82-0.17 was revealed as a clustered HMSFR with sources in different evolutionary stages, from starless cores to an ultracompact H II region \citep{Kim2020}. In G25.82-0.17 W1 core, a rotating signature has been found at 230.368 GHz CH$_{3}$OH thermal transition, and an outflow shock feature was also detected in the SiO line. The 229.759 GHz CH$_{3}$OH transition shows a mixture of thermal and maser emission. 
Hirota et al. (submitted to PASJ) investigated G24.33+0.14 region. From miltiple transitions of CH$_3$OH including the 229.759 GHz class I CH$_3$OH maser and the lines of 13 COMs, they detected the variability of line intensity responding to the accretion burst. 
In the observed spectral windows, a great number of unidentified lines of multiple COM candidates were also detected. 

In this paper, we present the chemical diversity of COMs in twelve HMSFRs using ALMA observation. We also explore the relation between the class II CH$_{3}$OH maser and COMs chemistry in hot cores. In Section \ref{sec:obs}, the observation is described, and the results are presented in Section \ref{sec:results} including core detection, line identification, class II CH$_{3}$OH maser search, and rotation diagram of COMs. In Section \ref{sec:discuss} we discuss the roles of physical properties and class II CH$_{3}$OH maser on the characteristics of COMs, the correlation between COMs, isotopic ratios and comparison with other regions. Finally, we summarize the contents in Section \ref{sec:summary}. 

%% file: s2_observation.tex
ALMA observations in Band 6 were carried out (2015.1.01571.S: P.I. M.-K. Kim) toward twelve HMSFRs.
The observation and target information is summarized in Table \ref{tab:obs_summary}. The data was calibrated using a CASA package (version 4.7; \citealt{McMullin2007}). 
Imaging was performed on each spectral window using the CASA 5.4.0 {\it tclean} task with a Briggs weighting of 0.5. 
The spectral resolution is $\sim$1.3 km s$^{-1}$ and the angular resolutions (beam size) range from 0.${''}$27$\times$0.${''}$24 to 0.${''}$38$\times$0.${''}$25 for continuum images and from 0.${''}$27$\times$0.${''}$24 to 0.${''}$40$\times$0.${''}$28 for line cubes.

Due to the crowded emission lines in some regions, which makes the continuum level estimation using line-free channels challenging, the STATCONT method \citep{SanchezMonge2018} was used for continuum subtraction. It statistically estimates the continuum level in each pixel to provide the continuum-subtracted cube (which only contains spectrum data) and pure continuum image.

%% file: s2_tab1_obstable.tex
\startlongtable
\begin{deluxetable*}{lccccccc}
\tablecaption{The summary of the observation and target information \label{tab:obs_summary}}
\tabletypesize{\scriptsize}
\tablehead{
\colhead{Target Name} & \multicolumn{2}{c}{Phase center of observation}	& \multicolumn{2}{c}{Position offset$^{a}$} & \colhead{RMS$^{b}$}&  \colhead{Distance}& \colhead{Reference}\\
\colhead{} & \colhead{R.A.($^{hms}$)} & \colhead{Decl.($^{\circ}$ $'$ $''$)} & \colhead{$\Delta$R.A.($''$)}& \colhead{$\Delta$Decl.,($''$)} & \colhead{Jy beam$^{-1}$}& \colhead{(Kpc)} & \colhead{(for distance)}}
\startdata
G10.32-0.26$^{*}$ & 18:09:24.20 & -20:08:07.0 & -13.80 & 0.15 & 4.5E-05 &2.0 & \citet{Wienen2015}\\
G10.34-0.14$^{*}$ & 18:09:00.76 & -20:03:35.0 & -12.30 & -4.15 &7.0E-05& 1.6 &  \citet{Towner2017} \\
G18.34+1.78 & 18:17:58.00 & -12:07:27.0 & +3.15 & +2.10 & 8.5E-05 & 2.6 & \citet{Urquhart2015}\\
G18.34+1.78SW & 18:17:50.40 & -12:07:55.0 & -1.65 & +0.50 &9.0E-05& 2.6 & \citet{Urquhart2015}$^{c}$ \\
G23.43-0.18 & 18:34:39.27 & -08:31:39.0 & -1.20 & +13.60$^{d}$ &7.5E-05& 5.9 & Chibueze et al. (submitted to ApJ) \\ 
G24.33+0.14 & 18:35:07.80 & -07:35:06.0 & +5.10 & +1.85 &2.2E-04& 7.2& Hirota et al. (submitted to PASJ) \\
G25.82-0.17 & 18:39:03.63 & -06:24:09.5 & +0.15 & -1.80 &3.0E-04& 5.0 & \citet{Kim2020}\\
G27.36-0.16 & 18:41:51.06 & -05:01:43.5 & +0.00 & +0.10 &4.5E-04& 8.0& \citet{Reid2016} \\
G28.37+0.07$^{*}$ & 18:42:52.10 & -03:59:45.0 & -1.65 &+9.05  &1.2E-04&5.0 &  \citet{Butler2012}\\
G29.91-0.03 & 18:46:05.30 & -02:42:26.9 & -1.35 & +2.40 &5.0E-05& 6.2 & \citet{Wienen2015} \\
G30.70-0.07$^{*}$ & 18:47:36.60 & -02:01:55.0 & -10.20 & +6.55 &5.5E-05& 5.6&  \citet{Saral2017}\\
G49.49-0.39 & 19:23:43.96 & +14:30:31.0 & +0.00 & +3.55 &3.0E-03& 5.4& \citet{Reid2014}\\
\enddata
\tablecomments{$^{a}$ Position offset from the phase center to the strongest continuum peak of the target. $^{b}$ RMS values of primary beam uncorrected continuum images are presented to estimate the sensitivity of the observation. $^{c}$ Distance is assumed as the same as G18.34+1.78. $^{d}$ G23.43-0.18 is observed toward the G23.43-0.18 C2 for the pointing center, southern from the strongest continuum peak at C1. 
$^{*}$ mark the regions observed from offset of the phase center.}
\end{deluxetable*}

%% file: s3_results.tex
\subsection{Dust Continuum} \label{sec:dustcont}
\begin{figure*}
  \centering
\includegraphics[clip,width=0.45\textwidth,keepaspectratio]{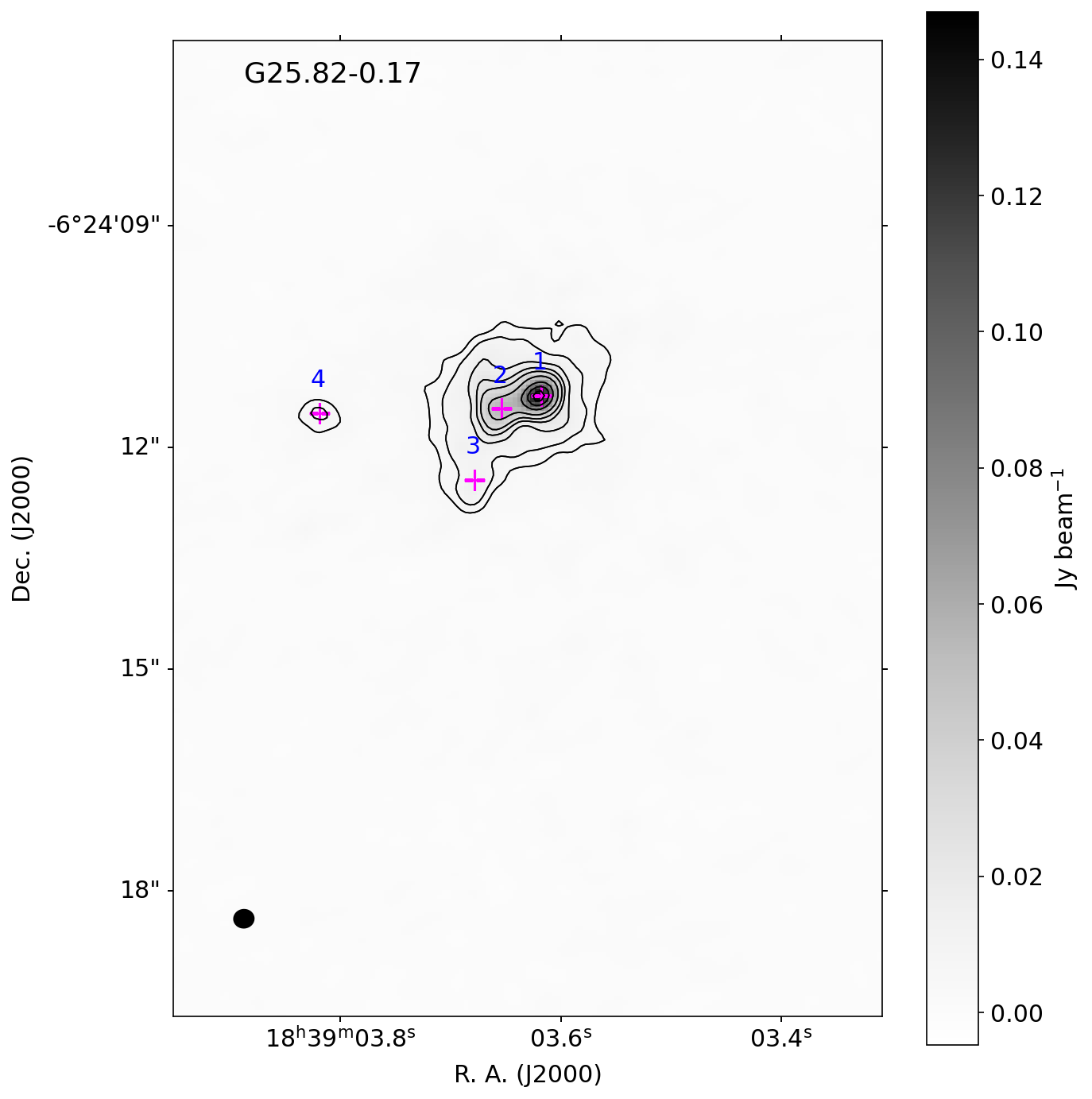}
\caption{The 1.3mm continuum images for the observed regions. Magenta plus symbols present the continuum intensity peaks, which indicates the locations of the resolved continuum cores.} For G25.82-0.17, contour levels are 0.0044, 0.0073, 0.0146, 0.0220, 0.0293, 0.0440, 0.0734, 0.1028, 0.1322 Jy beam$^{-1}$. \\ The complete figure set (12 images) is available in the online journal.
\label{fig:continuum}
\end{figure*}

We present 1.3 mm dust continuum images for the twelve HMSFRs.
For all regions, multiple continuum peaks were detected, and many of them have been resolved as multiple cores for the first time (Figure \ref{fig:continuum} and Figure \ref{cont:G10.32}--Figure \ref{cont:G49.49}). 
Although a few regions (G10.32-0.26, G10.34-0.14, G28.37+0.07) were not located at the phase center of observation, the cores were detected with the signal to noise ratios high enough to investigate their properties in primary beam uncorrected images (Table \ref{tab:cores}). 
As a result, a total of 68 continuum peaks were found in the 12 regions. We define the small regions that present their own continuum peaks as continuum cores, which cover the cores at different evolutionary stages from  
dense starless core, prestellar core, embedded protostellar cores at extremely early stage, hot cores with the COMs emission and cores associated with ultracompact HII region \citep{Kim2020}. Since this study focuses on the chemistry of hot cores, we identify cores associated with the COMs emission in Section \ref{sec:lineiden}.

Table \ref{tab:cores} summarizes the information of the detected cores and derived properties. 
Using the primary beam corrected continuum image, the peak position and intensity were found to estimate the H$_2$ column density (N(H$_{2}$)); it is given by $N(\rm H_{2})$=$S_{\nu}^{\rm beam}$/($\Omega B_{\nu}$($T_{\rm dust}$)$\kappa_{\nu}\mu_{\rm H_{2}}m_{\rm H}$, where $S_{\nu}^{\rm beam}$ is the dust continuum peak flux density in Jy beam$^{-1}$, $\Omega$ is the solid angle of the beam ($\Omega$=$\pi\theta_{maj}\theta_{min}$/4ln2), $\mu_{\rm H_{2}}$ is the mean molecular weight of 2.8 \citep{Kauffmann2008}, and m$_{\rm H}$ is the mass of the atomic hydrogen. 
$B_{\nu}$($T_{\rm dust}$) is the blackbody intensity at $T_{\rm dust}$.
For dust mass opacity, $\kappa_{\nu}$=0.1(250$\mu$m/$\lambda$)$^{\beta}$ cm$^{2}$g$^{-1}$ \citep{Hildebrand1983} with $\beta$=1.85 \citep{Ossenkopf1994} was adopted. A gas to dust mass ratio of 100 is assumed.

The uncertainty of the derived N(H$_{2}$) comes from the dust optical depth and temperature. To derive N(H$_{2}$), the optically thin dust emission is assumed. In the Rayleigh-Jeans limit ($h\nu/kT <<$ 1; $h$ and $k$ are planck and boltzmann constants, and $\nu$ is frequency) at 1.3 mm ($\nu\sim$230 GHz), a low optical depth is a valid assumption if the T$_{dust}$ is much higher than $\sim$10 K ($T >> h\nu/k$).
We adopted $T_{\rm dust}$ from the excitation temperature ($T_{\rm ex}$) of CH$_{3}$OH ($T_{ex}$(CH$_{3}$OH)$>$100 K; Section \ref{sec:rotdiagram}) at CH$_{3}$OH peak position, which are offset from the continuum peak position. The actual temperature at the protostellar position would be higher than the derived $T_{\rm ex}$, decreasing N(H$_{2}$). For some cores, the suppressed or missing COMs emission is observed at the continuum peak (Section \ref{sec:lineiden} and Section \ref{sec:contpeak}). This can be caused by the high dust optical depths at dust continuum peak position. In this case, the N(H$_{2}$) may be underestimated.  Multi-frequency continuum observation with comparable angular resolution will better estimate T$_{dust}$ and N(H$_{2}$).

For measuring the source size and integrated flux, the 2D gaussian fitting was performed for the observed continuum peaks using the $imfit$ task in CASA. A threshold of 5 $\sigma$ of RMS was applied. Masking was applied to constrain the fitting size in each core in the clustered regions. For cores surrounded by the diffuse cloud material, a zero-level offset fitting was applied to subtract diffuse emission. 

With the estimated source size and integrated flux density at 1.3 mm at each continuum core, the total enclosed mass was derived ($M_{\rm total}$=$S_{\nu}D^{2}$/$\kappa_{\nu}B_{\nu}$($T_{\rm dust}$),  \citealt{Hildebrand1983}). $S_{\nu}$ is the 1.3 mm integrated flux, and D is the adopted distance of each region presented in Table \ref{tab:obs_summary}. The optically thin dust emission and a gas to dust mass ratio of 100 are assumed. The most uncertain parameter for mass estimation is $T_{\rm dust}$. Since the continuum observation in multiple frequencies at comparable spatial resolution does not exist so far, we alternatively estimate the lower and upper limits of the mass in each continuum core. For all continuum cores, we assume 20 K as the lowest average temperature in the enclosed mass, giving the upper limit for $M_{\rm total}$. For lower-limits, toward the COMs-detected cores (see Table \ref{tab:216GHz} and Table \ref{tab:maserandCOMs}), the derived $T_{ex}$(CH$_{3}$OH) was adopted. Since the $T_{\rm ex}$ was derived at the CH$_{3}$OH position which is slightly off from the protostellar position, the actual average temperature of entire continuum core would be lower than T$_{ex}$(CH$_3$OH). Thus, the derived $M_{\rm total}$ can be considered as the lower limit. 
The COMs-undetected continuum cores may include the extremely embedded protostars, but they are probably cooler to have lower T$_{ex}$(CH$_3$OH) than COMs-detected cores in Table \ref{tab:rot_diagram1}. Thus, 100 K was adopted to give the lower limit of $M_{\rm total}$ for COMs-undetected continuum cores, which may be gravitationally bounded but chemically younger than COMs-detected cores. 
The enclosed masses of many COMs-undetected continuum cores are within the ranges of low ($M$ $\leqq$ 2$M_{\odot}$) or intermediate (2$M_{\odot}<$ $M$  $\leqq$ 8$M_{\odot}$) protostar. This is a natural consequence in the sense of the formation of YSOs in a cloud; high-mass stars are often born in a filamentary structure. In addition, inhomogeneous density distribution in a turbulent massive core may result in additional formation of low and intermediate mass protostars even during the monolithic collapse of the high-mass star formation. 

In-depth studies of G24.33+0.14 and G25.82-0.17 regions were conducted using the ALMA Cycle 3 data (\citealt{Kim2020}; Hirota et al., submitted to PASJ). The masses of G24.33+0.14 C1 -- C3 are estimated as 54.4, 8.7, and 4.0 M$_{\odot}$ (Hirota et al., submitted to PASJ), assuming T$_{dust}$ of 100 K. We adopted  T$_{ex}$(CH$_3$OH) (254.5 K) for G24.33+0.14 C1 as a low limit temperature, and 20--100 K for C2 and C3. The mass lower limits for C2 and C3 (Table \ref{tab:cores}) are larger than those of Hirota et al., (submitted to PASJ) by a factor of two in 100 K. However, this difference is caused by the different dust opacity $\kappa_{\nu}$ assumption by a factor of two (1.0 cm$^{2}$g$^{-1}$ in Hirota et al. (submitted to PASJ) and 0.047 cm$^{2}$g$^{-1}$ in this study). 
\citet{Kim2020} derived the masses of three cores in G25.82-0.17 (referred as W1--W3 in \citet{Kim2020}, and C1--C3 in this study). In the same $\kappa_{\nu}$, the derived mass for G25.82-0.17 C1 in \citet{Kim2020} is 20--84 M$_{\odot}$ assuming T$_{dust}$ of 75--300 K. We derived the mass lower limit of $\sim$26 M$_{\odot}$ adopting T$_{ex}$(CH$_3$OH) of 225.3 K, but it will be 19.1 M$_{\odot}$ if T$_{dust}$ is 300 K. For C2 and C3, \citet{Kim2020} derived mass upper limits which are higher (366 and 99 M$_{\odot}$ in 15 K) than ours (260 and 60 M$_{\odot}$ in 20 K) due to the different assumption of T$_{dust}$. Overall, the range of masses derived in this study is consistent with previous studies given the absence of the exact temperature information. To accurately constrain the physical properties of each cores, multi-band observation is essential.

\subsection{Millimeter Class II CH$_{3}$OH Maser Candidate} \label{sec:mmmaser}
The twelve high-mass star-forming regions covered by this study are associated with 44 and 95 GHz class I or 6.7 GHz class II CH$_3$OH masers \citep{Green2010,Matsumoto2014,Breen2015,Hu2016,Kang2016}. Since each region has been resolved into several continuum cores in our ALMA observation (Section \ref{sec:dustcont}), we specified the cores that are directly associated with previously detected 6.7 GHz class II CH$_3$OH maser sources by the Methanol Multibeam Survey (MMB; \citealt{Green2010,Breen2015}) and VLA \citep{Hu2016} observations (Figure \ref{fig:chnmap:G10.32}--Figure \ref{fig:chnmap:G49.49}). The association is marked in Table \ref{tab:216GHz}.

Two CH$_{3}$OH transitions, 1$_{1}$-0$_{0}$ E at 213.427~GHz and 5$_{1}$-4$_{2}$ E at 216.946~GHz, were theoretically predicted to be class~II CH$_{3}$OH masers \citep{Cragg2005}. Since our observation covers these two millimeter maser candidate transitions, we checked if the maser characteristics are detected in our data sets. We used three criteria to determine a maser: 1) narrower line widths near the protostellar position, 2) higher brightness temperature ($T_{\rm b}$), and 3) more spot-like features in the channel map compared to those of thermal lines.

In our observation, both the 213.427 and 216.946 GHz CH$_{3}$OH emission shows a combination of compact and extended emission structures. 
According to the masing mechanism, the compact structures are suspected to maser emitting candidate regions. The 213.427 GHz (1$_{1}$-0$_{0}$ E) transition shows a similar peak intensity to the 216.946 GHz (5$_{1}$-4$_{2}$ E) transition at compact structure but a stronger intensity at the extended structure. This might be due to more abundant population in a lower energy state. Thus, here we mainly investigate 216.946 GHz transition. We use the channel maps of the 216.946 GHz CH$_{3}$OH emission in each region (Figure \ref{fig:chnmap:G10.32} -- Figure \ref{fig:chnmap:G49.49}) to search for the class II CH$_3$OH maser signature. A more detailed analysis and discussion on the G24.33+0.14 region are conducted in a separate paper (Hirota et al., submitted to PASJ). 

Toward the cores emitting CH$_{3}$OH emission, the emission peak coordinates were found. 
At this position, the N(H$_{2}$) derived by a 1.3 mm continuum image, the peak intensity of CH$_{3}$OH emission, velocity and brightness temperatures were estimated. 
The information is summarized in Table \ref{tab:216GHz}. The estimated brightness temperatures of the 216.946 GHz CH$_{3}$OH lines range up to $\sim$140 K, which are not exceptional for thermal emission.

Given that the maser emission emerges in a very compact region and extremely narrow frequency range, we cannot rule out a possibility that the maser emission can be significantly diluted in our beam and frequency bin, considering our insufficient spatial and spectral resolutions. Since the emission shows both the extended and compact structures, this transition also has the characteristics of the thermal and maser emission mixture. 
As a result, the 216.946 GHz CH$_{3}$OH maser is not firmly confirmed but remains as a maser candidate. Higher spatial and spectral resolution observations will resolve the features and confirm their existence.

The 216.946 GHz CH$_{3}$OH maser candidates are listed in Table \ref{tab:maserandCOMs}.
In Table \ref{tab:maserandCOMs}, the cores marked with TH (Thermal) are cases where the emission is detected at the position. We determine the COMs association of a core when the morphology of the COMs emission is concentrated toward the core. In other cases, the COMs emission is detected at the position of a core but it is at the periphery of the extended emission associated with a nearby core.

In addition to the two Class II CH$_{3}$OH maser candidate transitions, the Class I CH$_{3}$OH maser at 229.759 GHz was also detected in many of our sources (Figure  \ref{fig:spectrum25}, Figure \ref{fig:spectrum:G10.32C1}--Figure \ref{fig:spectrum:G49.49C6}). 
The class I CH$_3$OH maser is pumped by molecular collisions in shocked region and has been observed toward several high-mass protostars \citep{Slysh2002,Cyganowski2011,Cyganowski2012,Hunter2014}. In our samples, the 229.759 GHz CH$_{3}$OH emission shows strong maser features; spot-like features were detected off from the protostellar position but coincident with the outflow shocked region (\citealt{Kim2020} for G25.82-0.17; Hirota et al., submitted to PASJ for G24.33+0.14 ). The detailed analysis of the class I maser emission will be presented in a separated paper (Baek et al., in prep).

\subsection{Line Identification} \label{sec:lineiden}
\begin{figure*}
  \centering
\includegraphics[clip,width=0.7\textwidth,keepaspectratio]{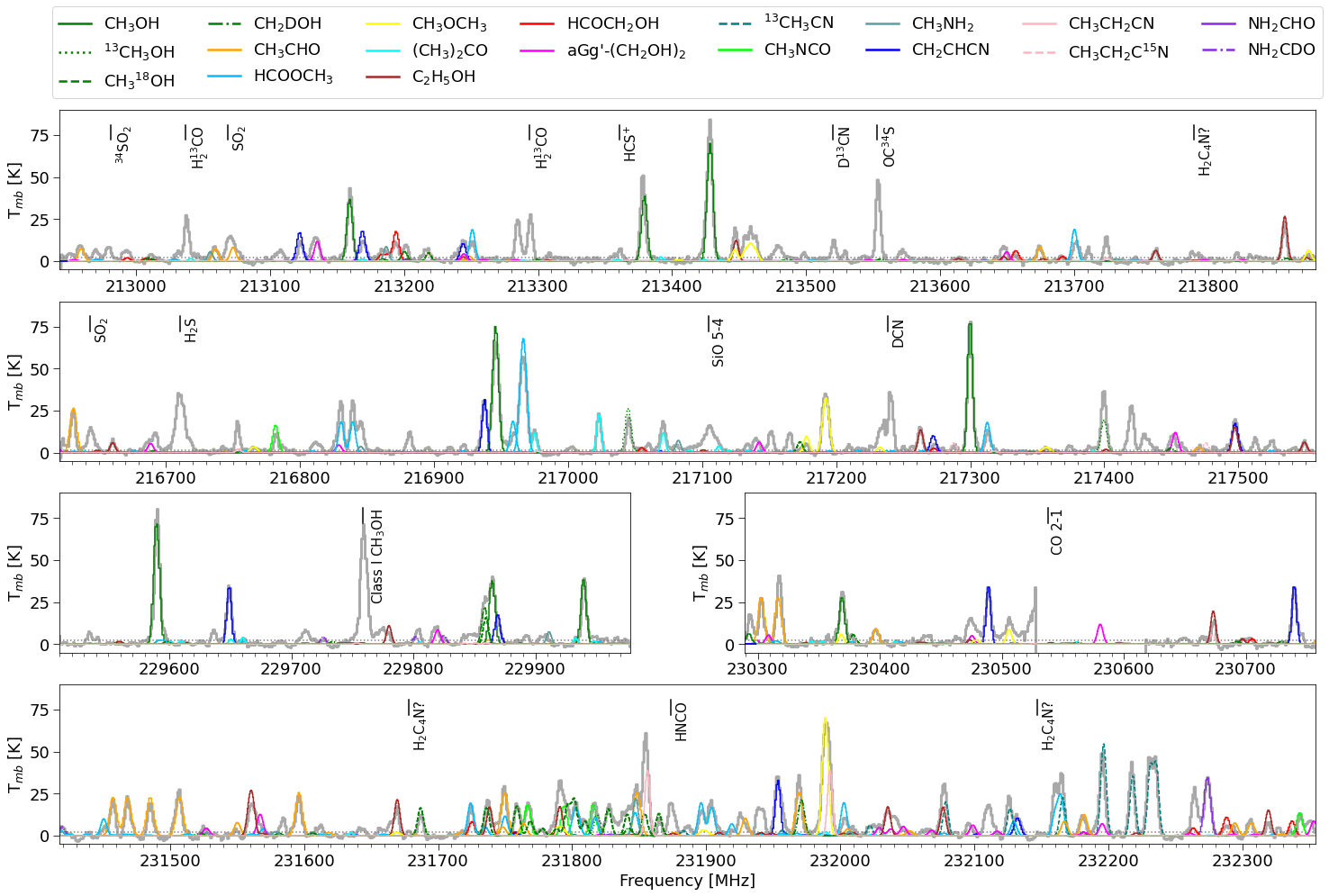}
\caption{The spectrum of G25.82-0.17 C1 at 216.946 GHz CH$_{3}$OH peak as marked with red triangles in Figure \ref{fig:chnmap:G25.82}.} The grey solid and dotted line presents the observed spectra and 3$\sigma$ noise level, respectively. The lines with different colors are the spectra of individual species simulated by the XCLASS. \\
The complete figure set (12 images) is available in the online journal.
\label{fig:spectrum25}
\end{figure*}

\begin{figure*}
  \centering
\includegraphics[clip,width=0.6\textwidth,keepaspectratio]{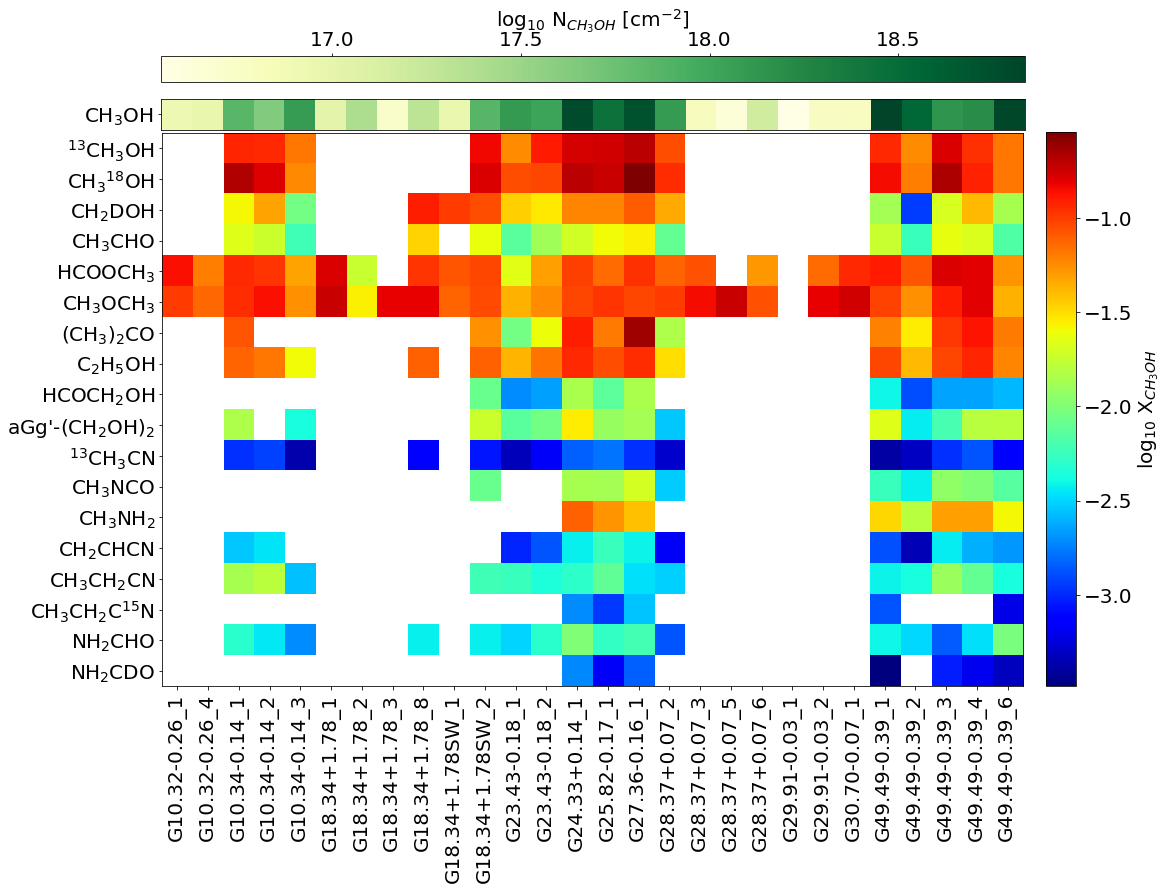}
\caption{Column density of CH$_{3}$OH and abundances of COMs with respect to CH$_{3}$OH obtained by XCLASS fitting.}
\label{fig:N_XCLASS}
\end{figure*}
In the five spectral windows in Band 6, many emission lines were detected in the vicinity of cores with chemical diversity. Toward hot cores, it has been known that the emission is dominated by COMs with relatively common and strong detections of CH$_3$OH, CH$_3$OCH$_3$, and HCOOCH$_3$, of which their intensity and relative abundances differ by cores \citep{Herbst2009}. Since our observation detected many other COMs, we define a COM-rich hot core if the core has more than 5 COMs, which include $^{13}$CH$_3$OH and CH$_2$DOH in addition to the three COMs.

While the emitting regions of COMs emission are compact toward the continuum sources, their peaks are slightly offset from the continuum peak positions. The reduced intensity of COMs at the continuum peak position could be caused by two reasons: 1) high dust opacity, obscuring the COMs emission at the continuum peaks \citep{Harsono2018,Lee2019} and 2) chemical effects, destroying COMs inside the water snow line \citep{Lee2020}. In addition to the reduced intensity, kinematic motions, such as rotation and infall, are detected at the continuum peak of some cores, producing blended line profiles or inverse P-Cygni profiles (Section \ref{sec:contpeak}). As a result, it is difficult to extract correct chemical properties at the continuum peak position. To identify the COMs in a consistent manner, we extract the spectrum of each COMs-detected core at the peak position of the 216.946 GHz CH$_{3}$OH emission, assuming that all COMs emission originates from the same region where the simplest COM, CH$_{3}$OH, exists.

From a total of 68 cores, 28 cores ($\sim$41\%) are associated with COMs emission. Assuming that the 68 continuum peaks are the site of high-mass star-formation, the ratio gives hot core emergence relative to the dense cores. If the total lifetime of HMYSOs is $\sim$10$^6$ years, the hot core lifetime is estimated $\sim$4.5$\times$10$^5$ years, slightly higher than previously derived statistical high mass protostar lifetime \citep[Table 2 of][]{Motte2018}. This can be because we targeted the actively accreting cores traced by the class I and II CH$_3$OH masers. Compared with low-mass counterparts, we derived slightly lower fraction, which is approximately 58\% \citep{Yang2021}.

To identify the detected emission lines in the extracted spectrum of each core, the eXtended CASA Line Analysis Software Suite (XCLASS; \citealt{Moller2017}) is used. 
XCLASS is a toolbox for modeling the spectra. It estimates physical properties of each species considering gas and dust attenuation in LTE conditions by searching for the best-fit model for observed spectra using MAGIX optimization algorithms and molecular databases from laboratory experiments (CDMS and JPL).

To synthesize the spectra, XCLASS mainly requires five parameters; source size, $T_{\rm ex}$, column density of the molecule (cm$^{-2}$), line velocity width (FHWM; km s$^{-1}$), and velocity offset relative to $V_{\rm LSR}$ (km s$^{-1}$). In addition, hydrogen column density ($N$(H)=2$\times N(\rm H_{2})$ in cm$^{-2}$) and dust property (dust mass opacity $\kappa_{\rm 1.3mm}$ in cm$^{2}$g$^{-1}$ and spectral index $\beta$) can be defined to consider the continuum optical depth. This may diminish the line intensity in the optically thick dusty disk/inner envelope. 
For the $N_{\rm H}$ used in XCLASS, we adopted 1.3 mm continuum intensity at the the CH$_{3}$OH peak position, where the spectrum is extracted, assuming that $T_{\rm dust}$ is the same as T$_{ex}$(CH$_3$OH) under the LTE condition.
Since we extract the spectrum in a pixel at the CH$_{3}$OH peak, which is close to the continuum peak, the gas and dust should be well coupled due to the high density.
Therefore, the remaining free parameter to fit the observed spectra is the column density of a given molecule. For line widths and offsets, since the spectra is extracted offset from the protostar position in a rotating hot core, ranges of line width (3 -- 10 km s$^{-1}$) and velocity offset from $V_{\rm LSR}$ (-8 -- +8 km s$^{-1}$) are given to match the observed spectra. 

To securely identify the lines, we followed the criteria of \citet{Herbst2009} and \citet{Jorgensen2020} using XCLASS; (1) synthesized spectrum of more than two transitions for the given species is systemically accounted for the observation, and (2) the predicted spectrum does not appear in any transitions that no observed signal is seen.
An exception is NH$_{2}$CHO (Formamide), of which one transition is covered in our spectrum; however, the strong intensity is expected and its deuterated species NH$_{2}$CDO is identified for this species. 
We identify the molecules in 3$\sigma$ detection, except for the following molecules which are detected as upper limits; CH$_{3}$NCO (Methyl isocyanate) in G18.34+1.78SW C2 Figure \ref{fig:spectrum:G18.34C2}), (CH$_{3}$)$_{2}$CO (Acetone) in G23.43-0.18 C1 (Figure \ref{fig:spectrum:G23.43C1}), CH$_{2}$CHCN (Vinyl Cyanide) in G28.37+0.07 C2 (Figure \ref{fig:spectrum:G28.37C2}), and CH$_{3}$OCH$_{3}$ (Dimethyl ether) in G28.37+0.07 C5 (Figure \ref{fig:spectrum:G28.37C5}).
Due to the heavily blended spectrum in some cores, the best fit parameters are found by fitting two or three molecules together. 

Figure \ref{fig:spectrum25} shows the observed and fitted spectra of G25.82-0.17 C1. Many COMs, including several oxygen and nitrogen bearing molecules, are identified. The rotational transitions in vibrationally and torsionally excited states are also detected, and isotopologues of carbon, oxygen and nitrogen molecules and their deuterated species are found. The column density derived by the XCLASS fitting is summarized in Table \ref{tab:N1}. Figure \ref{fig:N_XCLASS} shows the abundances of identified COMs relative to CH$_{3}$OH. Up to nineteen COMs are identified in a wide range of energy states (up to $\sim$800 K).  

Generally, O-bearing COMs are more abundant than N-bearing COMs in terms of greater detection rates and higher abundances. The most frequently detected COM is CH$_{3}$OCH$_{3}$, found in all regions except for G29.91-0.03 C1. 
The next abundant species are, in turn, HCOOCH$_{3}$ (Methyl formate), C$_{2}$H$_{5}$OH (Ethanol), and CH$_{3}$CHO (Acetaldehyde). For N-bearing molecules, besides CH$_{3}$CN (Methyl Cyanide), of which only its carbon isotope is covered in our observation, NH$_{2}$CHO, CH$_{3}$CH$_{2}$CN (Ethyl Cyanide) and CH$_{2}$CHCN are abundant.

We assume the filling factor is 1 (source size $>$ beam size). While this assumption is reasonable for most of COMs, some of the COMs and CH$_{3}$OH maser candidate transitions might have smaller emission distributions than the beam size. Thus, the derived beam-averaged values would be lower limit or underestimated compared with the actual values.

\subsection{Rotation diagram}\label{sec:rotdiagram}
To investigate the excitation conditions of COMs, we estimate the rotation temperature and column density toward the 216.946 GHz CH$_{3}$OH emission peak position. 
The integrated intensity was measured by gaussian fitting. The molecular data ($A_{\rm ij}$, $E_{\rm u}$, $g_{\rm u}$) is adopted from the $Splatalogue$\footnote{https://splatalogue.online/} database (Table \ref{tab:LID}). The best fit and uncertainty are calculated using the bootstrap method, which conducts linear fitting 10,000 times with randomly generated integrated intensity considering observation and gaussian fitting errors. 

The results are listed in Table \ref{tab:rot_diagram1} and Figure \ref{fig:rot_diagram}.
The T$_{ex}$(CH$_3$OH) range from 110 to 260 K; plausible for hot cores. The derived $T_{\rm ex}$ of $^{13}$CH$_{3}$OH tends to be lower than that of the main isotope. 

\begin{figure}
  \centering
\includegraphics[clip,width=0.9\columnwidth,keepaspectratio]{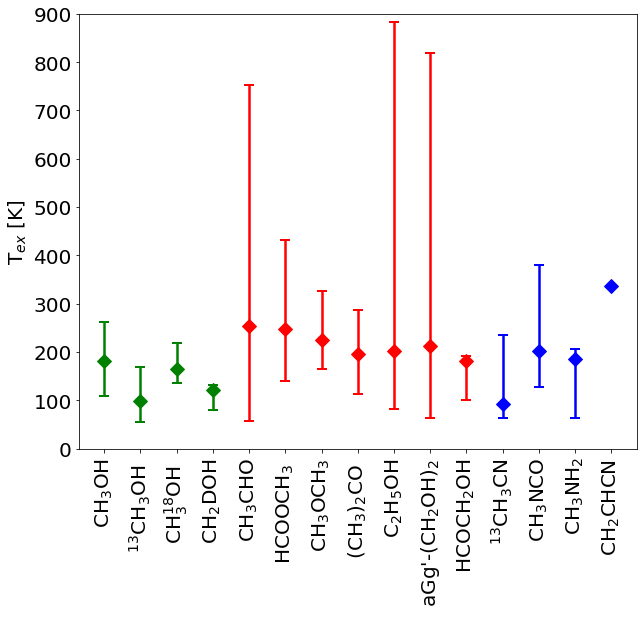}
\caption{T$_{ex}$ of COMs derived by the rotation diagram analysis. The medians of CH$_3$OH and isotopologues (green), O-bearing COMs (red), and N-bearing COMs (blue) are presented. Error bars depict the minimum and maximum T$_{ex}$ (Table \ref{tab:rot_diagram1}).} 
\label{fig:Texallcoms}
\end{figure}

We assume that the COMs coincide with CH$_{3}$OH. However, the $T_{\rm ex}$ derived for other COMs exhibit some scatters from that of CH$_{3}$OH, of which many are higher than that of  CH$_{3}$OH (Figure \ref{fig:Texallcoms} and Table \ref{tab:rot_diagram1}).
Together, given that the emitting region is smaller in the integrated intensity map, the COMs emission could be originated from the hotter and closer to the protostar compared to CH$_{3}$OH.
Larger COMs tend to show higher T$_{ex}$ than CH$_{3}$OH, implying that they have higher sublimation temperatures and trace the inner and hotter region. 
However, we note that the T$_{ex}$ of COMs are loosely constrained in the limited number of detected lines in our spectral coverage. 

%% file: s4_discussion.tex
\subsection{Detection rate of COMs}
We find significant diversity in the detection rates of COMs among the CH$_{3}$OH detected cores. Up to nineteen O- and N-bearing COMs were identified in each core, including rotational transitions in excited vibrational and tortional states. The isotopologues, including $^{13}$C, $^{15}$N, $^{18}$O, and deuterated species, were also detected. 
We note that the number of detected COMs is not dependent on the RMS and SNR of the 1.3 mm continuum image (Table \ref{tab:obs_summary} and Table \ref{tab:cores}), supporting that the observation sensitivity and continuum level do not affect the COMs detection rate.

As shown in Figure \ref{fig:N_XCLASS}, O-bearing COMs have higher detection rates compared to N-bearing COMs.
Among the 28 cores at which we found CH$_{3}$OH (representative O-bearing COMs), N-bearing COMs are detected in 16 cores ($\sim$57 \%).
The most frequently detected molecules following CH$_{3}$OH (detected in 28 cores) are CH$_{3}$OCH$_{3}$ (27 cores) and HCOOCH$_{3}$ (25 cores). The least frequently detected molecules are CH$_3$NH$_2$ (Methylamine; 8 cores), CH$_3$NCO (Methyl isocyanate; 10 cores), and HCOCH$_{2}$OH (Glycolaldehyde; 11 cores). Except for HCOCH$_{2}$OH, these higher detection rates of O-bearing COMs imply that the formation network for O-bearing COMs is more efficient than that of N-bearing COMs. 

\begin{figure}
  \centering
\includegraphics[clip,width=0.95\columnwidth,keepaspectratio]{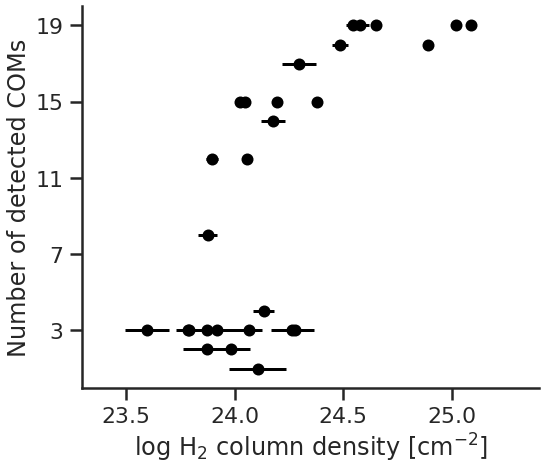}
\includegraphics[clip,width=0.95\columnwidth,keepaspectratio]{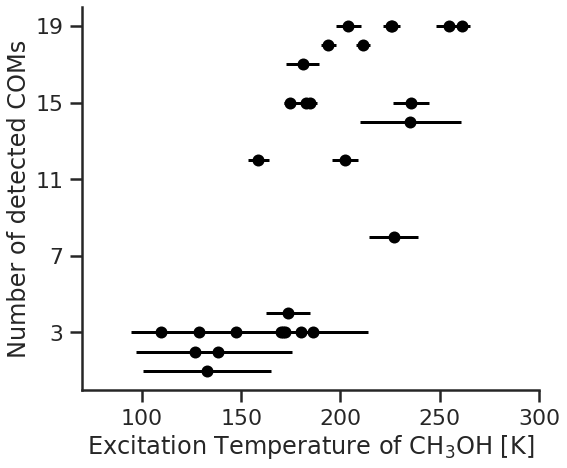}
\caption{The COMs detection rates in functions of (left) $N(\rm H_{2})$ column density and (right) the T$_{ex}$(CH$_3$OH).}
\label{fig:nh2_t_ncoms}
\end{figure}

Cores with only O-bearing species detected with no detection of N-bearing ones might be in an earlier stage of star formation, since nitrogen species are known to be formed via the surface and gas-phase chemistry at a longer timescale \citep{Lee2020}.
If a core experiences a longer timescale in the warm-up phase prior to the sublimation of COMs into the gas-phase, the formation of more complex COMs, including N-bearing species, can be promoted \citep{Garrod2013, Garrod2022}.

We note that the higher detection rate of O-bearing COMs could be caused by our spectrum extraction at the CH$_{3}$OH peak positions. 
We assume that O- and N-bearing COMs are present in the same gas parcel, but this might not always be the case.  
\citet{Csengeri2019} showed the spatial differentiation between the two groups toward the spatially resolved envelope of high-mass core within the G328.2551−0.5321 clump, interpreting that the O-bearing COMs trace the shocked region caused by the infall in addition to the protostellar heating, while N-bearing COMs are more concentrated toward the protostellar position. Higher angular-resolution and more sensitive observations are required to investigate the spatial distributions of the detected COMs and possible chemical differentiation. 

Figure \ref{fig:nh2_t_ncoms} illustrates the relation of the number of detected COMs with $N(\rm H_{2})$ and T$_{ex}$(CH$_3$OH). 
The COMs detection rate steeply grows with $N(\rm H_{2})$ above N(H$_{2}$) $>$ 2$\times$10$^{24}$cm$^{-2}$, implying that the formation of COMs is advanced in denser cores. This is consistent with the observational results of the low mass counterpart, hot corinos \citep{Yang2021}.

Although the cores with higher detection rates of COMs tend to have higher T$_{ex}$(CH$_3$OH), the COMs detection rate is less sensitive to T$_{ex}$(CH$_3$OH) than $N(\rm H_{2})$. For cores with the T$_{ex}$(CH$_3$OH) higher than 200 K, no clear trend is found between the number of detected COMs and the T$_{ex}$(CH$_3$OH).
This could be interpreted as the derived T$_{ex}$ is more related to the current physical condition rather than the conditions for the formation/sublimation of COMs.
It has been suggested that COMs mixed in the ice mantles tend to be sublimated when the main constituent of bulk ice, such as H$_{2}$O, evaporates \citep{oberg2021}. A detailed experiment of \citet{Garrod2013} demonstrates that the ice sublimation temperature largely depends on their bonding structures (CO-like or water-like species). However, the mixed ice with various constituents, including secondary or tertiary species to the dominant constituent water ice, can change the sublimation temperature \citep{Burke2015}, diluting the feasibility of the current temperature as the sublimation temperature tracer of individual species. 
Instead, a higher number of COMs appear to be detected because in a higher T$_{ex}$ the molecular desorption tends to be promoted at larger radii of hot cores, which allows the COMs exist in gas-phase, resulting in increasing the detection rate.

\begin{figure}
  \centering
\includegraphics[clip,width=0.8\columnwidth,keepaspectratio]{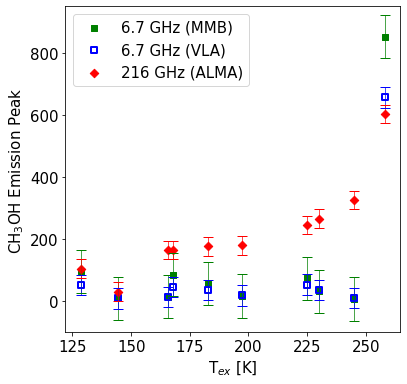}
\caption{CH$_{3}$OH emission peak flux versus T$_{ex}$(CH$_3$OH). The units of CH$_{3}$OH emission peak values are flux density (Jy) for 6.7 GHz CH$_{3}$OH maser emission by MMB and VLA observations and intensity (Jy beam$^{-1}$) for 216.946 GHz CH$_{3}$OH emission by ALMA observation.
}
\label{fig:maser}
\end{figure}

Among 28 cores with COMs, ten are associated with 6.7 GHz class II CH$_{3}$OH maser.
We also checked the COMs detection for those cores.
The 6.7 GHz class II CH$_{3}$OH maser flux density does not linearly respond to the T$_{ex}$(CH$_3$OH), whereas 216.946 GHz intensity which is thought to be a mixture of the thermal and maser emission tends to have a correlation with the T$_{ex}$(CH$_3$OH) (Figure \ref{fig:maser}). However, the emerging condition of the 6.7 GHz class II CH$_3$OH maser emission itself is very sensitive to the current physical conditions in both density and temperature (e.g., \citealt{Burns2020}).
The 6.7 GHz CH$_{3}$OH maser is also an indication for the evolutionary stages of the central objects. Survey observations have revealed that the sources detected in the 6.7 GHz CH$_{3}$OH maser are more evolved than quiescent IRDC cores but less evolved than ultracompact HII regions \citep{Paulson2022}. 

\begin{figure}
  \centering
\includegraphics[clip,width=0.45\textwidth,keepaspectratio]{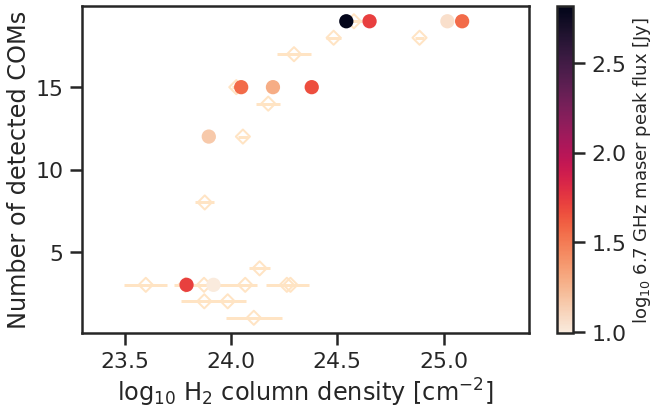}
\caption{The relaltion between the number of detected COMs and $N(\rm H_{2})$ column density. Filled and unfilled symbols present 6.7 GHz CH$_{3}$OH detected and undetected cores, respectively. Colors for the filled symbols depict the 6.7 GHz maser peak flux in VLA observation \citep{Hu2016}.}
\label{fig:nh2_ncoms_vla}
\end{figure}

In our targets, the cores associated with 6.7 GHz class II CH$_{3}$OH maser show a high COMs detection rate (Figure \ref{fig:nh2_ncoms_vla}). As shown in Figure \ref{fig:nh2_t_ncoms} and also discussed in Section \ref{sec:dis:comparison}, the number of detected COMs mainly depends on the density in the early stages. However, the thermal history and variation of UV radiation may also influence the COMs production and destruction rates. 
The 6.7 GHz class II CH$_{3}$OH maser emerges in a region with the active accretion process, which elevates the temperature and produce more UV radiations. Therefore, the high COMs detection rates in 6.7 GHz class II CH$_{3}$OH maser detected cores suggest that more COMs can be detected in actively accreting cores. 
This result implies that {\it 6.7 GHz class II CH$_{3}$OH maser could be used as a COMs-rich hot core tracer}. 

To sum up, the chemical diversity among hot cores could be originated by the degree of the accretion of each source as well as the different physical conditions of cores. 

\subsection{The effect of physical properties}\label{sec:dis:physical_properties}

\begin{figure*}
\includegraphics[clip,width=0.2\textwidth,keepaspectratio]{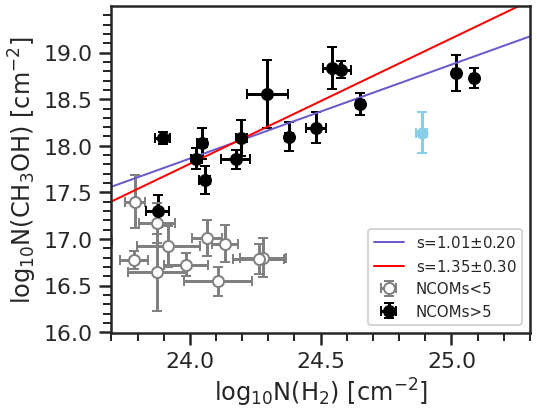}
\includegraphics[clip,width=0.2\textwidth,keepaspectratio]{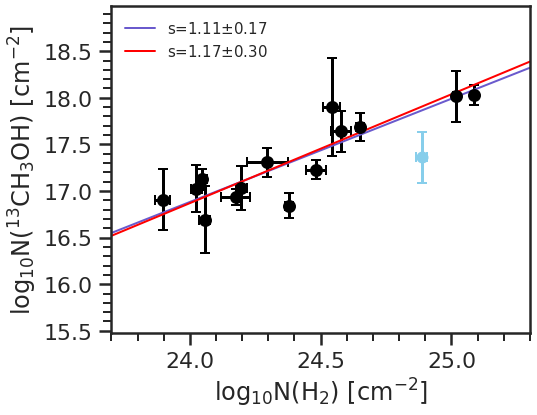}
\includegraphics[clip,width=0.2\textwidth,keepaspectratio]{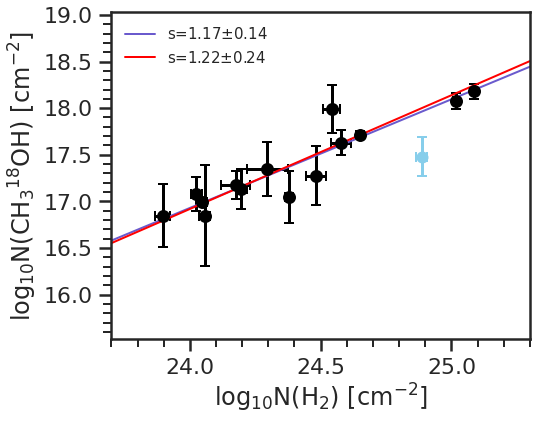}
\includegraphics[clip,width=0.2\textwidth,keepaspectratio]{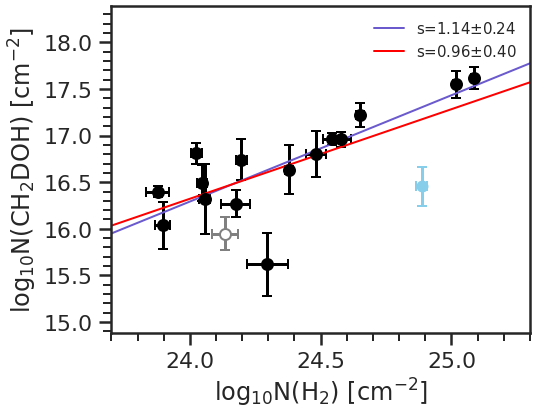}\\
\includegraphics[clip,width=0.2\textwidth,keepaspectratio]{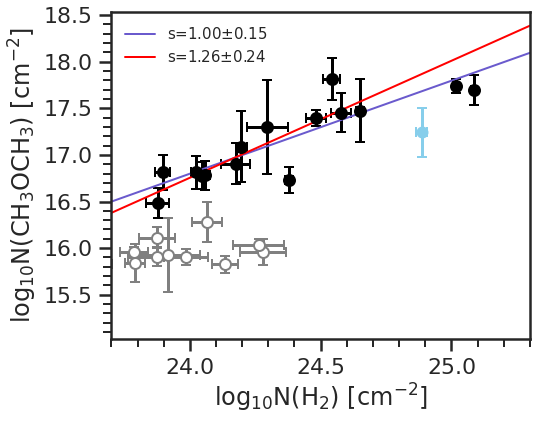}
\includegraphics[clip,width=0.2\textwidth,keepaspectratio]{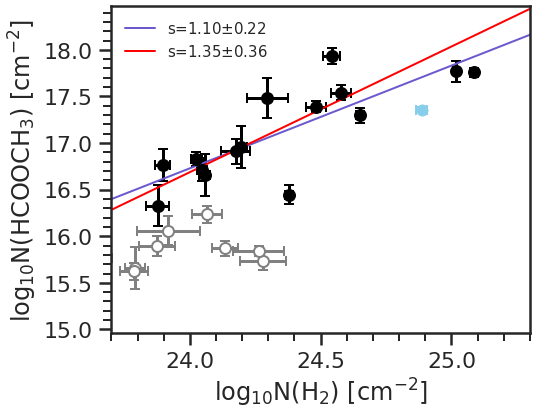}
\includegraphics[clip,width=0.2\textwidth,keepaspectratio]{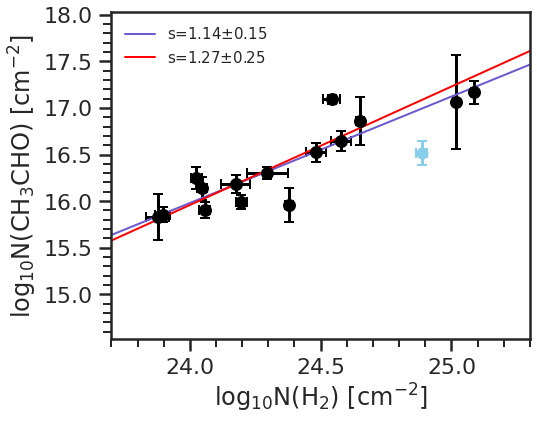}\\
\includegraphics[clip,width=0.2\textwidth,keepaspectratio]{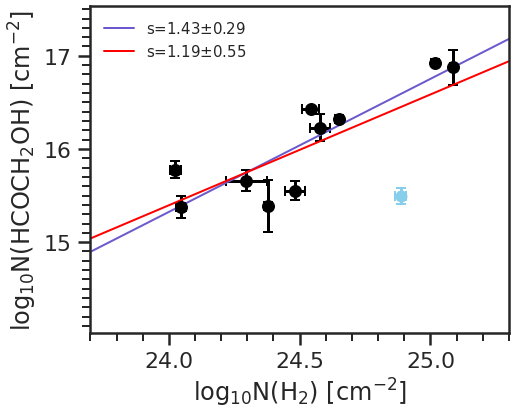}
\includegraphics[clip,width=0.2\textwidth,keepaspectratio]{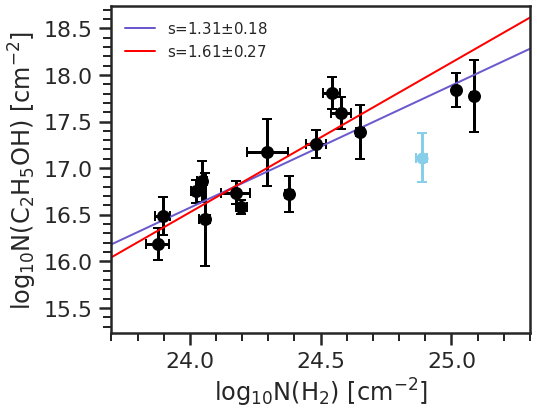}
\includegraphics[clip,width=0.2\textwidth,keepaspectratio]{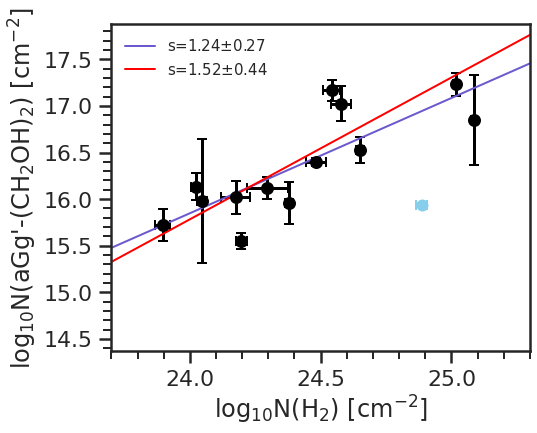}
\includegraphics[clip,width=0.2\textwidth,keepaspectratio]{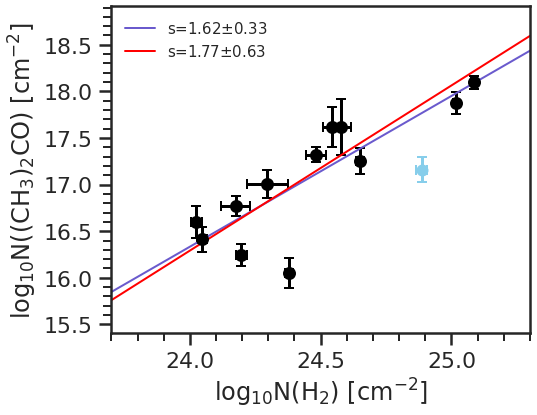}\\
\includegraphics[clip,width=0.2\textwidth,keepaspectratio]{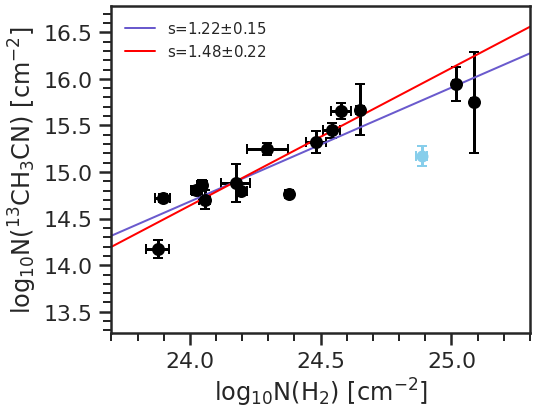}
\includegraphics[clip,width=0.2\textwidth,keepaspectratio]{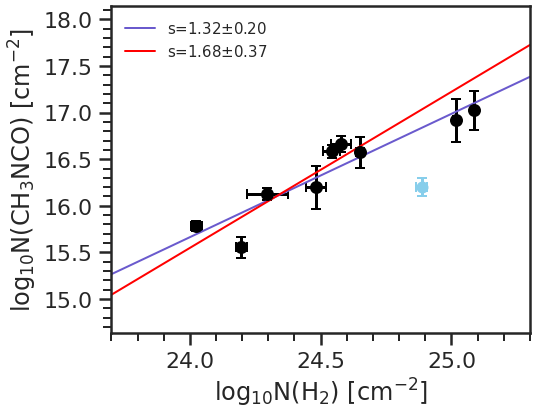}
\includegraphics[clip,width=0.2\textwidth,keepaspectratio]{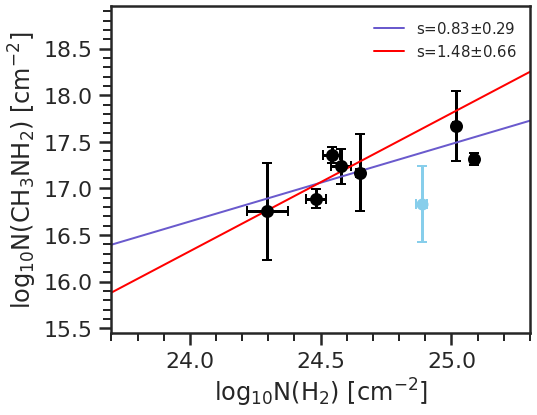}
\includegraphics[clip,width=0.2\textwidth,keepaspectratio]{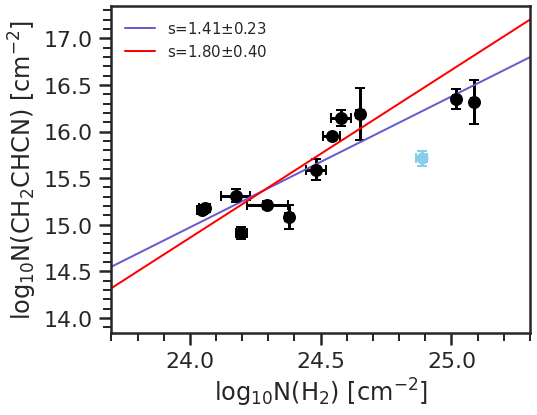}\\
\includegraphics[clip,width=0.2\textwidth,keepaspectratio]{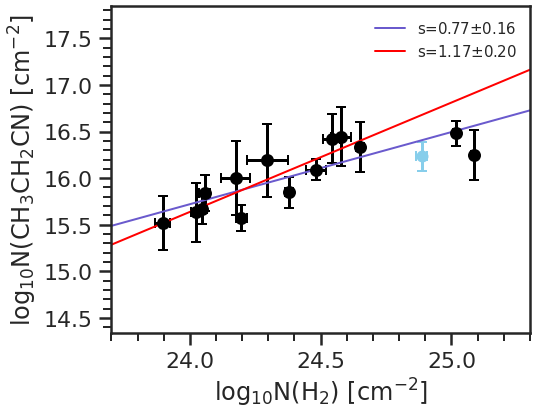}
\includegraphics[clip,width=0.2\textwidth,keepaspectratio]{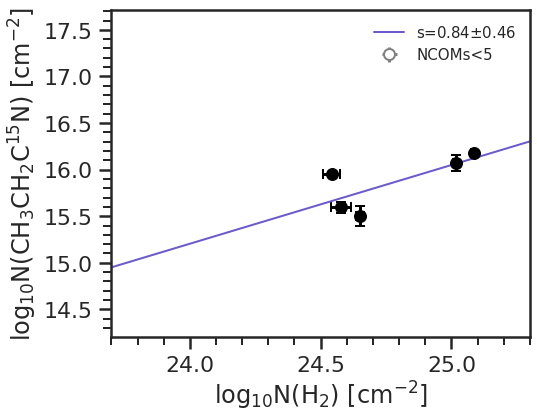}
\includegraphics[clip,width=0.2\textwidth,keepaspectratio]{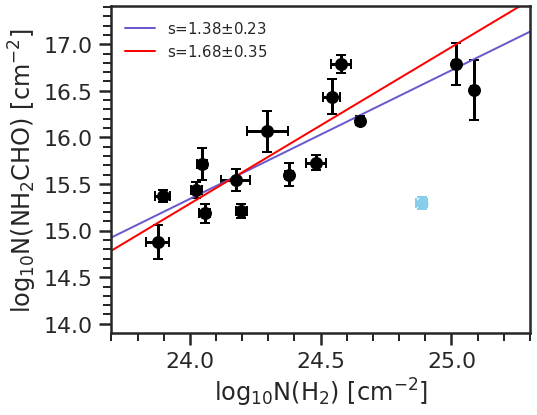}
\includegraphics[clip,width=0.2\textwidth,keepaspectratio]{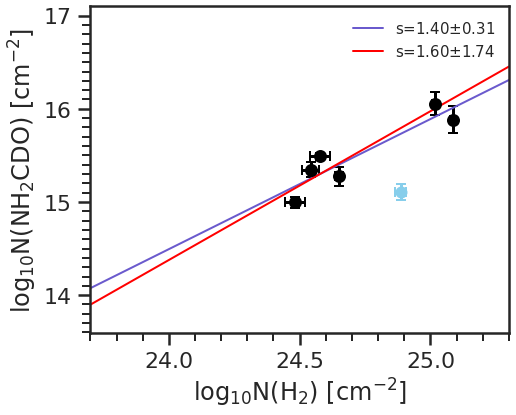}
\caption{Relation between the column densities of COMs and N(H$_{2}$), which is estimated using 1.3 mm continuum images. Grey empty and black filled circles represent the cores with the number of detected COMs smaller and greater than 5, respectively. Purple solid line is the linear fit in log-log space for the COM-rich cores, except for an outlier, G49.49-0.39 C3 (skyblue). Red solid line is the same as the purple line but fitted only at log$_{10}$N(H$_2) <$ 24.7 cm$^{-2}$, excluding the three densest cores.}
\label{fig:NCOMs_NH2_hue_NCOMs}
\end{figure*}

\begin{figure*}
\includegraphics[clip,width=0.2\textwidth,keepaspectratio]{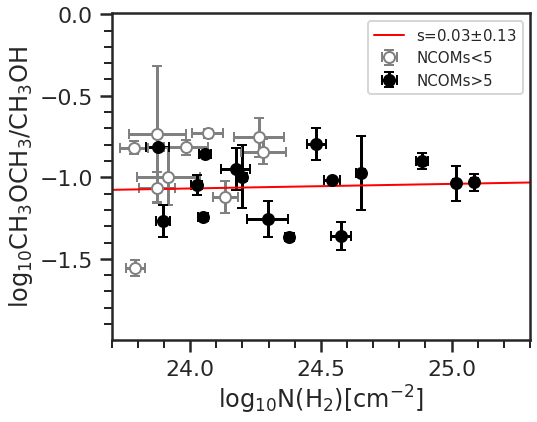}
\includegraphics[clip,width=0.2\textwidth,keepaspectratio]{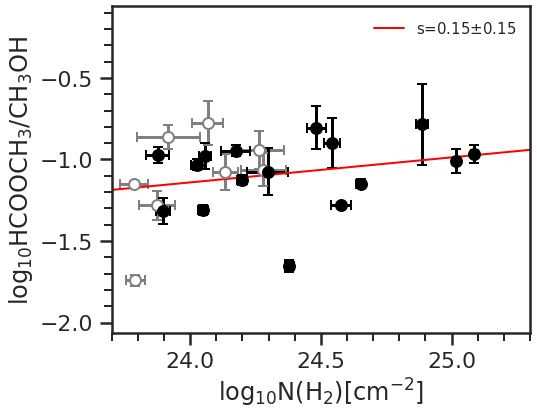}
\includegraphics[clip,width=0.2\textwidth,keepaspectratio]{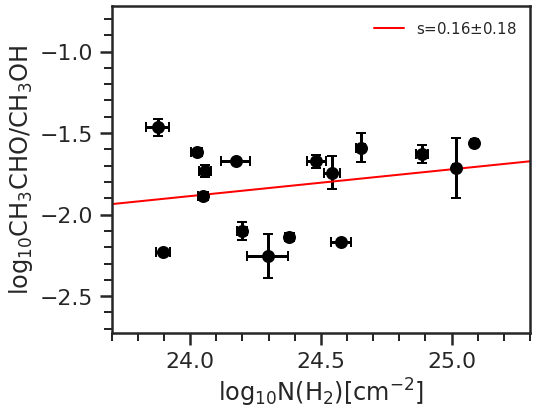}\\
\includegraphics[clip,width=0.2\textwidth,keepaspectratio]{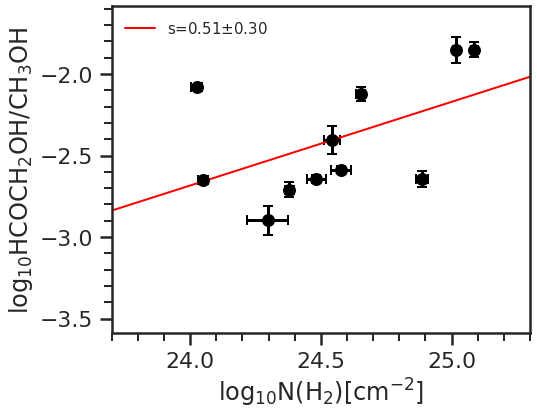}
\includegraphics[clip,width=0.2\textwidth,keepaspectratio]{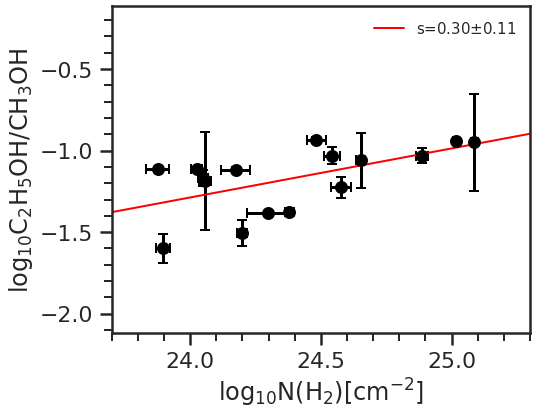}
\includegraphics[clip,width=0.2\textwidth,keepaspectratio]{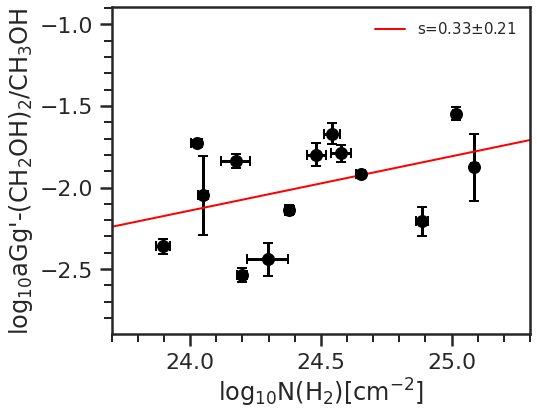}
\includegraphics[clip,width=0.2\textwidth,keepaspectratio]{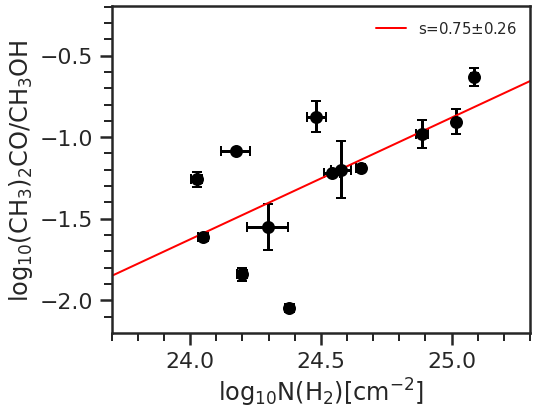}\\
\includegraphics[clip,width=0.2\textwidth,keepaspectratio]{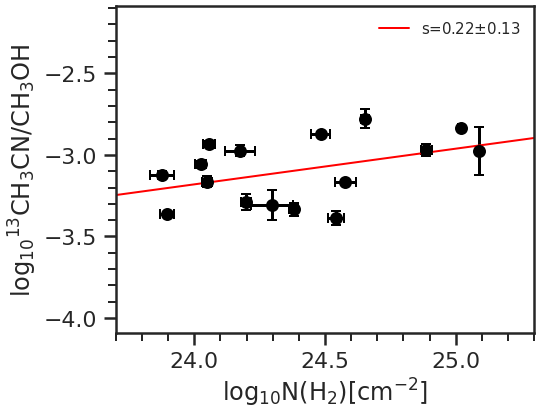}
\includegraphics[clip,width=0.2\textwidth,keepaspectratio]{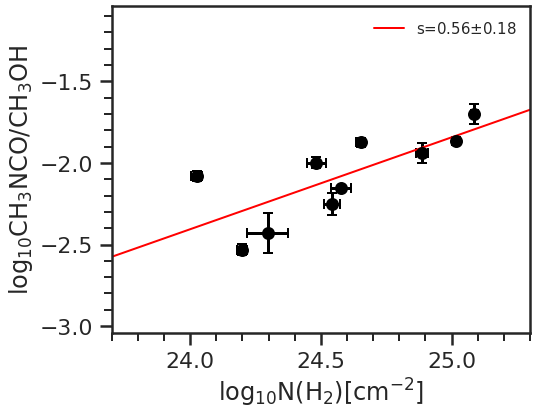}
\includegraphics[clip,width=0.2\textwidth,keepaspectratio]{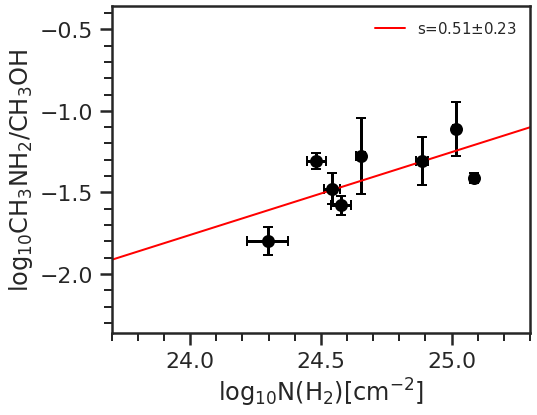}
\includegraphics[clip,width=0.2\textwidth,keepaspectratio]{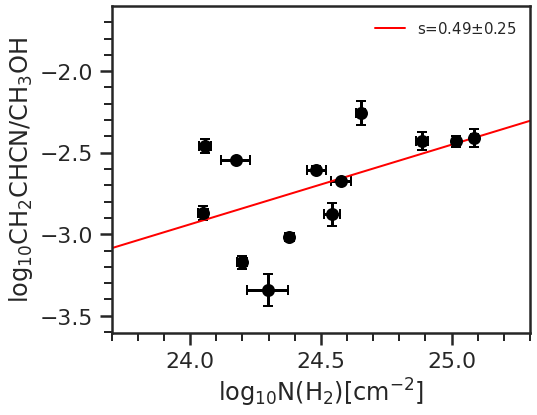}\\
\includegraphics[clip,width=0.2\textwidth,keepaspectratio]{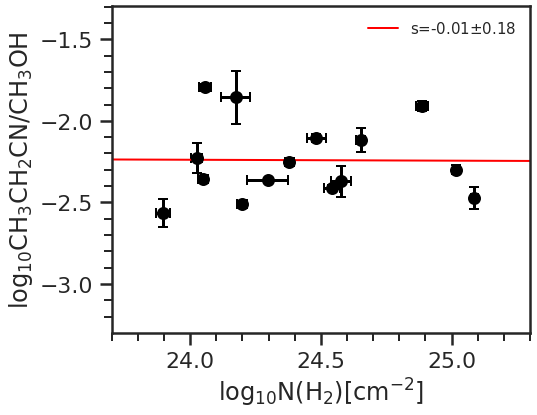}
\includegraphics[clip,width=0.2\textwidth,keepaspectratio]{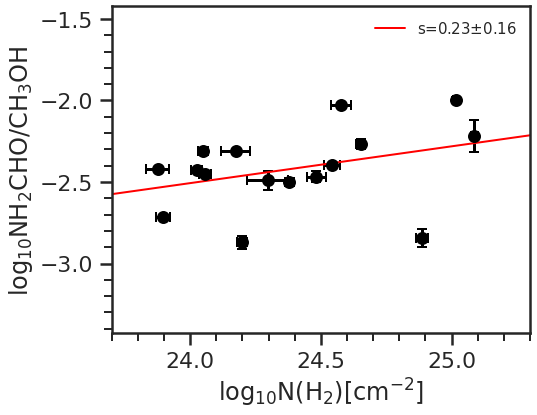}
\caption{Relation for the abundance ratios of COMs relative to CH$_{3}$OH versus N(H$_{2}$), which is estimated using 1.3 mm continuum images. Grey empty and black filled circles represent cores with the number of detected COMs smaller and greater than 5, respectively. Red solid line is the linear fit in log-log space for the COM-rich cores.}
\label{fig:corr_XCOMs_NH2_hue_NCOMs}
\end{figure*}

We test the effect of physical parameters on the chemical evolution. 
Figure \ref{fig:NCOMs_NH2_hue_NCOMs} and Figure \ref{fig:corr_NCOMs_Tex_hue_NCOMs} present the column density of each molecule (N(COMs)), while Figure \ref{fig:corr_XCOMs_NH2_hue_NCOMs} and Figure \ref{fig:corr_XCOMs_Tex_hue_NCOMs} show the COMs abundances with respect to the CH$_3$OH (X(COMs)) along with N(H$_{2}$) and T$_{ex}$(CH$_3$OH), respectively.

We divide the cores into two types by the number of detected COMs (i.e., $>$ 5 and $<$ 5). For all cores with $<$ 5 COMs, CH$_3$OH, $^{13}$CH$_3$OH, CH$_2$DOH, CH$_3$OCH$_3$, and HCOOCH$_3$ were detected, and for cores with $>$ 5 COMs (COM-rich cores), up to 19 COMs were detected. We focus on the COM-rich cores to consistently investigate the evolution of COMs complexity in the range of corresponding physical conditions.
In Figure \ref{fig:NCOMs_NH2_hue_NCOMs}, we measure the slopes of the N(COMs) with respective to the N(H$_{2}$) in the log-log space for COM-rich cores. For most of main isotopologues, the uptrend becomes slow down when log$_{10}$N(H$_2$) $>$ 24.7 cm$^{-2}$, indicative of the high dust optical depths, preventing the COMs emission from escaping the system, as well as the high line optical depths of the species. The cores with the largest N(H$_2$) are G24.33+0.14 C1 and G27.36-0.16 C1 (Table \ref{tab:216GHz}, Figure \ref{fig:spectrum:G24.33C1} and Figure \ref{fig:spectrum:G27.36C1}), of which the COMs emission is significantly suppressed at their continuum peak positions. The N(COMs) at the CH$_3$OH peaks might be also affected by the strong continuum emission, leading to a slightly flattened slope at the densest ends in the N(COMs) vs N(H$_{2}$) plots (Figure \ref{fig:NCOMs_NH2_hue_NCOMs}). An extreme case is G49.49-0.39 C3, which shows systematically lower column densities in all species compared to other COM-rich cores (Figure \ref{fig:spectrum:G49.49C3}). 
G49.49-0.39 C3 is located close to G49.49-0.39 clustered region with highest continuum levels (Table \ref{tab:cores}). Therefore, this core might be highly embedded in the cluster as well as its own envelope although the continuum emission is clearly resolved and makes a unique peak toward G49.49-0.39 C3.  As a result, we derived the slopes of N(COMs) with respect to N(H$_2$) only using the COM-rich cores and log$_{10}$ N(H$_2$) $<$ 24.7 cm$^{-2}$.

In the observed range of physical condition, for COM-rich cores, the characteristics of COMs are divided into two groups. The first group includes CH$_{3}$OH, CH$_{3}$OCH$_{3}$, HCOOCH$_{3}$, CH$_{3}$CHO, and CH$_{3}$CH$_{2}$CN, which were detected in most hot cores. The N(COMs) of this first group COMs are directly proportional to the N(H$_{2}$), and thus, X(COMs) are almost constant with the N(H$_{2}$).
Those X(COMs) also show relatively flat distribution of their abundance with T$_{ex}$(CH$_3$OH) except for CH$_{3}$CHO. 

The second group includes (CH$_{3}$)$_{2}$CO, HCOCH$_{2}$OH, C$_{2}$H$_{5}$OH, and aGg'-(CH$_{2}$OH)$_{2}$ for O-bearing COMs, and CH$_{3}$NCO, CH$_{3}$NH$_{2}$, CH$_{2}$CHCN, and NH$_{2}$CHO for N-bearing COMs. They show steeper gradients in N(COMs) versus N(H$_2$) and proportional relation in X(COMs) versus both N(H$_{2}$) and T$_{ex}$(CH$_3$OH), unlike the first group. 

This difference found in the two groups can be caused by the different formation timescales. At a low N(H$_{2}$) (e.g., $\sim$10$^{24}$ cm$^{-2}$), the second group species have low column densities, if detected, and grow with N(H$_2$) proportionally. The less efficient formation of the second group species than the first group species in the low density probably cause the difference in slope.
Moreover, the efficient destruction of the first group species at high densities or the high optical depth of the first group species lines may flatten the slope of the first species group (Figure \ref{fig:NCOMs_NH2_hue_NCOMs}).

Depending on the formation routes in both ice- and gas-phases, some of the second group species are formed from the destruction of the first group species, resulting in shallow gradients for the first group species at the high density regime. 
Our scenario is also supported by the comparison between this study and PEACHES. \citet{Yang2021} reported the elevated ratios of HCOOCH$_{3}$ and CH$_3$OCH$_3$ to CH$_3$OH in increasing continuum brightness temperature which traces the total gas column density (Figure 15 in \citealt{Yang2021}), while those in this study show almost constant levels. 
Compared with the PEACHES samples, the low mass hot corinos in Perseus molecular cloud, the cores in this study would be much denser because they are mostly high mass hot cores associated with class II CH$_3$OH maser (Table \ref{tab:cores} and Table \ref{tab:maserandCOMs}). Therefore, the differences between PEACHES survey and ours indicates that chemical timescale becomes shorter in denser cores, thus the reaction could proceed to form the other molecules resulting in the fast production of the second group species from HCOOCH$_{3}$ and CH$_3$OCH$_3$.

\subsection{Comparison with other sources}\label{sec:dis:comparison}

\begin{figure*}
  \centering
\includegraphics[clip,width=0.8\textwidth,keepaspectratio]{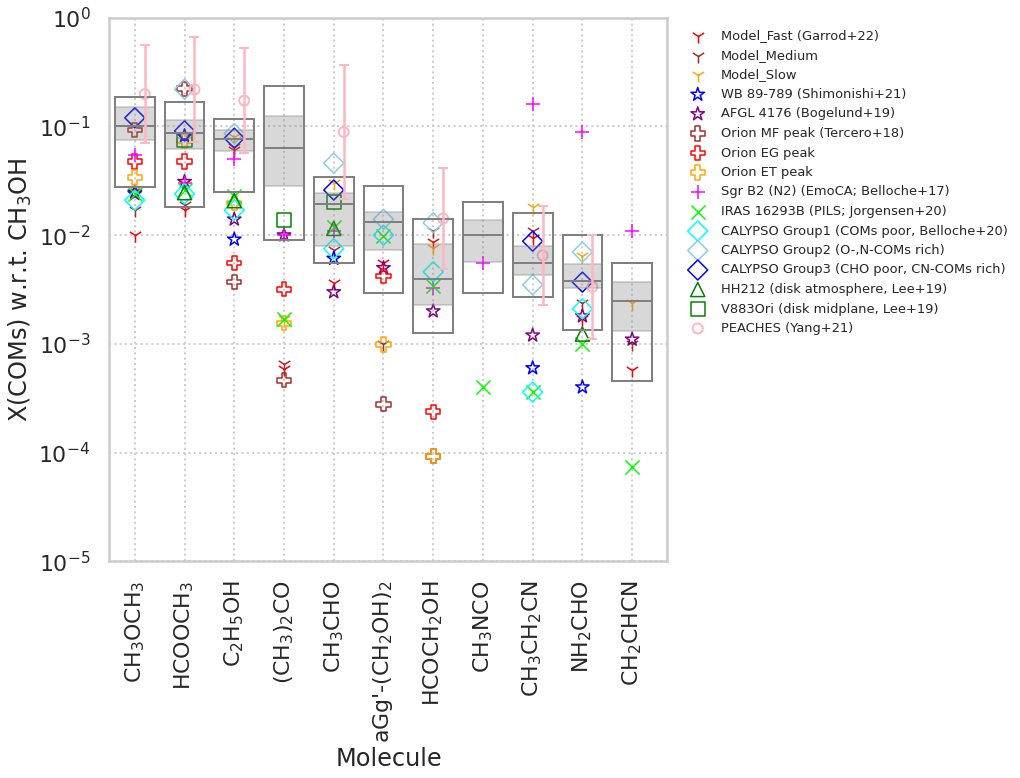}
\caption{Abundance relative to CH$_{3}$OH. Grey boxes indicate the abundance estimated in this work. Different symbols and colors are the adopted values in other regions.}
\label{fig:abundance}
\end{figure*}

We compare our abundance estimates of COMs with those found in various physical environments from recent interferometric observations.
Figure \ref{fig:abundance} shows the abundances of detected COMs relative to CH$_{3}$OH. To show their spreads, the grey boxes indicate the abundance ranges of our samples. The dark grey horizontal lines are the median and shades are the middle 25 - 75 \% values. 

The COMs detections are collected in various physical conditions, which are illustrated by the colored shapes in Figure \ref{fig:abundance}.
AFGL 4179 is taken as the high-mass hot core associated with an O-type star \citep{Bogelund2019}, Sgr B2(N2) as COMs-rich hot core at the Galactic center \citep{Belloche2016,Jorgensen2020}, WB 89-789 for a chemically rich hot core with low metallicity located in extremely outer region of our galaxy, \citep{Shimonishi2021}, Orion KL hot core and compact ridge regions as recently shocked regions \citep{Tercero2018}, IRAS 16293 B as a low mass hot conino in the Class 0 stage \citep{Jorgensen2018,Jorgensen2020}, CALYPSO survey results toward 11 Class 0 and I sources \citep{Belloche2020}, PEACHES survey results for 28 Class 0 and I protostars with COMs in Perseus molecular cloud \citep{Yang2021}, HH 212 observation to trace the COMs in low mass disk atmosphere \citep{LeeChin-Fei2019}, and V883 Ori to trace the COMs freshely evaporated from the disk midplane by the accretion burst \citep{Lee2019}. The recently updated astrochemical models are also included, considering non-diffusive mechanism \citep{Garrod2022}. 
Note that abundances of the hot core WB 89-789 are multiplied by 4. According to  \citet{Shimonishi2021}, when the COMs abundances are supplemented by a factor of 4, they show great similarity with those of Galactic hot core, which can be accounted for its low metalicity located in outer Galaxy.

From our samples, we sort the O- and N-bearing molecules in order of median of abundance. Except for the lower abundance molecules found in Orion KL, which cause different slopes from C$_{2}$H$_{5}$OH to HCOCH$_{2}$OH, overall trends agree within slightly more than an order of magnitude for both high and low mass sources. 

This similarity is very striking when we consider that the physical conditions of collected sources vary significantly, 0.16-47 L$_{\odot}$ and 0.12-32 L$_{\odot}$ for hot coninos from CALYPSO \citep{Belloche2020} and PEACHES \citep{Yang2021}, 3 L$_{\odot}$ for IRAS 16293B \citep{Jacobsen18}, 9 L$_{\odot}$ for low-mass disk harboring protostar in HH 212 \citep{LeeChin-Fei2019}, 400 L$_{\odot}$ for bursting protostar V883 Ori \citep{Lee2019}, and $\sim$2$\times$10$^5$L$_{\odot}$ for hot cores AFGL 4167 and Sgr B2(N2) \citep{Bogelund2019,Bonfand19}. The fact that over an order of magnitude range in abundance detected in huge luminosity differences over six orders of magnitude indicates the current luminosity is not a dominant factor affecting the characteristics of COMs. 

This similarity seems more robust in O-bearing COMs than N-bearing COMs, but we cannot confirm this due to the lack of the N-bearing COMs detection toward some regions (Orion KL, HH212 and V883 Ori).
This consistency with respect to CH$_{3}$OH abundance has been found in different star-forming regions, in various COMs \citep{Bianchi19,vanGelder2020,Belloche2020,Jorgensen2020} and the astrochemical model \citep{Aikawa2020}. 
The overall trend of COMs abundance suggests that the physical conditions and production mechanisms required for COMs are analogous, and this COMs chemistry is predominantly set by the ice chemistry in the cold prestellar core stage. 
One additional scenario is hinted by \citet{Shimonishi2021}. They compared the COMs abundances of a hot core in extremely outer Galaxy, Galactic hot core, and low metallicity hot core in Large Magellanic Cloud. The abundances between hot cores in outer Galaxy and inner Galactic region show a great similarity when the values of hot core in outer Galaxy are multiplied by a factor of 4, which corresponds to its poor metallicity level compared with near the solar neighborhood. In contrast, the abundances for the extreme hot core and that in Large Magellanic Cloud do not resemble, 
despite the similarity of the poor metallicity. Based on this, the found global COMs abundance agreement found in our collected samples may come from the coherent initial condition of our Galaxy which decides the overall COMs ratios.

On the other hand, in each species, scatters can reflect the roles of local conditions and subsequent physics, which control the desorption mechanisms and gas-phase chemistry following to the evaporation.

Our sources shows consistency with the CALYPSO group 3 samples which are classified as CN-COMs-rich hot corinos, PEACHES survey, and V883Ori for the detected COMs. One common characteristic for them is the burst accretion. Interestingly, the COMs emitting regions of CALYPSO group 3 sources are larger than their hot corino sizes with the temperature range of 100 K $<$ T $<$ 150 K estimated from their current luminosities (Figure 13 in \citealt{Belloche2020}). This is a strong indication of the prior burst accretion events that evaporates the COMs beyond the water snow line set by the current luminosity. For PEACHES, hot corino survey in Perseus molecular cloud known as one of the most active SFR, half (14 out of 28) of their sources have the evidence of prior or current burst accretion; the CO or H$_{2}$O snowlines were observed in larger radii than those can be explained by the current bolometric luminosity \citep{Hsieh2019,Yang2021}. V883 Ori is a low mass protostar currently in the FUor ourburst phase with a bolometric luminosity of $\sim$400 L$_{\odot}$ with an enhanced accretion rate by a few orders of magnitude than typical low mass protostars \citep{Cieza2016}. In addition, our samples include ten cores in active accretion, traced by the class II CH$_{3}$OH maser association. 

The systematically higher abundance compared with the hot core in AFGL 4176 is found in hot corino surveys (CALYPSO and PEACHES).
It may indicate no distinctive correlation between the occurrence of COMs and the current luminosity, which is consistent with the previous results  \citep{Taquet2015,vanGelder2020,Yang2021}. Instead, the strength of accretion events, which episodically varies during the star formation, may more affect the chemical characteristics.

The abundances of O-bearing COMs in Orion KL regions are exceptionally low. 
One can expect the density of Orion KL regions to be lower than other regions. It has been suggested that the COMs enhancement in Orion KL region is not caused by the internal heating source, but externally heated by a recent explosive shock within 550 years and/or outflow shocks driven by an HMYSO Orion source I \citep{Zapata2011}. 
In particular, the compact ridge region can be excited by the older outflows from Orion source I \citep{Tercero2018}. 
Unfortunately, the behavior of N-bearing COMs cannot be directly compared with our observation because of the absence of N-bearing COMs observations with high resolutions comparable to ours, toward Orion KL to date. 

The high (CH$_{3}$)$_{2}$CO abundance derived in our study stands out compared with other regions.
This may indicate that the chemical process of (CH$_{3}$)$_{2}$CO could be distinct from other O-bearing COMs. One suggestion is that (CH$_{3}$)$_{2}$CO is more sensitive to radiation or shocks compared to other O-bearing COMs. 
\citet{Peng2013} observed the (CH$_{3}$)$_{2}$CO toward Orion BN/KL, revealing that the distribution of (CH$_{3}$)$_{2}$CO is closer to N-bearing COMs than other O-bearing COMs. They suggested that the formation or destruction routes of (CH$_{3}$)$_{2}$CO can involve N-bearing species. 
This may be the case for our sources since the SiO emission, which traces the outflow shocks responding to recent burst accretions, is readily detected in many of our samples. 
The (CH$_{3}$)$_{2}$CO abundance range of our sources covers the values of Sgr B2(N2) and V883 Ori, while the abundances are systematically higher than all other samples. Sgr B2(N2) is a well known hot core in active star-forming region with a strong radiation field in the Galactic center \citep{Belloche2016}. V883 Ori is in the actively accreting phase with FUor type outburst.

Meanwhile, for N-bearing COMs, the abundance of our sources are systematically lower than those found in hot core Sgr B2(N2) but higher than low mass hot corino IRAS 16293 B. One exception is CH$_{3}$NCO. The abundance of our samples is comparable with Sgr B2(N2), while it is higher than that of IRAS 16293 B by more than an order of magnitude. Since N-bearing COMs are known to be sensitive to the strength of UV radiation field and/or shock, the abundance lower than that of Sgr B2(N2) and higher than that of IRAS 16293 B can be interpreted as intermediate strengths of UV radiation field and shock, induced by accretion and outflow, respectively. Therefore, the accretion rates of our sources might be in between the two regions. The other scenario is that our cores are chemically younger than Sgr B2(N2) but older than IRAS 16293 B. 
Given that most COMs formation occurs in the warm-up phase (T $\sim$ 20--40 K, \citealt{Garrod2013}) and N-bearing molecules require a longer timescale than O-bearing molecules to be synthesized in the cold cloud \citep{Lee2020}, the ratio of O-bearing and N-bearing COMs can be a tracer of the timescale for the warm-up phase. Our samples might have experienced the warm-up phase timescales shorter than Sgr B2(N2) but longer than IRAS 16293 B.

The model in \citet{Garrod2013}, where the chemical complexity operates efficiently via a diffusive mechanism in the warm-up phase, shows discrepancy with our results for the O-bearing COMs. However, their updated model \citep{Garrod2022}, which considered non-diffusive mechanisms, shows better agreement with our observations. 

\subsection{Correlation between COMs}

\begin{figure*}
  \centering
\includegraphics[clip,width=0.45\textwidth,keepaspectratio]{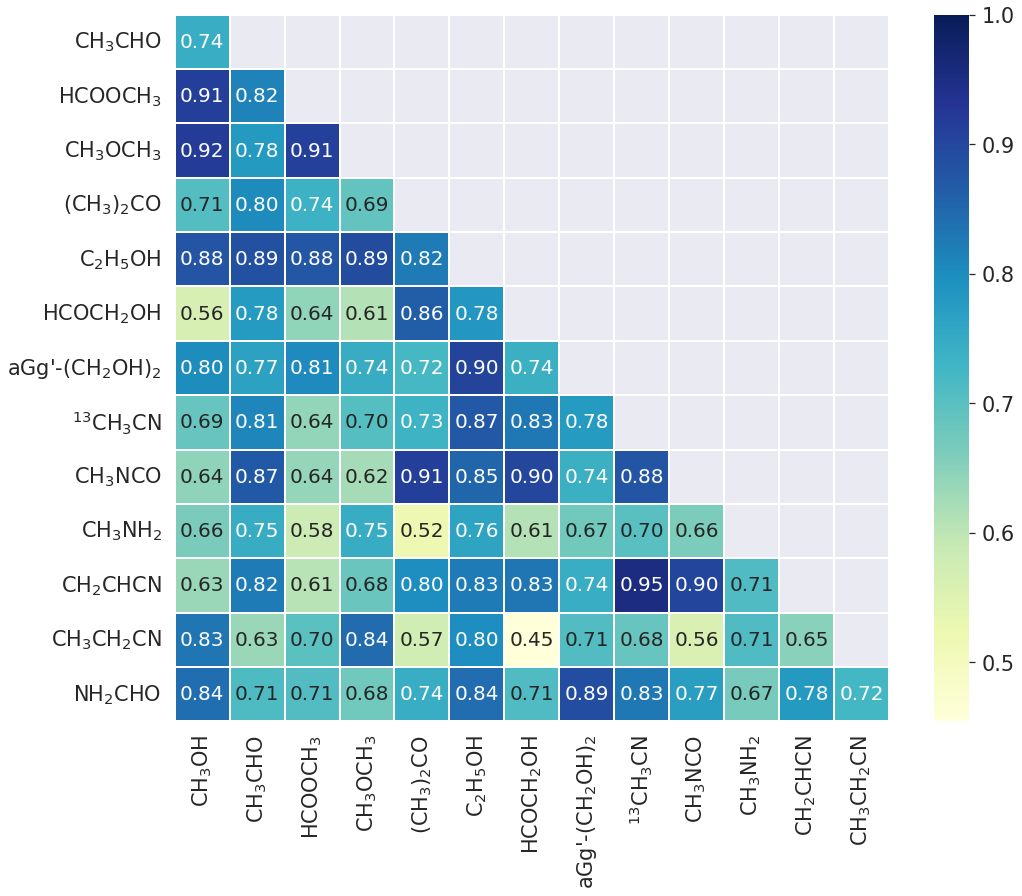}
\includegraphics[clip,width=0.45\textwidth,keepaspectratio]{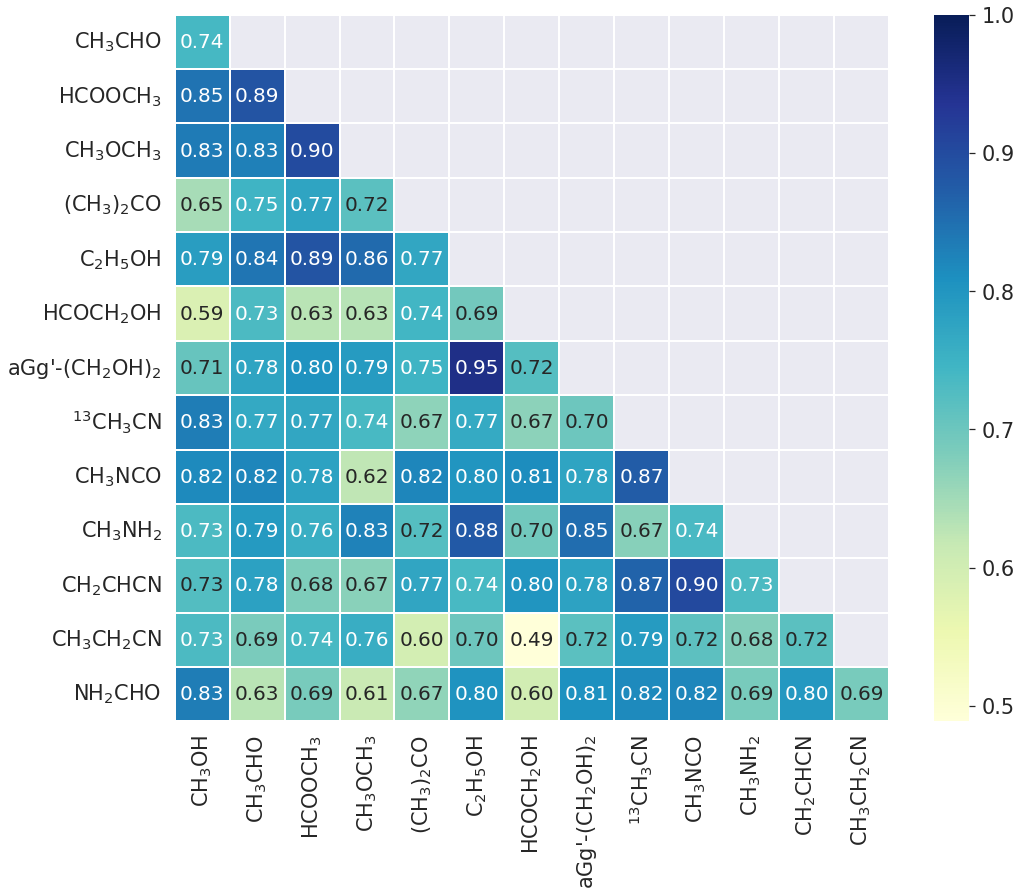}
\caption{Heat maps of the correlation using the column densities (left) and abundances to N(H$_2$) (right) between two species. The colors show the level of correlation which correspond to the Pearson $r$ coefficient shown inside in each box. The scatter plots with measured slopes are presented in Figure \ref{fig:corr_N_all} and \ref{fig:corr_N_NH2_all}.}.
\label{fig:heatmap}
\end{figure*}

We test the correlation of COMs to investigate how the molecules are chemically linked. The deduced molecular column densities span over 2 orders of magnitudes, which allows us to investigate their relative abundances in wide range of physical condition. The relative abundances of some species can provide the information on the branching ratio if they are indeed chemically related.
Figure \ref{fig:heatmap} displays the maps for correlation coefficients for all pairs of main isotopologues using the Pearson coefficient $r$. The correlation coefficient between two species in the column densities (left panel) and abundances relative to N(H$_2$) (right panel) ranges from 0.45 to 0.95 and from 0.49 to 0.95, respectively. For all pairs, the scatter plots with slopes measured in logarithmic scale are presented to infer the power-law index between two species (Figure \ref{fig:corr_N_all} and Figure \ref{fig:corr_N_NH2_all}).

CH$_3$OCH$_3$ and HCOOCH$_3$, the most abundant species following to CH$_3$OH in our observation, show a tight correlation. This is consistent with the previous observations in different physical conditions \citep{Brouillet13,Jaber14,Belloche2020,Yang2021}, implying their strong relevance in various chemical pathways. 
Under the grain-surface chemistry, their close relation is explained by a common precursor. CH$_3$OCH$_3$ and HCOOCH$_3$ are dominantly formed by the radical--radical reaction of CH$_3$O with CH$_3$ and HCO radicals, respectively \citep{Garrod2013,Oberg2016}. The gas-phase process has been also proposed, for CH$_3$OCH$_3$ being a precursor of HCOOCH$_3$ \citep{Balucani15}. In our observation, both species also show great correlations with CH$_3$OH (Figure \ref{fig:corr_N_all}), supporting that the origin of CH$_3$O radical is likely the photodissociation of CH$_3$OH both in grain and gas \citep{Garrod2013,Balucani15}.

A strong correlation between C$_2$H$_5$OH and aGg'-(CH$_2$OH)$_2$ is also explained by a shared precursor. On the grain-surface, C$_2$H$_5$OH, (CH$_2$OH)$_2$, and HCOCH$_2$OH can be efficiently formed via the radical-radical reactions involving CH$_2$OH radical, another product of the photodissociation of CH$_3$OH \citep{Vasyunin13,Garrod2013}. 

Moreover, in our analysis, C$_2$H$_5$OH (CH$_3$CH$_2$OH) tends to show high Pearson $r$ coefficients ($r >$ 0.8) with other O-bearing COMs containing CH$_3$, indicative of a common radical involving their formation routes.
Meanwhile, the coefficient in HCOCH$_2$OH and aGg'-(CH$_2$OH)$_2$ with others are relatively low. This can be due to the small samples with detection, or another nontrivial pathway which is less involved with the other molecules.
A recent laboratory experiment showed that HCOCH$_2$OH and (CH$_2$OH)$_2$ can be formed via a sequential atom addition reaction on the ice mantle started from CO deposition \citep{Coutens2018}. In their experiment, both species form with little CH$_2$OH radical, and HCOCH$_2$OH is produced with HCO radicals followed by hydrogenation and further forming (CH$_2$OH)$_2$. In addition to this formation network, gas-phase formation route is also opened (Section \ref{subsec:CH3CHO_HCOCH2OH}). 

The correlation coefficient between CH$_{2}$CHCN and CH$_{3}$CH$_{2}$CN is not as high as others, due to the plateau observed toward the CH$_{3}$CH$_{2}$CN column density above 10$^{16}$cm$^{-2}$, probably caused by the high optical depth of this molecule.
According to \citet{Garrod2017}, the appreciable amount of CH$_{2}$CHCN is not achieved on the grain-surfaces because the successive atom addition reaction for CH$_{2}$CHCN quickly proceeds to CH$_{3}$CH$_{2}$CN. Instead, after the ejection of CH$_{3}$CH$_{2}$CN from the dust grain, CH$_{2}$CHCN primarily survives in gas due to the CH$_{3}$CH$_{2}$CN destruction. In our analysis (Section \ref{sec:dis:physical_properties}), we grouped the detected COMs into two by the response to the physical condition, probably due to their different formation timescales. CH$_{3}$CH$_{2}$CN belongs to the first group, of which the abundance relative to CH$_3$OH maintains a constant level with respect to N(H$_2$) (Figure \ref{fig:corr_XCOMs_NH2_hue_NCOMs}). This indicates the co-existence of CH$_{3}$CH$_{2}$CN and CH$_3$OH, with similar formation timescale, primarily via grain-surface networks. In contrast, as belonging to the second group, CH$_{2}$CHCN abundance shows positive gradient, indicating a relatively longer formation timescale of CH$_{2}$CHCN than CH$_{3}$CH$_{2}$CN, which is consistent with the model prediction.

CH$_3$NH$_2$, CH$_3$NCO, and NH$_2$CHO contain peptide (–NH–(C=O)–) like bonds, which are regarded as potential compounds of amino acids. Thus, they are of interest to prebiotic chemistry because they are thought as precursors for more complex biomolecules. Their formation routes are still under debate, however, solid state reactions are suggested for their efficient formation starting with NH and NH$_2$ \citep{Garrod2008,Garrod2013,Barone2015,Kanuchov2016,Belloche2017,Ligterink2018,Quenard2018,Gorai2021}. 

From our study, one of the observed characteristics of the three species is their abundances, which are in between those of IRAS 16293B and Sgr B2(N2) (Figure \ref{fig:abundance}); the difference may be attributed to the degree of UV radiation. 
In correlation plots (Figure \ref{fig:heatmap}), among three, CH$_3$NCO and NH$_2$CHO have the closest correlation, but general correlation is not robust compared to other pairs. Perhaps this is due to the relatively small number of samples with detection ($\sim$10). 
 
We note that the column densities of some abundant species may be underestimated by the high optical depths of their lines. A similar response to the physical parameters such as temperature, density, or IR radiation field can give a high correlation by chance \citep{Brouillet13}. For the former issue, in our samples, the high optical depths of COMs seemingly affect only 3 out of 28 cores (Figure \ref{fig:NCOMs_NH2_hue_NCOMs}), thus they do not significantly affect the result. For the latter one, we distinguish a possible correlation using Figure \ref{fig:corr_NCOMs_Tex_hue_NCOMs}. Last, the single linear fitting may not be appropriate to account for the relation between two species, as reported by \cite{El-Abd2019}, which is not considered in this study.
 
\subsubsection{CH$_{3}$CHO and HCOCH$_{2}$OH}\label{subsec:CH3CHO_HCOCH2OH}
\begin{figure}
  \centering
\includegraphics[clip,width=0.9\columnwidth,keepaspectratio]{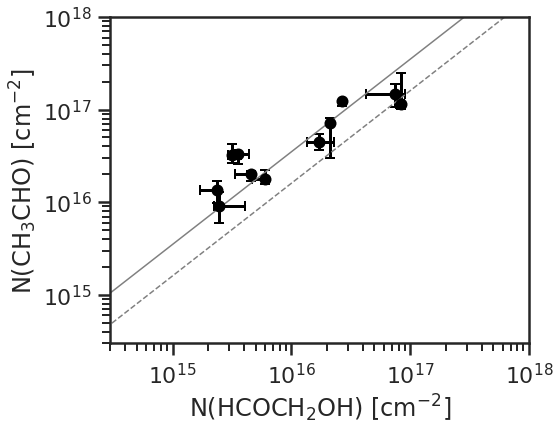}
\caption{The correlation between the CH$_{3}$CHO and HCOCH$_{2}$OH of eleven cores where the two COMs are detected. The sold and dashed lines are the abundance ratio of 3.5 and 1.6, respectively, which imply the branching ratio suggested gas-phase formation route including their uncertainty \citep{Vazart2020}.}
\label{fig:ch3cho_hcoch2oh}
\end{figure}

We compare our observations with theoretical computations. As one of the examples for key molecular species, we here compare correlation between HCOCH$_{2}$OH and CH$_{3}$CHO. 

Figure \ref{fig:ch3cho_hcoch2oh} shows that the correlation of column densities between CH$_{3}$CHO and HCOCH$_{2}$OH for eleven cores where both species were detected. The found proportional relation suggests that the two molecules may be chemically related. If they form from a sharing parent molecule, the slope of their proportional relation may provide their branching ratio. 

\citet{Skouteris2018} and \citet{Vazart2020} showed that the destruction of C$_{2}$H$_{5}$OH could efficiently form HCOCH$_{2}$OH and CH$_{3}$CHO with branching ratios of 3:1 in the gas-phase (solid and dashed lines in Figure \ref{fig:ch3cho_hcoch2oh} with the modeling uncertainty), which explains our data reasonably well. 
\citet{Skouteris2018} used the reaction coefficient measured at a sufficiently high temperature for the destruction of C$_{2}$H$_{5}$OH (100--300 K), and they reproduced the observation by using the gas-phase network followed by the sublimation of C$_{2}$H$_{5}$OH formed in ice. The uncertainty of their pathway is caused by the abundance of OH radicals for the destruction of C$_{2}$H$_{5}$OH at the first step of the gas-phase chemistry. 

On the other hand, the recently updated model by \citet{Garrod2022} adopts non-diffusive grain-surface and ice-chemistry.
Among three models with different warm-up timescales, the slow warm-up model better matches with our observation in COMs abundances (Figure \ref{fig:abundance}) and ratio between CH$_{3}$CHO and HCOCH$_2$OH. This model more concerns the relavance of HCOCH$_2$OH with HCOOCH$_3$ and CH$_3$COOH (than with CH$_{3}$CHO), as the three species are structural isomers. Nevertheless, the ratio between CH$_{3}$CHO and HCOCH$_2$OH in slow warm-up model is consistent with our data ($\sim$4 ; Table 18 of \citealt{Garrod2022}).

Therefore, both ice- and gas-phases chemistry reproduce the observed ratio between HCOCH$_{2}$OH and CH$_{3}$CHO \citep[e.g.,][]{Skouteris2018, Vazart2020,Garrod2022} in our samples. It suggests that the surface chemistry generally determine the COMs abundances, but the gas-phase production of aldehyde from the C$_{2}$H$_{5}$OH destruction also plays a role in hot cores.

\subsection{Isotope ratio} 
\begin{figure*}
  \centering
\includegraphics[clip,width=0.3\textwidth,keepaspectratio]{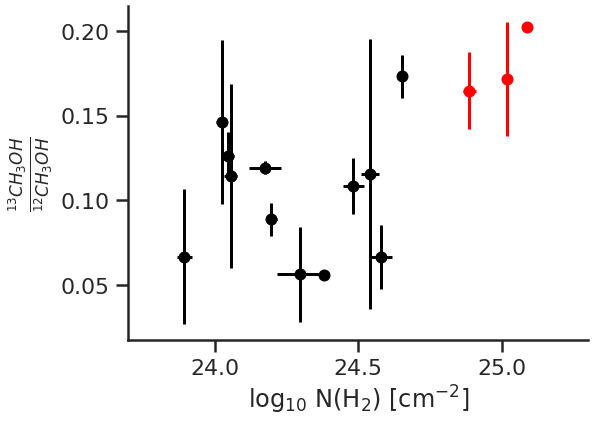}
\includegraphics[clip,width=0.3\textwidth,keepaspectratio]{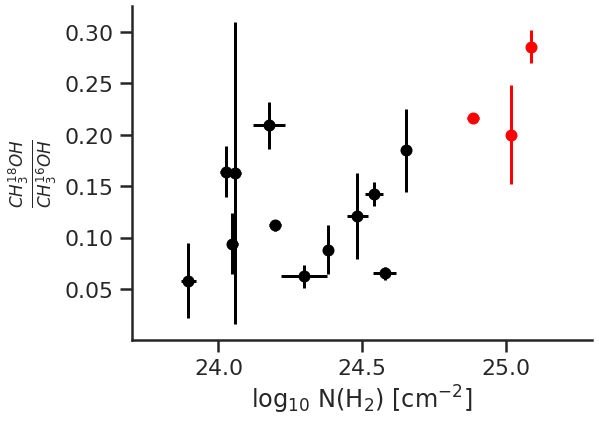}
\includegraphics[clip,width=0.3\textwidth,keepaspectratio]{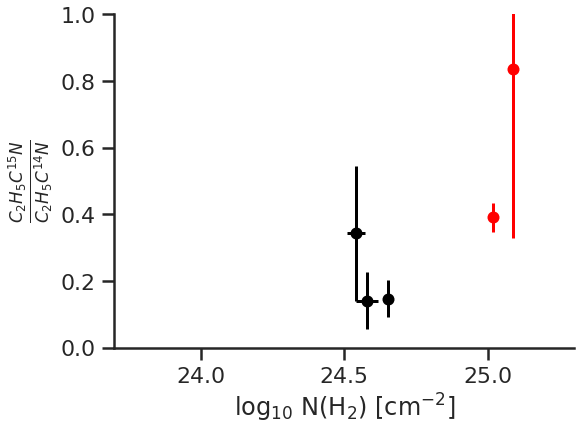}\\
\includegraphics[clip,width=0.3\textwidth,keepaspectratio]{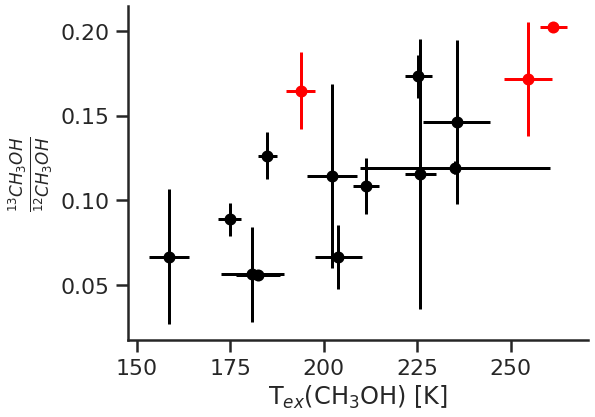}
\includegraphics[clip,width=0.3\textwidth,keepaspectratio]{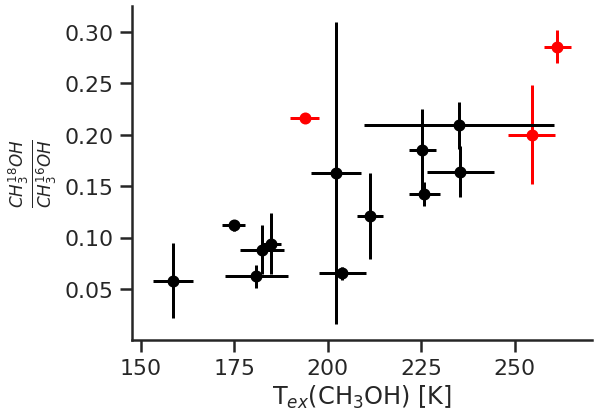}
\includegraphics[clip,width=0.3\textwidth,keepaspectratio]{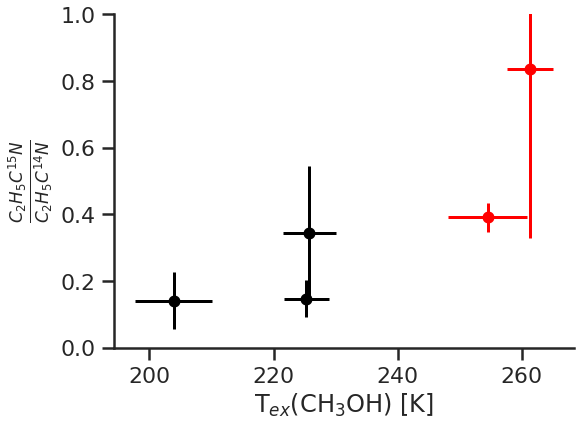}
\caption{The isotopic ratio of $^{13}$CH$_3$OH/$^{12}$CH$_3$OH, CH$_3^{18}$OH/CH$_3^{16}$OH, and C$_2$H$_5$C$^{15}$N/C$_2$H$_5$C$^{14}$N as functions of N(H$_2$) (upper panels) and T$_{ex}$(CH$_3$OH) (lower panels). Red and black dotts indicate the cores with N(H$_2$) $>$ 24.7 and N(H$_2$) $<$ 24.7 cm $^{-2}$, respectively.}
\label{fig:isotope_ratio}
\end{figure*}

The level of fractionations of COMs is sensitive to the density, temperature, and timescale so that they provide insight of when and where they are formed. There are a few mechanisms that can alter the isotopic ratios, including isotopic exchange reactions both in the gas- and ice-phases, selective photodissociations, and timescale spending at cold temperature \citep{Furuya2018,Jorgensen2020,oberg2021}. 
From our data sets, we examine the hydrogen, carbon, oxygen, nitrogen isotopic ratios. 

The isotopic ratios of carbon, oxygen and nitrogen are presented as functions of N(H$_2$) and T$_{ex}$(CH$_3$OH). As discussed in Section \ref{sec:dis:physical_properties}, column densities of main isotope species toward three highest column density cores (log$_{10}$N(H$_2$) $>$ 24.7 cm$^{-2}$; red dotts) might be underestimated due to the high optical depth, resulting in isotope ratios overestimated (upper panels of Figure \ref{fig:isotope_ratio}). Except for the measurements in the three densest cores, no clear trend is found in isotopic ratios with respective to N(H$_2$). 
Instead, the isotopic ratios are tentatively proportional to T$_{ex}$(CH$_3$OH) (lower panels of Figure \ref{fig:isotope_ratio}). Although we speculate the high optical depth for main isotopologues in the three densest cores, the detected linear trends are not significantly affected by the three cores. 

The selective photodissociation in COMs unlikely cause the proportional trends since the abundances of COMs are very low compared to CO and N$_2$, for which the self-shielding effect is important \citep{Visser2009,Heays2014}. 
More favored mechanism is the isotope change reactions. 
We found that the chemical processes become efficient in cores with high N(H$_2$) and T$_{ex}$(CH$_3$OH) (Figure \ref{fig:corr_XCOMs_NH2_hue_NCOMs} and Figure \ref{fig:corr_XCOMs_Tex_hue_NCOMs}). Based on more notable correlation between the isotopic ratio and T$_{ex}$(CH$_3$OH), the efficient gas and surface chemistry, which could be induced by the high gas temperature and strong radiation field, may promote isotopic fractionation, as discussed toward IRAS 16293 B \citep{Jorgensen2018}.
Nevertheless, the absolute isotope ratios of CH$_3$OH in our samples ($>$ 0.05) are much higher than those found in Sgr B2(N2) (0.04) and IRAS 16293B (0.015) \citep{Muller2016,Jorgensen2016}. 

We note that the levels of fractionation are much higher than the Galactic \citep{Wilson1994} and local interstellar medium (ISM) values (0.015 for $^{13}$C/$^{12}$C; 0.0036 for $^{15}$N/$^{14}$N; 0.0018 for $^{18}$O/$^{16}$O; \citealt{Nomura2022}). Since we observed interior region, close to the massive protostar, these unusually high isotopic ratios may come from significantly different physical and chemical conditions deviated from the Galactic scale environment. In addition, the isotope fractionations differ by species, but the values adopted for the Galactic and local ISM ratios are derived from atoms, and thus, they could be very different from the ratios in COMs \citep[e.g.,][]{Nomura2022,Drozdovskaya2022}.

\begin{figure*}
  \centering
\includegraphics[clip,width=0.3\textwidth,keepaspectratio]{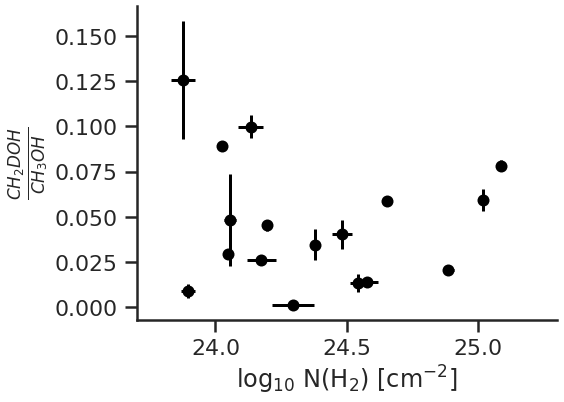}
\includegraphics[clip,width=0.3\textwidth,keepaspectratio]{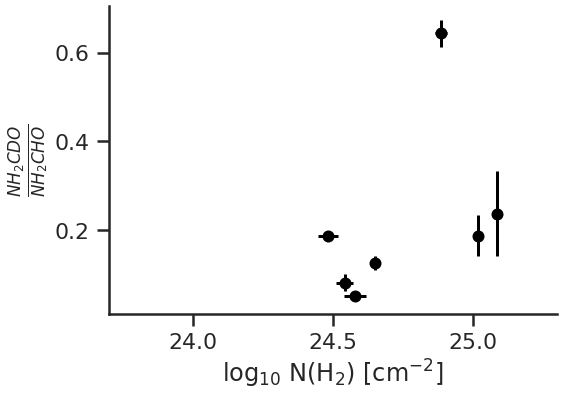}\\
\includegraphics[clip,width=0.3\textwidth,keepaspectratio]{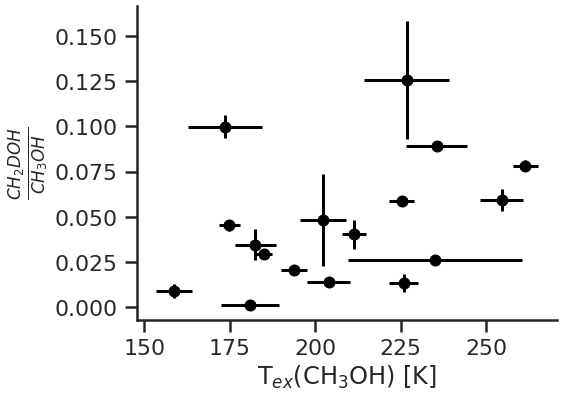}
\includegraphics[clip,width=0.3\textwidth,keepaspectratio]{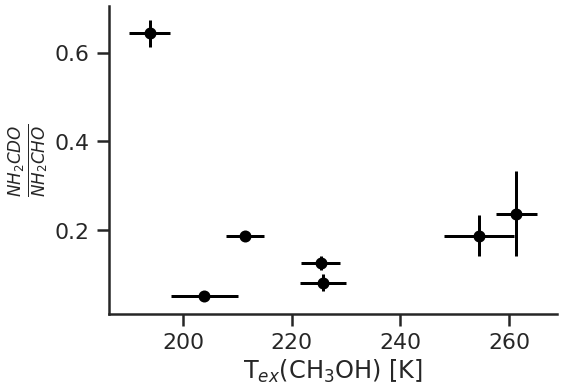}
\caption{The D/H ratios of CH$_3$OH and NH$_2$CHO, to N(H$_2$) (top) and T$_{ex}$(CH$_3$OH) (bottom).}
\label{fig:d_h_ratio}
\end{figure*}

Deuterium fractionation has been used to constrain how long the cores stayed at cold temperature. Since D atom is twice heavier than H atom, thus stronger in bond interactions, the deuterated ions or molecules are favorably formed from HD \citep{oberg2021}.
The deuteration of molecules begins with two reactions: (1) 
    $H_3^+ + HD \leftrightarrow H_2D^+ + H_2 + 232~K $ and 
  (2) $CH_3^+ + HD \leftrightarrow CH_2D^+ + H_2 + 654~K$.
In turn, the produced ions become the main reactants to form simple molecules and COMs. 
Since the forward reactions are exothermic, the reverse reaction is inactive at cold temperature ($<$ 30 K for $H_3^+$ and $<$ 300 K for $CH_3^+$), which results in increasing D/H ratio if a core stays in the cold-phase long. 

In Figure \ref{fig:d_h_ratio}, we could not find a clear trend of the D/H ratios with respective to N(H$_2$) within the range of observed properties: the D/H ratio found in CH$_{3}$OH tentatively proportional to T$_{ex}$(CH$_3$OH) and inversely proportional to N(H$_2$). We need to remind that $T_{\rm ex}$ does not represent the temperature when COMs were formed. Instead, the current temperature could be a measure of the efficiency of atomic exchange reaction after desorption.

Compared with other regions, in most of our cores, the D/H ratios (0.001--0.13 for CH$_3$OH) are higher than that of Sgr B2(N2) ($<$0.004 for O- and N-bearing COMs; \citealt{Belloche2016}). The low deuteration level of Sgr B2(N2) has been interpreted by the high temperature in prestellar stage in the extreme condition of Galactic Center region \citep{Belloche2016}.
Among our samples, G49.49-0.39 C2 and G10.34-0.14 C3 show the lowest D/H ratios of 0.001 and 0.009, respectively (Figure \ref{fig:spectrum:G49.49C2} and Figure \ref{fig:spectrum:G10.34C3}). Except for G49.49-0.39 C4, cores in G49.49-0.39 tend to show low D/H ratios ($\leq$0.02). Since deuterium fractionation increases if a core has stayed at the cold environment longer, the observed low ratios of G10.34-0.14 C3 and cores in G49.49-0.39 region indicate that the cores mostly stayed in the high temperature environment through
their formation history. The low deuteration level can be also caused by the shorter presellar core stage owing to more rapid collapse than other sources; in short timescale, the isotopic exchange reaction could not sufficiently occur \citep{Faure2015}.

This D/H ratio can be different by species. The fractionation of COMs tend to be higher than that of H$_2$O \citep{Persson2014,Taquet2019} because COMs would be formed on grain-surface after the CO freeze-out on grain surfaces: H$_2$O ice can be formed on grain surfaces even before the dense core is formed, while CO freeze-out occurs after the temperature drops below 18 K in cold and dense cores \citep{Bisschop2006}. Thus, ice structure can have two types of layers: inner H$_2$O-dominated ice layer with a lower D/H ratio and outer CO-based ice layer with a higher D/H ratio \citep{Jensen2021}. In addition, among COMs, the fractionation level also depends on the CO depletion timescale during core formation as well as core density \citep{Bergin2014,oberg2021} since the frozen CO on the grain surfaces can participate in the ice chemistry producing COMs.

The D/H ratios of CH$_3$OH and NH$_2$CHO measured in IRAS 16293B are $\sim$0.02, lower than those of most our sources. The ratios of different COMs range over 0.005--0.2 in IRAS 16293B \citep{Drozdovskaya2022}. Within a source, the varied deuteration of individual species imply the different formation timescales of those molecules \citep{Jorgensen2020}. The D/H ratios in our sources are generally in agreement with the COMs ratios found in IRAS 16293B. However, some cores (G18.34+1.78SW C1, G18.34+1.78SW C2, and G18.34+1.78 C8 for CH$_3$OH and G49.49-0.39 C3 for NH$_2$CHO) show notably high D/H ratios. The D/H ratio can be enhanced when core stay longer at the dense and cold environment shielded from the interstellar UV radiation field before protostars form at the core center. In addition, the level of deuteration can also change with time via gas-phase reactions induced by the photodesorption. 

\begin{figure}
  \centering
\includegraphics[clip,width=1.0\columnwidth,keepaspectratio]{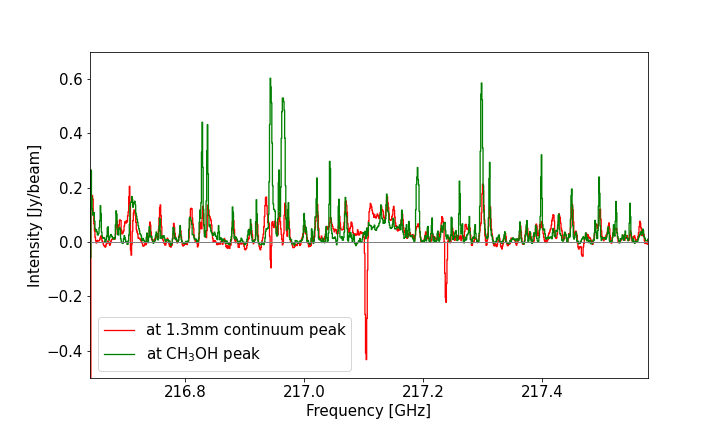}
\caption{Example spectrum of G49.49-0.39 C1. Red and green solid lines show the spectrum at the 1.3 mm continuum and CH$_{3}$OH peaks, respectively.}
\label{fig:sp_cont_ch3oh_peaks}
\end{figure}

\subsection{COMs intensity at the continuum peak position} \label{sec:contpeak}
In our observation, the CH$_{3}$OH emission lines are weakened at the continuum peak position (Figure \ref{fig:sp_cont_ch3oh_peaks}).
The suppressed CH$_{3}$OH emission at the continuum peak position might be caused by the dust obscuration due to the high dust opacity near the protostar. This has often been seen in the ALMA view toward early stages of both low- and high-mass YSOs (e.g., \citealt{Lee2019,Ko2020,Kim2020}). The observation toward low mass protostar TMC1A resolved the inner disk \citep{Harsono2018} with 1.3 mm continuum and isotopologues of CO. The emission of CO isotopologues is suppressed toward the central position of the protostar. They are even completely devoid at the dust continuum peak. This observation is best reproduced by the existence of CO emission at the central part, but obscured by the large (1 mm) size dust grains near the protostar. 

This effect of dust obscuration is also investigated in detail in the NGC 1333 IRAS 4A binary system. The 4A1, which showed a stronger continuum emission than 4A2, is free of COMs emission whereas 4A2 is associated with the rich spectrum at the submillimeter wavelengths \citep{DeSimone2020}. The polarization observation with ALMA (0.85--0.89 mm and 1.3 mm) and JVLA (6.3--16.7mm) showed a 90$^\circ$ angle offset between the two observations only toward 4A1. This is explained by the dust grains with different optical depths seen in two wavelengths; the high opacity foreground material of 4A1 causes high extinction. This leads to observed orthogonal polarization angles between two frequencies. In contrast to the ALMA observation, a similar brightness of CH$_{3}$OH emission is observed toward the two sources in the centimeter regime, at which the dust opacity is almost negligible \citep{DeSimone2020}.

This can be the case for some of our samples since they are in the early stage of star formation and embedded in thick envelope material. 
Different detection rates have been seen within a clustered region; a core with strong continuum emission shows weaker or no COMs emission, compared to other cores with weaker continuum emission.
One of such examples is the G29.91-0.03 region (Figure \ref{fig:spectrum:G29.91C1} and Figure \ref{fig:spectrum:G29.91C2}); 
C1 shows a higher $N(\rm H_{2})$ at both the CH$_{3}$OH peak and continuum peak positions than those of C2 by a factor of two. However, only the CH$_{3}$OH emission was detected in C1, while three COMs were detected in C2. The other example is G18.34+1.78SW; 15 COMs were detected in C2, while only three COMs were detected in C1, where $N(\rm H_{2})$ is higher than that of C2 by a factor of 1.3 (Figure \ref{fig:spectrum:G18.34SWC1}, Figure \ref{fig:spectrum:G18.34SWC2} and Table \ref{tab:maserandCOMs}).

Extended toward COMs non-detected cores with higher continuum peaks, the G28.37+0.07 region is particularly interesting. At G28.37+0.07 C1 the strongest continuum emission is observed, but no COM is detected (Table \ref{tab:216GHz} and Table \ref{tab:maserandCOMs}). 
In contrast, G28.37+0.07 C2 showed a rich spectra with 15 COMs (Figure \ref{fig:spectrum:G28.37C2}), associated with the 6.7 GHz class II CH$_{3}$OH maser.
 
However, the presence of any embedded protostar or gravitational boundness is not checked toward G28.37+0.07 C1.
If the G29.91-0.03 C1, G18.34+1.78SW C1, and G28.37+0.07 C1 cores are in the same situation with the aforementioned case toward NGC 1333 IRAS 4A system, they are still highly embedded and the density near the central protostar might be too high for the molecular emission to escape. As a result, to avoid the effect of thick dust material on the COMs lines, we extract the spectra from a position slightly off the continuum peak position.

To investigate the COMs richness and complexity without hinder of the high dust opacity, the more transparent windows at longer wavelengths can be useful to study the COMs emission close to the central protostar unless the non-thermal effect is high \citep{McGuire2020}. 

Another scenario for the offset of COMs peak from the central position is additional desorption mechanisms for COMs. We speculate the thermal evaporation of COMs due to their compact spatial structure around the protostar. There are several observations showing that COMs can be also evaporated by shocks by outflowing winds originated very near the protostar \citep{LeeChin-Fei2019} or infalling material \citep{Csengeri2019}. However, the portion of detected COMs desorbed by the non-thermal mechanism cannot be separated because it requires a detailed physical and chemical modeling toward individual cores, which is beyond the scope of this work. 

%% file: s3_tab1_1.3cont_peak.tex
\startlongtable
\begin{deluxetable*}{lcccccccccc}
\tabletypesize{\scriptsize}
\tablecaption{The summary of core detection in 1.3 mm continuum}
\tablehead{
\colhead{Name} & \colhead{R.A.} & \colhead{Decl.} & \colhead{SNR$^a$} & \colhead{Peak Intensity} & \colhead{T$_b$}& \colhead{$N_{H_2}$ $^b$} & \colhead{Source Size$^c$}& \colhead{Flux Density}& \colhead{$M_{total}$ $^d$} \\ 
\colhead{} & \colhead{($^{hms}$)} & \colhead{($^\circ$ $'$ $''$)}  & \colhead{}
& \colhead{(mJy beam$^{-1}$)} & \colhead{(K)} & \colhead{(cm$^{-2}$)} & \colhead{($''\times''$), P.A.$^\circ$} & \colhead{(mJy)} & \colhead{(M$_{\odot}$)} 
}
\startdata
G10.32-0.26 C1 & 18:09:23.28 & -20:08:07.2 & 216 & 18.4(0.1) & 5.41 & 1.58$^{+0.41}_{-0.27}\times$10$^{24}$ & [0.4$\times$0.3],  74.9 &   32(0.2) & 0.6--5.3 \\
G10.32-0.26 C2 & 18:09:23.22 & -20:08:07.4 & 75 &   7.0(0.1) & 2.06 & 0.89--5.68$\times 10^{24}$ & [0.7$\times$0.4],  93.1 &   26(0.3) & 0.7--4.5 \\
G10.32-0.26 C3 & 18:09:23.18 & -20:08:07.0 & 59 &   5.9(0.1) & 1.75 & 0.75--4.83$\times$10$^{24}$ & [0.4$\times$0.4],   2.1 &   11(0.2) & 0.3--1.8 \\
G10.32-0.26 C4 & 18:09:23.24 & -20:08:05.9 & 39 &  3.6(0.1) & 1.06 & 4.19$^{+0.84}_{-0.63}\times$10$^{23}$ & [0.5$\times$0.4],  56.1 &    8(0.2) & 0.2--1.4 \\
G10.32-0.26 C5 & 18:09:22.98 & -20:08:07.8 & 22 &   3.3(0.1) & 0.97 & 0.41--2.72$\times$10$^{24}$ & [0.4$\times$0.2], 103.0 &    4(0.2) & 0.1--0.7 \\
\tableline
G10.34-0.14 C1 & 18:08:59.98 & -20:03:39.1 & 275 & 32.2(0.2) & 9.40 & 1.69$^{+0.22}_{-0.18}\times$10$^{24}$ & [0.5$\times$0.4], 146.4 &   89(0.5) & 0.6--9.5 \\
G10.34-0.14 C2 & 18:09:00.01 & -20:03:38.8 & 170 & 19.1(0.2) & 5.58 & 1.17$^{+0.05}_{-0.05}\times$10$^{24}$ & [0.5$\times$0.4], 143.1 &   52(0.5) & 0.4--5.6 \\
G10.34-0.14 C3 & 18:08:59.98 & -20:03:35.7 & 144 & 15.8(0.2) & 4.60 & 1.24$^{+0.06}_{-0.06}\times$10$^{24}$ & [0.5$\times$0.4],   0.4 &   38(0.5) & 0.4--4.1 \\
G10.34-0.14 C4 & 18:09:00.00 & -20:03:37.1 & 41 &   4.6(0.2) & 1.34 & 0.56--3.76$\times$10$^{24}$ & [0.7$\times$0.7],  66.8 &   28(0.6) & 0.5--3.0 \\
\tableline
G18.34+1.78 C1 & 18:17:58.21 & -12:07:24.9 & 217 & 19.3(0.2) & 5.53 & 1.35$^{+0.14}_{-0.12}\times$10$^{24}$ & [0.4$\times$0.3], 108.9 &   31(0.4) & 0.8--8.9 \\
G18.34+1.78 C2 & 18:17:58.12 & -12:07:24.7 & 169 & 14.7(0.2) & 4.20 & 1.45$^{+0.11}_{-0.10}\times$10$^{24}$ & [0.4$\times$0.3],  84.9 &   21(0.4) & 0.7--6.0 \\
G18.34+1.78 C3 & 18:17:58.33 & -12:07:23.9 & 128 & 12.2(0.2) & 3.50 & 1.12$^{+0.45}_{-0.26}\times$10$^{24}$ & [0.4$\times$0.3], 105.3 &   19(0.4) & 0.6--5.4 \\
G18.34+1.78 C4 & 18:17:58.04 & -12:07:23.0 & 131 &  11.7(0.2) & 3.35 & 1.48--9.51$\times$10$^{24}$ & [0.4$\times$0.3],  84.3 &   16(0.4) & 0.7--4.6 \\
G18.34+1.78 C5 & 18:17:58.28 & -12:07:23.8 & 81 &   7.6(0.2) & 2.18 & 0.95--6.23$\times$10$^{24}$ & [0.5$\times$0.5],  24.1 &   21(0.5) & 0.9--5.9 \\
G18.34+1.78 C6 & 18:17:58.29 & -12:07:24.4 & 68 &   6.4(0.2) & 1.82 & 0.79--5.24$\times$10$^{24}$ & [0.5$\times$0.4],  48.8 &   17(0.5) & 0.8--5.0 \\
G18.34+1.78 C7 & 18:17:58.17 & -12:07:23.1 & 44 &   4.1(0.2) & 1.17 & 0.50--3.42$\times$10$^{24}$ & [0.7$\times$0.5], 144.3 &   15(0.6) & 0.7--4.4 \\
G18.34+1.78 C8 & 18:17:57.22 & -12:07:30.3 & 97 & 14.3(0.2) & 4.09 & 7.88$^{+0.59}_{-0.53}\times$10$^{23}$ & [0.5$\times$0.4], 102.4 &   30(0.4) & 0.6--8.5 \\
\tableline
G18.34+1.78SW C1 & 18:17:50.29 & -12:07:54.5 & 243 & 22.0(0.2) & 6.25 & 1.56$^{+0.12}_{-0.11}\times$10$^{24}$ & [0.4$\times$0.3],  96.4 &   33(0.4) & 0.8--9.4 \\
G18.34+1.78SW C2 & 18:17:50.31 & -12:07:53.0 & 225 & 20.5(0.2) & 5.82 & 1.07$^{+0.05}_{-0.05}\times$10$^{24}$ & [0.5$\times$0.4],  98.1 &   44(0.5) & 0.8--12.5 \\
G18.34+1.78SW C3 & 18:17:49.67 & -12:07:55.4 & 145 &  20.1(0.2) & 5.71 & 0.25--1.59$\times$10$^{25}$ & [0.4$\times$0.3], 107.5 &   30(0.4) & 1.4--8.5 \\
G18.34+1.78SW C4 & 18:17:50.31 & -12:07:59.1 & 94 &   9.1(0.2) & 2.57 & 1.12--7.24$\times$10$^{24}$ & [0.4$\times$0.3], 101.5 &   13(0.4) & 0.6--3.8 \\
G18.34+1.78SW C5 & 18:17:49.78 & -12:07:54.4 & 48 &   5.9(0.2) & 1.67 & 0.72--4.77$\times$10$^{24}$ & [0.4$\times$0.3],  86.8 &   10(0.4) & 0.4--2.8 \\
\tableline
G23.43-0.18 C1 & 18:34:39.19 &  -8:31:25.4 & 164 & 25.2(0.3) & 8.97 & 2.73$^{+0.13}_{-0.12}\times$10$^{24}$ & [0.3$\times$0.3],  79.4 &   34(0.5) & 4.3--50.3 \\
G23.43-0.18 C2 & 18:34:39.25 &  -8:31:39.3 & 181 & 13.6(0.3) & 4.85 & 1.46$^{+0.05}_{-0.05}\times$10$^{24}$ & [0.4$\times$0.3],  56.2 &   28(0.7) & 3.4--41.3 \\
G23.43-0.18 C3 & 18:34:39.18 &  -8:31:39.6 & 124 &   9.3(0.3) & 3.33 & 0.18--1.21$\times$10$^{25}$ & [0.3$\times$0.3], 104.6 &   15(0.6) & 3.5--21.9 \\
\tableline
G24.33+0.14 C1 & 18:35:08.14 &  -7:35:04.1 & 705 &171.1(0.7) & 61.45 & 1.35$^{+0.04}_{-0.04}\times$10$^{25}$ & [0.4$\times$0.4],  32.1 &  363(1.6) & 46.9--779.3 \\
G24.33+0.14 C2 & 18:35:08.18 &  -7:35:04.2 & 38 &   9.5(0.7) & 3.41 & 0.18--1.31$\times$10$^{25}$ & [0.6$\times$0.5],  45.9 &   45(2.1) & 15.4--97.2 \\
G24.33+0.14 C3 & 18:35:08.13 &  -7:35:04.7 & 29 &   7.0(0.7) & 2.52 & 1.31--9.91$\times$10$^{24}$ & [0.3$\times$0.3],  70.1 &    9(1.2) & 3.3--20.6 \\
\tableline
G25.82-0.17 C1 & 18:39:03.62 &  -6:24:11.3 & 501 &151.0(0.8) & 53.58 & 1.33$^{+0.03}_{-0.03}\times$10$^{25}$ & [0.4$\times$0.4], 147.5 &  364(1.9) & 25.6--376.2 \\
 G25.82-0.17 C2 & 18:39:03.65 &  -6:24:11.4 & 186 &  56.1(0.8) & 19.91 & 1.13--7.22$\times$10$^{25}$ & [0.6$\times$0.5], 173.0 &  251(2.3) & 41.1--259.8 \\
 G25.82-0.17 C3 & 18:39:03.68 &  -6:24:12.5 & 45 &  13.8(0.8) & 4.90 & 0.27--1.85$\times$10$^{25}$ & [0.6$\times$0.4], 155.9 &   58(2.3) & 9.5--60.1 \\
 G25.82-0.17 C4 & 18:39:03.82 &  -6:24:11.5 & 27 &   8.4(0.8) & 2.98 & 0.16--1.17$\times$10$^{25}$ & [0.6$\times$0.4],  78.6 &   32(2.3) & 5.3--33.5 \\
\tableline
 G27.36-0.16 C1 & 18:41:51.06 &  -5:01:43.4 & 431 &194.0(0.8) & 68.26 & 1.44$^{+0.03}_{-0.03}\times$10$^{25}$ & [0.4$\times$0.4],  73.7 &  494(1.9) & 76.5--1307.9 \\
 G27.36-0.16 C2 & 18:41:51.07 &  -5:01:43.1 & 280 & 126.3(0.8) & 44.44 & 0.25--1.58$\times$10$^{26}$ & [0.3$\times$0.2],   8.4 &   76(0.8) & 31.8--201.1 \\
 G27.36-0.16 C3 & 18:41:51.49 &  -5:01:36.9 & 27 &  16.8(0.8) & 5.93 & 0.32--2.19$\times$10$^{25}$ & [0.3$\times$0.3],  55.8 &   29(1.6) & 12.5--79.1 \\
 G27.36-0.16 C4 & 18:41:51.55 &  -5:01:36.4 & 13 &   9.1(0.8) & 3.20 & 0.17--1.23$\times$10$^{25}$ & [0.4$\times$0.3], 106.1 &   13(1.5) & 5.8--36.4 \\
\tableline
 G28.37+0.07 C1 & 18:42:51.99 &  -3:59:54.0 & 431 &  70.5(0.4) & 24.67 & 1.40--8.79$\times$10$^{25}$ & [0.4$\times$0.4], 174.9 &  153(0.9) & 25.0--158.1 \\
 G28.37+0.07 C2 & 18:42:51.98 &  -3:59:54.6 & 119 & 20.8(0.4) & 7.26 & 2.32$^{+0.09}_{-0.09}\times$10$^{24}$ & [0.3$\times$0.3], 105.0 &   29(0.7) & 2.7--30.6 \\
 G28.37+0.07 C3 & 18:42:52.14 &  -3:59:54.1 & 107 & 17.5(0.4) & 6.12 & 1.83$^{+0.37}_{-0.28}\times$10$^{24}$ & [0.4$\times$0.4], 111.0 &   47(1.0) & 4.0--48.8 \\
 G28.37+0.07 C4 & 18:42:52.08 &  -3:59:53.4 & 85 &  13.3(0.4) & 4.64 & 0.26--1.69$\times$10$^{25}$ & [0.5$\times$0.5],  56.4 &   46(1.1) & 7.6--48.1 \\
 G28.37+0.07 C5 & 18:42:52.00 &  -3:59:53.0 & 68 & 10.4(0.4) & 3.64 & 1.62$^{+0.61}_{-0.37}\times$10$^{24}$ & [0.5$\times$0.3],  44.9 &   25(0.9) & 3.2--26.1 \\
 G28.37+0.07 C6 & 18:42:52.17 &  -3:59:51.3 & 49 &  6.9(0.4) & 2.41 & 7.86$^{+1.27}_{-1.07}\times$10$^{23}$ & [0.4$\times$0.3], 101.0 &   15(0.9) & 1.4--15.8 \\
 G28.37+0.07 C7 & 18:42:51.97 &  -3:59:46.8 & 40 &   5.0(0.4) & 1.74 & 0.91--6.66$\times$10$^{24}$ & [0.3$\times$0.3], 100.3 &    6(0.6) & 1.0--6.2 \\
 G28.37+0.07 C8 & 18:42:52.10 &  -3:59:54.5 & 54 &   9.0(0.4) & 3.16 & 0.17--1.17$\times$10$^{25}$ & [0.5$\times$0.3],  50.6 &   23(1.0) & 3.8--23.9 \\
 G28.37+0.07 C9 & 18:42:52.49 &  -3:59:40.9 & 36 &   5.2(0.4) & 1.80 & 0.95--6.88$\times$10$^{24}$ & [0.4$\times$0.3],  54.6 &    9(0.8) & 1.5--9.4 \\
\tableline
 G29.91-0.03 C1 & 18:46:05.21 &  -2:42:24.5 & 233 & 11.9(0.1) & 4.11 & 1.76$^{+0.62}_{-0.37}\times$10$^{24}$ & [0.4$\times$0.3],  93.5 &   19(0.3) & 3.7--31.7 \\
 G29.91-0.03 C2 & 18:46:04.94 &  -2:42:25.3 & 91 &  5.1(0.1) & 1.76 & 5.77$^{+0.72}_{-0.61}\times$10$^{23}$ & [0.4$\times$0.4], 130.5 &   12(0.4) & 1.8--20.2 \\
 G29.91-0.03 C3 & 18:46:04.98 &  -2:42:24.5 & 62 &   3.4(0.1) & 1.19 & 0.66--4.44$\times$10$^{24}$ & [0.5$\times$0.3], 119.3 &    7(0.3) & 1.9--12.2 \\
 G29.91-0.03 C4 & 18:46:05.25 &  -2:42:24.1 & 66 &   3.4(0.1) & 1.17 & 0.65--4.37$\times$10$^{24}$ & [0.8$\times$0.5],  29.0 &   20(0.5) & 5.2--32.5 \\
 G29.91-0.03 C5 & 18:46:05.18 &  -2:42:23.9 & 62 &   3.3(0.1) & 1.14 & 0.63--4.26$\times$10$^{24}$ & [0.4$\times$0.3], 131.2 &    6(0.3) & 1.6--10.0 \\
 G29.91-0.03 C6 & 18:46:04.72 &  -2:42:19.6 & 20 &   1.7(0.1) & 0.58 & 0.30--2.25$\times$10$^{24}$ & [0.3$\times$0.3],  78.1 &    2(0.3) & 0.5--3.4 \\
 G29.91-0.03 C7 & 18:46:04.19 &  -2:42:27.2 & 55 &   8.4(0.1) & 2.91 & 0.16--1.06$\times$10$^{25}$ & [0.4$\times$0.3],  96.6 &   14(0.3) & 3.6--22.5 \\
 G29.91-0.03 C8 & 18:46:06.12 &  -2:42:31.7 & 37 &   3.7(0.1) & 1.29 & 0.71--4.77$\times$10$^{24}$ & [0.3$\times$0.3], 102.3 &    4(0.2) & 1.1--6.9 \\
\tableline
 G30.70-0.07 C1 & 18:47:35.92 &  -2:01:48.4 & 165 & 15.8(0.2) & 5.47 & 1.80$^{+0.46}_{-0.31}\times$10$^{24}$ & [0.5$\times$0.4], 138.9 &   45(0.5) & 5.4--58.9 \\
 G30.70-0.07 C2 & 18:47:35.96 &  -2:01:49.0 & 203 &  18.1(0.2) & 6.26 & 0.35--2.24$\times$10$^{25}$ & [0.4$\times$0.3], 123.5 &   36(0.4) & 7.5--47.2 \\
\tableline
 G49.49-0.39 C1 & 19:23:43.96 &  14:30:34.5 & 193 &600.5(4.0) & 143.80 & 2.61$^{+0.07}_{-0.07}\times$10$^{25}$ & [0.6$\times$0.5],   4.7 & 2118(11.0) & 173.5--2553.2 \\
 G49.49-0.39 C2 & 19:23:43.91 &  14:30:34.5 & 141 &439.8(4.0) & 105.32 & 2.40$^{+0.14}_{-0.13}\times$10$^{25}$ & [0.6$\times$0.4], 148.7 & 1121(9.7) & 115.4--1352.1 \\
 G49.49-0.39 C3 & 19:23:43.88 &  14:30:35.9 & 53 &174.6(4.0) & 41.81 & 8.88$^{+0.39}_{-0.37}\times$10$^{24}$ & [0.7$\times$0.5], 142.3 &  625(10.9) & 59.9--753.7 \\
 G49.49-0.39 C4 & 19:23:43.93 &  14:30:36.5 & 25 & 84.8(4.0) & 20.31 & 3.95$^{+0.26}_{-0.25}\times$10$^{24}$ & [0.7$\times$0.4], 145.9 &  268(10.5) & 23.5--323.1 \\
 G49.49-0.39 C5 & 19:23:43.99 &  14:30:33.1 & 19 &  59.2(4.0) & 14.18 & 0.56--3.97$\times$10$^{25}$ & [0.5$\times$0.4],  29.0 &  108(8.3) & 20.7--130.7 \\
 G49.49-0.39 C6 & 19:23:43.90 &  14:30:28.2 & 161 &495.6(4.0) & 118.68 & 2.39$^{+0.10}_{-0.09}\times$10$^{25}$ & [0.6$\times$0.6], 153.5 & 1700(10.9) & 154.7--2049.8 \\
 G49.49-0.39 C7 & 19:23:43.86 &  14:30:26.5 & 39 &128.6(4.0) & 30.80 & 8.56$^{+0.85}_{-0.75}\times$10$^{24}$ & [0.7$\times$0.5],  38.9 &  537(11.5) & 67.5--648.3 \\
 G49.49-0.39 C8 & 19:23:43.83 &  14:30:25.1 & 14 & 48.3(4.0) & 11.57 & 3.66$^{+0.78}_{-0.63}\times$10$^{24}$ & [0.7$\times$0.5],  41.1 &  150(10.4) & 21.5--181.2 \\
 G49.49-0.39 C9 & 19:23:43.84 &  14:30:24.7 & 11 &  40.6(4.0) & 9.73 & 0.37--2.80$\times$10$^{25}$ & [0.8$\times$0.4],  40.1 &  115(9.8) & 22.1--139.6 \\
G49.49-0.39 C10 & 19:23:43.82 &  14:30:23.4 & 11 &  43.3(4.0) & 10.38 & 0.40--2.97$\times$10$^{25}$ & [0.6$\times$0.4],   6.9 &  112(9.7) & 21.5--135.7 \\
G49.49-0.39 C11 & 19:23:43.79 &  14:30:22.3 & 11 &  46.7(4.0) & 11.17 & 0.43--3.18$\times$10$^{25}$ & [0.5$\times$0.4], 126.4 &   89(8.5) & 17.0--107.3 \\
G49.49-0.39 C12 & 19:23:43.74 &  14:30:21.4 & 8 & 35.3(4.0) & 8.46 & 1.68$^{+0.36}_{-0.30}\times$10$^{24}$ & [0.5$\times$0.3], 140.1 &   48(7.0) & 4.3--58.7 \\
G49.49-0.39 C13 & 19:23:43.79 &  14:30:19.7 & 5 & 28.0(4.0) & 6.70 & 1.56$^{+0.42}_{-0.35}\times$10$^{24}$ & [0.4$\times$0.3], 140.6 &   32(6.3) & 3.4--39.2
\enddata
\tablecomments{$^{a}$ SNR is measured in primary beam uncorrected images. \\
$^{b}$For N(H$_{2}$), we adopt the T$_{ex}$(CH$_3$OH) which is derived from the rotation diagram of CH$_3$OH  for COMs-detected cores. Note that T$_{ex}$(CH$_3$OH) is derived at the CH$_{3}$OH emission peak located off from the continuum peak where the high dust optical depths affects the line intensity. The actual T$_{ex}$(CH$_3$OH) would be higher than the adopted temperature.
The T$_{dust}$ and $T_{gas}$ are assumed to be the same. For COMs-undetected continuum peaks, a range of T$_{dust}$ temperature of 20 to 100 K are adopted for prestellar to protostellar cores, assuming that the cores without COMs emission are cooler than those with COMs emission. \\
$^{c}$ Convolved source size. \\
$^{d}$ The lower and upper limits are derived. For all components, the lowest average temperature is assumed as 20 K to provide the upper limit of the total mass. For COMs detected cores (Table \ref{tab:maserandCOMs}), the T$_{ex}$(CH$_3$OH) is adopted (Table \ref{tab:rot_diagram1}). The actual average temperature of each core would be lower than the derived T$_{ex}$(CH$_3$OH), which is measured at the CH$_3$OH peak position, thus the derived mass is the lower limit. For COMs undetected continuum components, the highest temperature of 100 K is assumed for mass lower limit.
}
\label{tab:cores}
\end{deluxetable*}

%% file: s3_tab2_216GHz_info.tex
\startlongtable
\begin{deluxetable*}{l|cc|c|cccc}
\tablecaption{The summary of information at the peak of 216.946 GHz CH$_{3}$OH emission \label{tab:216GHz}}
\tabletypesize{\scriptsize}
\tablehead{
\colhead{Source Name} & \multicolumn{2}{c}{Equatorial coordinates}	& \colhead{N(H$_{2}$)}  & \multicolumn{4}{c}{Peak intensity}\\
\colhead{} & \colhead{R.A.} & \colhead{Decl.} &\colhead{N(H$_{2}$)$^a$}   & \colhead{$v_{\rm peak}$} &\colhead{$\Delta v_{\rm offset}$}& \colhead{I$_{\rm peak}$}   &\colhead{T$_{\rm b}$}\\
\colhead{} & \colhead{($^{hms}$)} & \colhead{($^{\circ}$ $'$ $''$)} &  \colhead{(cm$^{-2}$)} &  \colhead{(km s$^{-1}$)} &  \colhead{(km s$^{-1}$)}& \colhead{(mJy beam$^{-1}$)} & \colhead{(K)}
}
\startdata
G10.32-0.26 C1 & 18:09:23.27 & -20:08:7.24 & 8.28$^{+2.22}_{-1.47}\times$10$^{23}$ & 32.7 & 0.2 & 31.4 & 8.5 \\
G10.32-0.26 C4 & 18:09:23.24 & -20:08:5.92 & 3.97$^{+0.80}_{-0.60}\times$10$^{23}$ & 32.7 & 0.2 & 36.5 & 9.8 \\
\tableline
G10.34-0.14 C1 & 18:08:59.98 & -20:03:39.08 & 1.58$^{+0.21}_{-0.17}\times$10$^{24}$ & 13.8 & -0.8 & 59.9 & 16.2 \\
G10.34-0.14 C2 & 18:09:00.00 & -20:03:38.78 & 1.14$^{+0.05}_{-0.05}\times$10$^{24}$ & 13.8 & -0.8 & 67.5 & 18.2 \\
G10.34-0.14 C3 & 18:08:59.98 & -20:03:35.54 & 7.98$^{+0.45}_{-0.42}\times$10$^{23}$ & 9.8 & -4.8 & 161.6 & 43.7 \\
\tableline
G18.34+1.78 C1 & 18:17:58.22 & -12:07:24.90 & 1.26$^{+0.13}_{-0.11}\times$10$^{24}$ & 34.4 & 4.1 & 25.8 & 6.9 \\
G18.34+1.78 C2 & 18:17:58.12 & -12:07:24.90 & 6.18$^{+0.59}_{-0.52}\times$10$^{23}$ & 30.4 & 0.1 & 104.9 & 28.0 \\
G18.34+1.78 C3 & 18:17:58.34 & -12:07:24.00 & 1.04$^{+0.42}_{-0.24}\times$10$^{24}$ & 31.7 & 1.4 & 14.8 & 4.0 \\
G18.34+1.78 C8 & 18:17:57.21 & -12:07:30.36 & 7.53$^{+0.57}_{-0.51}\times$10$^{23}$ & 33.1 & 2.8 & 32.8 & 8.7 \\
\tableline
G18.34+1.78SW C1 & 18:17:50.29 & -12:07:54.46 & 1.38$^{+0.11}_{-0.10}\times$10$^{24}$ & 29.0 & -1.3 & 20.1 & 5.3 \\
G18.34+1.78SW C2 & 18:17:50.31 & -12:07:53.02 & 1.07$^{+0.05}_{-0.05}\times$10$^{24}$ & 30.4 & 0.1 & 76.6 & 20.2 \\
\tableline
G23.43-0.18 C1 & 18:34:39.19 &  -08:31:25.44 & 2.33$^{+0.11}_{-0.11}\times$10$^{24}$ & 96.3 & -6.0 & 164.9 & 41.3 \\
G23.43-0.18 C2 & 18:34:39.25 &  -08:31:39.30 & 1.12$^{+0.05}_{-0.05}\times$10$^{24}$ & 100.3 & -2.0 & 177.5 & 44.5 \\
\tableline
G24.33+0.14 C1 & 18:35:8.15 &  -07:35:4.20 & 1.01$^{+0.03}_{-0.03}\times$10$^{25}$ & 112.8 & -0.7 & 326.3 & 113.2 \\
\tableline
G25.82-0.17 C1 & 18:39:3.62 &  -06:24:11.06 & 4.13$^{+0.14}_{-0.14}\times$10$^{24}$ & 89.7 & -4.0 & 245.9 & 65.6 \\
\tableline
G27.36-0.16 C1 & 18:41:51.06 &  -5:01:43.50 & 1.23$^{+0.02}_{-0.02}\times$10$^{25}$ & 94.6 & 0.5 & 266.1 & 88.6 \\
\tableline
G28.37+0.07 C2 & 18:42:51.98 &  -3:59:54.42 & 1.56$^{+0.07}_{-0.07}\times$10$^{24}$ & 75.5 & -1.0 & 180.7 & 43.8 \\
G28.37+0.07 C3 & 18:42:52.14 &  -3:59:54.12 & 1.62$^{+0.34}_{-0.25}\times$10$^{24}$ & 76.9 & 0.4 & 23.3 & 5.6 \\
G28.37+0.07 C5 & 18:42:51.98 &  -3:59:53.04 & 7.90$^{+3.38}_{-2.05}\times$10$^{23}$ & 76.9 & 0.4 & 19.6 & 4.8 \\
G28.37+0.07 C6 & 18:42:52.17 &  -3:59:51.24 & 7.67$^{+1.26}_{-1.05}\times$10$^{23}$ & 75.5 & -1.0 & 25.9 & 6.3 \\
\tableline
G29.91-0.03 C1 & 18:46:5.21 &  -2:42:24.62 & 1.24$^{+0.45}_{-0.27}\times$10$^{24}$ & 97.8 & -0.4 & 10.3 & 3.3 \\
G29.91-0.03 C2 & 18:46:4.94 &  -2:42:25.34 & 5.43$^{+0.69}_{-0.58}\times$10$^{23}$ & 97.8 & -0.4 & 11.7 & 3.7 \\
\tableline
G30.70-0.07 C1 & 18:47:35.93 &  -2:01:48.46 & 1.67$^{+0.43}_{-0.29}\times$10$^{24}$ & 90.1 & 0.1 & 15.9 & 3.6 \\
\tableline
G49.49-0.39 C1 & 19:23:44.00 &  14:30:34.36 & 3.47$^{+0.24}_{-0.24}\times$10$^{24}$ & 58.6 & 2.9 & 602.9 & 142.3 \\
G49.49-0.39 C2 & 19:23:43.88 &  14:30:34.30 & 1.92$^{+0.33}_{-0.30}\times$10$^{24}$ & 51.8 & -3.9 & 519.2 & 122.6 \\
G49.49-0.39 C3 & 19:23:43.88 &  14:30:35.86 & 7.58$^{+0.36}_{-0.35}\times$10$^{24}$ & 54.5 & -1.2 & 170.4 & 40.2 \\
G49.49-0.39 C4 & 19:23:43.92 &  14:30:36.64 & 3.06$^{+0.24}_{-0.23}\times$10$^{24}$ & 51.8 & -3.9 & 180.4 & 42.6 \\
G49.49-0.39 C6 & 19:23:43.87 &  14:30:27.88 & 3.82$^{+0.32}_{-0.30}\times$10$^{24}$ & 57.2 & 1.5 & 432.2 & 102.1
\enddata
\tablecomments{$^{a}$ N(H$_{2}$) is obtained by 1.3mm dust continuum image at which the spectra is extracted.}
\end{deluxetable*}

%% file: s3_tab3_maser_association.tex
\startlongtable
\begin{deluxetable*}{lcccc}
\tablecaption{Methanol Maser emission and hot core associations\label{tab:maserandCOMs}}
\tabletypesize{\scriptsize}
\tablehead{
\colhead{Name} & \colhead{216.946 GHz CH$_{3}$OH emission\tablenotemark{a}} & \colhead{6.7 GHz class II CH$_{3}$OH maser\tablenotemark{b}} & \colhead{detected COMs\tablenotemark{c}}}
\startdata
G10.32-0.26 C1 & TH & Y & 3 \\
G10.32-0.26 C2 & TH & N& - \\
G10.32-0.26 C3 & N & N& - \\
G10.32-0.26 C4 & TH & N&3 \\
G10.32-0.26 C5 & TH &N& - \\
\tableline
G10.34-0.14 C1 & TH & N&14\\
G10.34-0.14 C2 & TH & N&12\\
G10.34-0.14 C3 & TH+M? &Y&12\\
G10.34-0.14 C4 & N &N&-\\
\tableline
G18.34+1.78 C1 & TH &N&3\\
G18.34+1.78 C2 & TH &Y&3\\
G18.34+1.78 C3 & TH &N&2\\
G18.34+1.78 C4 & N &N&-\\
G18.34+1.78 C5 & N &N&-\\
G18.34+1.78 C6 & N &N&-\\
G18.34+1.78 C7 & N &N&-\\
G18.34+1.78 C8 & TH &N&8\\
\tableline
G18.34+1.78SW C1 & TH &N& 4\\
G18.34+1.78SW C2 & TH & N&15 \\
G18.34+1.78SW C3 & N &N&- \\
G18.34+1.78SW C4 & N &N&-\\
G18.34+1.78SW C5 & N &N&- \\
\tableline
G23.43-0.18 C1 & TH+M? &Y&15 \\
G23.43-0.18 C2 & TH+M? &Y&15\\
\tableline 
G24.33+0.14 C1 & TH+M? &Y&19\\
G24.33+0.14 C2 & N &N&-\\
G24.33+0.14 C3 & N &N&-\\
\tableline
G25.82-0.17 C1 & TH+M? &Y& 19\\
G25.82-0.17 C2 & TH & N&-\\
G25.82-0.17 C3 & TH & N&-\\
G25.82-0.17 C4 & TH & N&-\\
\tableline
G27.36-0.16 C1 & TH+M?&Y& 19\\
G27.36-0.16 C2 & TH & N& - \\
G27.36-0.16 C3 & TH & N& - \\
G27.36-0.16 C4 & TH & N& - \\
\tableline
G28.37+0.07 C1 & TH& N & - \\
G28.37+0.07 C2 & TH+M? &Y & 15\\
G28.37+0.07 C3 & TH &N& 3 \\
G28.37+0.07 C4 & TH &N& - \\
G28.37+0.07 C5 & TH &N& 2 \\
G28.37+0.07 C6 & TH &N& 3\\
G28.37+0.07 C7 & N &N& -\\
G28.37+0.07 C8 & N &N& -\\
\tableline
G29.91-0.03 C1 & TH &N&1\\
G29.91-0.03 C2 & TH &N&3\\
G29.91-0.03 C3 & TH &N&-\\
G29.91-0.03 C4 & TH &N&-\\
G29.91-0.03 C5 & N &N&-\\
\tableline
G30.70-0.07 C1 & TH & N&3\\
G30.70-0.07 C2 & N & N&- \\
\tableline
G49.49-0.39 C1 & TH+M? & Y & 19\\
G49.49-0.39 C2 & TH+M? & N & 17\\
G49.49-0.39 C3 & TH+M? & N & 18\\
G49.49-0.39 C4 & TH &N& 18 \\
G49.49-0.39 C5 & TH &N& - \\
G49.49-0.39 C6 & TH &N& 19\\
G49.49-0.39 C7 & TH & N&-\\
G49.49-0.39 C8 & TH &N& -\\
G49.49-0.39 C9 & TH &N& -\\
G49.49-0.39 C10 & TH &N& -\\
G49.49-0.39 C11 & TH & N& -\\
G49.49-0.39 C12 & TH & N&-\\
G49.49-0.39 C13 & TH &N&-\\
\enddata
\tablenotetext{a}{TH: Thermal emission. M?: 216.946 GHz methanol maser candidate; maser detection was not firmly confirmed despite evident spot emission features. N: no CH$_3$OH emission detected.}
\tablenotetext{b}{\citet{Green2010,Breen2015} and \citet{Hu2016}.}
\tablenotetext{c}{ The number of COMs identified using XCLASS. There are detected but unidentified lines, thus the number is lower limit.}
\end{deluxetable*}

%% file: s3_tab4_rotation_diagram_all.tex
\clearpage 
\begin{longrotatetable}
\begin{deluxetable}{lcccccccc}
\tablecaption{Rotation Diagram Analysis\label{tab:rot_diagram1}}
\tabletypesize{\scriptsize}
\tablehead{\colhead{Source} & \multicolumn{8}{c}{Molecules}}
\startdata
	&	\multicolumn{2}{c}{CH$_{3}$OH} &	\multicolumn{2}{c}{$^{13}$CH$_{3}$OH}	&	\multicolumn{2}{c}{CH$_{3}^{18}$OH}	&	\multicolumn{2}{c}{CH$_{2}$DOH}\\
&	 T$_{\rm rot}$[K] 	&	 N[$\rm cm^{-2}$] &  T$_{\rm rot}$[K]	&	 N[$\rm cm^{-2}$] &  T$_{\rm rot}$[K] 	&	 N[$\rm cm^{-2}$] &  T$_{\rm rot}$[K] 	&	 N[$\rm cm^{-2}$]\\
\tableline
G10.32-0.26 C1   &    147.2$\pm$28.9 &  1.15$\pm$0.39E+17   &  $\cdots$  &  $\cdots$   & $\cdots$ &$\cdots$  & $\cdots$ &$\cdots$ \\ 
G10.32-0.26 C4   &    109.8$\pm$15.1 &  1.19$\pm$0.24E+17   &  $\cdots$  &  $\cdots$   & $\cdots$ &$\cdots$ & $\cdots$ &$\cdots$  \\ 
G10.34-0.14 C1   &    235.1$\pm$25.5 &  9.13$\pm$0.51E+17 &  85.2$\pm$19.1 &  2.75$\pm$0.53E+16  & $\cdots$ &$\cdots$  & $\cdots$ &$\cdots$ \\ 
G10.34-0.14 C2   &    202.3$\pm$6.7 &  6.46$\pm$0.21E+17 & 61.0$\pm$19.0 &  2.62$\pm$1.51E+16  & $\cdots$ &$\cdots$  & $\cdots$ &$\cdots$ \\ 
G10.34-0.14 C3   &    158.6$\pm$5.3 &  1.30$\pm$0.06E+18 &94.9$\pm$72.5 &  2.75$\pm$1.22E+16   & $\cdots$ &$\cdots$ & $\cdots$ &$\cdots$  \\ 
G18.34+1.78 C1   &    179.9$\pm$15.0 &  9.19$\pm$0.98E+16  &  $\cdots$  &  $\cdots$   & $\cdots$ &$\cdots$  & $\cdots$ &$\cdots$ \\ 
G18.34+1.78 C2   &    128.7$\pm$7.1 &  2.67$\pm$0.10E+17  &  $\cdots$  &  $\cdots$   & $\cdots$ &$\cdots$  & $\cdots$ &$\cdots$ \\ 
G18.34+1.78 C3   &    138.5$\pm$36.9 &  5.75$\pm$3.42E+16  &  $\cdots$  &  $\cdots$   & $\cdots$ &$\cdots$  & $\cdots$ &$\cdots$ \\ 
G18.34+1.78 C8   &    226.7$\pm$12.5 &  2.92$\pm$0.19E+17  &  $\cdots$  &  $\cdots$   & $\cdots$ &$\cdots$  & $\cdots$ &$\cdots$ \\ 
G18.34+1.78SW C1   &    173.6$\pm$10.9 &  1.15$\pm$0.10E+17  &  $\cdots$  &  $\cdots$   & $\cdots$ &$\cdots$  & $\cdots$ &$\cdots$ \\ 
G18.34+1.78SW C2   &    235.5$\pm$9.0 &  9.66$\pm$0.35E+17 &  159.9$\pm$79.4 &  3.36$\pm$1.08E+16  & $\cdots$ &$\cdots$ &  120.5$\pm$82.0 &  8.10$\pm$4.15E+15 \\ 
G23.43-0.18 C1   &    182.5$\pm$5.9 &  1.55$\pm$0.06E+18 &  82.0$\pm$30.8 &  2.71$\pm$0.68E+16  & $\cdots$ &$\cdots$ & $\cdots$ &$\cdots$  \\ 
G23.43-0.18 C2   &    184.9$\pm$2.5 &  1.29$\pm$0.02E+18 & 54.8$\pm$2.4 &  8.47$\pm$0.85E+16 & $\cdots$ &$\cdots$  & $\cdots$ &$\cdots$ \\ 
G24.33+0.14 C1   &    254.5$\pm$6.4 &  4.96$\pm$0.11E+18 &  73.1$\pm$3.8 &  2.53$\pm$0.14E+17  & $\cdots$ &$\cdots$ & $\cdots$ &$\cdots$  \\ 
G25.82-0.17 C1   &    225.3$\pm$3.6 &  3.57$\pm$0.05E+18 & 61.9$\pm$2.9 &  2.05$\pm$0.13E+17  & 153.2$\pm$176.9 & 15.97$\pm$66.43E+15 & 130.8$\pm$10.7 &  2.50$\pm$0.24E+16 \\
G27.36-0.16 C1   &    261.4$\pm$3.7 &  5.15$\pm$0.08E+18 &  98.0$\pm$6.8 &  2.27$\pm$0.09E+17 & 178.2$\pm$91.3 &  5.38$\pm$8.10E+16 & 80.0$\pm$6.6 &  5.56$\pm$0.41E+16 \\
G28.37+0.07 C2   &    174.8$\pm$3.1 &  1.71$\pm$0.04E+18 &  104.4$\pm$16.2 &  4.33$\pm$0.45E+16  & $\cdots$ &$\cdots$  & $\cdots$ &$\cdots$ \\ 
G28.37+0.07 C3   &    186.3$\pm$27.3 &  1.07$\pm$0.21E+17  &  $\cdots$  &  $\cdots$   & $\cdots$ &$\cdots$  & $\cdots$ &$\cdots$ \\ 
G28.37+0.07 C5   &    127.0$\pm$29.8 &  5.33$\pm$2.16E+16  &  $\cdots$  &  $\cdots$   & $\cdots$ &$\cdots$  & $\cdots$ &$\cdots$ \\ 
G28.37+0.07 C6   &    171.3$\pm$14.9 &  2.02$\pm$0.24E+17  &  $\cdots$  &  $\cdots$   & $\cdots$ &$\cdots$  & $\cdots$ &$\cdots$ \\ 
G29.91-0.03 C1   &    132.9$\pm$32.3 &  3.63$\pm$1.74E+16  &  $\cdots$  &  $\cdots$   & $\cdots$ &$\cdots$  & $\cdots$ &$\cdots$ \\ 
G29.91-0.03 C2   &    171.9$\pm$14.2 &  7.45$\pm$0.89E+16  &  $\cdots$  &  $\cdots$   & $\cdots$ &$\cdots$  & $\cdots$ &$\cdots$ \\ 
G30.70-0.07 C1   &    169.9$\pm$31.7 &  1.13$\pm$0.26E+17  &  $\cdots$  &  $\cdots$   & $\cdots$ &$\cdots$ & $\cdots$ &$\cdots$ \\ 
G49.49-0.39 C1   &    225.8$\pm$4.2 &  8.20$\pm$0.14E+18 & 168.8$\pm$13.5 &  2.56$\pm$0.15E+17  & 218.1$\pm$25.2 &  9.04$\pm$1.64E+16 & $\cdots$ &$\cdots$ \\
G49.49-0.39 C2   &    180.9$\pm$8.4 &  5.38$\pm$0.31E+18 &  108.2$\pm$14.6 &  9.34$\pm$1.89E+16  & $\cdots$ &$\cdots$  & $\cdots$ &$\cdots$ \\ 
G49.49-0.39 C3   &    193.8$\pm$3.8 &  2.53$\pm$0.06E+18 &  130.5$\pm$15.6 &  8.83$\pm$0.61E+16  & 135.4$\pm$43.8 &  1.20$\pm$0.78E+16 & $\cdots$ &$\cdots$ \\
G49.49-0.39 C4   &    211.3$\pm$3.5 &  2.25$\pm$0.04E+18 &  112.5$\pm$15.1 &  5.16$\pm$0.32E+16  & $\cdots$ &$\cdots$  & $\cdots$ &$\cdots$ \\ 
G49.49-0.39 C6   &    203.9$\pm$6.2 & 10.16$\pm$0.33E+18 & 144.1$\pm$9.2 &  1.90$\pm$0.11E+17  & $\cdots$ &$\cdots$  & $\cdots$ &$\cdots$ \\ 
G49.49-0.39 C7  &    149.3$\pm$9.0 &  6.98$\pm$0.70E+17  &  $\cdots$  &  $\cdots$   & $\cdots$ &$\cdots$  & $\cdots$ &$\cdots$ \\ 
G49.49-0.39 C8   &    131.8$\pm$13.5 &  1.99$\pm$0.27E+17  &  $\cdots$  &  $\cdots$   & $\cdots$ &$\cdots$  & $\cdots$ &$\cdots$ \\ 
G49.49-0.39 C12  &    207.5$\pm$16.9 &  7.07$\pm$0.60E+17  &  $\cdots$  &  $\cdots$   & $\cdots$ &$\cdots$  & $\cdots$ &$\cdots$ \\ 
G49.49-0.39 C13   &    176.9$\pm$17.4 &  3.80$\pm$0.43E+17  &  $\cdots$  &   $\cdots$  & $\cdots$ &$\cdots$ & $\cdots$ &$\cdots$\\
\tableline
&	\multicolumn{2}{c}{CH$_{3}$CHO} &	\multicolumn{2}{c}{HCOOCH$_{3}$}	&		\multicolumn{2}{c}{CH$_{3}$OCH$_{3}$}	&	\multicolumn{2}{c}{(CH$_{3}$)$_{2}$CO}\\
&	 T$_{\rm rot}$[K] 	&	 N[$\rm cm^{-2}$] &  T$_{\rm rot}$[K]	&	 N[$\rm cm^{-2}$] &  T$_{\rm rot}$[K] 	&	 N[$\rm cm^{-2}$] &  T$_{\rm rot}$[K] 	&	 N[$\rm cm^{-2}$]\\
\tableline
G10.34-0.14 1   &    $\cdots$  &  $\cdots$   &179.7$\pm$22.5 &  3.18$\pm$0.35E+16  &    225.2$\pm$68.1 &  7.86$\pm$2.53E+16 & $\cdots$ &$\cdots$  \\ 
G10.34-0.14 2   &    $\cdots$  &  $\cdots$   &216.0$\pm$57.5 &  3.12$\pm$0.98E+16 &    164.0$\pm$17.5 &  3.91$\pm$0.32E+16 & $\cdots$ &$\cdots$  \\ 
G10.34-0.14 3   &    $\cdots$  &  $\cdots$   &140.3$\pm$17.6 &  2.90$\pm$0.30E+16 &    219.3$\pm$45.7 &  6.63$\pm$1.88E+16 & $\cdots$ &$\cdots$  \\ 
G18.34+1.78 8   &    $\cdots$  &  $\cdots$   &297.3$\pm$147.4 &  2.04$\pm$1.06E+16 &    325.5$\pm$230.7 &  0.28$\pm$0.27E+17 & $\cdots$ &$\cdots$  \\ 
G18.34+1.78SW 2   &   $\cdots$  &  $\cdots$   & 284.5$\pm$36.0 &  6.02$\pm$0.90E+16&    273.1$\pm$92.6 &  7.23$\pm$3.45E+16 & $\cdots$ &$\cdots$  \\ 
G23.43-0.18 1   &    $\cdots$  &  $\cdots$   &326.1$\pm$127.0 &  6.18$\pm$2.93E+16 &    238.6$\pm$66.5 &  6.32$\pm$2.11E+16 & $\cdots$ &$\cdots$  \\ 
G23.43-0.18 2   &   $\cdots$  &  $\cdots$   & 160.9$\pm$26.2 &  2.51$\pm$0.42E+16  &    216.4$\pm$33.7 &  4.32$\pm$0.86E+16 & $\cdots$ &$\cdots$  \\ 
G24.33+0.14 1   &    752.3$\pm$123.3 &  2.53$\pm$0.39E+17  &    308.3$\pm$21.2 &  3.40$\pm$0.33E+17  &    313.3$\pm$47.9 &  4.40$\pm$0.82E+17 & $\cdots$ &$\cdots$  \\ 
G25.82-0.17 1   &    214.9$\pm$16.5 &  3.41$\pm$0.35E+16 &    238.0$\pm$14.8 &  1.63$\pm$0.09E+17  &    191.8$\pm$13.7 &  2.06$\pm$0.12E+17 &    113.0$\pm$9.1 &  4.74$\pm$0.68E+17 \\
G27.36-0.16 1   &    509.0$\pm$86.5 &  1.93$\pm$0.36E+17  &    432.3$\pm$46.9 &  5.92$\pm$0.63E+17 &$\cdots$ &$\cdots$ &    213.0$\pm$31.5 &  0.43$\pm$0.31E+19 \\
G28.37+0.07 2   &    $\cdots$  &  $\cdots$   &149.1$\pm$18.8 &  4.68$\pm$0.55E+16 & $\cdots$ &$\cdots$  &$\cdots$ &$\cdots$\\ 
G28.37+0.07 6   &    $\cdots$  &  $\cdots$   &241.9$\pm$214.1 &  1.48$\pm$1.84E+16 & $\cdots$ &$\cdots$ &$\cdots$ &$\cdots$ \\ 
G49.49-0.39 1   &    252.9$\pm$33.7 &  4.92$\pm$0.89E+16  &    258.2$\pm$17.5 &  4.71$\pm$0.24E+17 &    189.4$\pm$6.5 &  2.92$\pm$0.22E+17 &    177.6$\pm$56.2 &  1.80$\pm$188.36E+18 \\
G49.49-0.39 2   &     $\cdots$  &  $\cdots$   &187.3$\pm$18.6 &  2.19$\pm$0.15E+17  &    167.5$\pm$119.9 &  2.66$\pm$0.95E+17  &    287.0$\pm$202.7 &  0.47$\pm$171.18E+19 \\
G49.49-0.39 3   &     $\cdots$  &  $\cdots$   &248.5$\pm$12.9 &  2.39$\pm$0.11E+17  &    257.3$\pm$35.2 &  2.51$\pm$0.43E+17 &    146.0$\pm$129.9 &  0.84$\pm$735.80E+18 \\
G49.49-0.39 4   &     $\cdots$  &  $\cdots$   &265.2$\pm$27.6 &  2.08$\pm$0.21E+17 &    239.1$\pm$43.3 &  2.38$\pm$0.45E+17  &    238.8$\pm$27.2 &  1.69$\pm$0.73E+18 \\
G49.49-0.39 6   &     57.1$\pm$16.9 &  1.36$\pm$0.20E+16 &    279.6$\pm$71.2 &  4.59$\pm$1.26E+17 &$\cdots$ &$\cdots$  &$\cdots$ &$\cdots$\\
\tableline
 &		\multicolumn{2}{c}{C$_{2}$H$_{5}$OH} &	\multicolumn{2}{c}{aGg'-(CH$_{2}$OH)$_{2}$}		&	\multicolumn{2}{c}{HCOCH$_{2}$OH} & &	\\
&	 T$_{\rm rot}$[K] 	&	 N[$\rm cm^{-2}$] &  T$_{\rm rot}$[K]	&	 N[$\rm cm^{-2}$] &  T$_{\rm rot}$[K] 	&	 N[$\rm cm^{-2}$] & 	&\\
\tableline
G10.34-0.14 1   &    118.3$\pm$114.1 &  8.63$\pm$11.96E+15 &    156.8$\pm$213.3 &  0.49$\pm$0.71E+16 &$\cdots$ &$\cdots$  \\
G10.34-0.14 2   &    186.0$\pm$196.8 &  0.77$\pm$1.39E+16 & $\cdots$ &$\cdots$  &$\cdots$ &$\cdots$  \\
G10.34-0.14 3   &     82.1$\pm$32.8 &  6.40$\pm$2.53E+15  &     62.6$\pm$94.0 &  2.18$\pm$1.83E+15 &$\cdots$ &$\cdots$  \\
G18.34+1.78 8   &    202.7$\pm$164.2 &  6.34$\pm$7.56E+15 & $\cdots$ &$\cdots$ & $\cdots$ &$\cdots$  \\
G18.34+1.78SW 2   &    131.5$\pm$26.2 &  7.30$\pm$1.21E+15 &    269.2$\pm$205.3 &  8.62$\pm$7.97E+15 &$\cdots$ &$\cdots$  \\
G23.43-0.18 1   &    106.3$\pm$72.7 &  7.53$\pm$6.30E+15 &    307.2$\pm$226.9 &  8.77$\pm$8.65E+15 &$\cdots$ &$\cdots$  \\
G23.43-0.18 2   &    103.6$\pm$28.6 &  1.01$\pm$0.22E+16  &    116.5$\pm$48.0 &  4.90$\pm$1.12E+15 &$\cdots$ &$\cdots$  \\
G24.33+0.14 1   &    882.0$\pm$65.9 &  7.74$\pm$0.77E+17  &    628.3$\pm$183.3 &  2.10$\pm$0.75E+17 &    181.8$\pm$113.1 &  1.30$\pm$1.62E+16 \\
G25.82-0.17 1   &     $\cdots$ &$\cdots$ &    143.6$\pm$33.3 &  1.74$\pm$0.25E+16 &$\cdots$ &$\cdots$  \\
G27.36-0.16 1   &    430.6$\pm$54.6 &  3.23$\pm$0.49E+17  &    818.4$\pm$115.7 &  2.54$\pm$0.46E+17& 191.4$\pm$21.2 &  2.47$\pm$0.28E+16 \\
G28.37+0.07 2   &    202.8$\pm$180.9 &  1.12$\pm$1.30E+16 & $\cdots$ &$\cdots$ & $\cdots$ &$\cdots$  \\
G49.49-0.39 1   &    825.0$\pm$112.9 & 10.14$\pm$1.73E+17 & $\cdots$ &$\cdots$ & $\cdots$ &$\cdots$  \\
G49.49-0.39 2   &    510.0$\pm$149.0 &  1.91$\pm$0.71E+17 & $\cdots$ &$\cdots$ & $\cdots$ &$\cdots$  \\
G49.49-0.39 4   &    655.5$\pm$165.7 & 21.72$\pm$6.72E+16 & $\cdots$ &$\cdots$ & $\cdots$ &$\cdots$  \\
G49.49-0.39 6   &    683.3$\pm$159.0 &  7.14$\pm$2.06E+17& $\cdots$ &$\cdots$ &    100.5$\pm$12.3 &  1.66$\pm$0.16E+16 \\
\tableline
&		\multicolumn{2}{c}{$^{13}$CH$_{3}$CN} &	\multicolumn{2}{c}{CH$_{3}$NCO}	&		\multicolumn{2}{c}{CH$_{3}$NH$_{2}$}	&		\multicolumn{2}{c}{CH$_{2}$CHCN}\\
&	 T$_{\rm rot}$[K] 	&	 N[$\rm cm^{-2}$] &  T$_{\rm rot}$[K]	&	 N[$\rm cm^{-2}$] &  T$_{\rm rot}$[K] 	&	 N[$\rm cm^{-2}$] &  T$_{\rm rot}$[K] 	&	 N[$\rm cm^{-2}$]\\
\tableline
G10.34-0.14 1   &    141.7$\pm$34.5 &  1.56$\pm$0.32E+14 & $\cdots$ &$\cdots$  & $\cdots$ &$\cdots$  & $\cdots$ &$\cdots$  \\
G10.34-0.14 2   &    119.3$\pm$143.4 & 10.13$\pm$11.73E+13 & $\cdots$ &$\cdots$  & $\cdots$ &$\cdots$  & $\cdots$ &$\cdots$  \\
G10.34-0.14 3   &    122.5$\pm$56.1 &  1.09$\pm$0.48E+14 & $\cdots$ &$\cdots$  & $\cdots$ &$\cdots$  & $\cdots$ &$\cdots$  \\
G18.34+1.78SW 2   &     90.3$\pm$51.8 &  9.18$\pm$3.79E+13 & $\cdots$ &$\cdots$  & $\cdots$ &$\cdots$  & $\cdots$ &$\cdots$  \\
G23.43-0.18 1   &     87.6$\pm$44.9 &  0.86$\pm$0.27E+14 & $\cdots$ &$\cdots$  & $\cdots$ &$\cdots$  & $\cdots$ &$\cdots$  \\
G23.43-0.18 2   &    112.2$\pm$128.0 &  9.37$\pm$11.93E+13 & $\cdots$ &$\cdots$  & $\cdots$ &$\cdots$  & $\cdots$ &$\cdots$  \\
G24.33+0.14 1   &    117.4$\pm$77.6 &  9.54$\pm$6.77E+14 & $\cdots$ &$\cdots$ &    205.2$\pm$182.4 &  3.39$\pm$4.25E+17 & $\cdots$ &$\cdots$  \\
G25.82-0.17 1   &     64.8$\pm$24.7 &  6.19$\pm$0.92E+14 & $\cdots$ &$\cdots$  &    184.6$\pm$213.1 &  2.14$\pm$3.09E+17  &    335.7$\pm$19.6 &  2.14$\pm$0.14E+16 \\
G27.36-0.16 1   &     91.3$\pm$49.8 &  8.47$\pm$3.81E+14  &    380.6$\pm$230.2 &  0.95$\pm$20.72E+17 & $\cdots$ &$\cdots$  & $\cdots$ &$\cdots$  \\
G28.37+0.07 2   &     63.7$\pm$21.6 &  1.21$\pm$0.23E+14 & $\cdots$ &$\cdots$  & $\cdots$ &$\cdots$  & $\cdots$ &$\cdots$  \\
G49.49-0.39 2   &     80.4$\pm$166.5 &  4.27$\pm$6.91E+14 & $\cdots$ &$\cdots$  & $\cdots$ &$\cdots$  & $\cdots$ &$\cdots$  \\
G49.49-0.39 3 & $\cdots$ &$\cdots$ &    127.5$\pm$170.5 &  0.00$\pm$0.19E+18  &     62.9$\pm$134.0 &  8.34$\pm$11.86E+16 & $\cdots$ &$\cdots$  \\
G49.49-0.39 4   &     91.6$\pm$8.4 &  3.88$\pm$0.12E+14 & $\cdots$ &$\cdots$  & $\cdots$ &$\cdots$  & $\cdots$ &$\cdots$  \\
G49.49-0.39 6   &    234.8$\pm$205.4 &  1.49$\pm$1.82E+15  &    201.0$\pm$150.9 &  1.23$\pm$40.63E+16 & $\cdots$ &$\cdots$  & $\cdots$ &$\cdots$
\enddata
\end{deluxetable}
\end{longrotatetable}

%% file: s3_tab5_columndensity_XCLASS.tex
\startlongtable
\begin{deluxetable*}{lccccc}
\tablecaption{Column densities of the molecules from XCLASS fitting}
\tabletypesize{\scriptsize}
\tablehead{\colhead{Source} & \multicolumn{5}{c}{Molecules}}
\startdata
Source &  CH$_{3}$OH &  $^{13}$CH$_{3}$OH  &  CH$_{3}^{18}$OH &  CH$_{2}$DOH &  CH$_{3}$CHO\\
\tableline
G10.32-0.26 C1  &  8.36$_{-4.23}^{+2.40}\times$10$^{16}$ & --& --&--& --\\
G10.32-0.26 C4  &  9.27$_{-3.37}^{+2.61}\times$10$^{16}$ & --& --&--& --\\
G10.34-0.14 C1  &  7.17$_{-1.34}^{+1.65}\times$10$^{17}$ &  8.55$_{-1.18}^{+1.70}\times$10$^{16}$ &  1.50$_{-0.51}^{+0.11}\times$10$^{17}$ & 1.87$_{-0.61}^{+0.02}\times$10$^{16}$ &  1.53$_{-0.32}^{+0.33}\times$10$^{16}$ \\
G10.34-0.14 C2  &  4.32$_{-1.49}^{+0.86}\times$10$^{17}$ &  4.94$_{-0.50}^{+4.05}\times$10$^{16}$ &  7.03$_{-2.59}^{+8.79}\times$10$^{16}$ & 2.08$_{-0.30}^{+1.81}\times$10$^{16}$ &  7.99$_{-1.59}^{+0.09}\times$10$^{15}$ \\
G10.34-0.14 C3  &  1.21$_{-0.18}^{+0.17}\times$10$^{18}$ &  8.05$_{-1.55}^{+6.03}\times$10$^{16}$ &  7.00$_{-4.60}^{+5.47}\times$10$^{16}$ & 1.09$_{-0.29}^{+0.64}\times$10$^{16}$ &  7.15$_{-0.51}^{+1.36}\times$10$^{15}$ \\
G18.34+1.78 C1  &  1.02$_{-0.46}^{+0.17}\times$10$^{17}$ & --& --&--& --\\
G18.34+1.78 C2  &  2.50$_{-0.49}^{+1.64}\times$10$^{17}$ & --& --&--& --\\
G18.34+1.78 C3  &  5.32$_{-1.50}^{+0.43}\times$10$^{16}$ & --& --&--& --\\
G18.34+1.78 C8  &  1.99$_{-0.67}^{+0.79}\times$10$^{17}$ & --& --& 2.50$_{-0.34}^{+0.02}\times$10$^{16}$ &  6.83$_{-1.40}^{+3.89}\times$10$^{15}$ \\
G18.34+1.78SW C1  &  8.88$_{-4.11}^{+1.03}\times$10$^{16}$ & --& --& 8.88$_{-3.54}^{+0.52}\times$10$^{15}$ & --\\
G18.34+1.78SW C2  &  7.27$_{-1.90}^{+1.29}\times$10$^{17}$ &  1.06$_{-0.46}^{+0.63}\times$10$^{17}$ &  1.19$_{-0.49}^{+0.13}\times$10$^{17}$ & 6.46$_{-0.91}^{+1.65}\times$10$^{16}$ &  1.76$_{-0.21}^{+0.47}\times$10$^{16}$ \\
G23.43-0.18 C1  &  1.25$_{-0.40}^{+0.46}\times$10$^{18}$ &  6.98$_{-2.24}^{+0.84}\times$10$^{16}$ &  1.11$_{-0.71}^{+0.06}\times$10$^{17}$ & 4.33$_{-2.66}^{+1.16}\times$10$^{16}$ &  9.16$_{-3.19}^{+3.84}\times$10$^{15}$ \\
G23.43-0.18 C2  &  1.06$_{-0.04}^{+0.40}\times$10$^{18}$ &  1.34$_{-0.31}^{+0.35}\times$10$^{17}$ &  9.96$_{-0.58}^{+0.40}\times$10$^{16}$ & 3.11$_{-1.41}^{+1.20}\times$10$^{16}$ &  1.37$_{-0.38}^{+0.34}\times$10$^{16}$ \\
G24.33+0.14 C1  &  6.02$_{-0.37}^{+2.64}\times$10$^{18}$ &  1.03$_{-0.05}^{+0.66}\times$10$^{18}$ &  1.21$_{-0.24}^{+0.21}\times$10$^{18}$ & 3.57$_{-1.19}^{+0.29}\times$10$^{17}$ &  1.16$_{-0.14}^{+1.35}\times$10$^{17}$ \\
G25.82-0.17 C1  &  2.80$_{-0.74}^{+0.42}\times$10$^{18}$ &  4.86$_{-1.63}^{+1.20}\times$10$^{17}$ &  5.18$_{-0.24}^{+0.17}\times$10$^{17}$ & 1.66$_{-0.50}^{+0.15}\times$10$^{17}$ &  7.20$_{-4.23}^{+0.93}\times$10$^{16}$ \\
G27.36-0.16 C1  &  5.32$_{-0.47}^{+1.27}\times$10$^{18}$ &  1.08$_{-0.26}^{+0.21}\times$10$^{18}$ &  1.52$_{-0.19}^{+0.28}\times$10$^{18}$ & 4.16$_{-0.78}^{+1.16}\times$10$^{17}$ &  1.47$_{-0.39}^{+0.41}\times$10$^{17}$ \\
G28.37+0.07 C2  &  1.22$_{-0.53}^{+0.19}\times$10$^{18}$ &  1.08$_{-0.59}^{+0.33}\times$10$^{17}$ &  1.36$_{-0.67}^{+0.11}\times$10$^{17}$ & 5.53$_{-2.09}^{+2.81}\times$10$^{16}$ &  9.71$_{-1.70}^{+1.32}\times$10$^{15}$ \\
G28.37+0.07 C3  &  6.30$_{-2.38}^{+3.00}\times$10$^{16}$ & --& --&--& --\\
G28.37+0.07 C5  &  4.39$_{-0.89}^{+4.15}\times$10$^{16}$ & --& --&--& --\\
G28.37+0.07 C6  &  1.49$_{-0.74}^{+0.32}\times$10$^{17}$ & --& --&--& --\\
G29.91-0.03 C1  &  3.50$_{-0.18}^{+1.24}\times$10$^{16}$ & --& --&--& --\\
G29.91-0.03 C2  &  5.93$_{-1.25}^{+1.27}\times$10$^{16}$ & --& --&--& --\\
G30.70-0.07 C1  &  6.07$_{-0.99}^{+2.29}\times$10$^{16}$ & --& --&--& --\\
G49.49-0.39 C1  &  6.85$_{-3.53}^{+1.01}\times$10$^{18}$ &  7.91$_{-3.77}^{+9.55}\times$10$^{17}$ &  9.77$_{-5.83}^{+2.49}\times$10$^{17}$ & 9.19$_{-1.31}^{+1.23}\times$10$^{16}$ &  1.24$_{-0.14}^{+0.04}\times$10$^{17}$ \\
G49.49-0.39 C2  &  3.60$_{-1.01}^{+3.04}\times$10$^{18}$ &  2.02$_{-0.70}^{+0.71}\times$10$^{17}$ &  2.23$_{-1.50}^{+0.35}\times$10$^{17}$ & 4.17$_{-1.57}^{+3.25}\times$10$^{15}$ &  2.01$_{-0.30}^{+0.19}\times$10$^{16}$ \\
G49.49-0.39 C3  &  1.39$_{-0.33}^{+0.70}\times$10$^{18}$ &  2.29$_{-0.64}^{+1.46}\times$10$^{17}$ &  3.01$_{-0.20}^{+1.45}\times$10$^{17}$ & 2.85$_{-1.37}^{+1.20}\times$10$^{16}$ &  3.25$_{-0.61}^{+0.97}\times$10$^{16}$ \\
G49.49-0.39 C4  &  1.57$_{-0.36}^{+0.60}\times$10$^{18}$ &  1.70$_{-0.39}^{+0.02}\times$10$^{17}$ &  1.89$_{-1.38}^{+0.29}\times$10$^{17}$ & 6.35$_{-2.33}^{+3.68}\times$10$^{16}$ &  3.34$_{-0.77}^{+0.16}\times$10$^{16}$ \\
G49.49-0.39 C6  &  6.54$_{-1.39}^{+1.13}\times$10$^{18}$ &  4.36$_{-0.30}^{+2.17}\times$10$^{17}$ &  4.26$_{-1.33}^{+0.95}\times$10$^{17}$ & 9.11$_{-1.67}^{+1.05}\times$10$^{16}$ &  4.41$_{-0.77}^{+1.09}\times$10$^{16}$ \\
\tableline
Source  &  HCOOCH$_{3}$  &  CH$_{3}$OCH$_{3}$ &  (CH$_{3}$)$_{2}$CO  &  C$_{2}$H$_{5}$OH  &  HCOCH$_{2}$OH\\
\tableline
G10.32-0.26 C1  &  1.15$_{-0.37}^{+0.42}\times$10$^{16}$ &  8.46$_{-2.96}^{+7.69}\times$10$^{15}$ & -- &-- & -- \\
G10.32-0.26 C4  &  5.81$_{-0.52}^{+1.88}\times$10$^{15}$ &  6.76$_{-0.73}^{+1.26}\times$10$^{15}$ & -- &-- & -- \\
G10.34-0.14 C1  &  8.13$_{-2.50}^{+0.55}\times$10$^{16}$ &  8.07$_{-4.14}^{+2.33}\times$10$^{16}$ &  5.92$_{-1.51}^{+1.29}\times$10$^{16}$ & 5.45$_{-1.53}^{+0.20}\times$10$^{16}$ & -- \\
G10.34-0.14 C2  &  4.55$_{-1.11}^{+2.38}\times$10$^{16}$ &  6.03$_{-1.61}^{+2.14}\times$10$^{16}$ & -- & 2.82$_{-0.49}^{+3.27}\times$10$^{16}$ & -- \\
G10.34-0.14 C3  &  5.82$_{-2.27}^{+0.76}\times$10$^{16}$ &  6.54$_{-2.82}^{+0.42}\times$10$^{16}$ & -- & 3.04$_{-0.83}^{+1.44}\times$10$^{16}$ & -- \\
G18.34+1.78 C1  &  1.72$_{-0.36}^{+0.23}\times$10$^{16}$ &  1.91$_{-0.49}^{+0.95}\times$10$^{16}$ & -- &-- & -- \\
G18.34+1.78 C2  &  4.54$_{-0.33}^{+2.39}\times$10$^{15}$ &  6.93$_{-0.97}^{+3.32}\times$10$^{15}$ & -- &-- & -- \\
G18.34+1.78 C3  & -- &  8.12$_{-1.61}^{+0.62}\times$10$^{15}$ & -- &-- & -- \\
G18.34+1.78 C8  &  2.12$_{-0.38}^{+1.07}\times$10$^{16}$ &  3.05$_{-1.14}^{+0.27}\times$10$^{16}$ & -- & 1.54$_{-0.61}^{+0.15}\times$10$^{16}$ & -- \\
G18.34+1.78SW C1  &  7.43$_{-1.38}^{+1.36}\times$10$^{15}$ &  6.72$_{-1.16}^{+1.40}\times$10$^{15}$ & -- &-- & -- \\
G18.34+1.78SW C2  &  6.77$_{-1.19}^{+1.02}\times$10$^{16}$ &  6.50$_{-2.68}^{+1.24}\times$10$^{16}$ &  4.00$_{-0.66}^{+1.60}\times$10$^{16}$ & 5.62$_{-1.65}^{+0.51}\times$10$^{16}$ &  6.02$_{-1.32}^{+0.13}\times$10$^{15}$ \\
G23.43-0.18 C1  &  2.79$_{-0.66}^{+0.34}\times$10$^{16}$ &  5.37$_{-1.74}^{+0.78}\times$10$^{16}$ &  1.12$_{-0.42}^{+0.38}\times$10$^{16}$ & 5.26$_{-2.33}^{+1.50}\times$10$^{16}$ &  2.45$_{-0.21}^{+1.60}\times$10$^{15}$ \\
G23.43-0.18 C2  &  5.20$_{-0.96}^{+1.48}\times$10$^{16}$ &  6.02$_{-1.72}^{+1.96}\times$10$^{16}$ &  2.59$_{-0.72}^{+0.80}\times$10$^{16}$ & 7.20$_{-1.86}^{+3.70}\times$10$^{16}$ &  2.36$_{-0.65}^{+0.04}\times$10$^{15}$ \\
G24.33+0.14 C1  &  5.87$_{-1.57}^{+0.83}\times$10$^{17}$ &  5.52$_{-0.30}^{+0.99}\times$10$^{17}$ &  7.47$_{-2.08}^{+1.12}\times$10$^{17}$ & 6.88$_{-2.88}^{+0.69}\times$10$^{17}$ &  8.46$_{-0.81}^{+0.48}\times$10$^{16}$ \\
G25.82-0.17 C1  &  1.99$_{-0.37}^{+0.02}\times$10$^{17}$ &  2.97$_{-2.30}^{+1.34}\times$10$^{17}$ &  1.80$_{-0.57}^{+0.41}\times$10$^{17}$ & 2.44$_{-0.48}^{+1.64}\times$10$^{17}$ &  2.12$_{-0.10}^{+0.02}\times$10$^{16}$ \\
G27.36-0.16 C1  &  5.76$_{-0.65}^{+0.34}\times$10$^{17}$ &  4.92$_{-2.07}^{+1.73}\times$10$^{17}$ &  1.25$_{-0.17}^{+0.20}\times$10$^{18}$ & 4.96$_{-1.82}^{+1.04}\times$10$^{17}$ &  7.49$_{-3.25}^{+1.57}\times$10$^{16}$ \\
G28.37+0.07 C2  &  9.10$_{-3.43}^{+4.74}\times$10$^{16}$ &  1.23$_{-0.18}^{+1.08}\times$10$^{17}$ &  1.76$_{-0.48}^{+0.42}\times$10$^{16}$ & 3.82$_{-0.54}^{+0.63}\times$10$^{16}$ & -- \\
G28.37+0.07 C3  &  5.46$_{-0.56}^{+1.22}\times$10$^{15}$ &  9.05$_{-2.98}^{+2.76}\times$10$^{15}$ & -- &-- & -- \\
G28.37+0.07 C5  & -- &  8.10$_{-1.92}^{+1.44}\times$10$^{15}$ & -- &-- & -- \\
G28.37+0.07 C6  &  7.81$_{-1.88}^{+1.57}\times$10$^{15}$ &  1.29$_{-0.32}^{+0.33}\times$10$^{16}$ & -- &-- & -- \\
G29.91-0.03 C1  & -- & -- & -- &-- & -- \\
G29.91-0.03 C2  &  4.22$_{-0.73}^{+0.88}\times$10$^{15}$ &  9.01$_{-1.10}^{+1.25}\times$10$^{15}$ & -- &-- & -- \\
G30.70-0.07 C1  &  6.94$_{-0.85}^{+0.05}\times$10$^{15}$ &  1.07$_{-0.18}^{+0.08}\times$10$^{16}$ & -- &-- & -- \\
G49.49-0.39 C1  &  8.64$_{-1.76}^{+1.68}\times$10$^{17}$ &  6.56$_{-3.39}^{+1.92}\times$10$^{17}$ &  4.14$_{-1.52}^{+2.03}\times$10$^{17}$ & 6.38$_{-2.48}^{+2.48}\times$10$^{17}$ &  2.69$_{-0.08}^{+0.06}\times$10$^{16}$ \\
G49.49-0.39 C2  &  3.03$_{-1.12}^{+1.49}\times$10$^{17}$ &  1.99$_{-1.07}^{+2.32}\times$10$^{17}$ &  1.01$_{-0.18}^{+0.35}\times$10$^{17}$ & 1.48$_{-0.34}^{+1.23}\times$10$^{17}$ &  4.56$_{-1.18}^{+0.07}\times$10$^{15}$ \\
G49.49-0.39 C3  &  2.28$_{-0.01}^{+0.13}\times$10$^{17}$ &  1.74$_{-0.14}^{+1.05}\times$10$^{17}$ &  1.45$_{-0.29}^{+0.46}\times$10$^{17}$ & 1.30$_{-0.79}^{+0.57}\times$10$^{17}$ &  3.14$_{-0.19}^{+0.63}\times$10$^{15}$ \\
G49.49-0.39 C4  &  2.46$_{-0.19}^{+0.33}\times$10$^{17}$ &  2.51$_{-0.51}^{+0.15}\times$10$^{17}$ &  2.09$_{-0.33}^{+0.40}\times$10$^{17}$ & 1.83$_{-0.64}^{+0.48}\times$10$^{17}$ &  3.56$_{-0.24}^{+0.82}\times$10$^{15}$ \\
G49.49-0.39 C6  &  3.46$_{-0.64}^{+0.63}\times$10$^{17}$ &  2.84$_{-1.36}^{+0.69}\times$10$^{17}$ &  4.14$_{-2.90}^{+1.29}\times$10$^{17}$ & 3.89$_{-1.55}^{+1.23}\times$10$^{17}$ &  1.70$_{-0.36}^{+0.57}\times$10$^{16}$ \\
\tableline 
Source  &  aGg'-(CH$_{2}$OH)$_{2}$  &  $^{13}$CH$_{3}$CN  &  CH$_{3}$NCO  &  CH$_{3}$NH$_{2}$  &  CH$_{2}$CHCN\\
\tableline
G10.32-0.26 C1  & -- & -- & -- &-- & -- \\
G10.32-0.26 C4  & -- & -- & -- &-- & -- \\
G10.34-0.14 C1  &  1.04$_{-0.42}^{+0.07}\times$10$^{16}$ &  7.62$_{-3.56}^{+1.78}\times$10$^{14}$ & -- &-- &  2.05$_{-0.03}^{+0.32}\times$10$^{15}$ \\
G10.34-0.14 C2  & -- &  5.06$_{-1.13}^{+1.16}\times$10$^{14}$ & -- &-- &  1.50$_{-0.14}^{+0.13}\times$10$^{15}$ \\
G10.34-0.14 C3  &  5.26$_{-2.13}^{+1.01}\times$10$^{15}$ &  5.25$_{-0.23}^{+0.57}\times$10$^{14}$ & -- &-- & -- \\
G18.34+1.78 C1  & -- & -- & -- &-- & -- \\
G18.34+1.78 C2  & -- & -- & -- &-- & -- \\
G18.34+1.78 C3  & -- & -- & -- &-- & -- \\
G18.34+1.78 C8  & -- &  1.51$_{-0.12}^{+0.33}\times$10$^{14}$ & -- &-- & -- \\
G18.34+1.78SW C1  & -- & -- & -- &-- & -- \\
G18.34+1.78SW C2  &  1.36$_{-0.45}^{+0.35}\times$10$^{16}$ &  6.37$_{-0.72}^{+0.33}\times$10$^{14}$ &  6.03$_{-0.51}^{+0.73}\times$10$^{15}$ &-- & -- \\
G23.43-0.18 C1  &  9.14$_{-1.50}^{+4.74}\times$10$^{15}$ &  5.81$_{-0.47}^{+0.03}\times$10$^{14}$ & -- &-- &  1.21$_{-0.36}^{+0.08}\times$10$^{15}$ \\
G23.43-0.18 C2  &  9.47$_{-1.41}^{+14.48}\times$10$^{15}$ &  7.24$_{-0.15}^{+0.92}\times$10$^{14}$ & -- &-- &  1.43$_{-0.06}^{+0.14}\times$10$^{15}$ \\
G24.33+0.14 C1  &  1.71$_{-0.34}^{+0.49}\times$10$^{17}$ &  8.79$_{-3.62}^{+2.25}\times$10$^{15}$ &  8.24$_{-3.91}^{+4.42}\times$10$^{16}$ & 4.66$_{-2.05}^{+4.05}\times$10$^{17}$ &  2.24$_{-0.56}^{+0.47}\times$10$^{16}$ \\
G25.82-0.17 C1  &  3.40$_{-0.98}^{+1.09}\times$10$^{16}$ &  4.67$_{-1.80}^{+2.94}\times$10$^{15}$ &  3.74$_{-1.44}^{+0.48}\times$10$^{16}$ & 1.48$_{-0.47}^{+1.40}\times$10$^{17}$ &  1.56$_{-0.18}^{+1.00}\times$10$^{16}$ \\
G27.36-0.16 C1  &  7.09$_{-4.02}^{+7.95}\times$10$^{16}$ &  5.61$_{-1.98}^{+7.01}\times$10$^{15}$ &  1.07$_{-0.51}^{+0.19}\times$10$^{17}$ & 2.06$_{-0.17}^{+0.29}\times$10$^{17}$ &  2.07$_{-1.11}^{+0.06}\times$10$^{16}$ \\
G28.37+0.07 C2  &  3.55$_{-0.49}^{+0.70}\times$10$^{15}$ &  6.24$_{-0.43}^{+0.11}\times$10$^{14}$ &  3.59$_{-0.92}^{+0.04}\times$10$^{15}$ &-- &  8.18$_{-0.68}^{+1.14}\times$10$^{14}$ \\
G28.37+0.07 C3  & -- & -- & -- &-- & -- \\
G28.37+0.07 C5  & -- & -- & -- &-- & -- \\
G28.37+0.07 C6  & -- & -- & -- &-- & -- \\
G29.91-0.03 C1  & -- & -- & -- &-- & -- \\
G29.91-0.03 C2  & -- & -- & -- &-- & -- \\
G30.70-0.07 C1  & -- & -- & -- &-- & -- \\
G49.49-0.39 C1  &  1.47$_{-0.17}^{+0.39}\times$10$^{17}$ &  2.80$_{-0.50}^{+0.37}\times$10$^{15}$ &  3.85$_{-0.62}^{+0.60}\times$10$^{16}$ & 2.28$_{-0.36}^{+0.43}\times$10$^{17}$ &  9.05$_{-0.23}^{+0.38}\times$10$^{15}$ \\
G49.49-0.39 C2  &  1.31$_{-0.10}^{+0.36}\times$10$^{16}$ &  1.78$_{-0.13}^{+0.27}\times$10$^{15}$ &  1.34$_{-0.06}^{+0.21}\times$10$^{16}$ & 5.70$_{-1.32}^{+6.83}\times$10$^{16}$ &  1.63$_{-0.17}^{+0.12}\times$10$^{15}$ \\
G49.49-0.39 C3  &  8.61$_{-0.22}^{+0.43}\times$10$^{15}$ &  1.48$_{-0.25}^{+0.37}\times$10$^{15}$ &  1.60$_{-0.04}^{+0.36}\times$10$^{16}$ & 6.86$_{-4.32}^{+6.46}\times$10$^{16}$ &  5.15$_{-1.00}^{+0.06}\times$10$^{15}$ \\
G49.49-0.39 C4  &  2.50$_{-0.23}^{+0.16}\times$10$^{16}$ &  2.10$_{-0.59}^{+0.52}\times$10$^{15}$ &  1.57$_{-0.84}^{+0.22}\times$10$^{16}$ & 7.76$_{-0.63}^{+1.80}\times$10$^{16}$ &  3.91$_{-0.29}^{+1.00}\times$10$^{15}$ \\
G49.49-0.39 C6  &  1.06$_{-0.12}^{+0.45}\times$10$^{17}$ &  4.50$_{-0.38}^{+0.87}\times$10$^{15}$ &  4.61$_{-0.20}^{+0.88}\times$10$^{16}$ & 1.72$_{-0.44}^{+0.74}\times$10$^{17}$ &  1.39$_{-0.04}^{+0.27}\times$10$^{16}$ \\
\tableline 
Source  & CH$_{3}$CH$_{2}$CN &  CH$_{3}$CH$_{2}$C$^{15}$N &  NH$_{2}$CHO  & NH$_{2}$CDO & \\
\tableline
G10.32-0.26 C1  & -- & -- & -- &-- \\
G10.32-0.26 C4  & -- & -- & -- &-- \\
G10.34-0.14 C1  &  1.00$_{-0.25}^{+0.92}\times$10$^{16}$ & -- &  3.50$_{-0.19}^{+0.95}\times$10$^{15}$ &-- \\
G10.34-0.14 C2  &  6.93$_{-3.03}^{+1.46}\times$10$^{15}$ & -- &  1.53$_{-0.35}^{+0.23}\times$10$^{15}$ &-- \\
G10.34-0.14 C3  &  3.30$_{-0.94}^{+2.17}\times$10$^{15}$ & -- &  2.33$_{-0.24}^{+0.34}\times$10$^{15}$ &-- \\
G18.34+1.78 C1  & -- & -- & -- &-- \\
G18.34+1.78 C2  & -- & -- & -- &-- \\
G18.34+1.78 C3  & -- & -- & -- &-- \\
G18.34+1.78 C8  & -- & -- &  7.54$_{-0.84}^{+3.19}\times$10$^{14}$ &-- \\
G18.34+1.78SW C1  & -- & -- & -- &-- \\
G18.34+1.78SW C2  &  4.29$_{-1.55}^{+3.14}\times$10$^{15}$ & -- &  2.73$_{-0.37}^{+0.55}\times$10$^{15}$ &-- \\
G23.43-0.18 C1  &  7.02$_{-2.64}^{+0.88}\times$10$^{15}$ & -- &  3.96$_{-1.11}^{+0.77}\times$10$^{15}$ &-- \\
G23.43-0.18 C2  &  4.63$_{-1.68}^{+0.99}\times$10$^{15}$ & -- &  5.16$_{-0.09}^{+2.06}\times$10$^{15}$ &-- \\
G24.33+0.14 C1  &  3.03$_{-0.84}^{+0.92}\times$10$^{16}$ &  1.18$_{-0.23}^{+0.03}\times$10$^{16}$ &  6.07$_{-1.48}^{+3.16}\times$10$^{16}$ & 1.14$_{-0.07}^{+0.31}\times$10$^{16}$ \\
G25.82-0.17 C1  &  2.15$_{-1.10}^{+1.34}\times$10$^{16}$ &  3.15$_{-0.78}^{+0.41}\times$10$^{15}$ &  1.51$_{-0.12}^{+0.16}\times$10$^{16}$ & 1.88$_{-0.44}^{+0.11}\times$10$^{15}$ \\
G27.36-0.16 C1  &  1.79$_{-0.20}^{+1.10}\times$10$^{16}$ &  1.49$_{-0.02}^{+0.01}\times$10$^{16}$ &  3.23$_{-1.18}^{+2.40}\times$10$^{16}$ & 7.65$_{-2.59}^{+1.00}\times$10$^{15}$ \\
G28.37+0.07 C2  &  3.77$_{-1.22}^{+1.13}\times$10$^{15}$ & -- &  1.64$_{-0.29}^{+0.11}\times$10$^{15}$ &-- \\
G28.37+0.07 C3  & -- & -- & -- &-- \\
G28.37+0.07 C5  & -- & -- & -- &-- \\
G28.37+0.07 C6  & -- & -- & -- &-- \\
G29.91-0.03 C1  & -- & -- & -- &-- \\
G29.91-0.03 C2  & -- & -- & -- &-- \\
G30.70-0.07 C1  & -- & -- & -- &-- \\
G49.49-0.39 C1  &  2.65$_{-1.61}^{+0.97}\times$10$^{16}$ &  9.08$_{-0.15}^{+0.13}\times$10$^{15}$ &  2.74$_{-0.70}^{+1.18}\times$10$^{16}$ & 2.23$_{-0.02}^{+0.42}\times$10$^{15}$ \\
G49.49-0.39 C2  &  1.55$_{-0.23}^{+1.40}\times$10$^{16}$ & -- &  1.16$_{-0.05}^{+0.60}\times$10$^{16}$ &-- \\
G49.49-0.39 C3  &  1.71$_{-0.61}^{+0.59}\times$10$^{16}$ & -- &  1.99$_{-0.30}^{+0.08}\times$10$^{15}$ & 1.28$_{-0.03}^{+0.25}\times$10$^{15}$ \\
G49.49-0.39 C4  &  1.24$_{-0.05}^{+0.32}\times$10$^{16}$ & -- &  5.35$_{-0.57}^{+0.99}\times$10$^{15}$ & 9.95$_{-1.37}^{+0.22}\times$10$^{14}$ \\
G49.49-0.39 C6  &  2.79$_{-2.03}^{+0.43}\times$10$^{16}$ &  3.94$_{-0.52}^{+0.31}\times$10$^{15}$ &  6.17$_{-1.40}^{+1.16}\times$10$^{16}$ & 3.09$_{-0.25}^{+0.22}\times$10$^{15}$ 
\enddata
\tablecaption{Column densities of the molecules from XCLASS fitting}
\tabletypesize{\scriptsize}
\label{tab:N1}
\end{deluxetable*}

%% file: s3_f1_figure1_continuum_figureset.tex
\newpage

\begin{figure*}
  \centering
\includegraphics[clip,width=0.95\columnwidth,keepaspectratio]{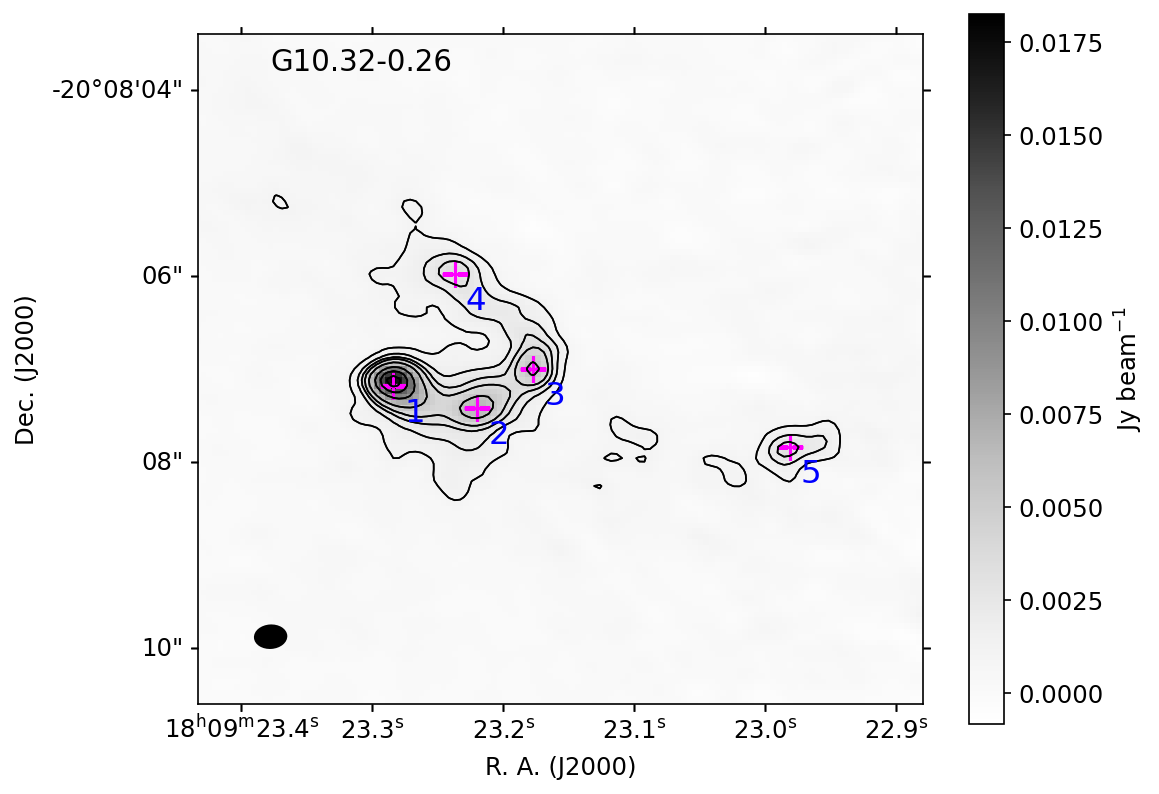}
\caption{The 1.3mm continuum image. Magenta plus symbols indicate the continuum intensity peaks.} For G10.32-0.26, contour levels are 0.0009, 0.0018, 0.0027, 0.0036, 0.0054, 0.0091, 0.0127, 0.0164 Jy beam$^{-1}$. The coordinates of continuum peaks are summarized in Table \ref{tab:cores}.
\label{cont:G10.32}
\end{figure*}

\begin{figure*}
  \centering
\includegraphics[clip,width=0.95\columnwidth,keepaspectratio]{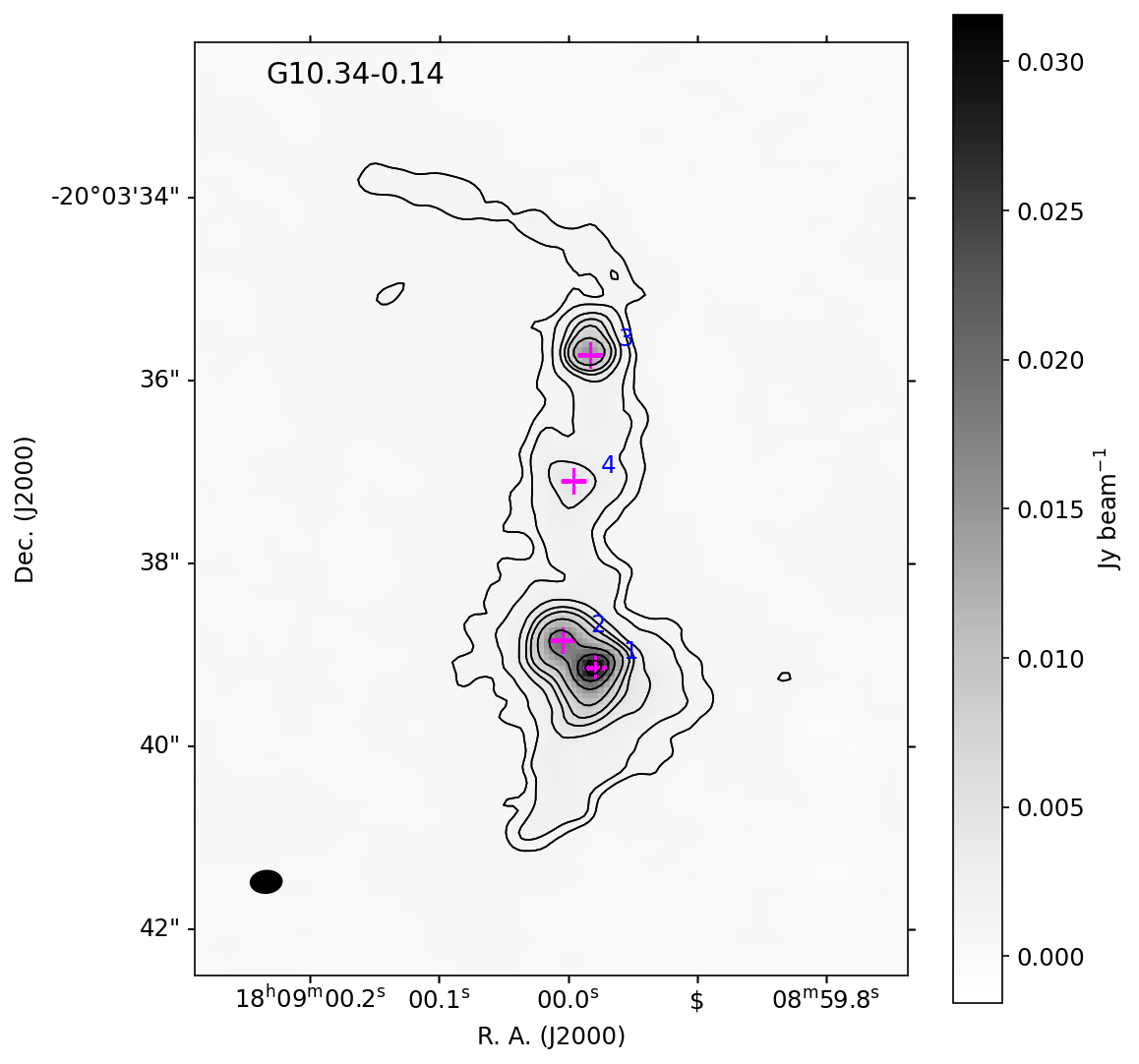}
\caption{The same as Figure \ref{cont:G10.32} except for G10.34-0.14. Contour levels are 0.001, 0.0015, 0.0031, 0.0047, 0.0063, 0.0094, 0.0157, 0.0220, 0.0284 Jy beam$^{-1}$. }
\end{figure*}

\begin{figure*}
  \centering
\includegraphics[clip,width=0.95\columnwidth,keepaspectratio]{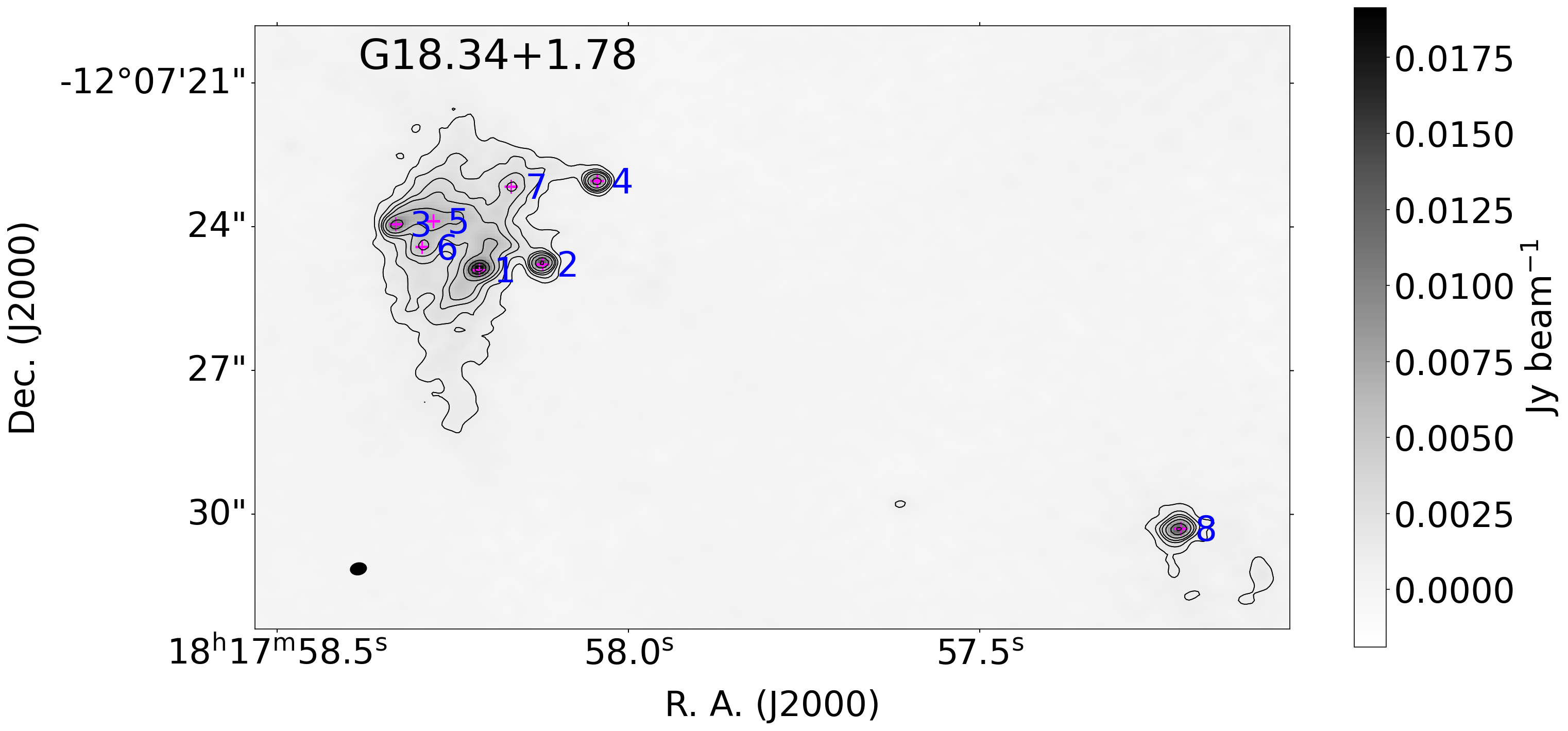}
\caption{The same as Figure \ref{cont:G10.32} except for G18.34+1.78. Contour levels are 0.001, 0.0019, 0.0028, 0.0038, 0.0057, 0.0095, 0.0133, 0.0172 Jy beam$^{-1}$.}
\end{figure*}

\begin{figure*}
  \centering
\includegraphics[clip,width=0.95\columnwidth,keepaspectratio]{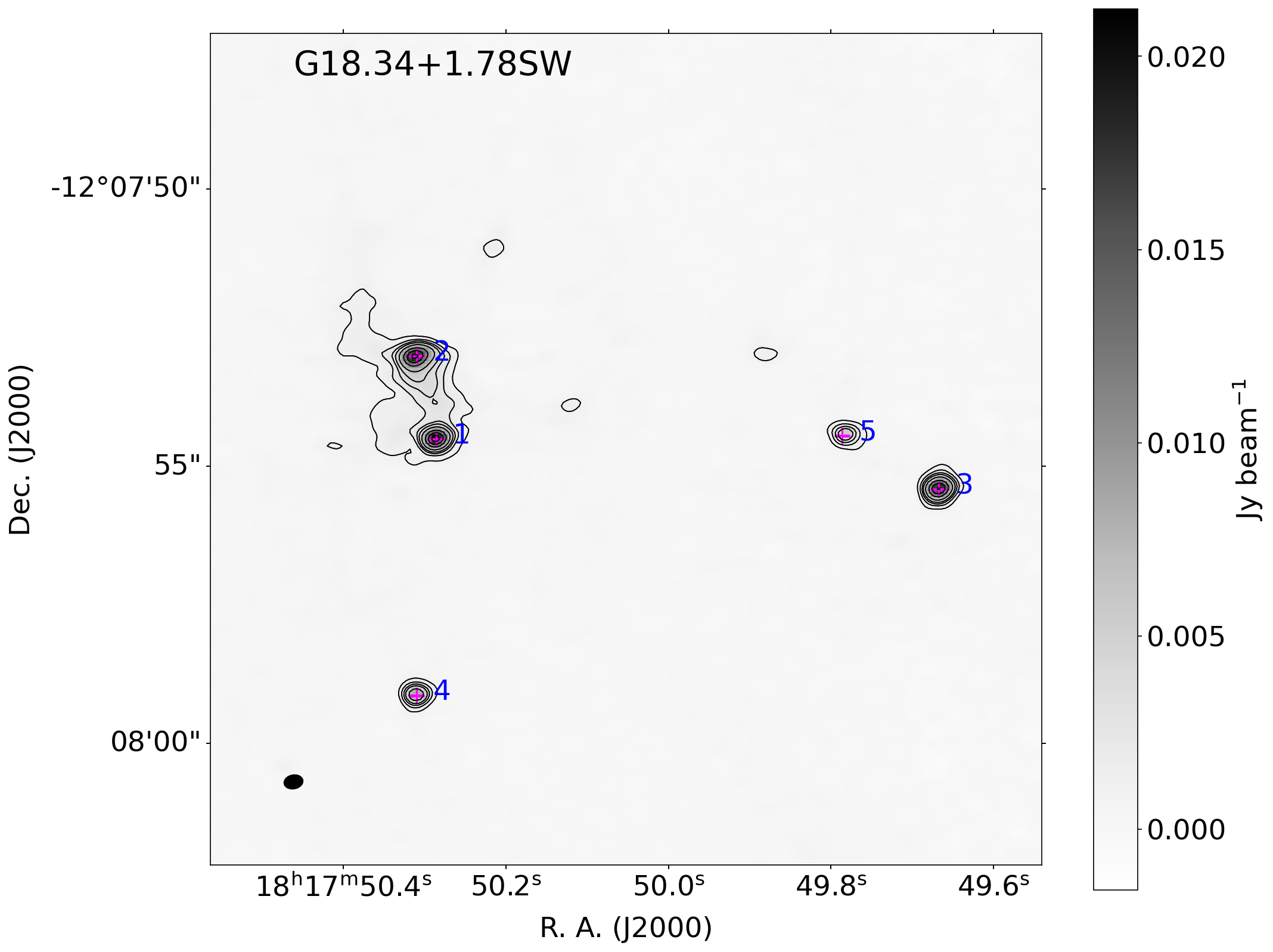}
\caption{The same as Figure \ref{cont:G10.32} except for G18.34+1.78SW. Contour levels are 0.0010, 0.0021, 0.0031, 0.0042, 0.0063, 0.0106, 0.0148, 0.0191 Jy beam$^{-1}$.}
\label{fig:continuum1}
\end{figure*}

\begin{figure*}
  \centering
\includegraphics[clip,width=0.95\columnwidth,keepaspectratio]{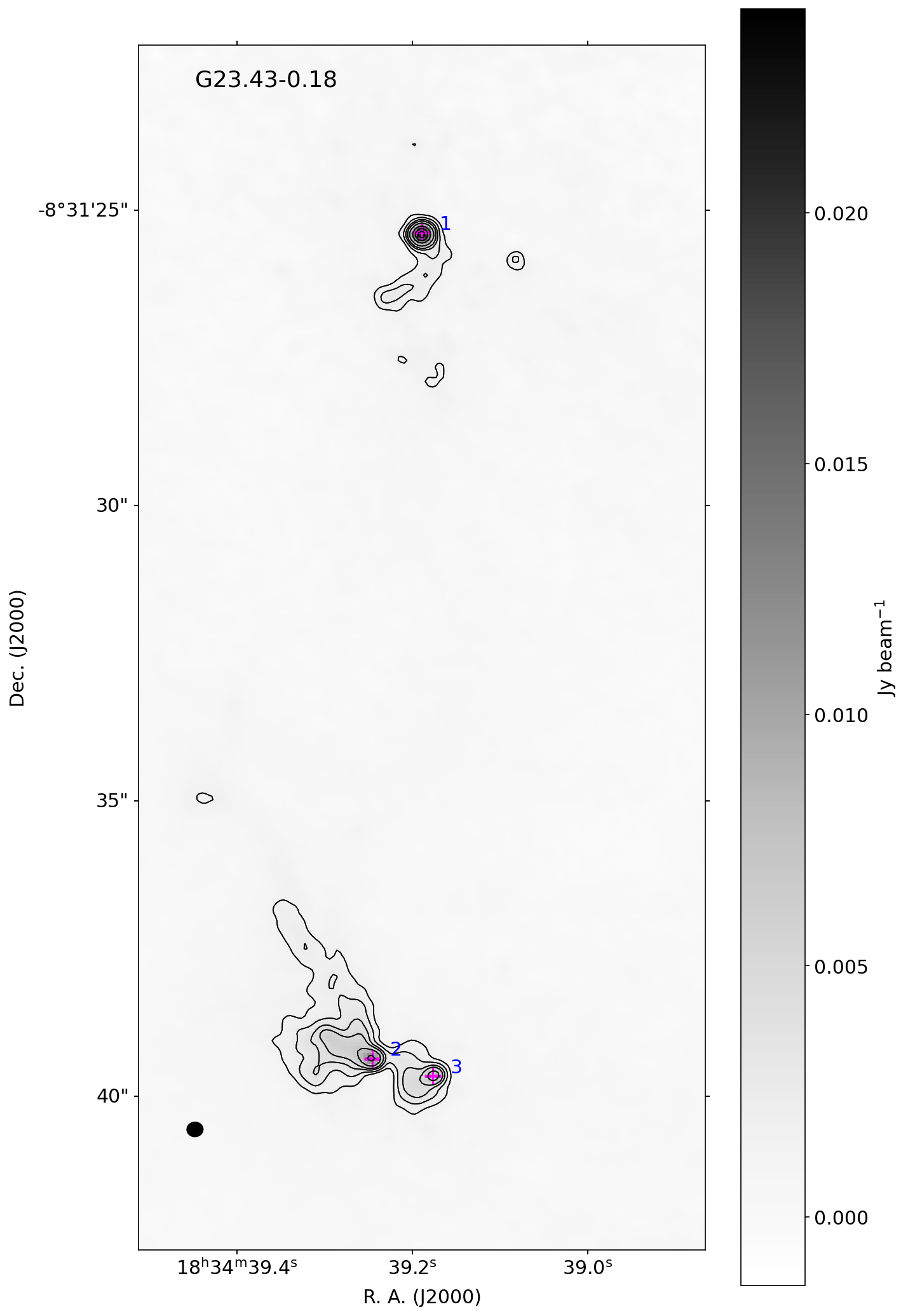}
\caption{The same as Figure \ref{cont:G10.32} except for G23.43-0.18. Contour levels are 0.0014, 0.0024, 0.0036, 0.0048, 0.0072, 0.0120, 0.0168, 0.0216 Jy beam$^{-1}$.}
\end{figure*}

\begin{figure*}
  \centering
\includegraphics[clip,width=0.95\columnwidth,keepaspectratio]{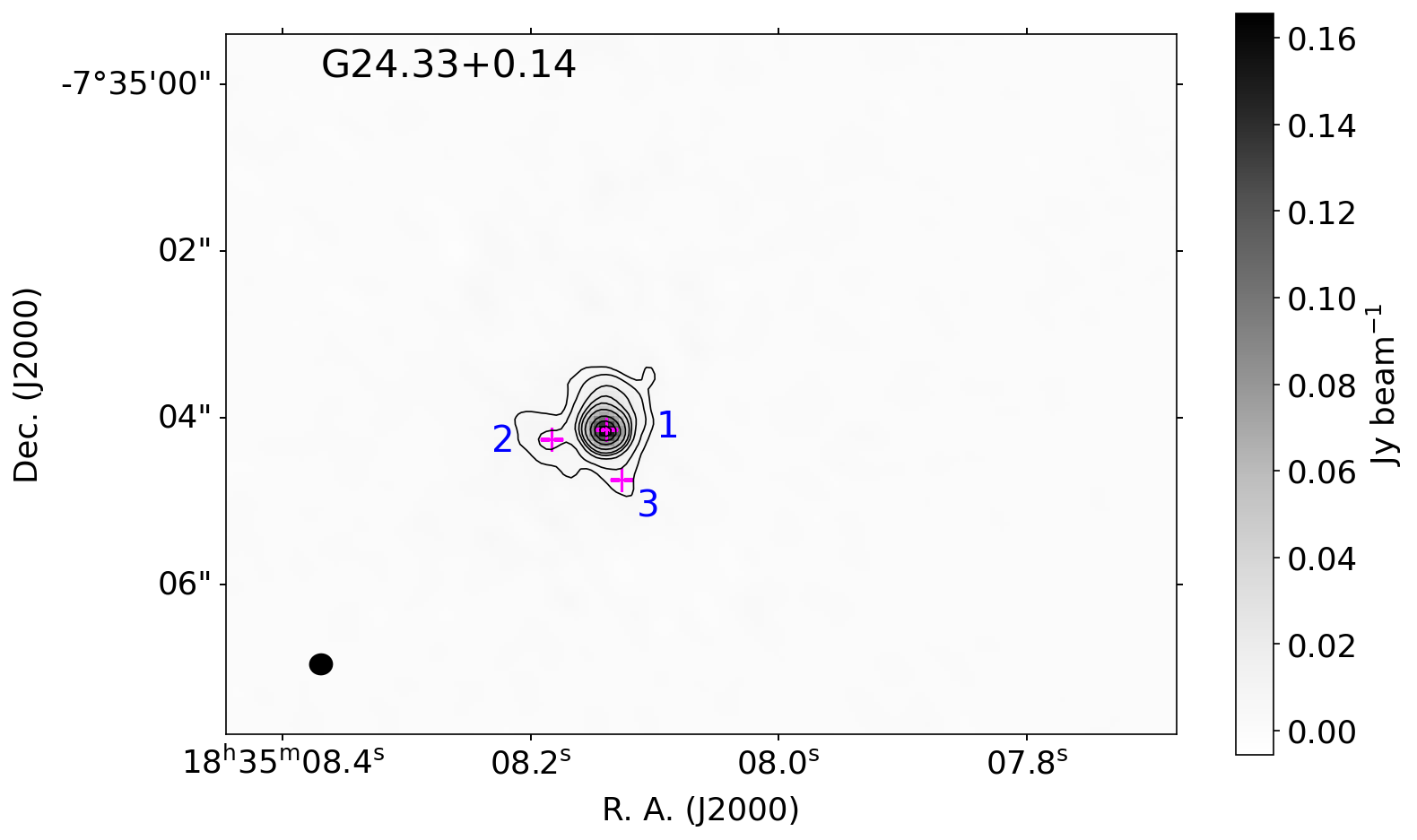}
\caption{The same as Figure \ref{cont:G10.32} except for G24.33+0.14. Contour levels are 0.0049, 0.0082, 0.0165, 0.0248, 0.0331, 0.0496, 0.0828, 0.1159, 0.1490 Jy beam$^{-1}$.}
\end{figure*}

\begin{figure*}
  \centering
\includegraphics[clip,width=0.95\columnwidth,keepaspectratio]{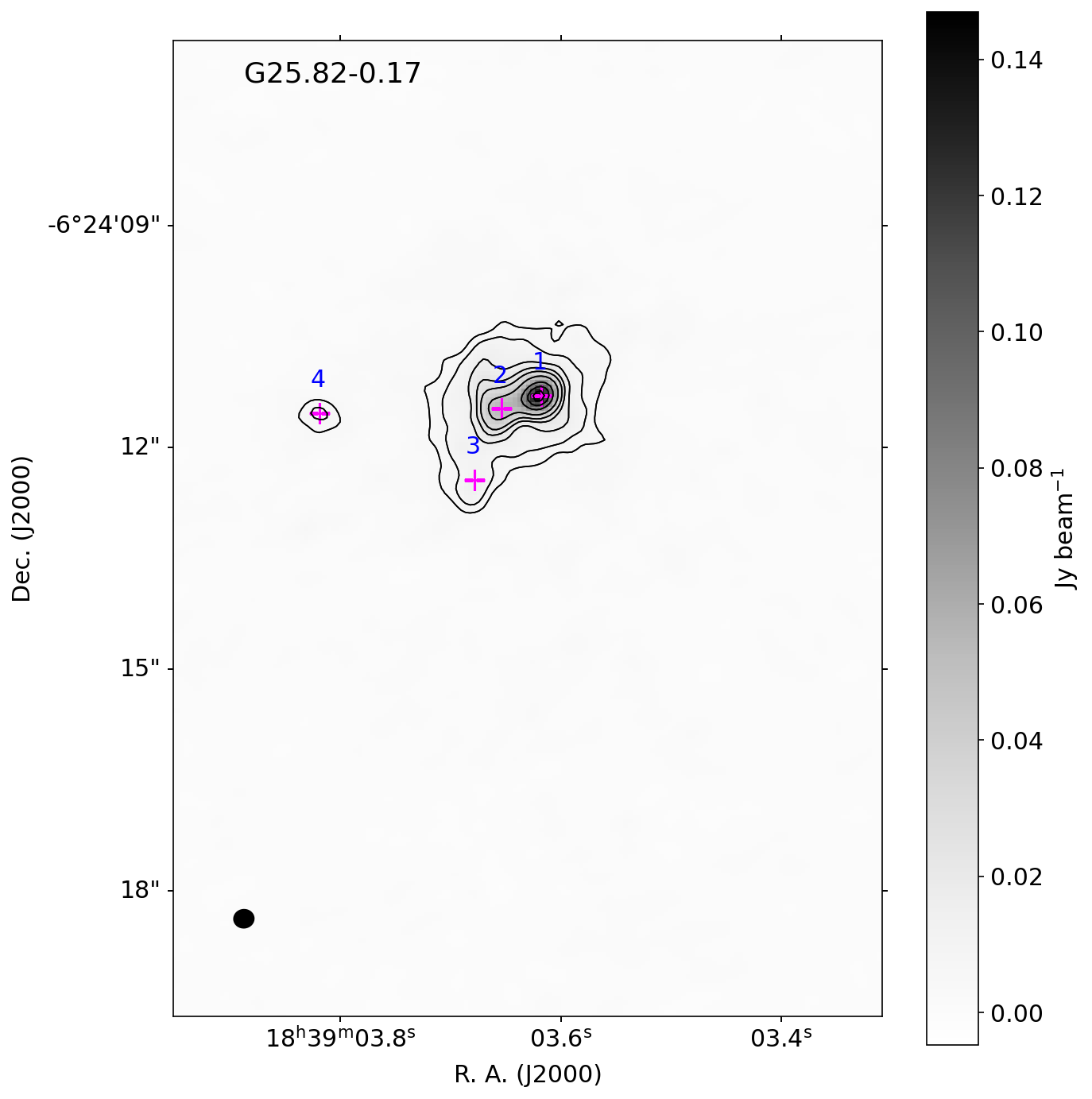}
\caption{The same as Figure \ref{cont:G10.32} except for G25.82-0.17. Contour levels are 0.0044, 0.0073, 0.0146, 0.0220, 0.0293, 0.0440, 0.0734, 0.1028, 0.1322 Jy beam$^{-1}$.}
\end{figure*}

\begin{figure*}
  \centering
\includegraphics[clip,width=0.95\columnwidth,keepaspectratio]{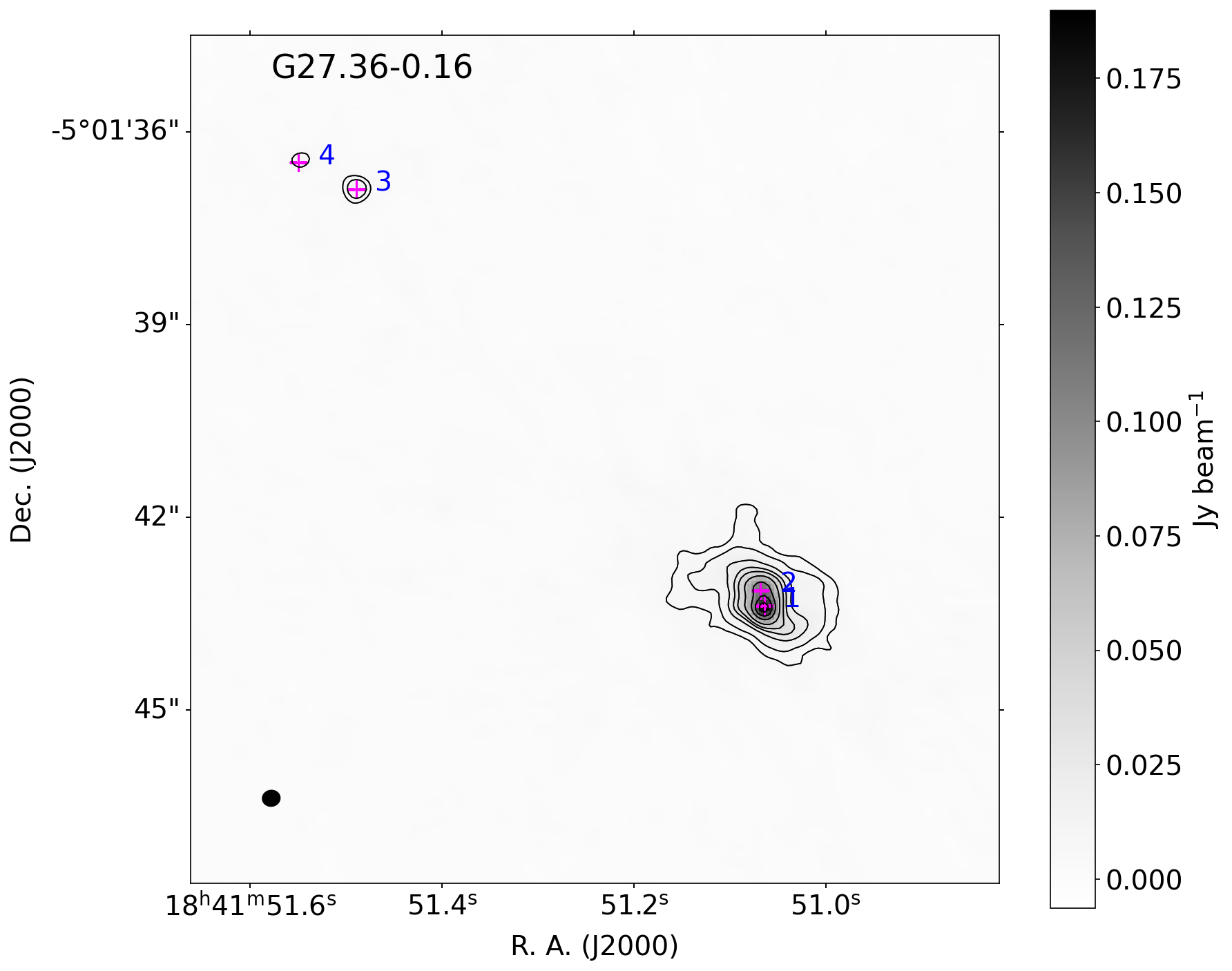}
\caption{The same as Figure \ref{cont:G10.32} except for G27.36-0.16. Contour levels are 0.0056, 0.0094, 0.0189, 0.0284, 0.0379, 0.0569, 0.0949, 0.1329, 0.1709 Jy beam$^{-1}$.}
\end{figure*}

\begin{figure*}
  \centering
\includegraphics[clip,width=0.95\columnwidth,keepaspectratio]{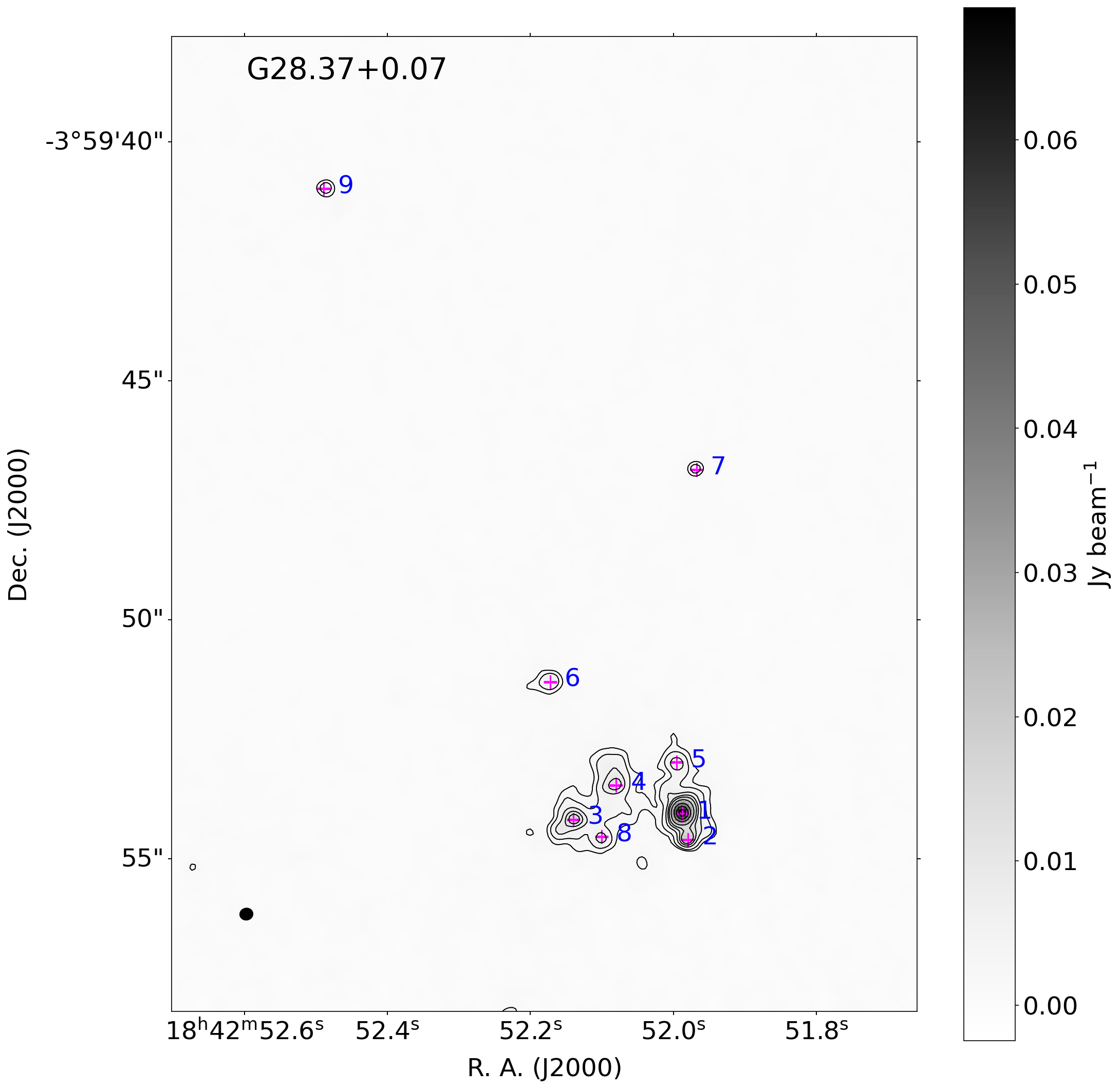}
\caption{The same as Figure \ref{cont:G10.32} except for G28.37+0.07. Contour levels are 0.0020, 0.0034, 0.0069, 0.0103, 0.0138, 0.0207, 0.0345, 0.0484, 0.0622 Jy beam$^{-1}$.}
\end{figure*}

\begin{figure*}
  \centering
\includegraphics[clip,width=0.95\columnwidth,keepaspectratio]{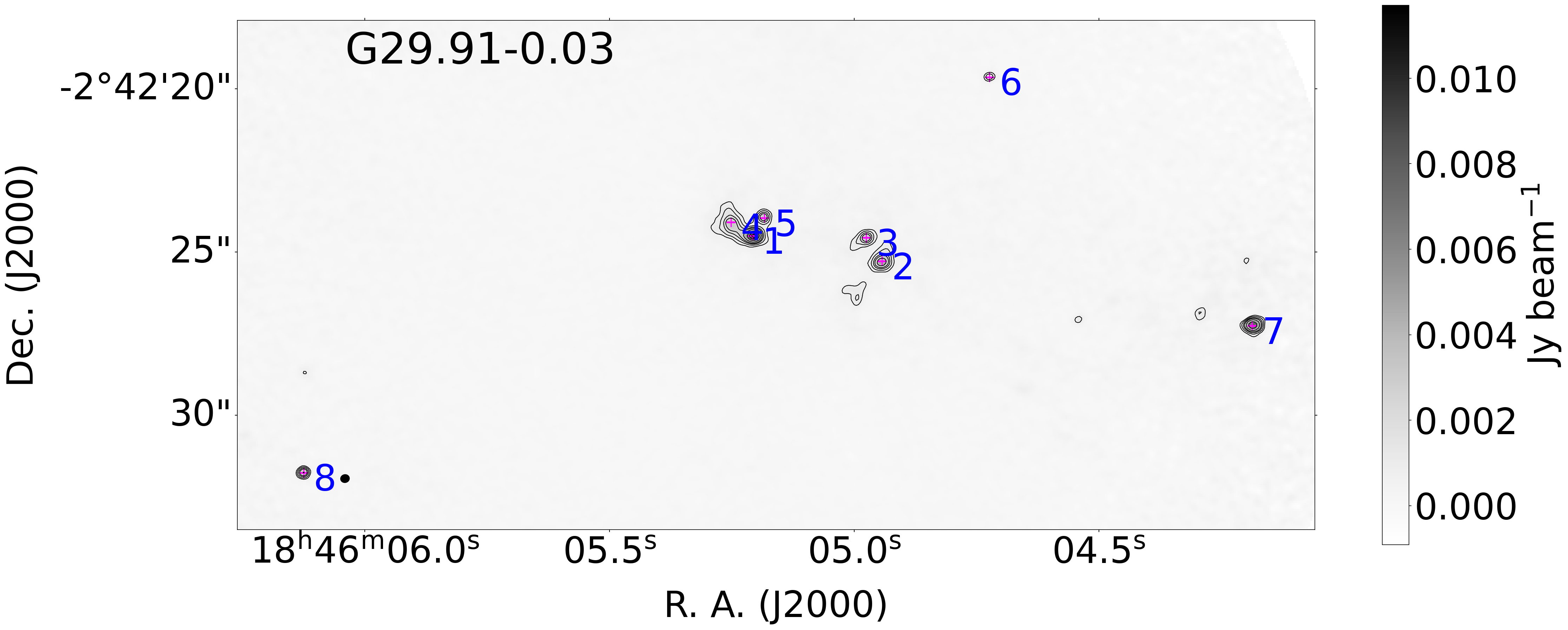}
\caption{The same as Figure \ref{cont:G10.32} except for G29.91-0.03. Contour levels are 0.0007, 0.0011, 0.0017, 0.0023, 0.0035, 0.0058, 0.0081, 0.0105 Jy beam$^{-1}$.}
\end{figure*}

\begin{figure*}
  \centering
\includegraphics[clip,width=0.95\columnwidth,keepaspectratio]{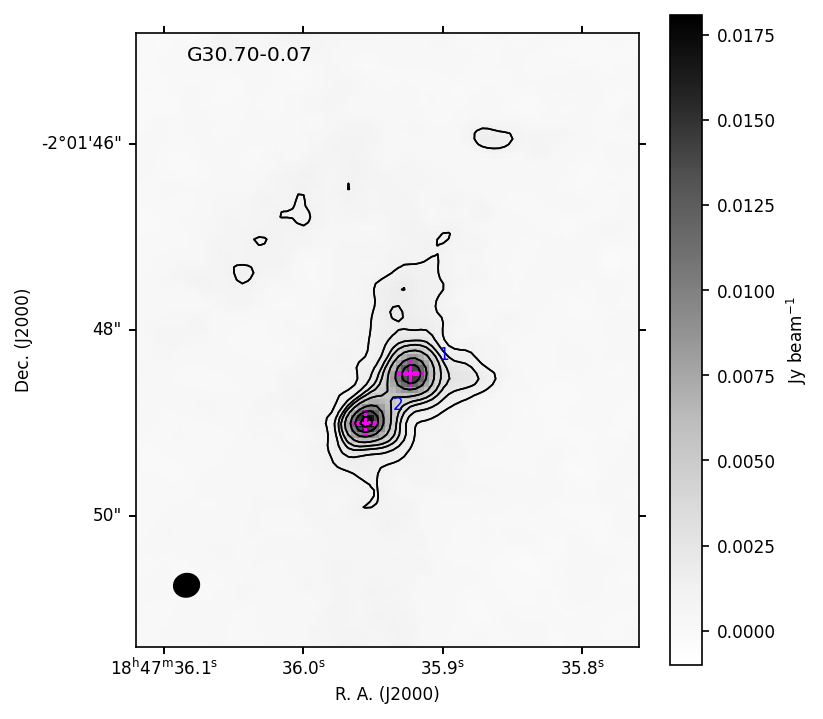}
\caption{The same as Figure \ref{cont:G10.32} except for G30.70-0.07. Contour levels are 0.001, 0.0018, 0.0027, 0.0036, 0.0054, 0.0090, 0.0126, 0.0162 Jy beam$^{-1}$.}
\end{figure*}

\begin{figure*}
  \centering
\includegraphics[clip,width=0.95\columnwidth,keepaspectratio]{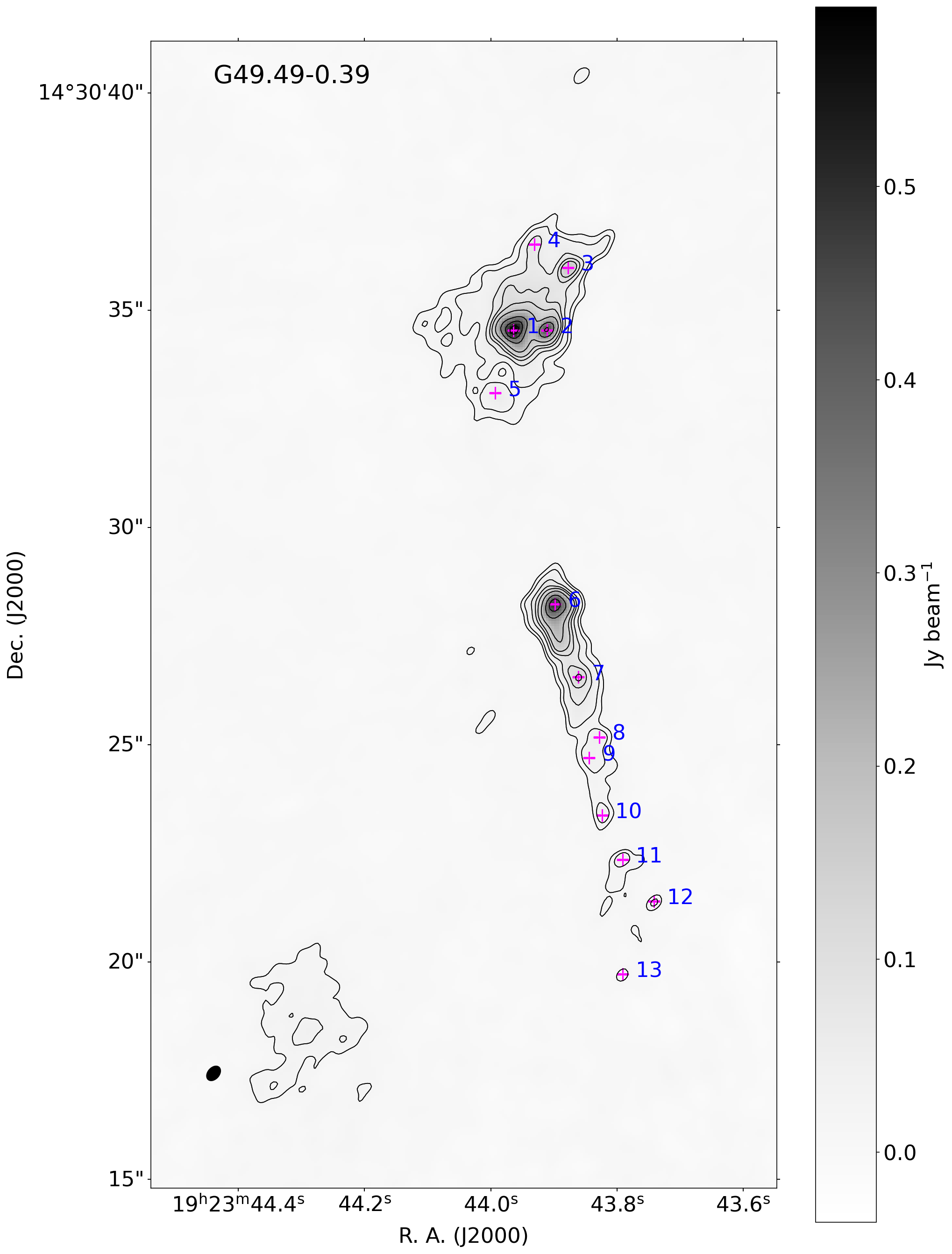}
\caption{The same as Figure \ref{cont:G10.32} except for G49.49-0.39. Contour levels are 0.0177, 0.0296, 0.0593, 0.0889, 0.1186, 0.1779, 0.2965, 0.4151, 0.5337 Jy beam$^{-1}$.}
\label{cont:G49.49}
\end{figure*}

%% file: s3_f2_spectrum_figureset.tex
\begin{figure*}
  \centering
\includegraphics[clip,width=0.7\textwidth,keepaspectratio]{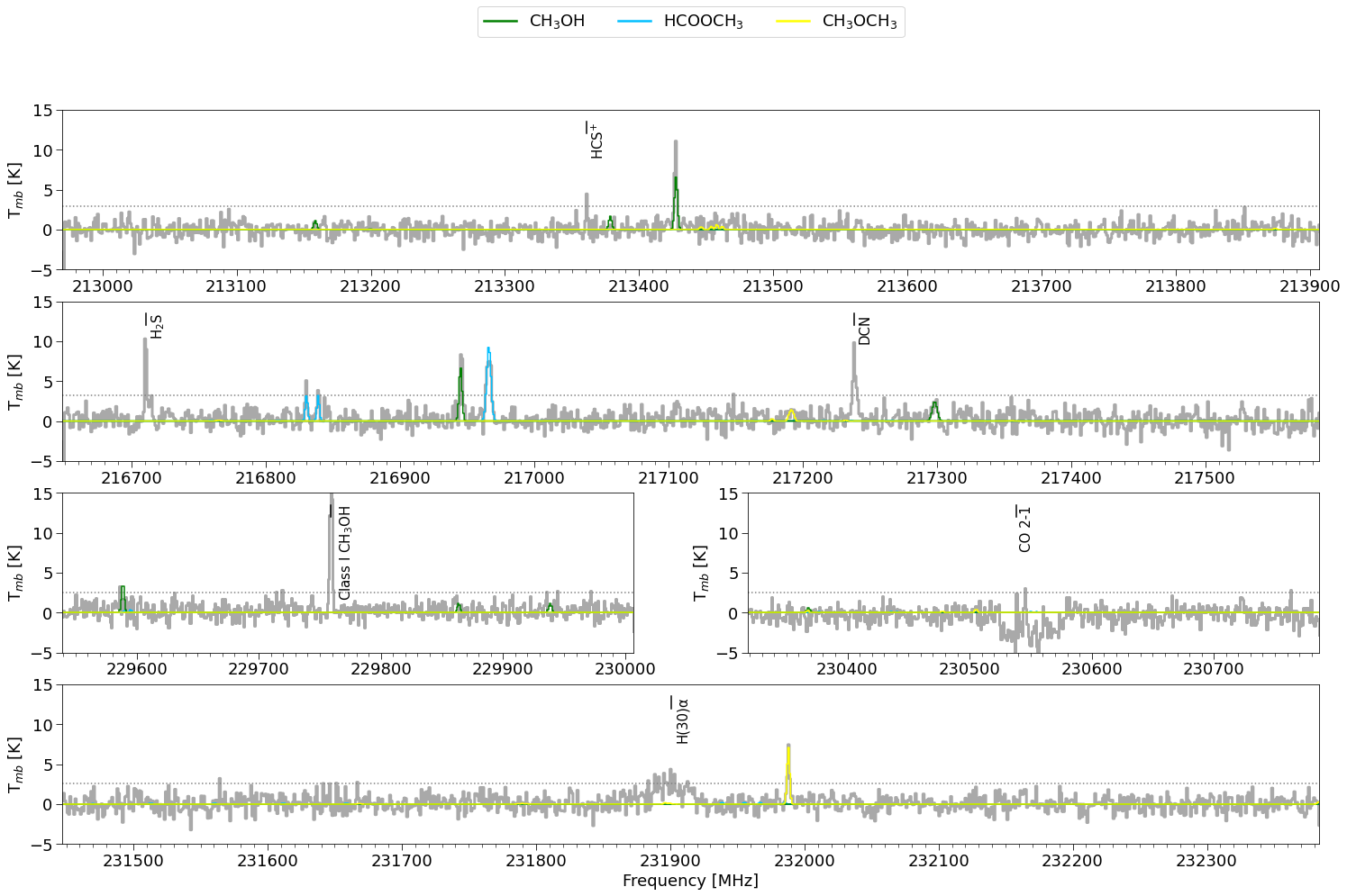}
\caption{The spectra of G10.32-0.26 C1 at the 216.946 GHz CH$_3$OH emission peak {\bf as marked with red triangles in Figure \ref{fig:chnmap:G10.32}.} The grey solid and dotted line presents the observed spectra and 3$\sigma$ noise level, respectively. The lines with different colors are the spectra of individual species simulated by the XCLASS.}
\label{fig:spectrum:G10.32C1}
\end{figure*}

\begin{figure*}
  \centering
\includegraphics[clip,width=0.7\textwidth,keepaspectratio]{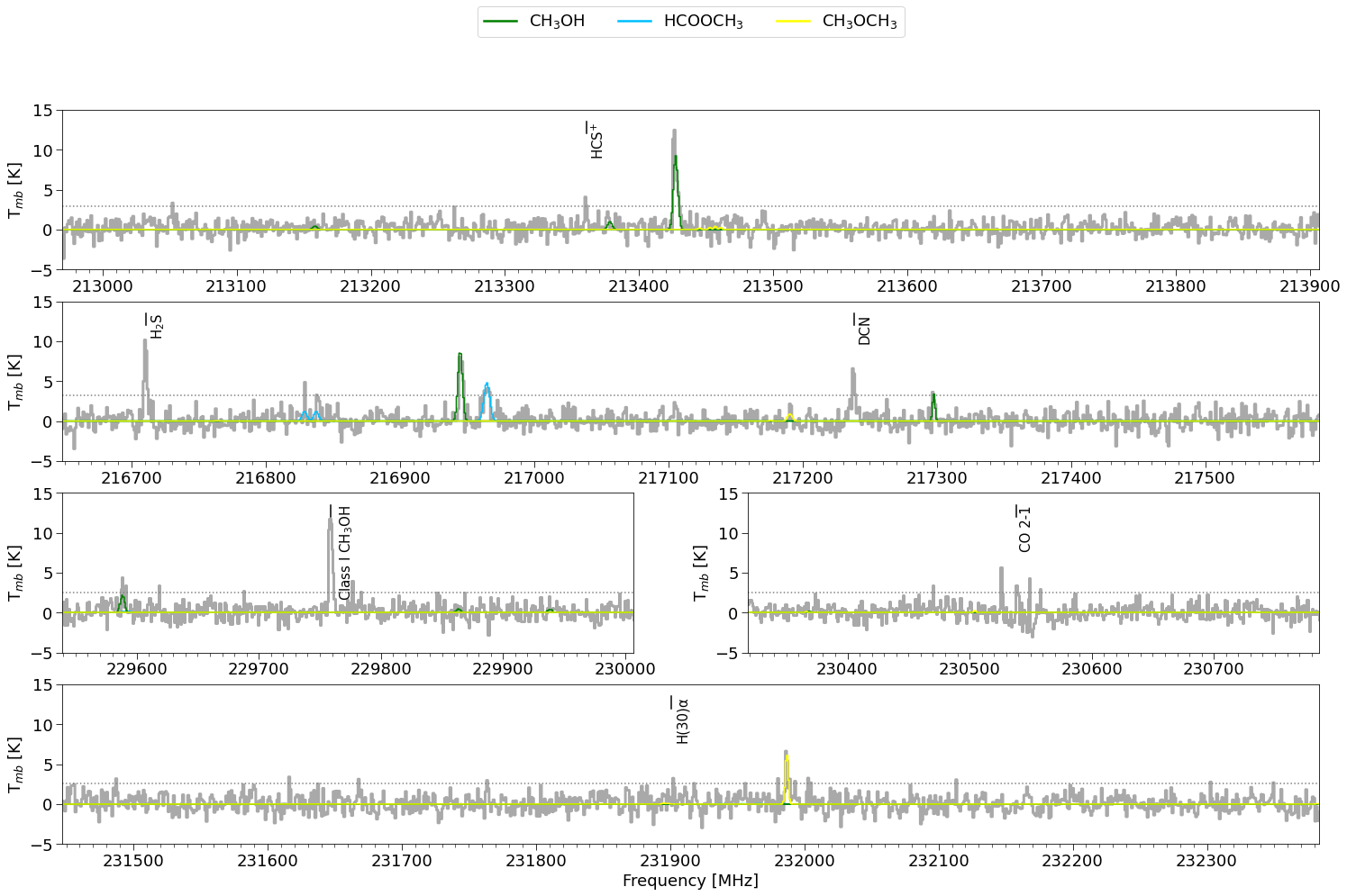}
\caption{The same as Figure \ref{fig:spectrum:G10.32C1} except for G10.32-0.26 C4. Check Figure \ref{fig:chnmap:G10.32} for the CH$_3$OH emission peak position from which the spectra were extracted.}
\label{fig:spectrum:G10.32C4}
\end{figure*}

\begin{figure*}
  \centering
\includegraphics[clip,width=0.7\textwidth,keepaspectratio]{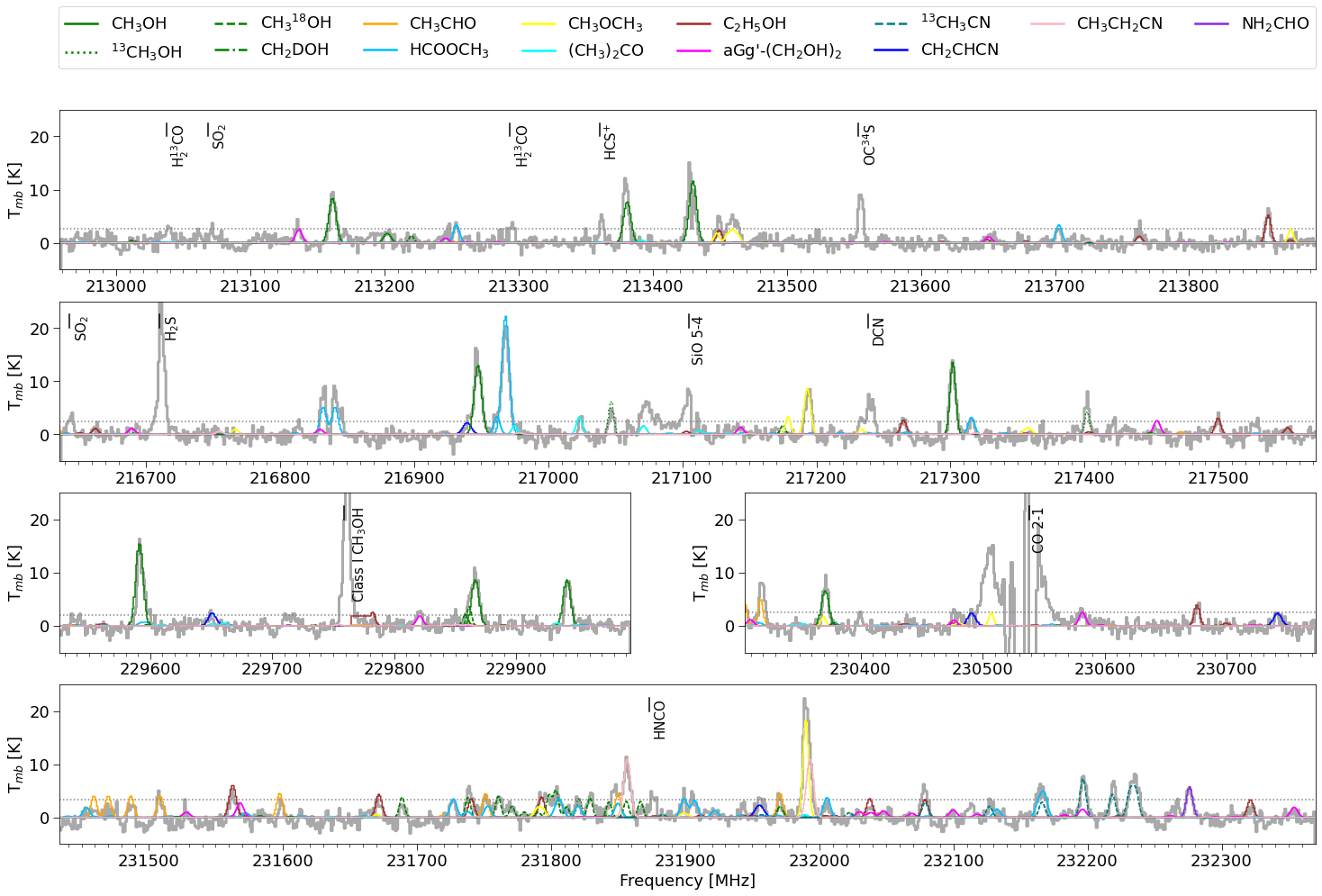}
\caption{The same as Figure \ref{fig:spectrum:G10.32C1} except for G10.34-0.14 C1. Check Figure \ref{fig:chnmap:G10.34} for the CH$_3$OH emission peak position from which the spectra were extracted.}
\label{fig:spectrum:G10.34C1}
\end{figure*}

\begin{figure*}
  \centering
\includegraphics[clip,width=0.7\textwidth,keepaspectratio]{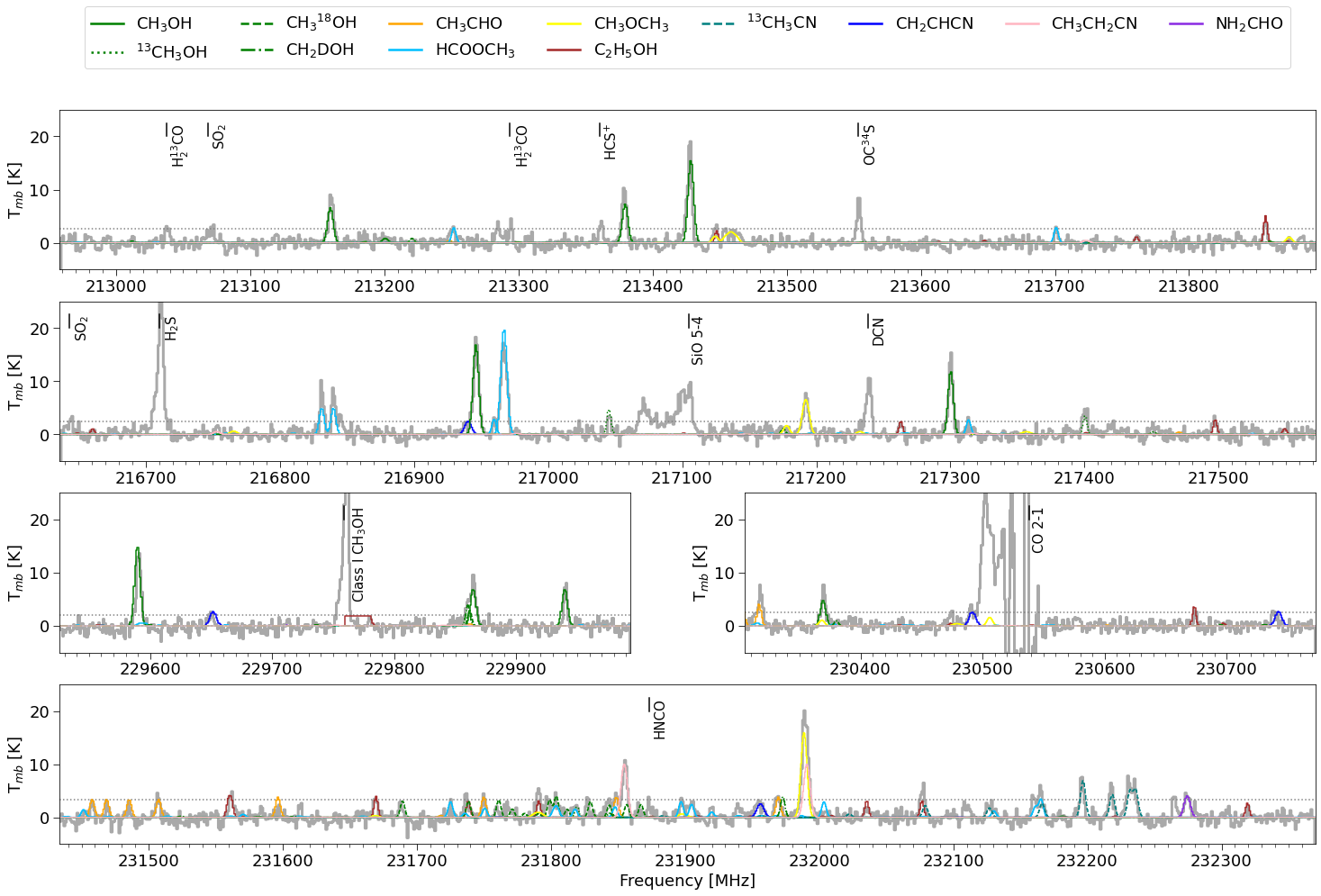}
\caption{The same as Figure \ref{fig:spectrum:G10.32C1} except for G10.34-0.14 C2. Check Figure \ref{fig:chnmap:G10.34} for the CH$_3$OH emission peak position from which the spectra were extracted.}
\label{fig:spectrum:G10.34C2}
\end{figure*}

\begin{figure*}
  \centering
\includegraphics[clip,width=0.7\textwidth,keepaspectratio]{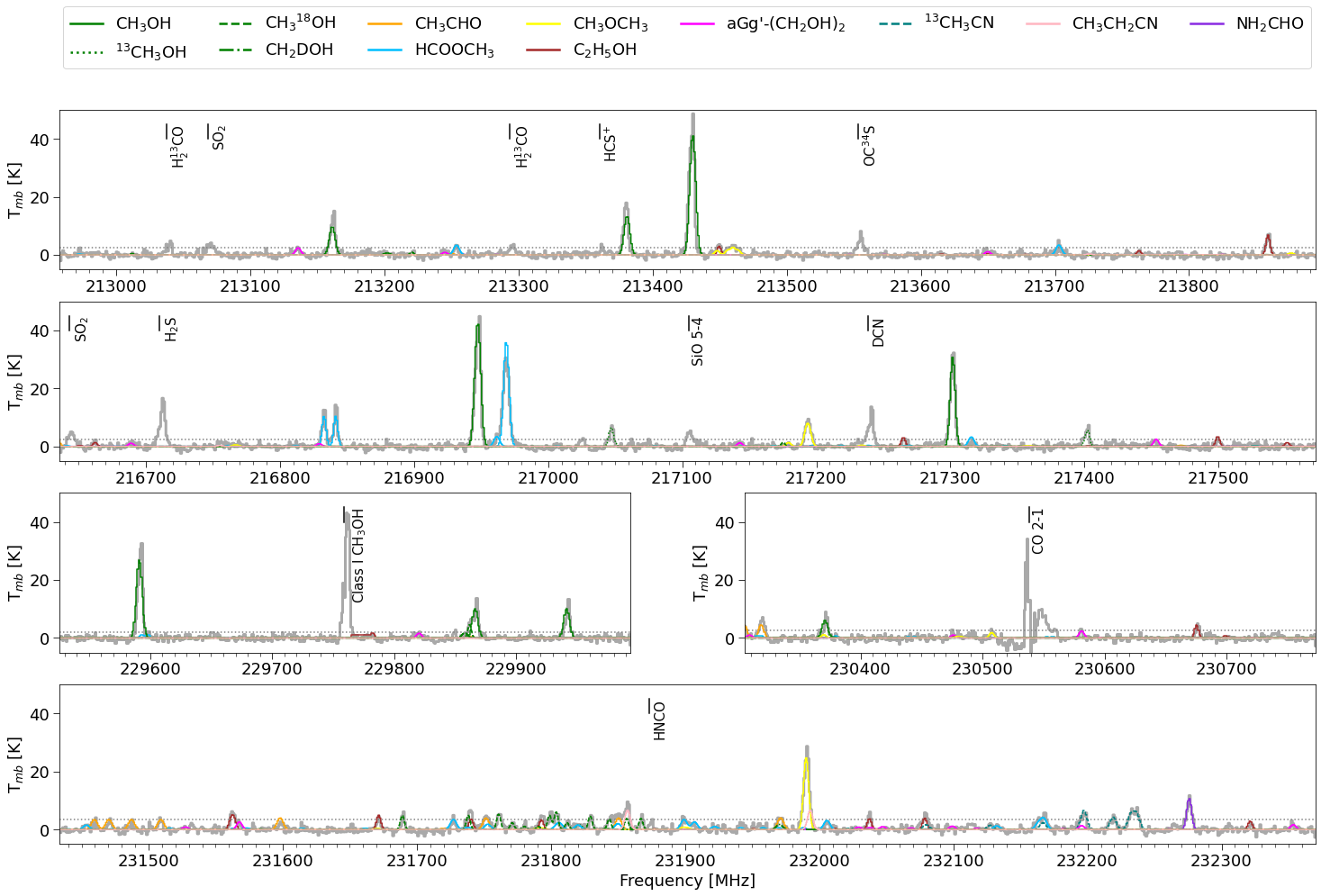}
\caption{The same as Figure \ref{fig:spectrum:G10.32C1} except for G10.34-0.14 C3. Check Figure \ref{fig:chnmap:G10.34} for the CH$_3$OH emission peak position from which the spectra were extracted.}
\label{fig:spectrum:G10.34C3}
\end{figure*}

\begin{figure*}
  \centering
\includegraphics[clip,width=0.7\textwidth,keepaspectratio]{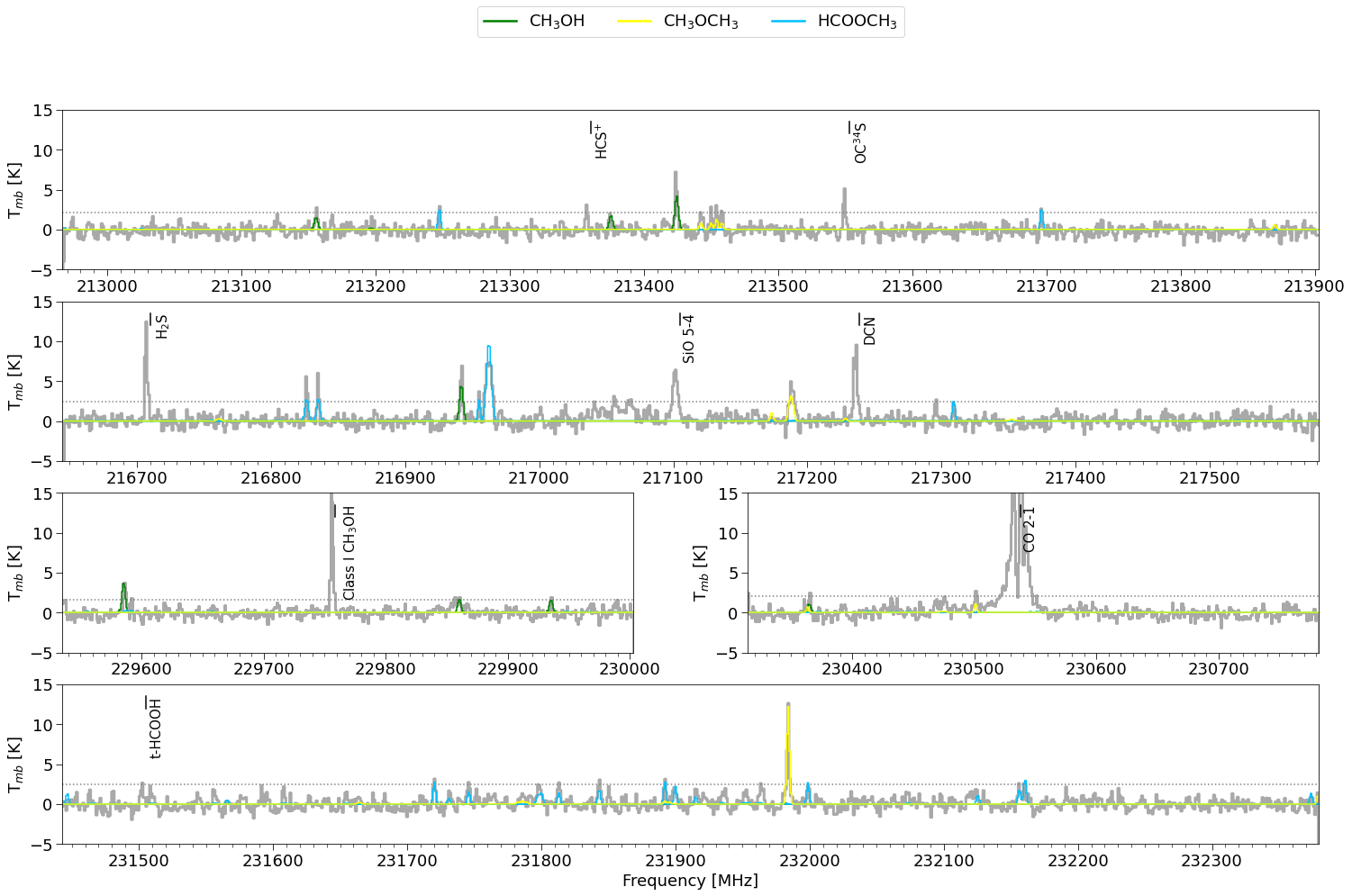}
\caption{The same as Figure \ref{fig:spectrum:G10.32C1} except for G18.34+1.78 C1. Check Figure \ref{fig:chnmap:G18.34} for the CH$_3$OH emission peak position from which the spectra were extracted.}
\label{fig:spectrum:G18.34C1}
\end{figure*}

\begin{figure*}
  \centering
\includegraphics[clip,width=0.7\textwidth,keepaspectratio]{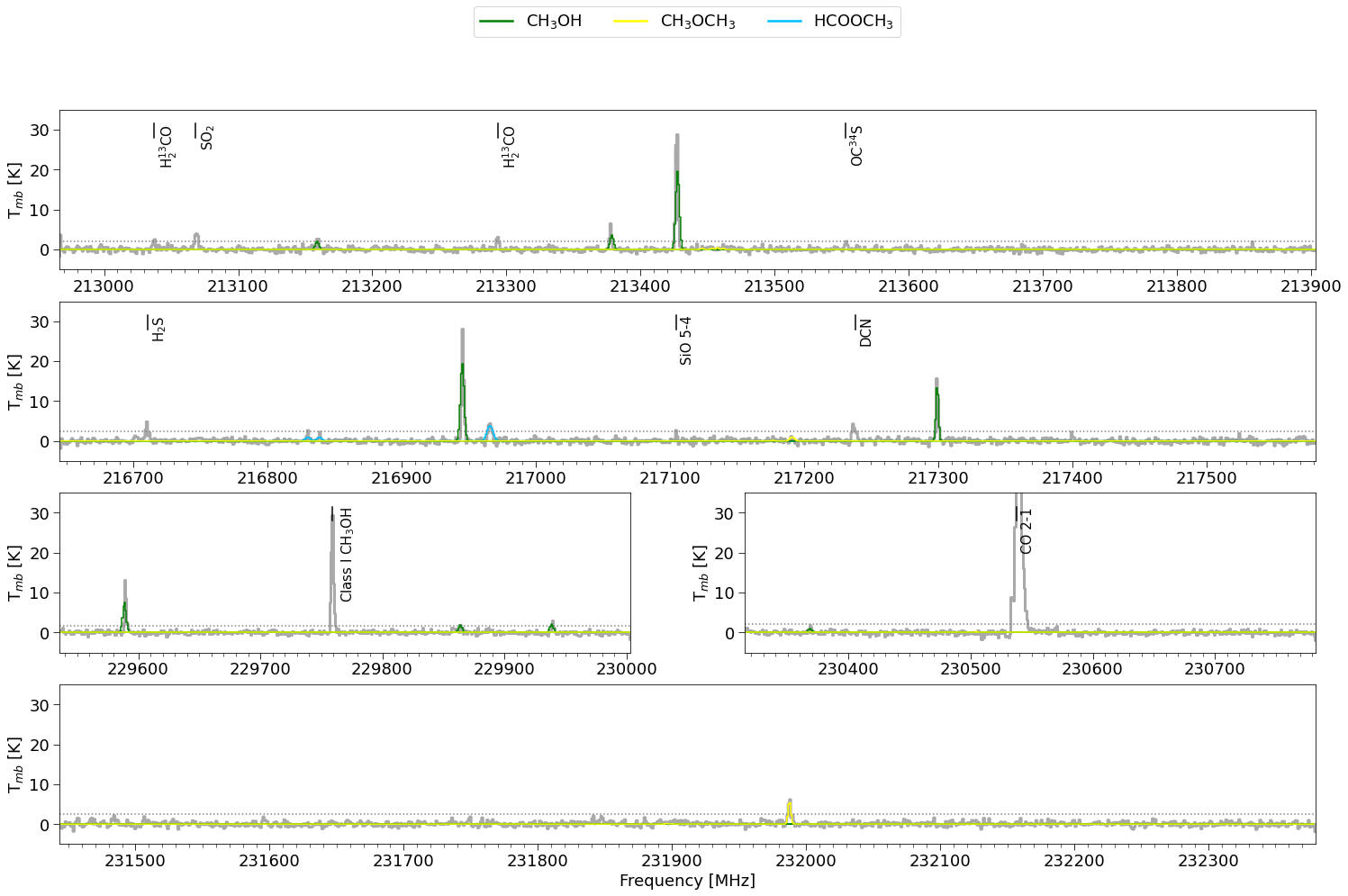}
\caption{The same as Figure \ref{fig:spectrum:G10.32C1} except for G18.34+1.78 C2. Check Figure \ref{fig:chnmap:G18.34} for the CH$_3$OH emission peak position from which the spectra were extracted.}
\label{fig:spectrum:G18.34C2}
\end{figure*}

\begin{figure*}
  \centering
\includegraphics[clip,width=0.7\textwidth,keepaspectratio]{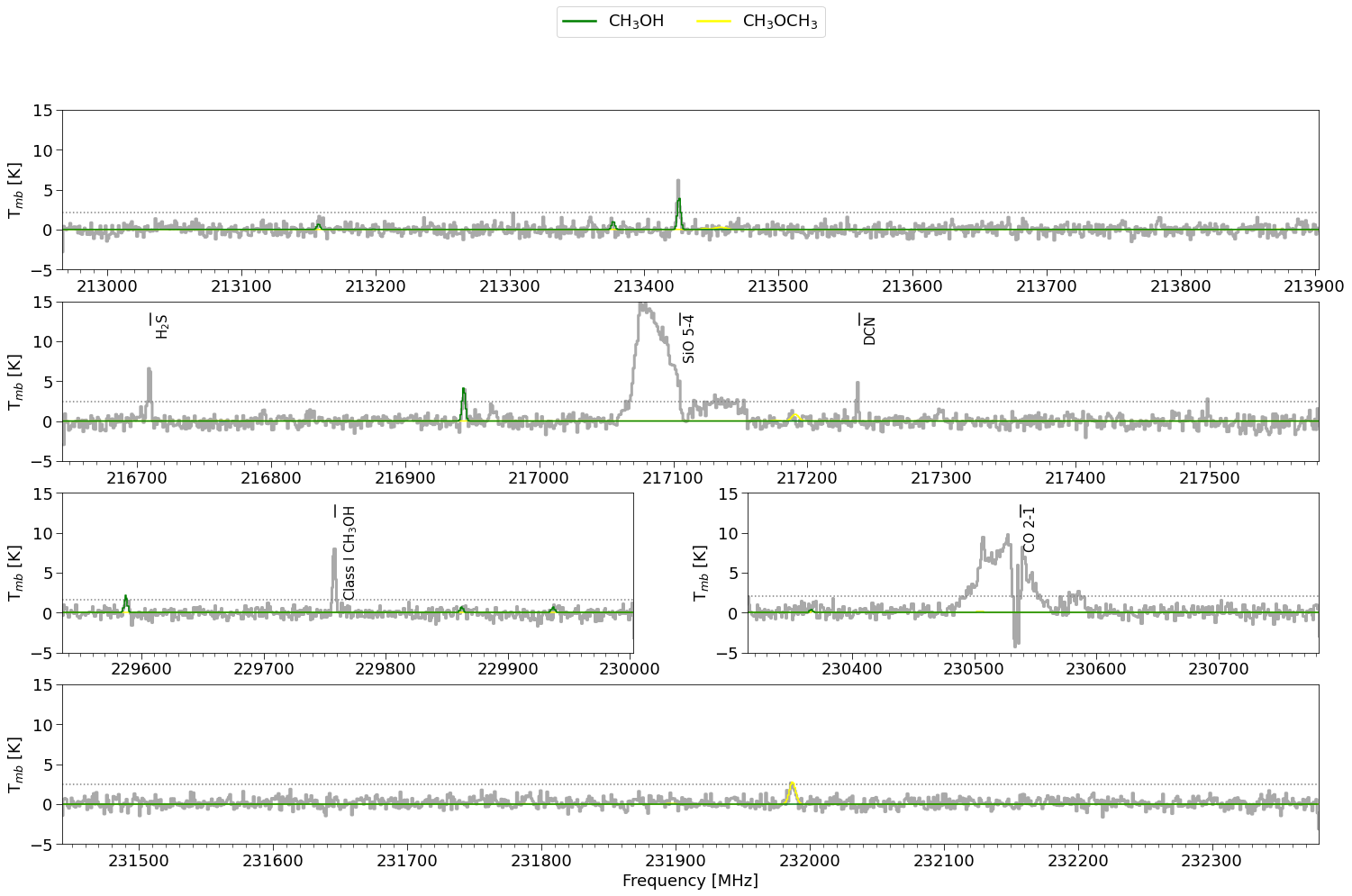}
\caption{The same as Figure \ref{fig:spectrum:G10.32C1} except for G18.34+1.78 C3. Check Figure \ref{fig:chnmap:G18.34} for the CH$_3$OH emission peak position from which the spectra were extracted.}
\label{fig:spectrum:G18.34C3}
\end{figure*}

\begin{figure*}
  \centering
\includegraphics[clip,width=0.7\textwidth,keepaspectratio]{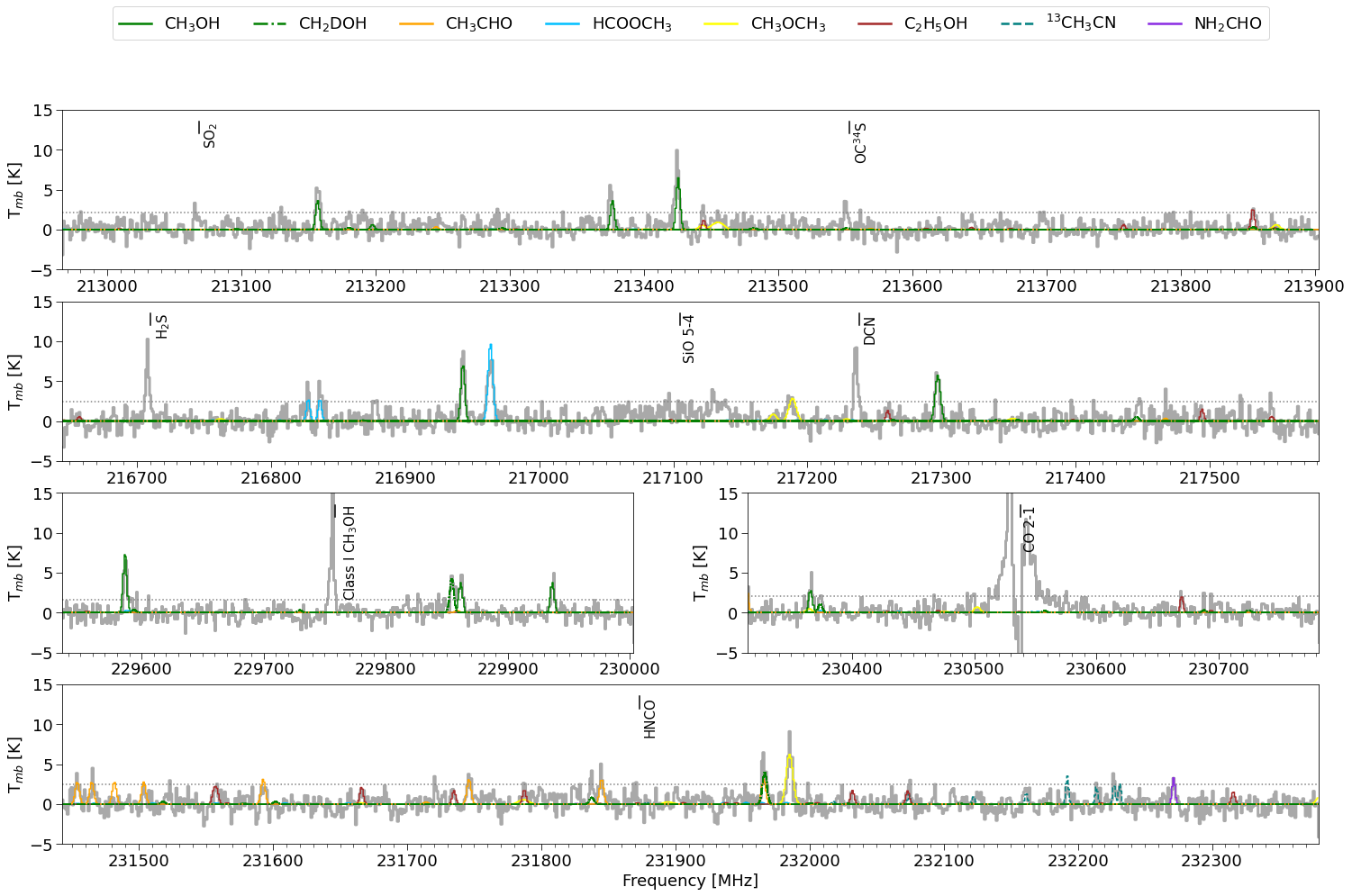}
\caption{The same as Figure \ref{fig:spectrum:G10.32C1} except for G18.34+1.78 C8. Check Figure \ref{fig:chnmap:G18.34} for the CH$_3$OH emission peak position from which the spectra were extracted.}
\label{fig:spectrum:G18.34C8}
\end{figure*}

\begin{figure*}
  \centering
\includegraphics[clip,width=0.7\textwidth,keepaspectratio]{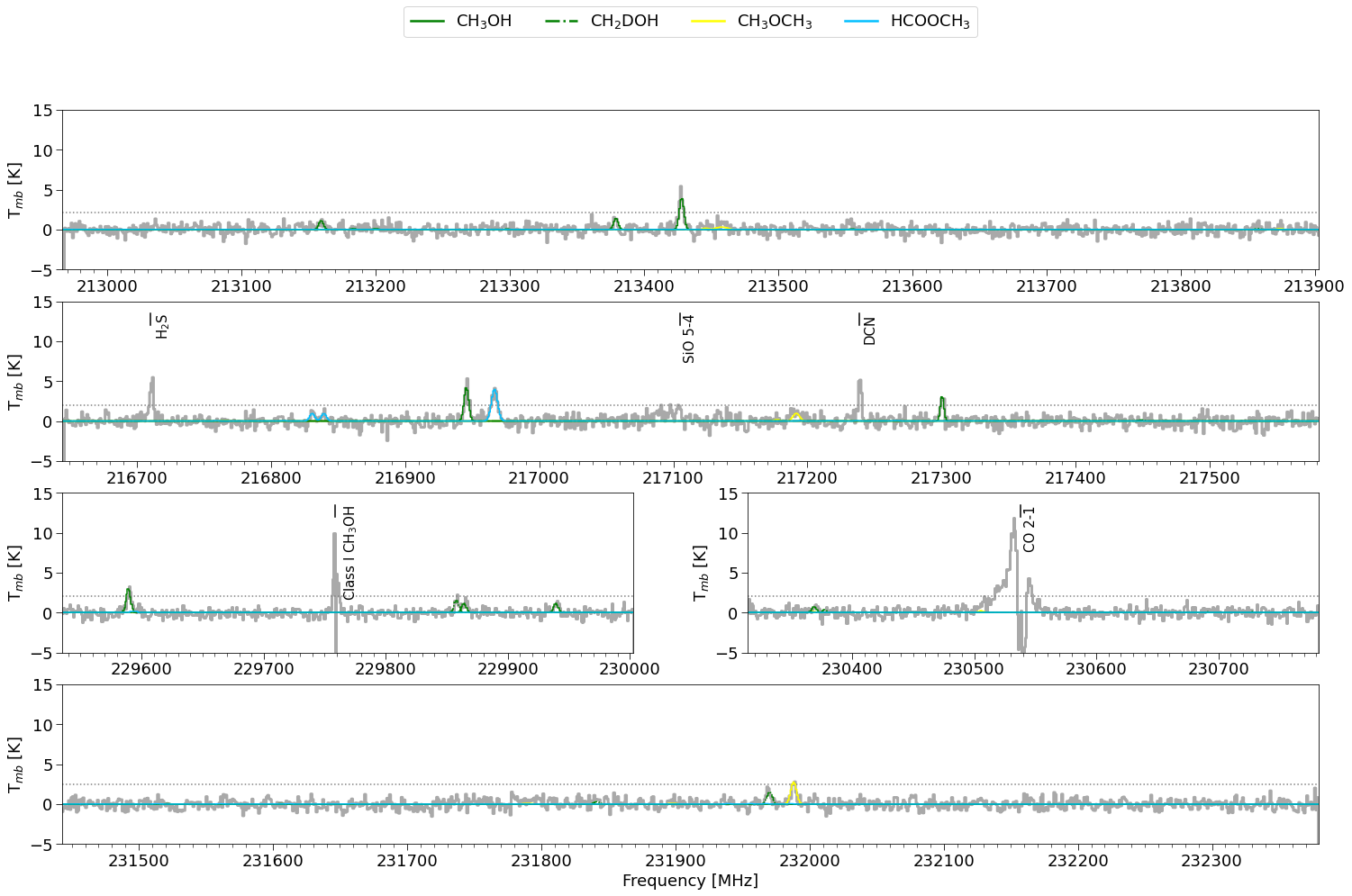}
\caption{The same as Figure \ref{fig:spectrum:G10.32C1} except for G18.34+1.78SW C1. Check Figure \ref{fig:chnmap:G18.34SW} for the CH$_3$OH emission peak position from which the spectra were extracted.}
\label{fig:spectrum:G18.34SWC1}
\end{figure*}

\begin{figure*}
  \centering
\includegraphics[clip,width=0.7\textwidth,keepaspectratio]{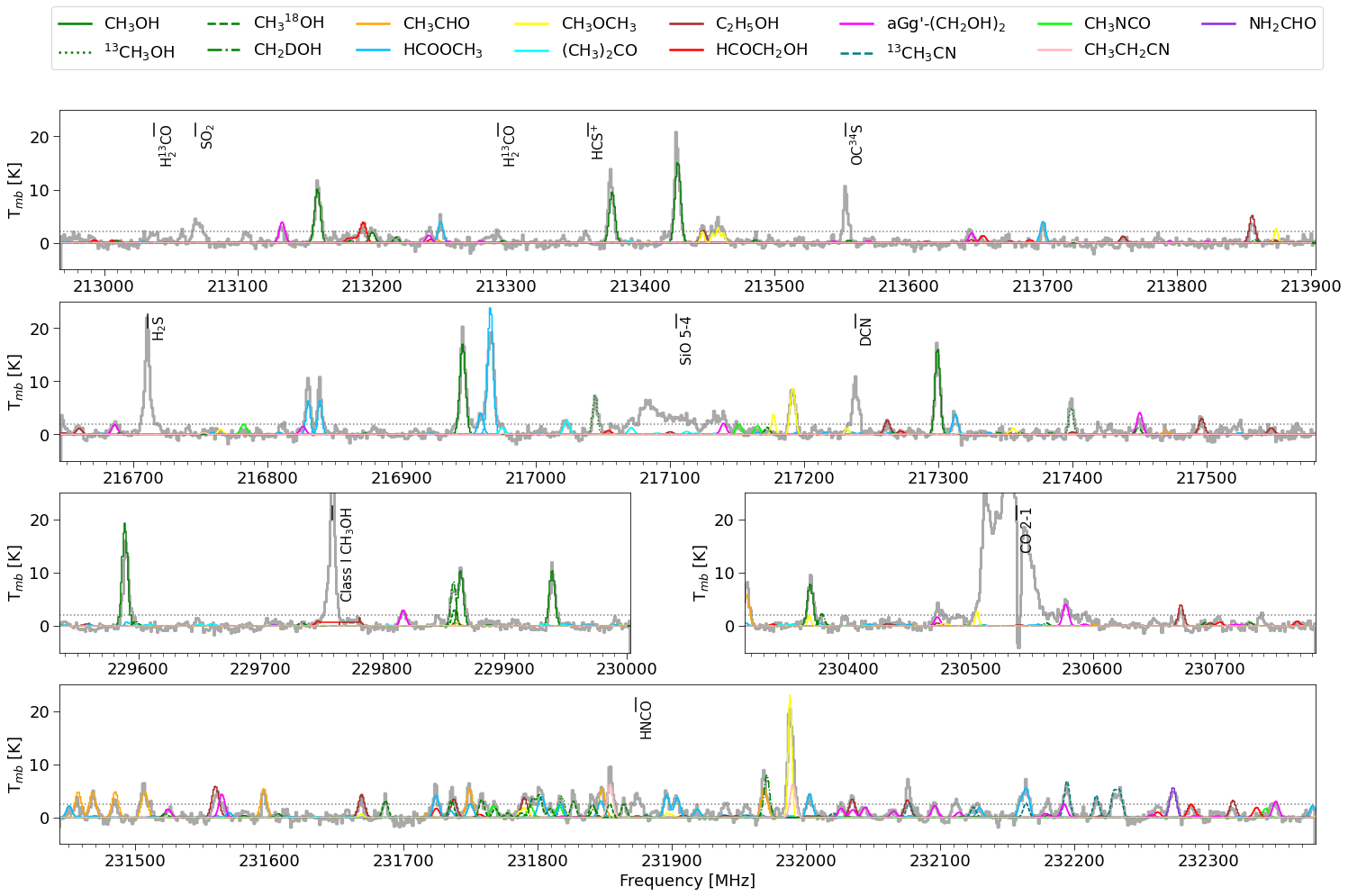}
\caption{The same as Figure \ref{fig:spectrum:G10.32C1} except for G18.34+1.78SW C2. Check Figure \ref{fig:chnmap:G18.34SW} for the CH$_3$OH emission peak position from which the spectra were extracted.}
\label{fig:spectrum:G18.34SWC2}
\end{figure*}

\begin{figure*}
  \centering
\includegraphics[clip,width=0.7\textwidth,keepaspectratio]{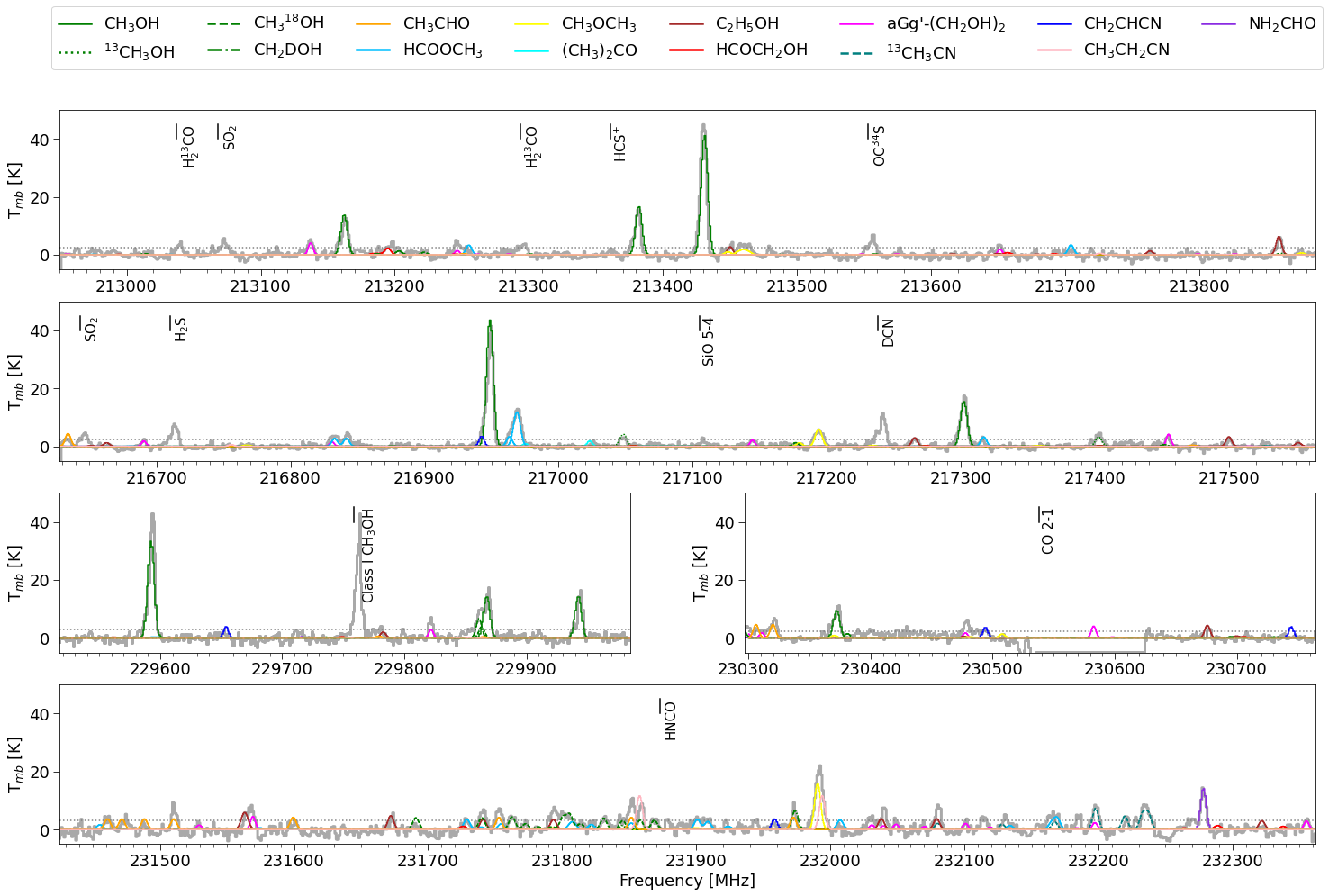}
\caption{The same as Figure \ref{fig:spectrum:G10.32C1} except for G23.43-0.18 C1. Check Figure \ref{fig:chnmap:G23.43} for the CH$_3$OH emission peak position from which the spectra were extracted.}
\label{fig:spectrum:G23.43C1}
\end{figure*}

\begin{figure*}
  \centering
\includegraphics[clip,width=0.7\textwidth,keepaspectratio]{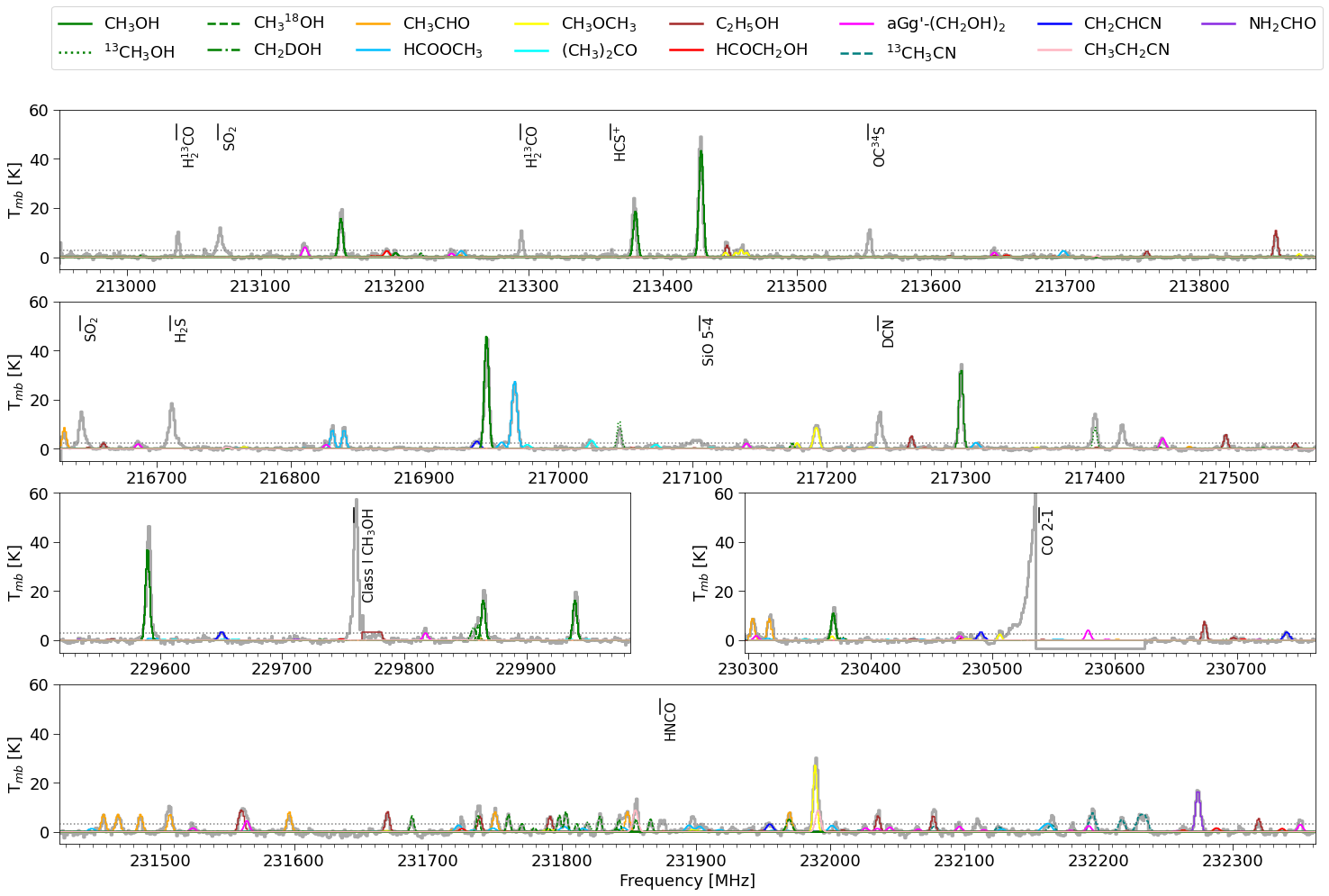}
\caption{The same as Figure \ref{fig:spectrum:G10.32C1} except for G23.43-0.18 C2. Check Figure \ref{fig:chnmap:G23.43} for the CH$_3$OH emission peak position from which the spectra were extracted.}
\label{fig:spectrum:G23.43C2}
\end{figure*}

\begin{figure*}
  \centering
\includegraphics[clip,width=0.7\textwidth,keepaspectratio]{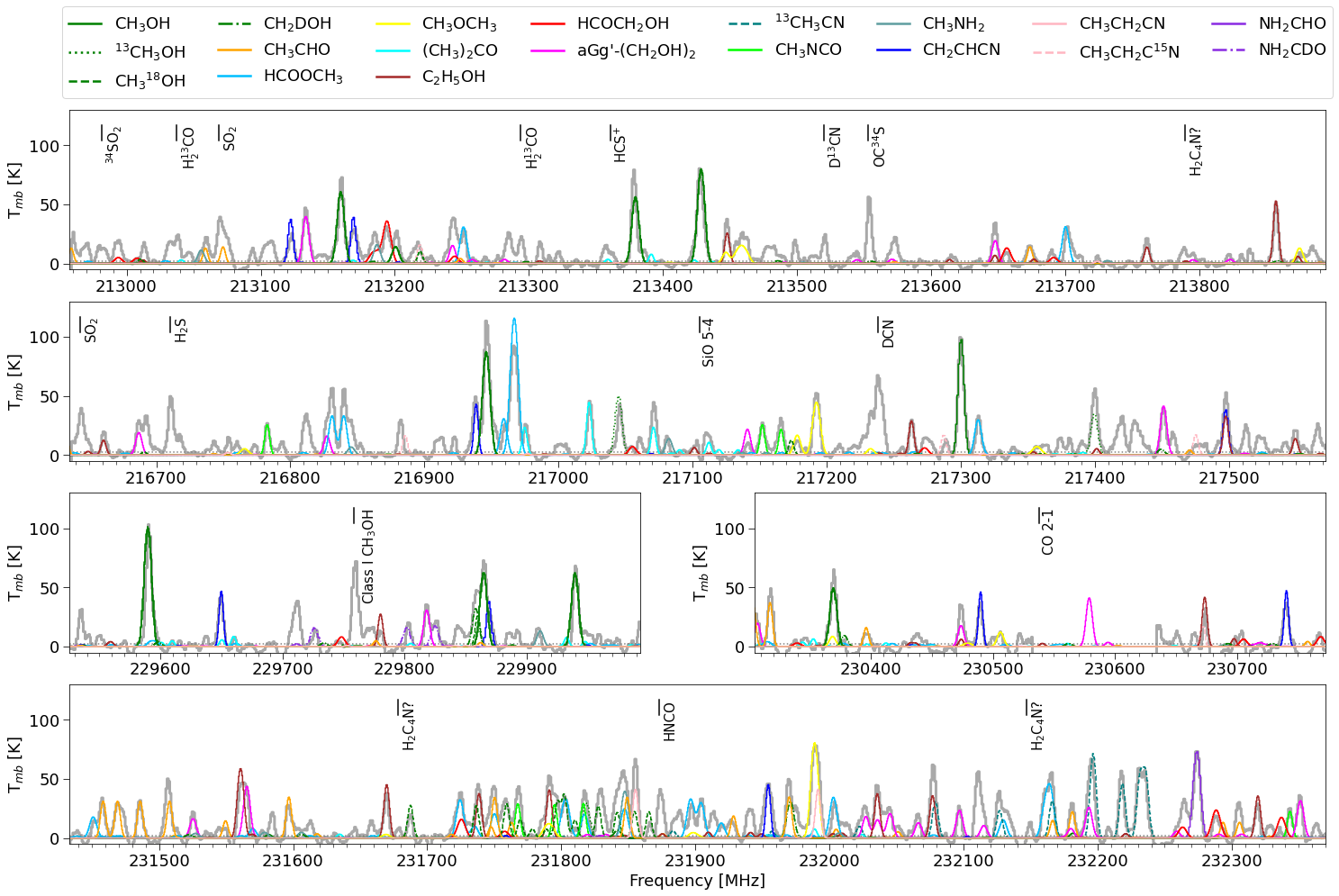}
\caption{The same as Figure \ref{fig:spectrum:G10.32C1} except for G24.33+0.14 C1. Check Figure \ref{fig:chnmap:G24.33} for the CH$_3$OH emission peak position from which the spectra were extracted.}
\label{fig:spectrum:G24.33C1}
\end{figure*}

\begin{figure*}
  \centering
\includegraphics[clip,width=0.7\textwidth,keepaspectratio]{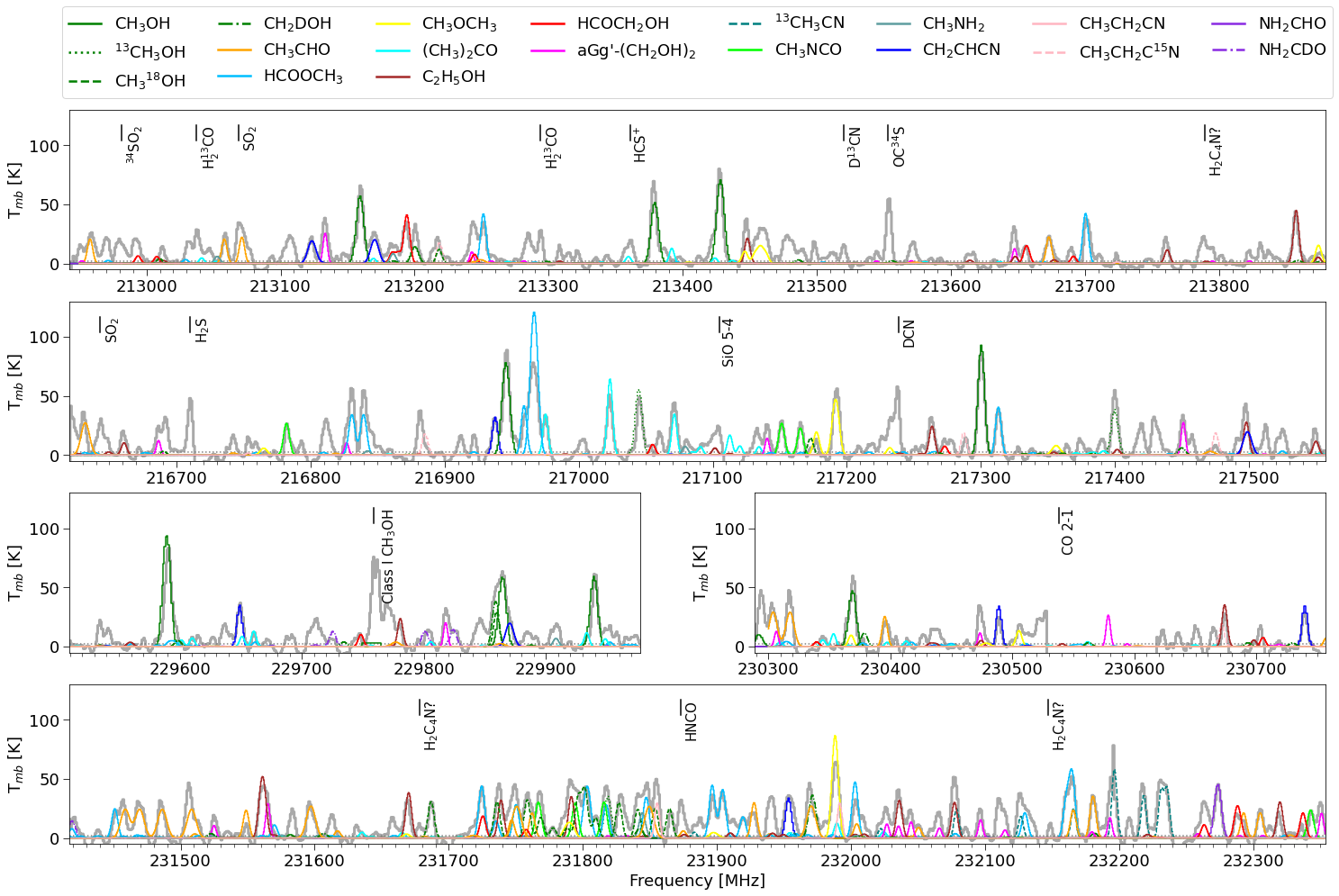}
\caption{The same as Figure \ref{fig:spectrum:G10.32C1} except for G27.36-0.16 C1. Check Figure \ref{fig:chnmap:G27.36} for the CH$_3$OH emission peak position from which the spectra were extracted.}
\label{fig:spectrum:G27.36C1}
\end{figure*}

\begin{figure*}
  \centering
\includegraphics[clip,width=0.7\textwidth,keepaspectratio]{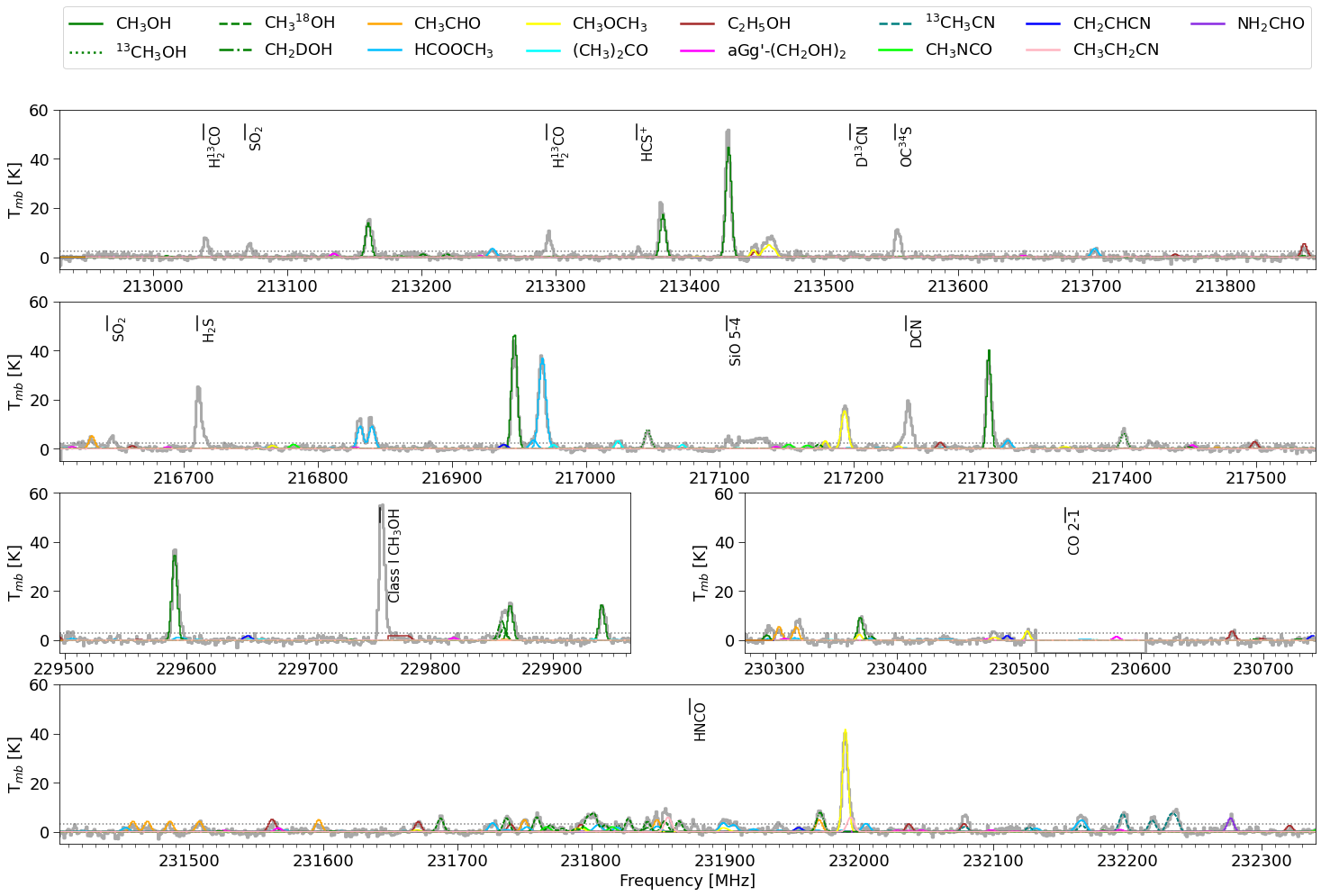}
\caption{The same as Figure \ref{fig:spectrum:G10.32C1} except for G28.37+0.07 C2. Check Figure \ref{fig:chnmap:G28.37} for the CH$_3$OH emission peak position from which the spectra were extracted.}
\label{fig:spectrum:G28.37C2}
\end{figure*}

\begin{figure*}
  \centering
\includegraphics[clip,width=0.7\textwidth,keepaspectratio]{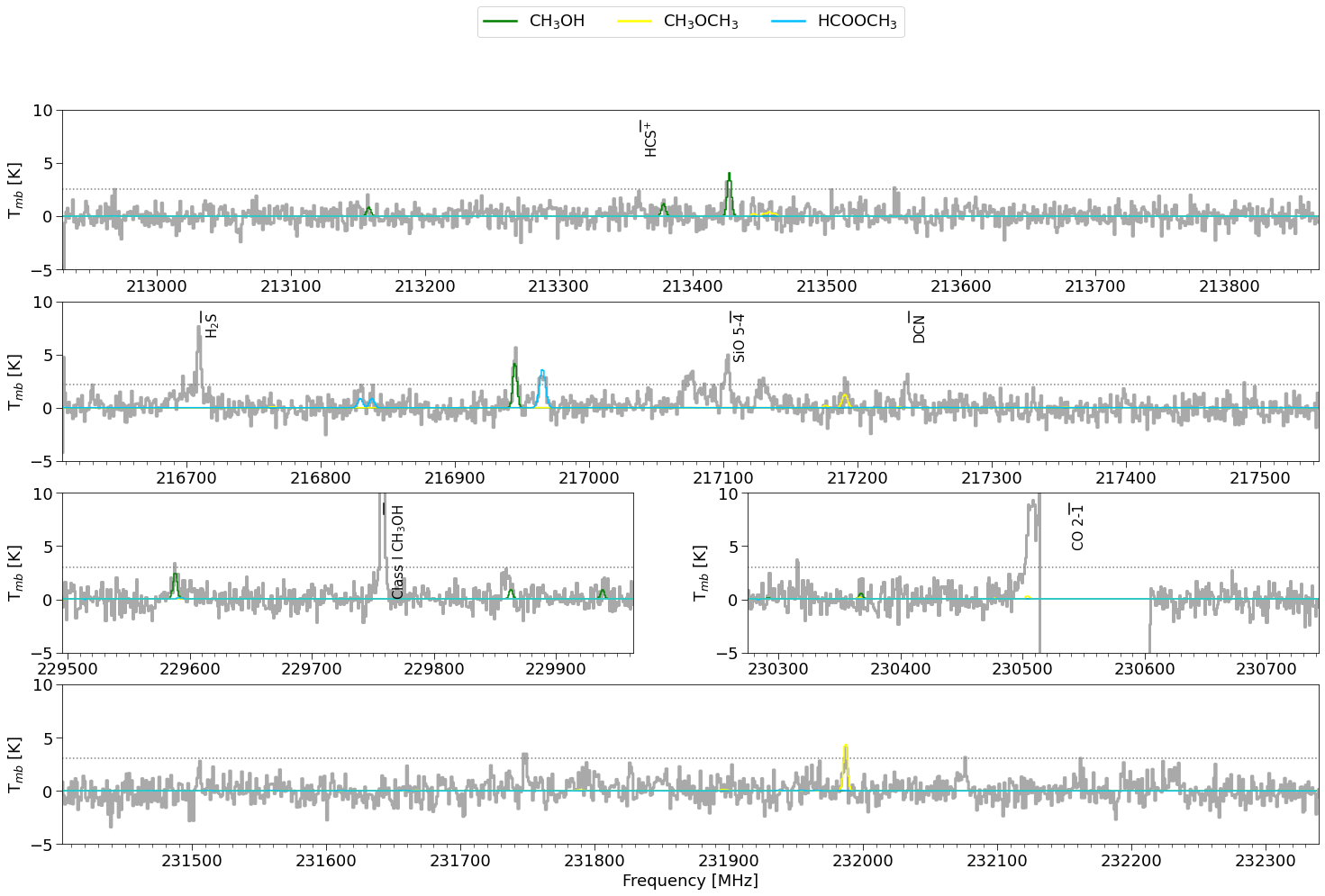}
\caption{The same as Figure \ref{fig:spectrum:G10.32C1} except for G28.37+0.07 C3. Check Figure \ref{fig:chnmap:G28.37} for the CH$_3$OH emission peak position from which the spectra were extracted.}
\label{fig:spectrum:G28.37C3}
\end{figure*}

\begin{figure*}
  \centering
\includegraphics[clip,width=0.7\textwidth,keepaspectratio]{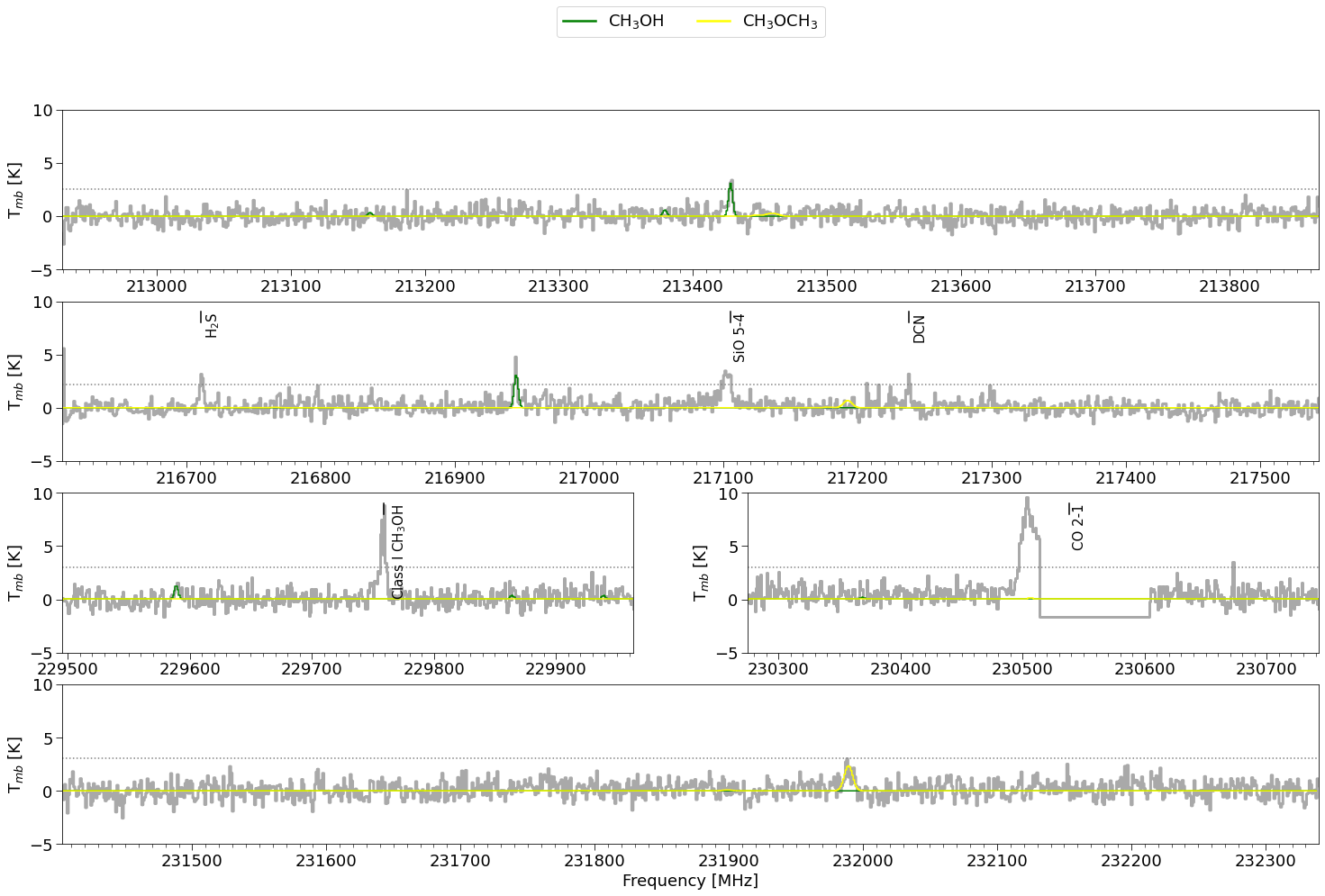}
\caption{The same as Figure \ref{fig:spectrum:G10.32C1} except for G28.37+0.07 C5. Check Figure \ref{fig:chnmap:G28.37} for the CH$_3$OH emission peak position from which the spectra were extracted.}
\label{fig:spectrum:G28.37C5}
\end{figure*}

\begin{figure*}
  \centering
\includegraphics[clip,width=0.7\textwidth,keepaspectratio]{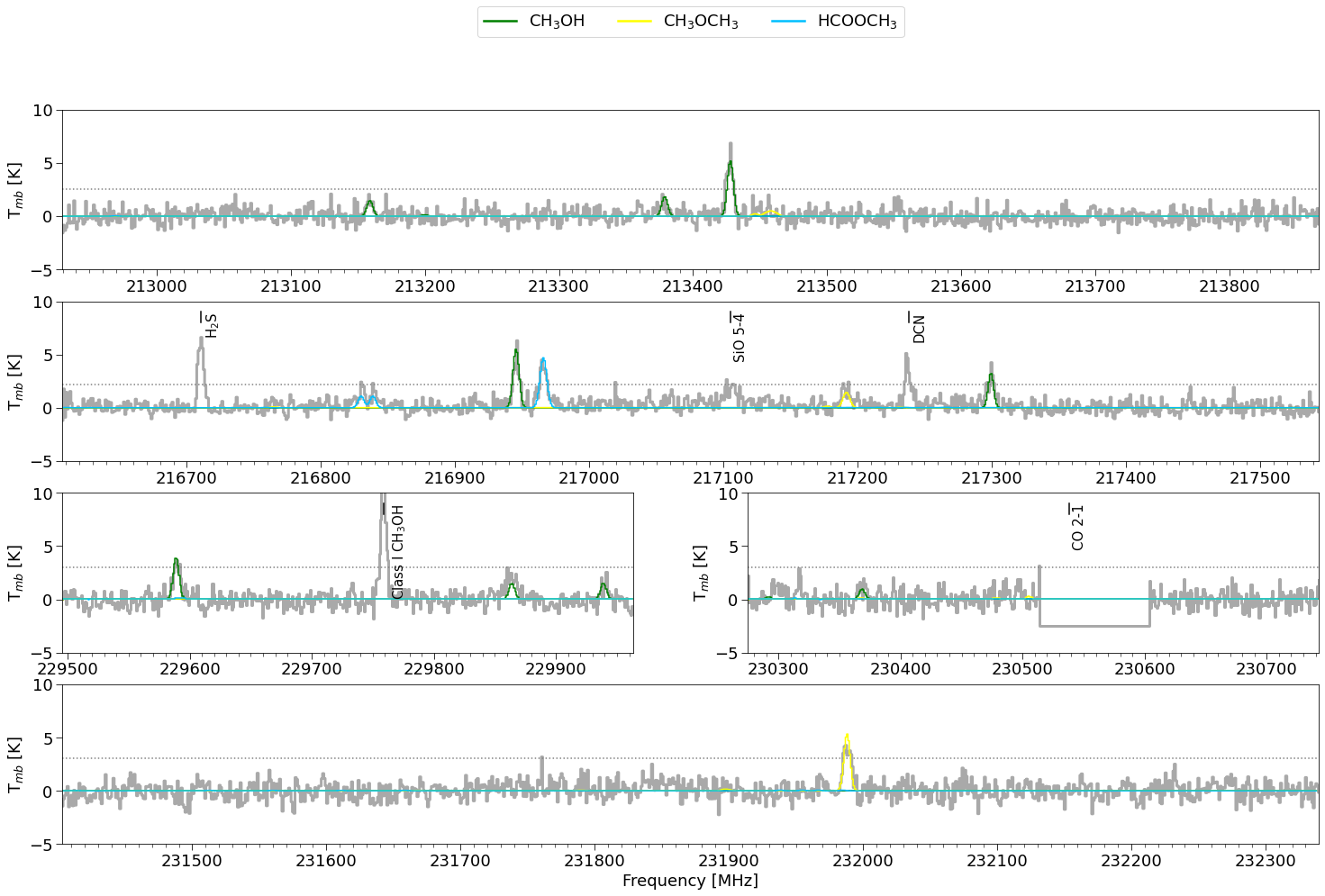}
\caption{The same as Figure \ref{fig:spectrum:G10.32C1} except for G28.37+0.07 C6. Check Figure \ref{fig:chnmap:G28.37} for the CH$_3$OH emission peak position from which the spectra were extracted.}
\label{fig:spectrum:G28.37C6}
\end{figure*}

\begin{figure*}
  \centering
\includegraphics[clip,width=0.7\textwidth,keepaspectratio]{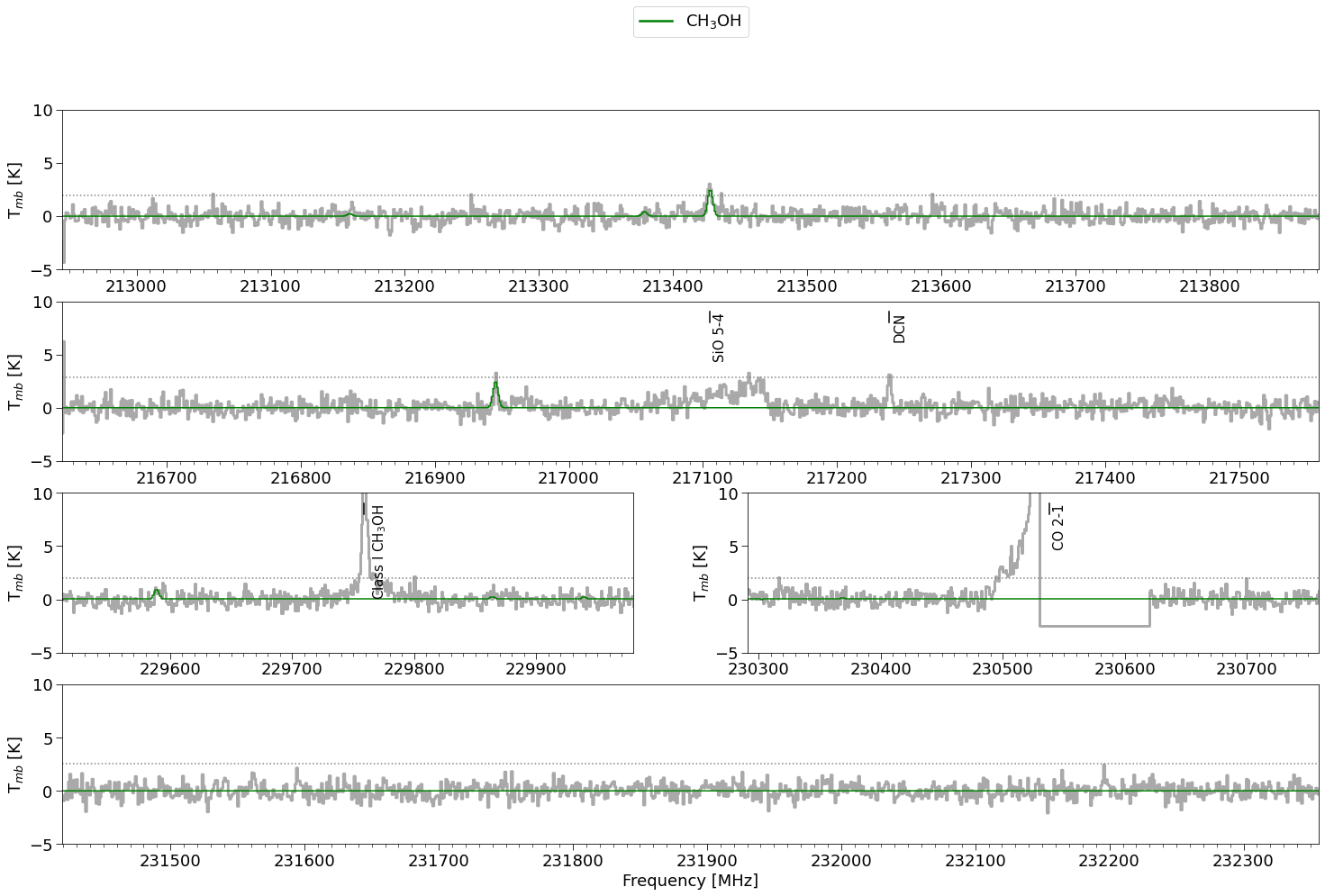}
\caption{The same as Figure \ref{fig:spectrum:G10.32C1} except for G29.91-0.03 C1. Check Figure \ref{fig:chnmap:G29.91} for the CH$_3$OH emission peak position from which the spectra were extracted.}
\label{fig:spectrum:G29.91C1}
\end{figure*}

\begin{figure*}
  \centering
\includegraphics[clip,width=0.7\textwidth,keepaspectratio]{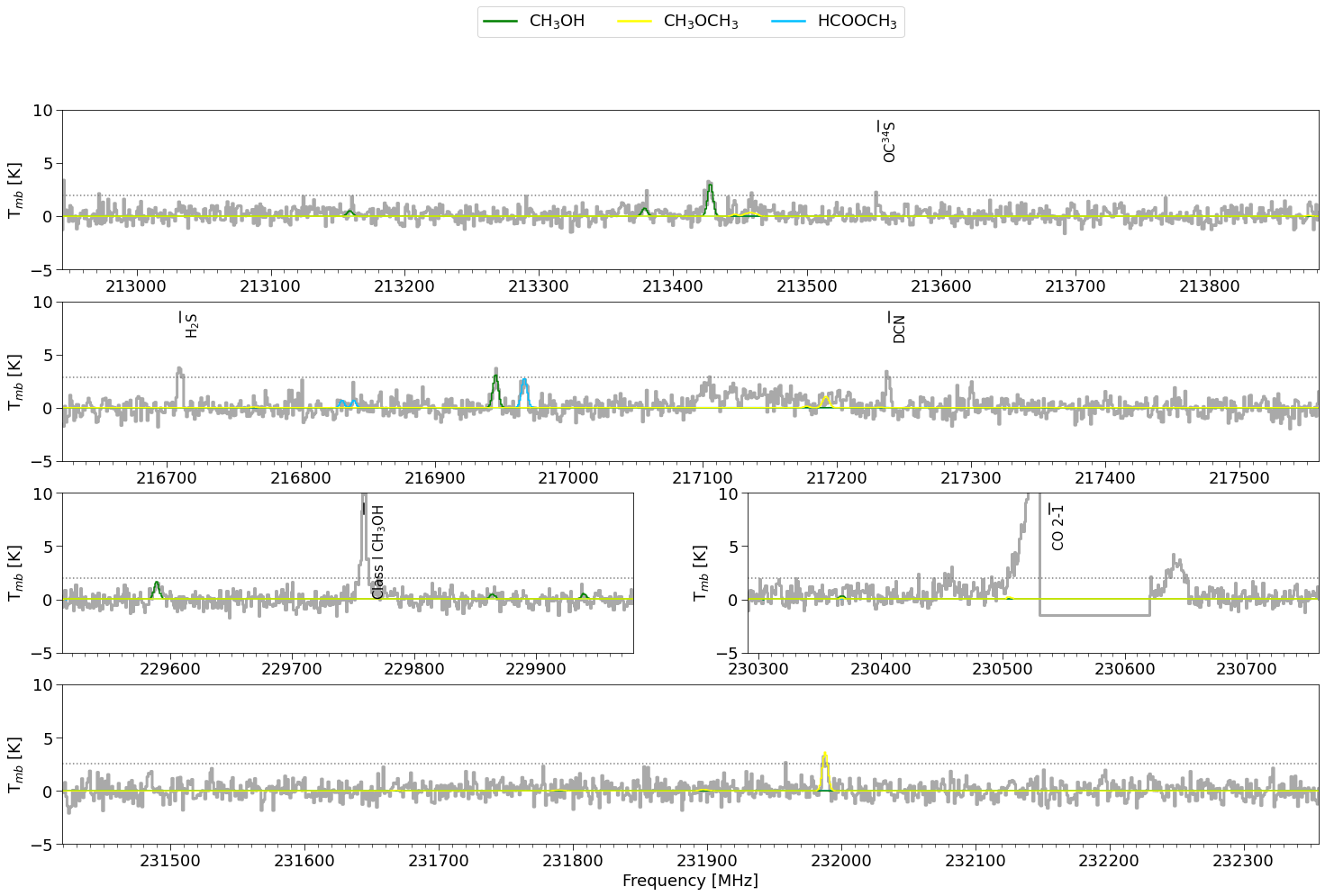}
\caption{The same as Figure \ref{fig:spectrum:G10.32C1} except for G29.91-0.03 C2. Check Figure \ref{fig:chnmap:G29.91} for the CH$_3$OH emission peak position from which the spectra were extracted.}
\label{fig:spectrum:G29.91C2}
\end{figure*}

\begin{figure*}
  \centering
\includegraphics[clip,width=0.7\textwidth,keepaspectratio]{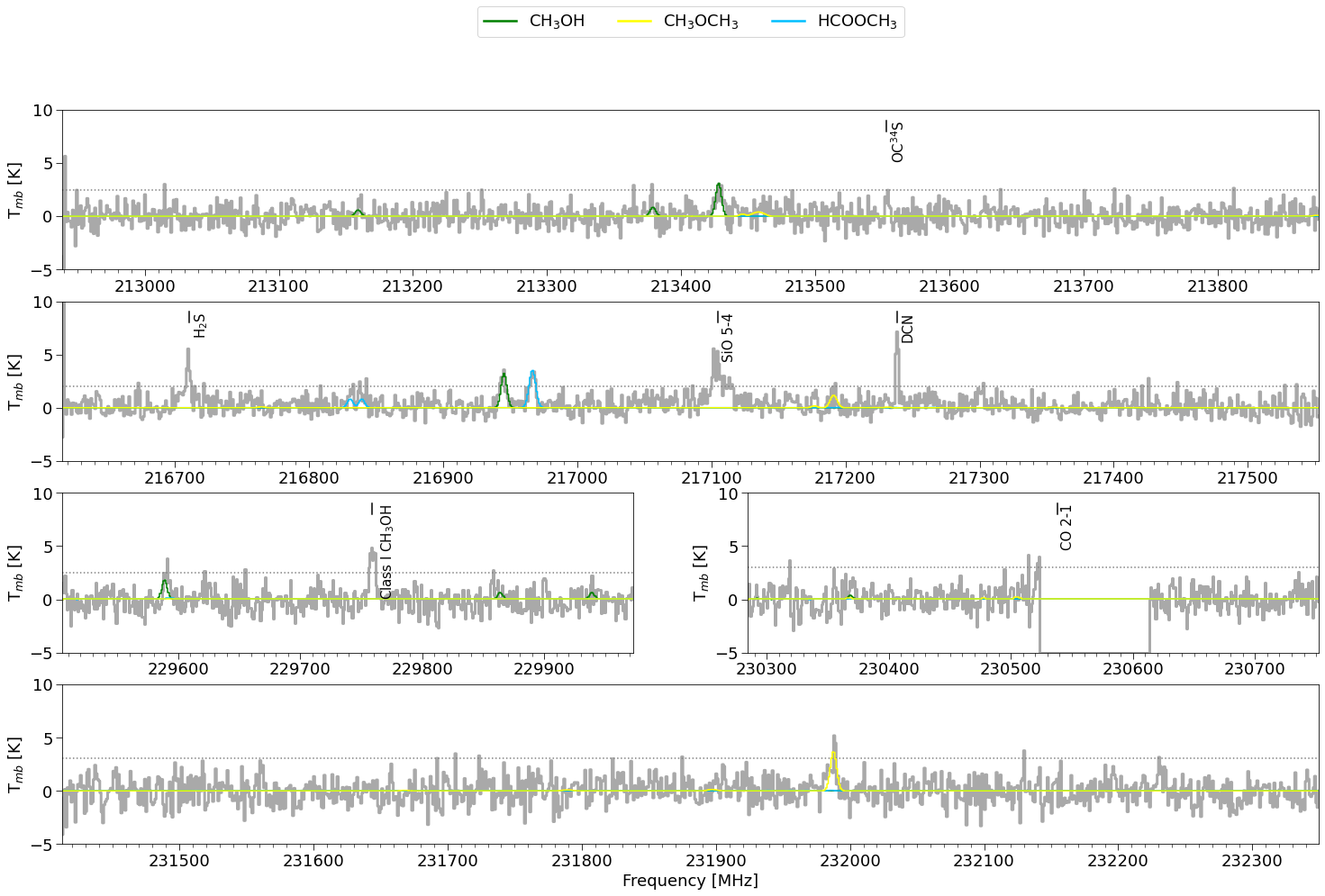}
\caption{The same as Figure \ref{fig:spectrum:G10.32C1} except for G30.70-0.07 C1. Check Figure \ref{fig:chnmap:G30.70} for the CH$_3$OH emission peak position from which the spectra were extracted.}
\label{fig:spectrum:G30.70C1}
\end{figure*}

\begin{figure*}
  \centering
\includegraphics[clip,width=0.7\textwidth,keepaspectratio]{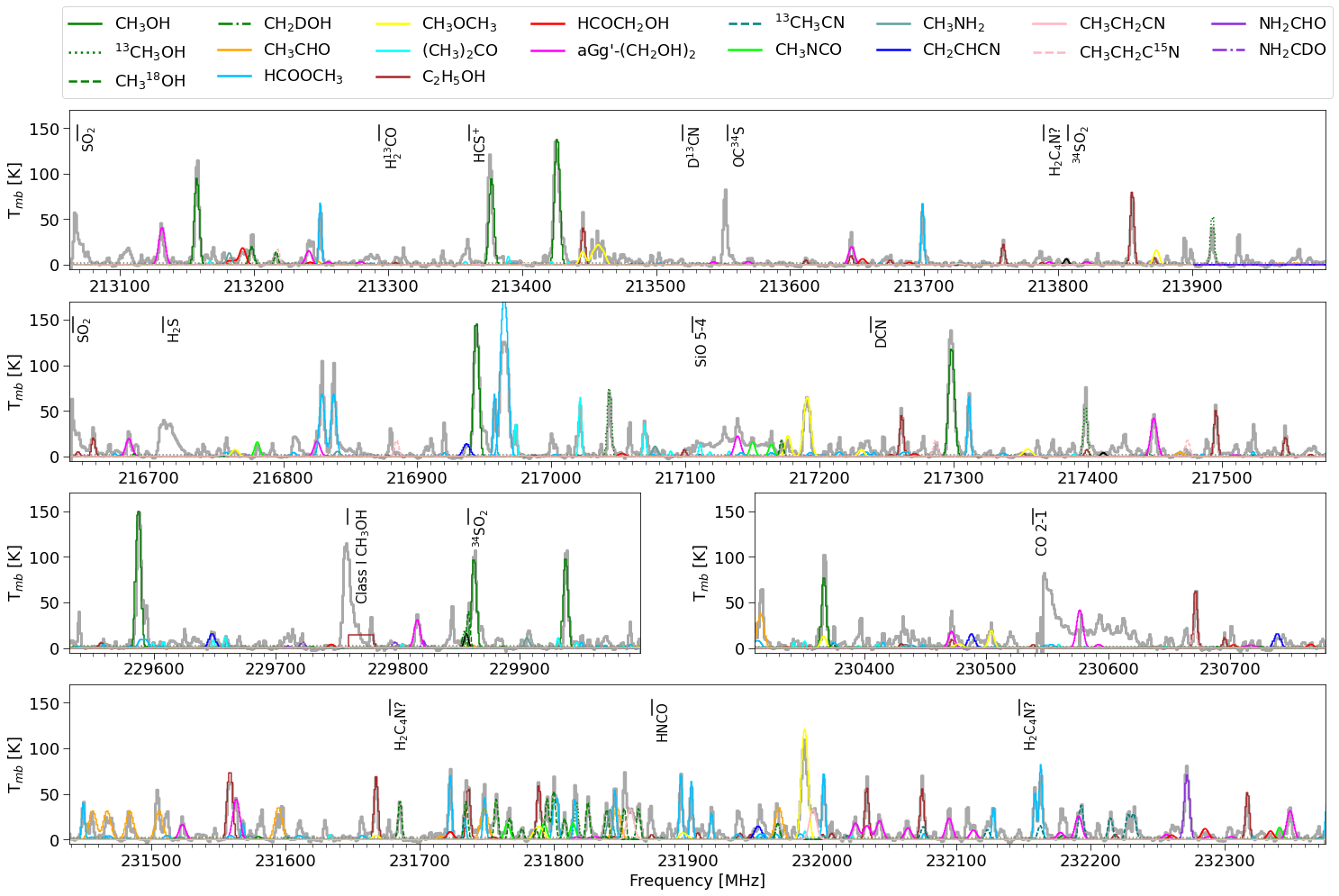}
\caption{The same as Figure \ref{fig:spectrum:G10.32C1} except for G49.49-0.39 C1. Check Figure \ref{fig:chnmap:G49.49} for the CH$_3$OH emission peak position from which the spectra were extracted.}
\label{fig:spectrum:G49.49C1}
\end{figure*}

\begin{figure*}
  \centering
\includegraphics[clip,width=0.7\textwidth,keepaspectratio]{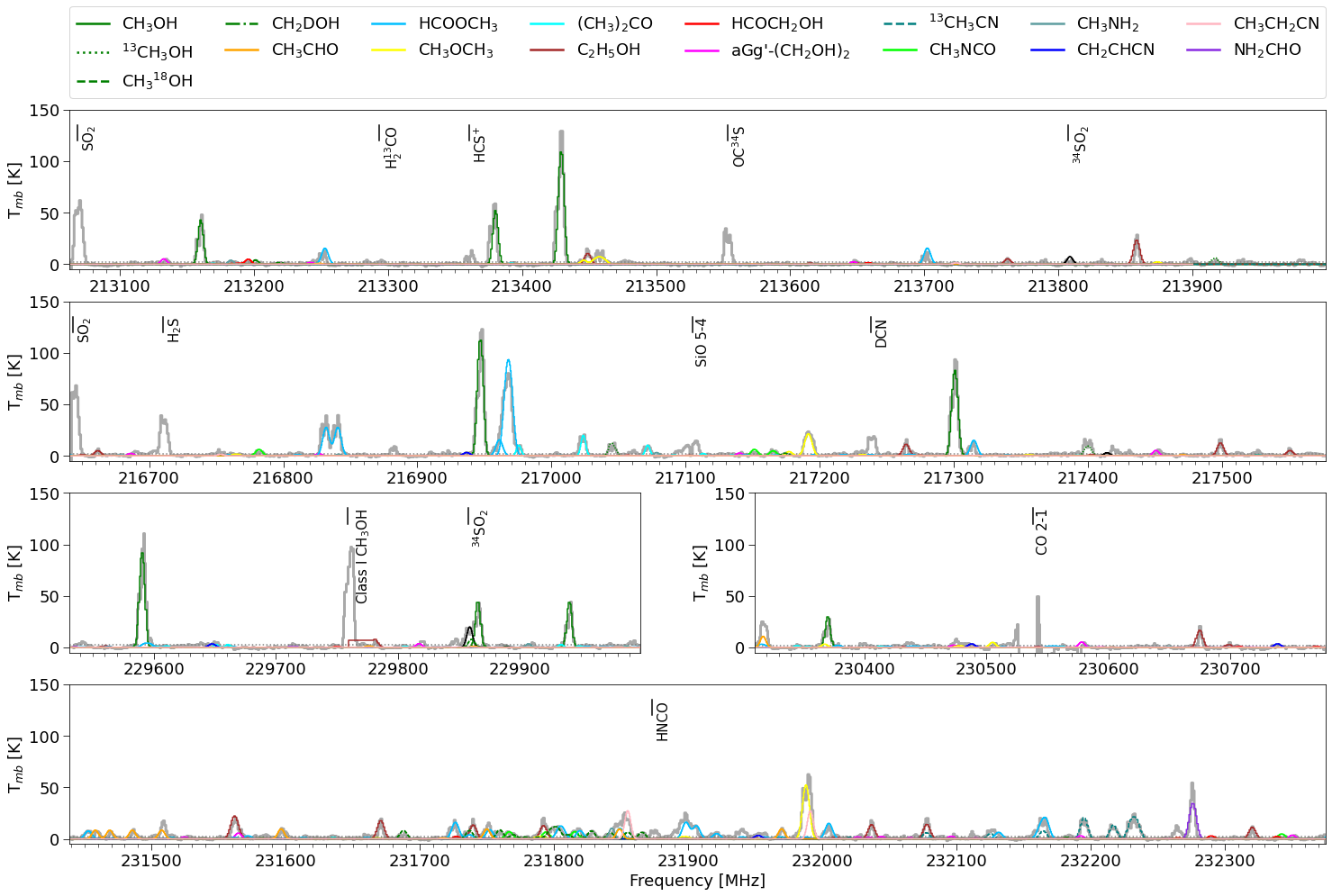}
\caption{The same as Figure \ref{fig:spectrum:G10.32C1} except for G49.49-0.39 C2. Check Figure \ref{fig:chnmap:G49.49} for the CH$_3$OH emission peak position from which the spectra were extracted.}
\label{fig:spectrum:G49.49C2}
\end{figure*}

\begin{figure*}
  \centering
\includegraphics[clip,width=0.7\textwidth,keepaspectratio]{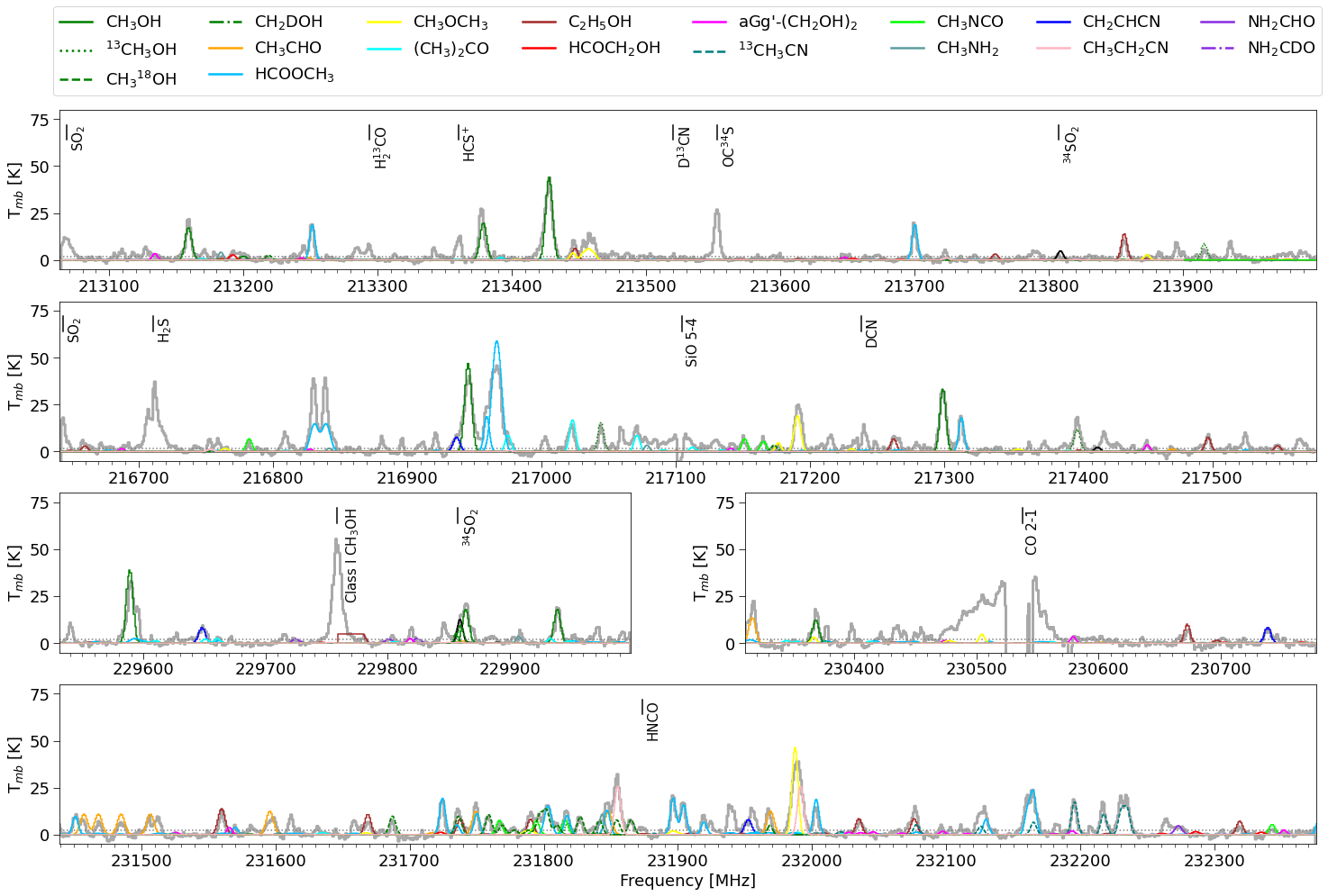}
\caption{The same as Figure \ref{fig:spectrum:G10.32C1} except for G49.49-0.39 C3. Check Figure \ref{fig:chnmap:G49.49} for the CH$_3$OH emission peak position from which the spectra were extracted.}
\label{fig:spectrum:G49.49C3}
\end{figure*}

\begin{figure*}
  \centering
\includegraphics[clip,width=0.7\textwidth,keepaspectratio]{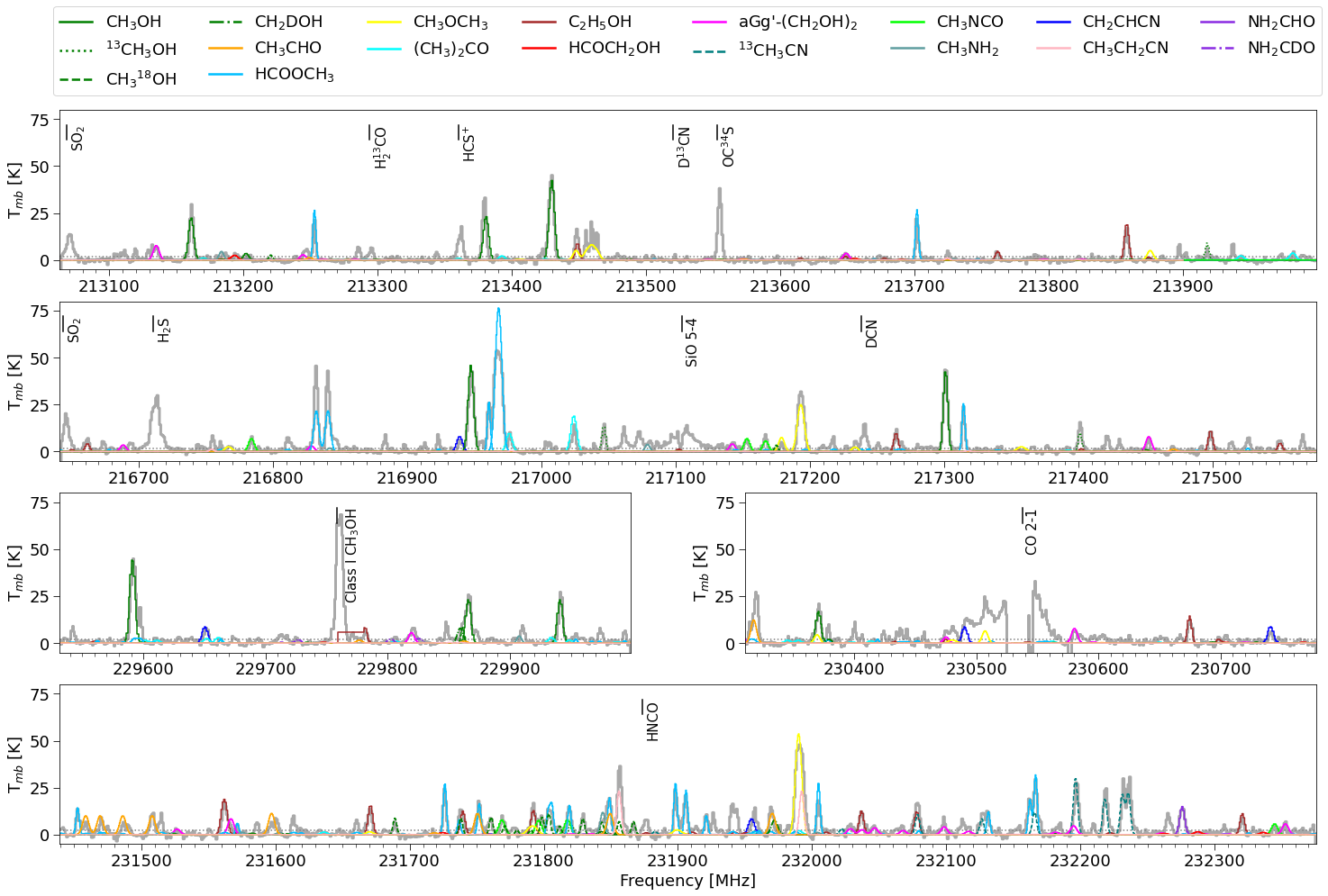}
\caption{The same as Figure \ref{fig:spectrum:G10.32C1} except for G49.49-0.39 C4. Check Figure \ref{fig:chnmap:G49.49} for the CH$_3$OH emission peak position from which the spectra were extracted.}
\label{fig:spectrum:G49.49C4}
\end{figure*}

\begin{figure*}
  \centering
\includegraphics[clip,width=0.7\textwidth,keepaspectratio]{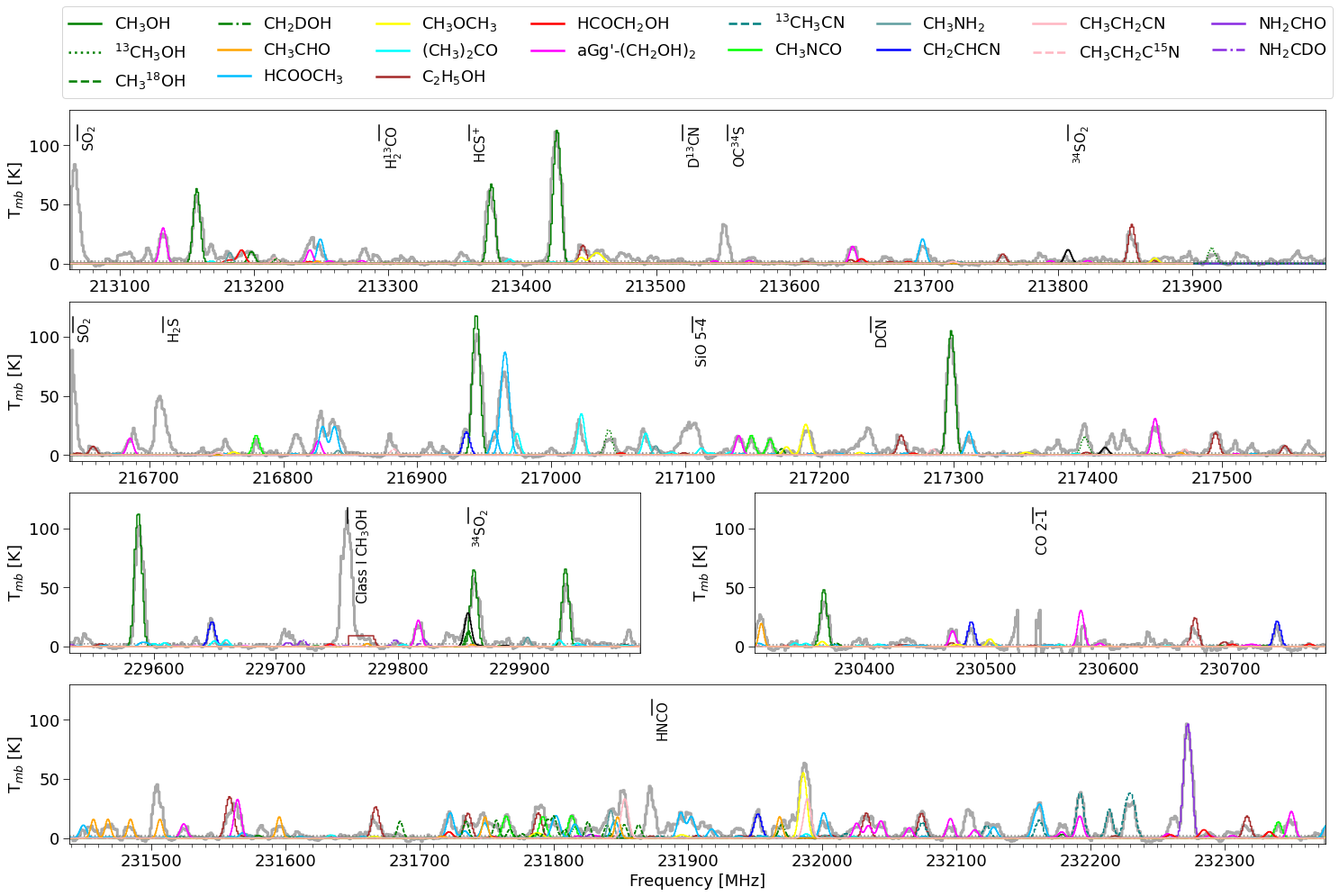}
\caption{The same as Figure \ref{fig:spectrum:G10.32C1} except for G49.49-0.39 C6. Check Figure \ref{fig:chnmap:G49.49} for the CH$_3$OH emission peak position from which the spectra were extracted.}
\label{fig:spectrum:G49.49C6}
\end{figure*}

%% file: s5_appendix.tex
\clearpage
\appendix{}
\restartappendixnumbering

\section{Line identification}
\input{s5_tab1_line_identification}
\clearpage

\section{216.946 GHz Channel maps}
\input{s5_216GHz_channel_maps_figureset}

\clearpage
\section{Rotation diagram}
\input{s3_f4_rotation_diagram}

\clearpage
\section{The effect of physical properties}

\begin{figure*}[htb!]
\includegraphics[clip,width=0.2\textwidth,keepaspectratio]{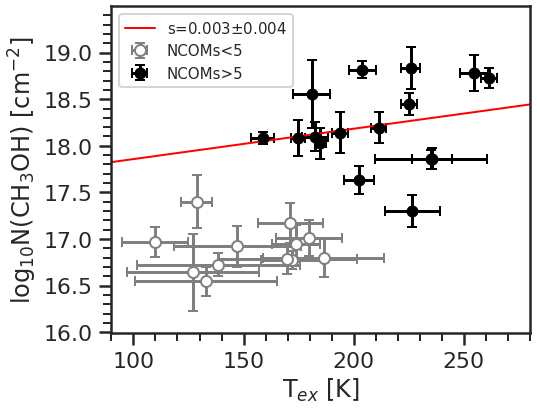}
\includegraphics[clip,width=0.2\textwidth,keepaspectratio]{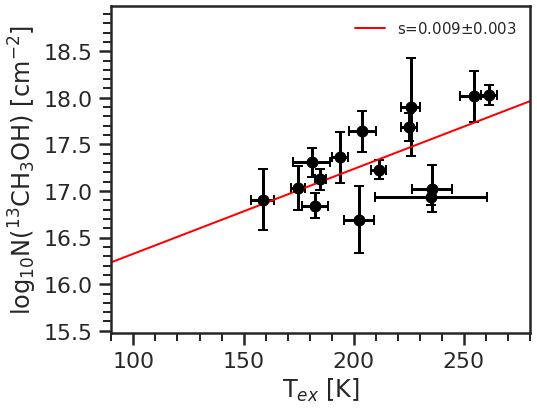}
\includegraphics[clip,width=0.2\textwidth,keepaspectratio]{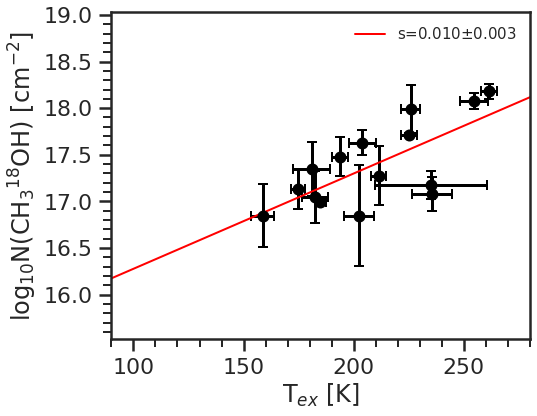}
\includegraphics[clip,width=0.2\textwidth,keepaspectratio]{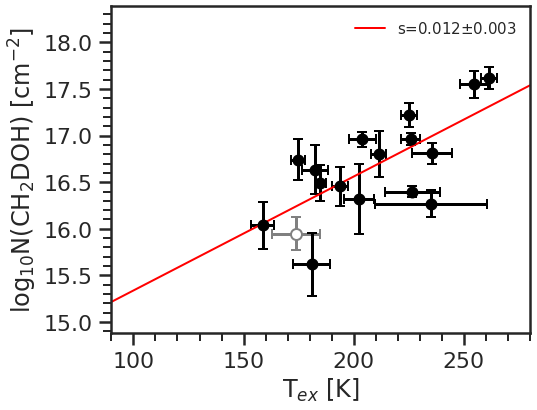}\\
\includegraphics[clip,width=0.2\textwidth,keepaspectratio]{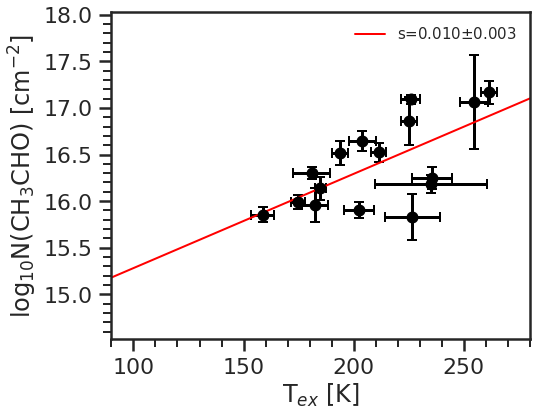}
\includegraphics[clip,width=0.2\textwidth,keepaspectratio]{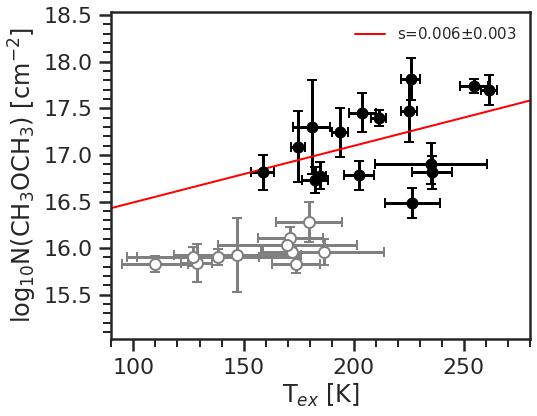}
\includegraphics[clip,width=0.2\textwidth,keepaspectratio]{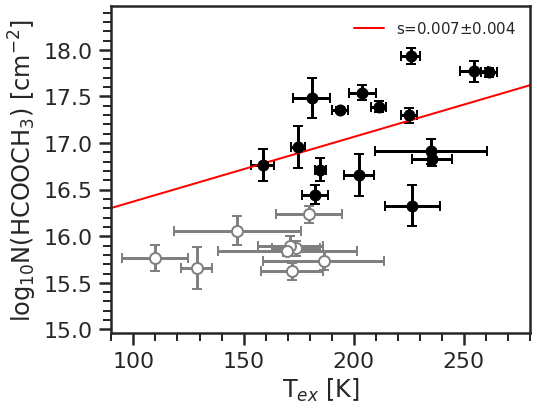}\\
\includegraphics[clip,width=0.2\textwidth,keepaspectratio]{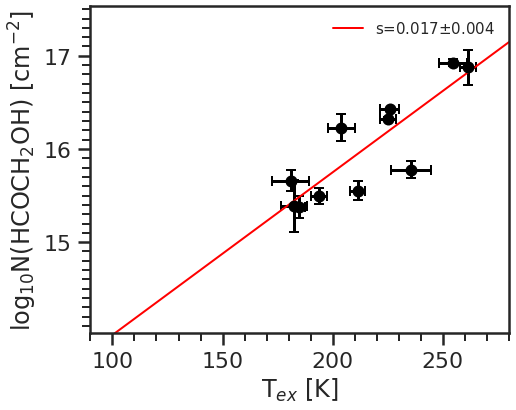}
\includegraphics[clip,width=0.2\textwidth,keepaspectratio]{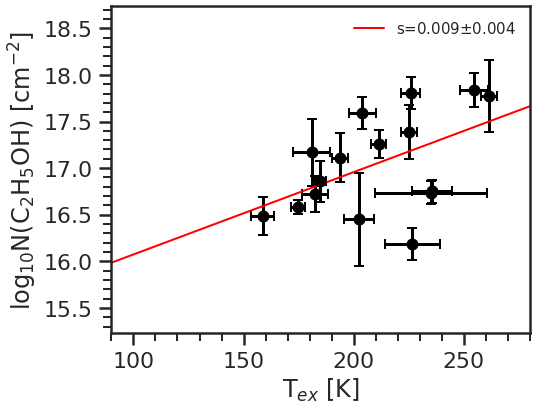}
\includegraphics[clip,width=0.2\textwidth,keepaspectratio]{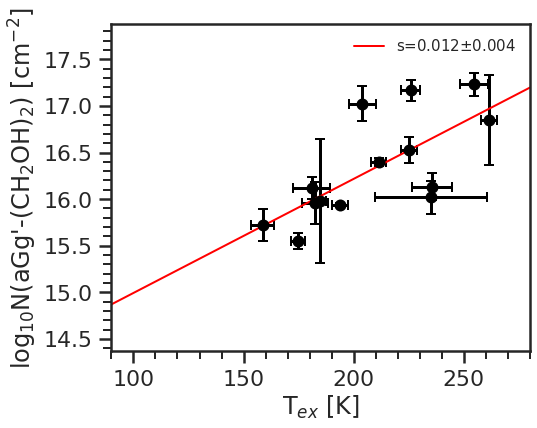}
\includegraphics[clip,width=0.2\textwidth,keepaspectratio]{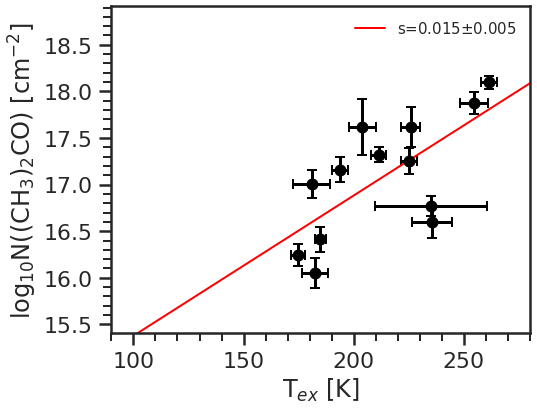}\\
\includegraphics[clip,width=0.2\textwidth,keepaspectratio]{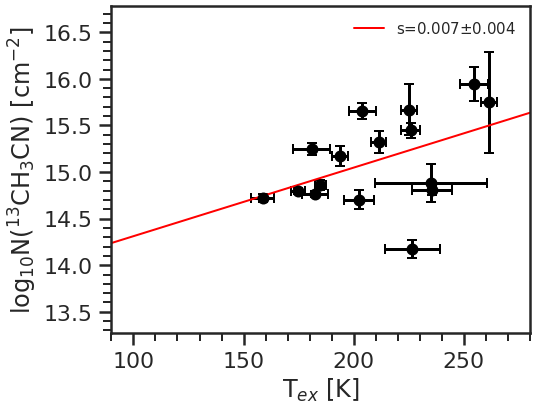}
\includegraphics[clip,width=0.2\textwidth,keepaspectratio]{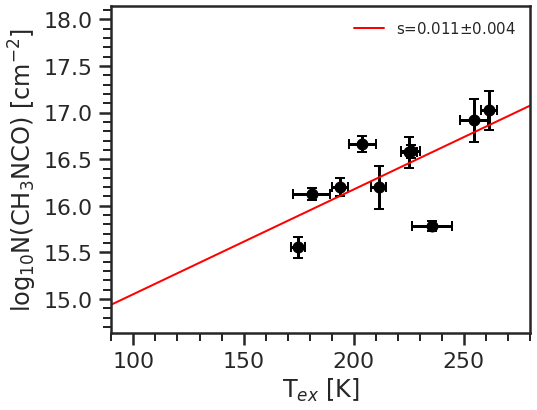}
\includegraphics[clip,width=0.2\textwidth,keepaspectratio]{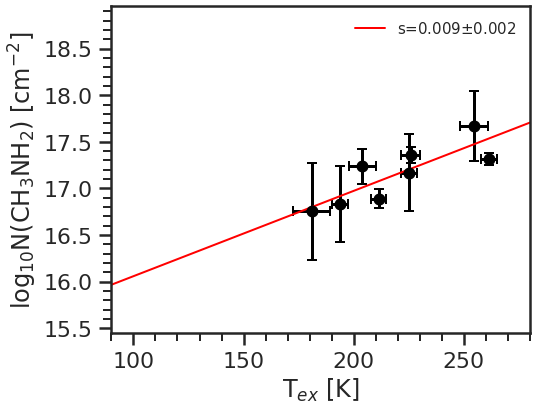}
\includegraphics[clip,width=0.2\textwidth,keepaspectratio]{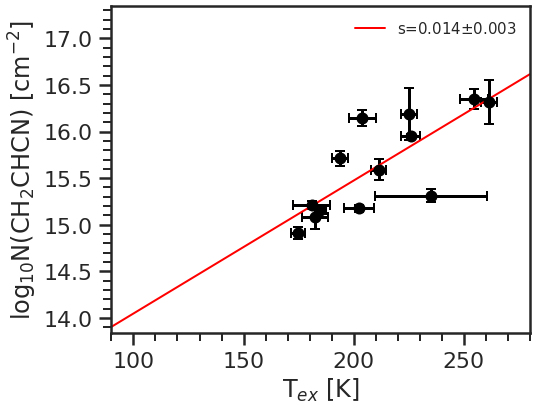}\\
\includegraphics[clip,width=0.2\textwidth,keepaspectratio]{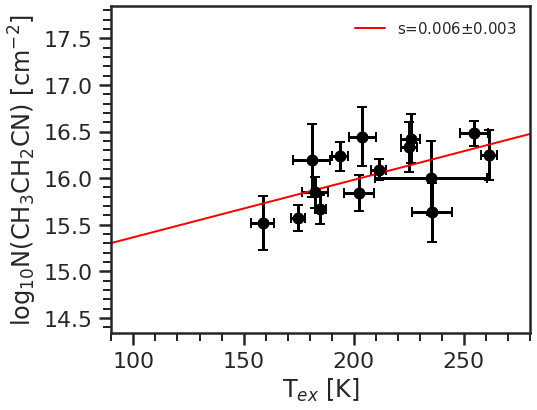}
\includegraphics[clip,width=0.2\textwidth,keepaspectratio]{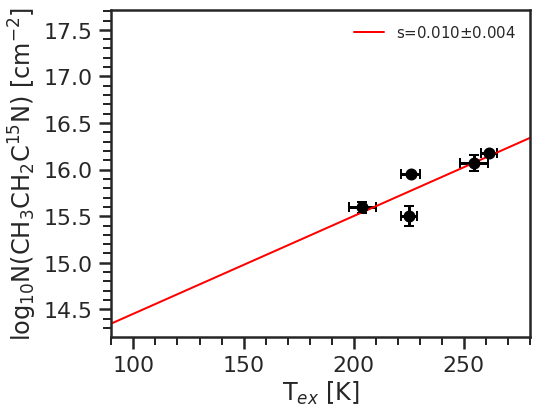}
\includegraphics[clip,width=0.2\textwidth,keepaspectratio]{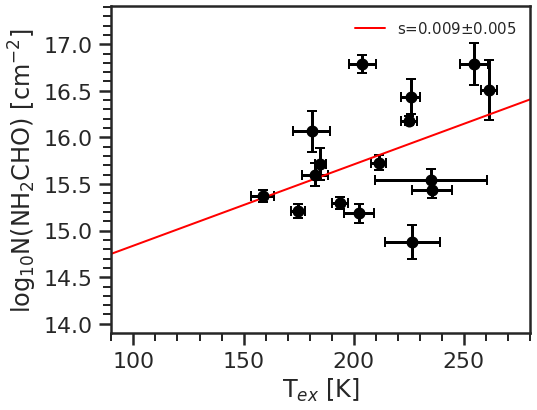}
\includegraphics[clip,width=0.2\textwidth,keepaspectratio]{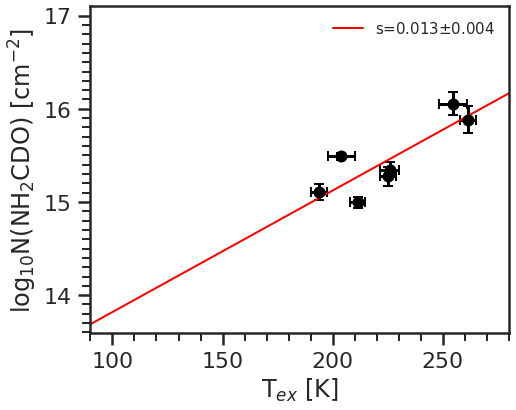}
\caption{Relation between the N(COMs) versus T$_{ex}$(CH$_3$OH). Symbols indicate the same as Figure \ref{fig:NCOMs_NH2_hue_NCOMs}. Red solid lines are the linear fits in log-linear space for all cores with COM-rich cores.}
\label{fig:corr_NCOMs_Tex_hue_NCOMs}
\end{figure*}

\begin{figure*}[htb!]
\includegraphics[clip,width=0.2\textwidth,keepaspectratio]{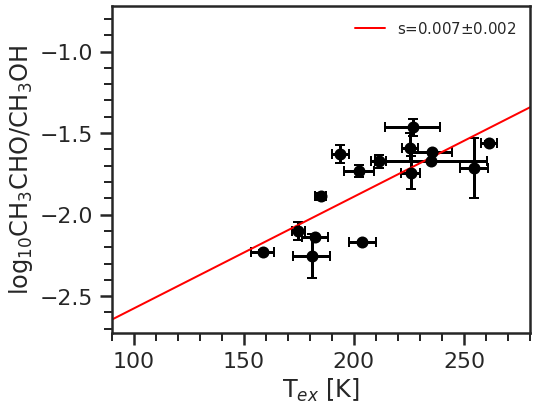}
\includegraphics[clip,width=0.2\textwidth,keepaspectratio]{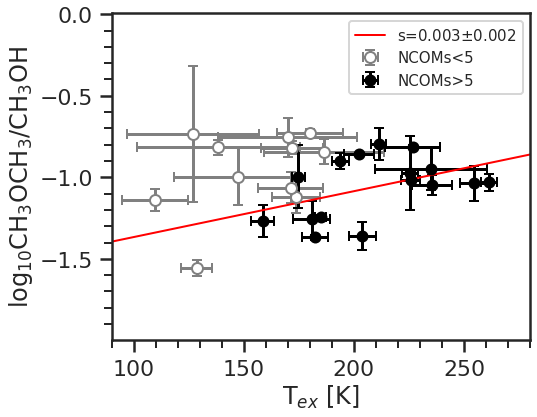}
\includegraphics[clip,width=0.2\textwidth,keepaspectratio]{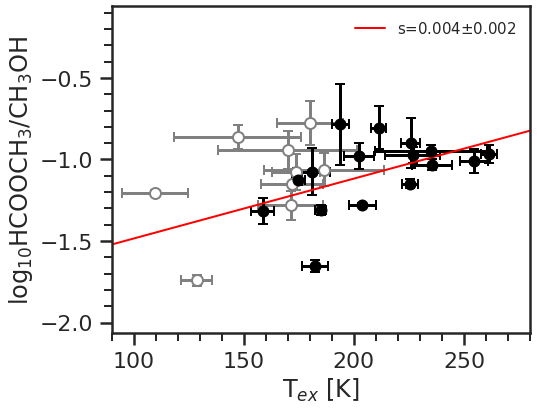}\\
\includegraphics[clip,width=0.2\textwidth,keepaspectratio]{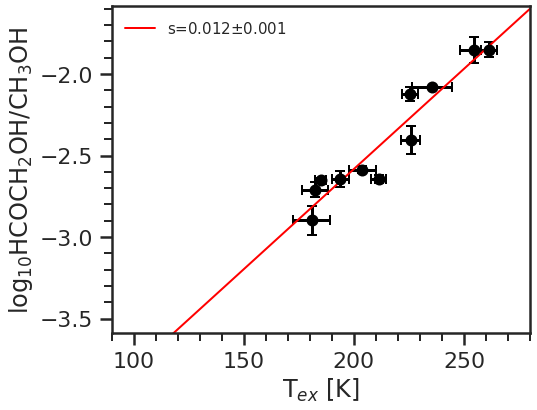}
\includegraphics[clip,width=0.2\textwidth,keepaspectratio]{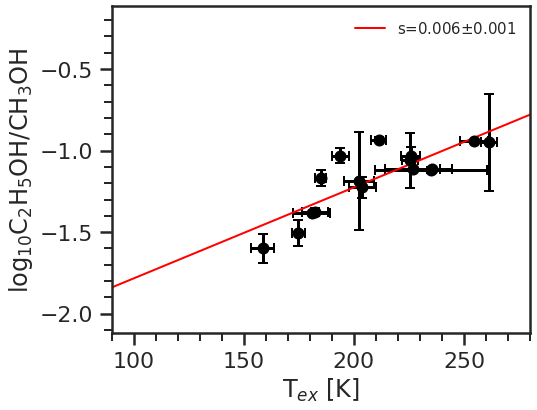}
\includegraphics[clip,width=0.2\textwidth,keepaspectratio]{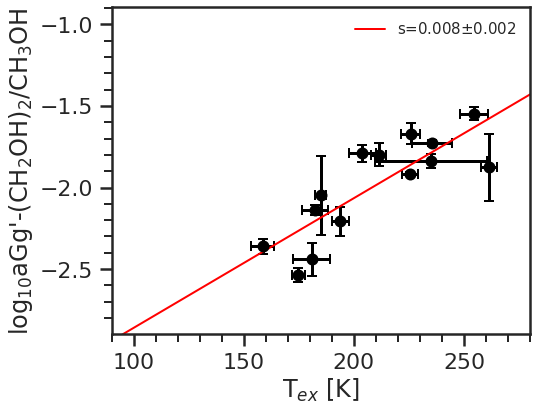}
\includegraphics[clip,width=0.2\textwidth,keepaspectratio]{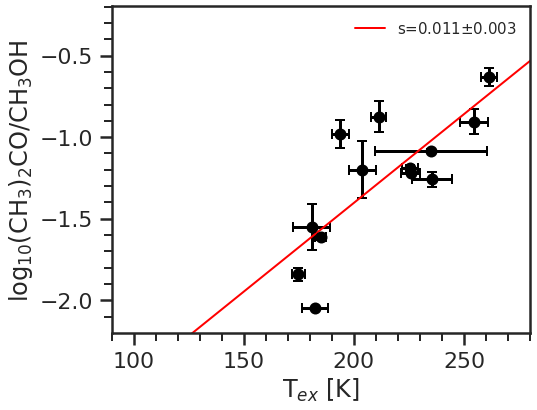}\\
\includegraphics[clip,width=0.2\textwidth,keepaspectratio]{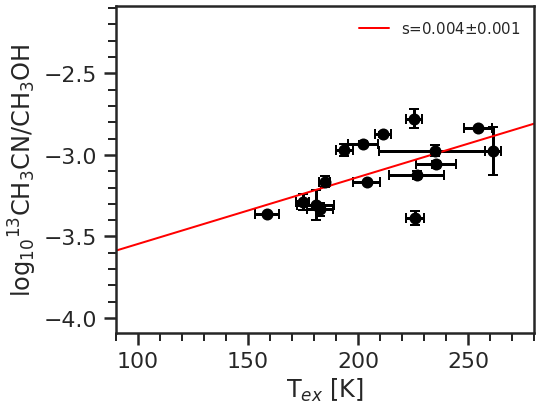}
\includegraphics[clip,width=0.2\textwidth,keepaspectratio]{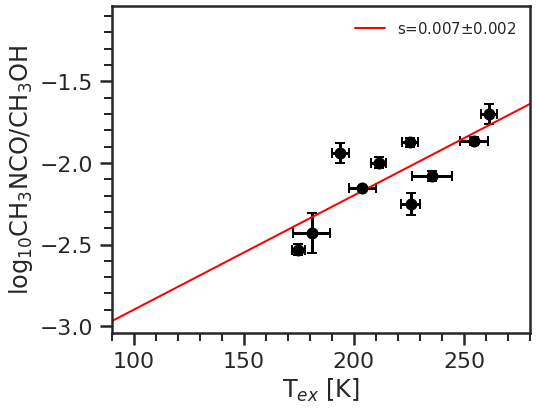}
\includegraphics[clip,width=0.2\textwidth,keepaspectratio]{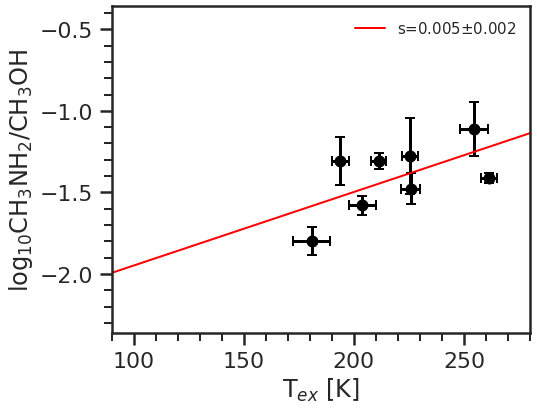}
\includegraphics[clip,width=0.2\textwidth,keepaspectratio]{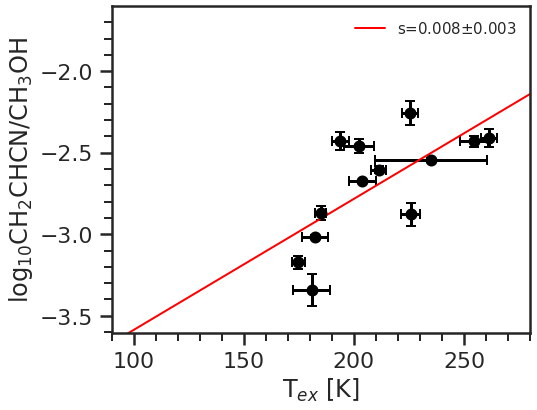}\\
\includegraphics[clip,width=0.2\textwidth,keepaspectratio]{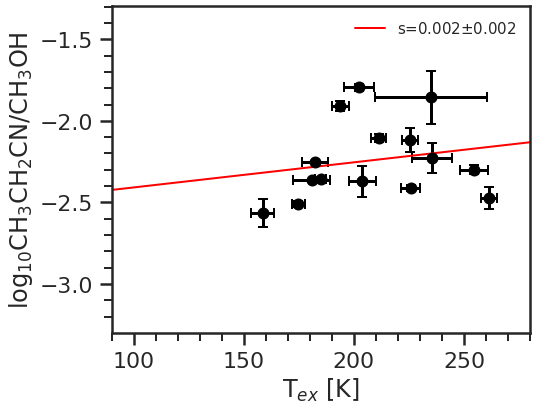}
\includegraphics[clip,width=0.2\textwidth,keepaspectratio]{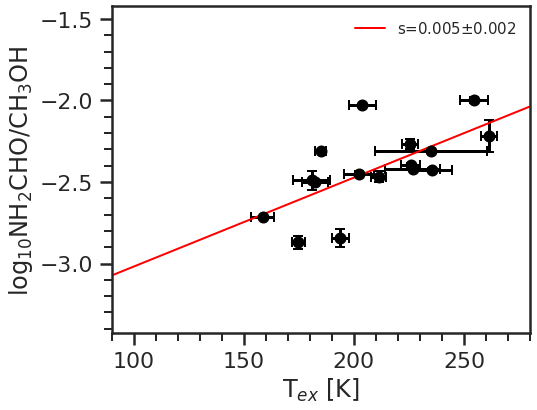}
\caption{Relation between the X(COMs) versus T$_{ex}$(CH$_3$OH). The convention of symbols and colors are identical to Figure \ref{fig:corr_NCOMs_Tex_hue_NCOMs}.}
\label{fig:corr_XCOMs_Tex_hue_NCOMs}
\end{figure*}

We present the relation between the N(COMs) (Figure \ref{fig:corr_NCOMs_Tex_hue_NCOMs}) and X(COMs) (Figure \ref{fig:corr_XCOMs_Tex_hue_NCOMs}) versus T$_{ex}$(CH$_3$OH). Similar tendency is found with Figure \ref{fig:NCOMs_NH2_hue_NCOMs} and Figure \ref{fig:corr_XCOMs_NH2_hue_NCOMs}, except for the increasing tendency in CH$_3$CHO abundance in increasing T$_{ex}$(CH$_3$OH). The details are discussed in Section \ref{sec:dis:physical_properties}. 

\clearpage
\section{Correlation of column densities and abuncances of COMs}

\begin{figure*}
\includegraphics[clip,width=1.0\textwidth,keepaspectratio]{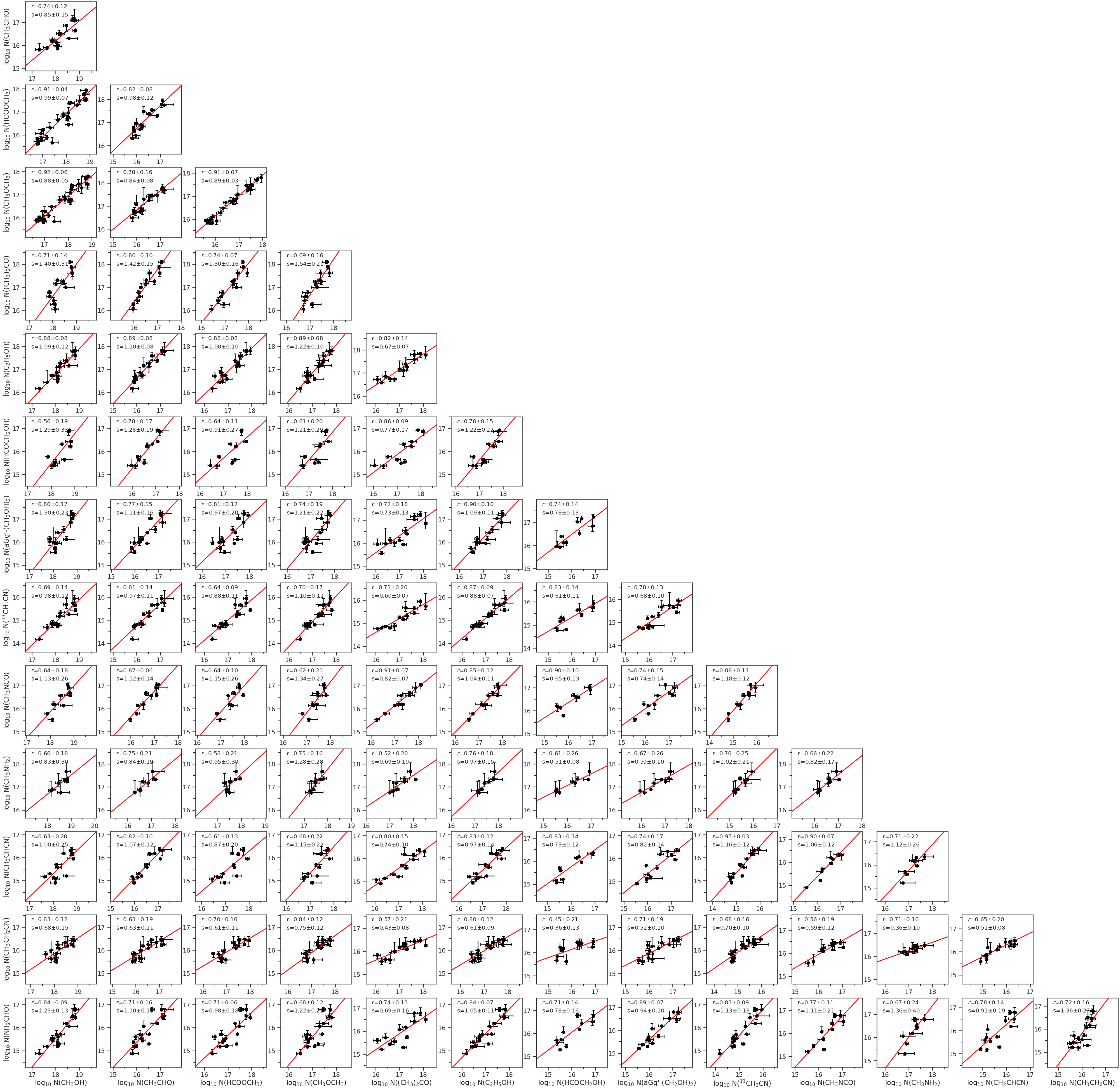}
\caption{Correlation between the column density of COMs. Pearson $r$ coefficient of each pair is presented. The slope measured in log-log space are presented as $s$ to give the power-law indicies of two given species.}
\label{fig:corr_N_all}
\end{figure*}

\begin{figure*}
\includegraphics[clip,width=1.0\textwidth,keepaspectratio]{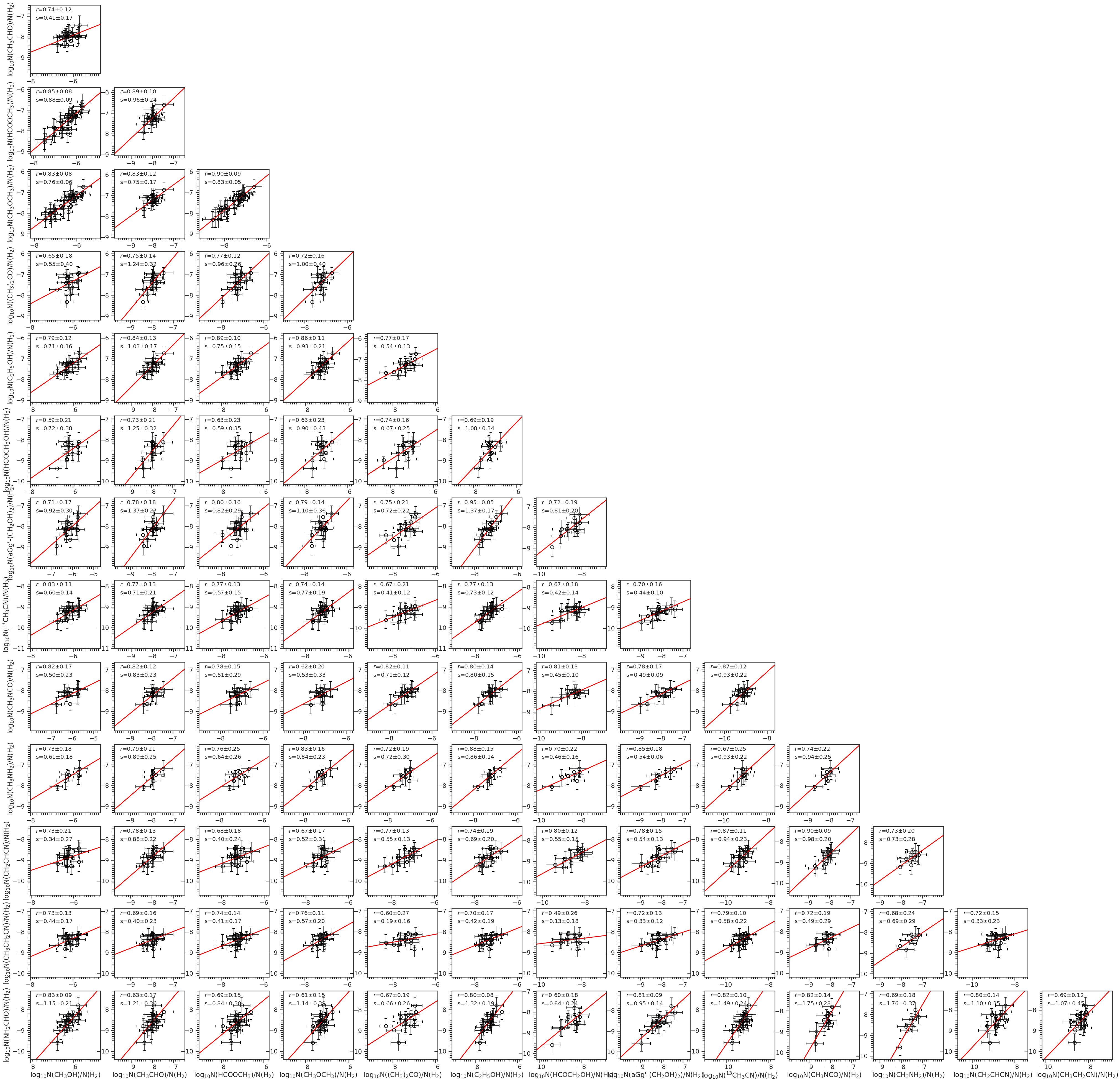}
\caption{Correlation between the abundances of COMs to N(H$_2$). Pearson $r$ coefficient of each pair is presented. The slope measured in log-log space are presented as $s$ to give the power-law indicies of two given species.}
\label{fig:corr_N_NH2_all}
\end{figure*}

Figure \ref{fig:corr_N_all} and Figure \ref{fig:corr_N_NH2_all} depict the pair plots of column density of COMs and their abundances relative to N(H$_2$) for all main species detected in this work. Pearson r coefficients are estimated in all pairs, and slopes are measured in log-log space to give power-law indices. The red solid lines are the fitted line with linear function with the corresponding slope.  

%% file: s5_tab1_line_identification.tex
\startlongtable
\begin{deluxetable*}{llcccc} 
\tablecaption{Line identification\label{tab:LID}}
\tabletypesize{\scriptsize}
\tablehead{									
\colhead{Frequency [MHz]} & \colhead{$trans$ition} & \colhead{log$_{10}$(Einstein-A)} &	\colhead{E$_u$ [K]} &	\colhead{g$_u$}  & \colhead{Reference}}
\startdata
\multicolumn{6}{c}{Methanol (CH$_{3}$OH) } \\
\tableline
213159.15 (0.01) &             20(-4)-19(-5)E2, $v_t$=0  &     -4.80 &    574.94 &  41  &    JPL \\
213377.53 (0.02) &               13(6)-14(5)E1, $v_t$=0  &     -4.97 &    389.92 &  27  &    JPL \\
213427.06 (0.01) &                 1(1)-0(0)E1, $v_t$=0  &     -4.47 &     23.37 &  3  &    JPL \\
216945.52 (0.01) &                 5(1)-4(2)E1, $v_t$=0  &     -4.92 &     55.87 &  11  &    JPL \\
217299.20 (0.02) &                 6(1)--7(2)-, $v_t$=1  &     -4.37 &    373.92 &  13  &    JPL \\
229589.06 (0.01) &               15(4)-16(3)E1, $v_t$=0  &     -4.68 &    374.44 &  31  &    JPL \\
229864.12 (0.01) &               19(5)+-20(4)+, $v_t$=0  &     -4.68 &    578.60 &  39  &    JPL \\
229939.10 (0.01) &               19(5)--20(4)-, $v_t$=0  &     -4.68 &    578.60 &  39  &    JPL \\
230368.76 (0.02) &               22(4)-21(5)E1, $v_t$=0  &     -4.68 &    682.74 &  45  &    JPL \\
\tableline
\multicolumn{6}{c}{Methanol ($^{13}$CH$_{3}$OH  $v_t$=0) } \\
\tableline
213916.35 (0.55) &           20(1,19)-20(0,20)  &     -4.48 &    496.63 &  41  &   CDMS \\
217044.62 (0.06) &         14(1,13)-13(2,12)-- &     -4.62 &    254.25 &  29  &   CDMS \\
217399.55 (0.05) &            10(2,8)-9(3,7)++  &     -4.82 &    162.41 &  21  &   CDMS \\
\tableline
\multicolumn{6}{c}{Methanol (CH$_{3}^{18}$OH  $v_t$=0) } \\
\tableline
213218.17 (0.05) &          16(2,14)-15(3,12)E  &     -4.79 &    325.14 &  132  &   CDMS \\
229859.25 (0.03) &              5(1,5)-4(1,4)A  &     -4.30 &     47.44 &  44  &   CDMS \\
231686.68 (0.03) &            5(-0,5)-4(-0,4)E &     -4.27 &     46.21 &  44  &   CDMS \\
231758.45 (0.03) &              5(0,5)-4(0,4)A &     -4.27 &     33.37 &  44  &   CDMS \\
231758.45 (0.03) &              5(0,5)-4(0,4)A  &     -4.27 &     33.37 &  44  &   CDMS \\
231768.54 (0.03) &              5(4,2)-4(4,1)A &     -4.72 &    114.23 &  44  &   CDMS \\
231777.71 (0.03) &            5(-4,2)-4(-4,1)E  &     -4.72 &    121.80 &  44  &   CDMS \\
231788.11 (0.03) &              5(4,1)-4(4,0)E  &     -4.71 &    129.53 &  44  &   CDMS \\
231796.52 (0.03) &              5(3,2)-4(3,1)A  &     -4.47 &     83.47 &  44  &   CDMS \\
231801.47 (0.03) &              5(2,4)-4(2,3)A  &     -4.34 &     70.85 &  44  &   CDMS \\
231809.48 (0.03) &            5(-3,3)-4(-3,2)E  &     -4.46 &     96.00 &  44  &   CDMS \\
231826.74 (0.03) &              5(1,4)-4(1,3)E  &     -4.27 &     54.13 &  44  &   CDMS \\
231840.92 (0.03) &              5(2,3)-4(2,2)A  &     -4.34 &     70.85 &  44  &   CDMS \\
231853.85 (0.03) &            5(-2,4)-4(-2,3)E  &     -4.35 &     59.25 &  44  &   CDMS \\
231864.50 (0.03) &              5(2,3)-4(2,2)E  &     -4.36 &     55.76 &  44  &   CDMS \\
\tableline
\multicolumn{6}{c}{Methanol (CH$_{2}$DOH $v_t$=0) } \\
\tableline
229856.75 (0.01) &           10(2,8)-10(1,9),e0  &     -4.06 &    134.73 &  21  &    JPL \\
230376.57 (0.01) &         15(2,13)-15(1,14),o1 &     -4.51 &    292.77 &  31  &    JPL \\
231969.23 (0.01) &             9(2,7)-9(1,8),e0  &     -4.06 &    113.11 &  19  &    JPL \\
\tableline
\multicolumn{6}{c}{Acetaldehyde (CH$_{3}$CHO) } \\
\tableline
212957.77 (0.01) &            11(3,9)-10(3,8)E, $v_t$=1  &     -3.50 &    285.62 &  46  &    JPL \\
213057.90 (0.01) &            11(3,8)-10(3,7)A, $v_t$=1  &     -3.50 &    285.74 &  46  &    JPL \\
213071.02 (0.01) &            11(2,9)-10(2,8)E, $v_t$=1  &     -3.48 &    276.51 &  46  &    JPL \\
213672.88 (0.01) &           11(2,10)-10(2,9)E, $v_t$=1  &     -3.48 &    273.95 &  46  &    JPL \\
216630.23 (0.01) &           11(1,10)-10(1,9)A, $v_t$=0  &     -3.45 &     64.81 &  46  &    JPL \\
217469.28 (0.01) &          14(3,11)-14(2,12)A, $v_t$=0  &     -4.42 &    117.70 &  58  &    JPL \\
230301.92 (0.01) &          12(2,11)-11(2,10)A, $v_t$=0  &     -3.38 &     81.04 &  50  &    JPL \\
230315.79 (0.01) &          12(2,11)-11(2,10)E, $v_t$=0  &     -3.38 &     81.06 &  50  &    JPL \\
230395.16 (0.01) &          12(2,11)-11(2,10)A, $v_t$=1  &     -3.37 &    286.41 &  50  &    JPL \\
231456.74 (0.01) &            12(4,9)-11(4,8)A, $v_t$=0  &     -3.41 &    108.35 &  50  &    JPL \\
231467.50 (0.01) &            12(4,8)-11(4,7)A, $v_t$=0  &     -3.41 &    108.35 &  50  &    JPL \\
231484.37 (0.01) &            12(4,8)-11(4,7)E, $v_t$=0  &     -3.41 &    108.29 &  50  &    JPL \\
231506.29 (0.01) &            12(4,9)-11(4,8)E, $v_t$=0  &     -3.41 &    108.25 &  50  &    JPL \\
231548.63 (0.01) &            12(7,5)-11(7,4)A, $v_t$=1  &     -3.54 &    387.01 &  50  &    JPL \\
231595.27 (0.01) &           12(3,10)-11(3,9)A, $v_t$=0  &     -3.39 &     92.58 &  50  &    JPL \\
231748.72 (0.01) &           12(3,10)-11(3,9)E, $v_t$=0  &     -3.39 &     92.51 &  50  &    JPL \\
231847.58 (0.01) &            12(3,9)-11(3,8)E, $v_t$=0  &     -3.39 &     92.61 &  50  &    JPL \\
232180.27 (0.01) &            12(5,7)-11(5,6)A, $v_t$=1  &     -3.44 &    331.84 &  50  &    JPL \\
232292.60 (0.01) &            12(4,9)-11(4,8)A, $v_t$=1  &     -3.41 &    311.80 &  50  &    JPL \\
232304.93 (0.01) &            12(4,8)-11(4,7)A, $v_t$=1  &     -3.41 &    311.81 &  50  &    JPL \\
\tableline
\multicolumn{6}{c}{Methyl formate (HCOOCH$_{3}$) } \\
\tableline
213250.92 (0.10) &           18(3,16)-17(3,15)A $v_t$=1  &     -3.85 &    292.89 &  74  &    JPL \\
213700.01 (0.10) &           18(3,16)-17(3,15)E $v_t$=1  &     -3.85 &    292.44 &  74  &    JPL \\
216830.20 (0.10) &           18(2,16)-17(2,15)E $v_t$=0  &     -3.83 &    105.68 &  74  &    JPL \\
216838.89 (0.10) &           18(2,16)-17(2,15)A $v_t$=0  &     -3.83 &    105.67 &  74  &    JPL \\
216958.83 (0.10) &           17(3,14)-16(3,13)E $v_t$=1  &     -3.83 &    286.21 &  70  &    JPL \\
216967.42 (0.10) &           20(0,20)-19(0,19)A $v_t$=0  &     -3.81 &    111.48 &  82  &    JPL \\
217312.63 (0.10) &           17(4,13)-16(4,12)A $v_t$=1  &     -3.84 &    289.96 &  70  &    JPL \\
229661.38 (0.10) &           25(9,16)-25(8,17)A $v_t$=1  &     -4.77 &    432.59 &  102  &    JPL \\
231414.45 (0.10) &          35(10,25)-35(9,26)A $v_t$=0  &     -4.72 &    440.93 &  142  &    JPL \\
231450.50 (0.10) &           19(11,8)-18(11,7)E $v_t$=1  &     -3.90 &    380.13 &  78  &    JPL \\
231724.16 (0.20) &           18(4,14)-17(4,13)E $v_t$=1  &     -3.75 &    300.80 &  74  &    JPL \\
231749.76 (0.10) &           19(10,9)-18(10,8)E $v_t$=1  &     -3.87 &    366.05 &  78  &    JPL \\
231896.06 (0.10) &           19(4,16)-18(4,15)E $v_t$=1  &     -3.75 &    309.75 &  78  &    JPL \\
231903.90 (0.10) &           19(12,7)-18(12,6)A $v_t$=1  &     -3.95 &    395.09 &  78  &    JPL \\
231918.95 (0.10) &           19(13,7)-18(13,6)E $v_t$=1  &     -4.00 &    411.79 &  78  &    JPL \\
232002.60 (0.10) &           19(11,9)-18(11,8)A $v_t$=1  &     -3.90 &    379.73 &  78  &    JPL \\
232129.22 (0.10) &           19(12,8)-18(12,7)E $v_t$=1  &     -3.94 &    394.99 &  78  &    JPL \\
232160.19 (0.10) &            19(9,10)-18(9,9)E $v_t$=1  &     -3.83 &    353.32 &  78  &    JPL \\
232164.44 (0.10) &          19(10,10)-18(10,9)A $v_t$=1  &     -3.86 &    365.73 &  78  &    JPL \\
\tableline
\multicolumn{6}{c}{Dimethyl ether (CH$_{3}$OCH$_{3}$) } \\
\tableline
213445.99 (0.01) &               25(4,22)-24(5,19)EE     &     -4.87 &    319.22 &  816  &   CDMS \\
213453.66 (0.01) &               21(1,20)-21(0,21)AE     &     -4.65 &    212.92 &  258  &   CDMS \\
213457.67 (0.01) &               21(1,20)-21(0,21)EE     &     -4.65 &    212.92 &  688  &   CDMS \\
213457.67 (0.01) &               21(1,20)-21(0,21)EE     &     -4.65 &    212.92 &  688  &   CDMS \\
213461.68 (0.01) &               21(1,20)-21(0,21)AA     &     -4.65 &    212.92 &  430  &   CDMS \\
213873.15 (0.01) &               41(5,36)-41(4,37)AA     &     -4.16 &    829.35 &  830  &   CDMS \\
216764.54 (0.01) &             23(11,13)-24(12,13)EE     &     -5.15 &    453.35 &  752  &   CDMS \\
217177.49 (0.01) &               36(4,32)-36(3,33)AA     &     -4.20 &    638.63 &  438  &   CDMS \\
217191.40 (0.01) &               22(4,19)-22(3,20)EE     &     -4.27 &    253.41 &  720  &   CDMS \\
217352.58 (0.01) &               42(6,36)-42(5,37)AA     &     -4.11 &    878.95 &  510  &   CDMS \\
230505.08 (0.01) &               26(4,23)-25(5,20)EE     &     -4.78 &    342.96 &  848  &   CDMS \\
231987.93 (0.01) &               13(0,13)-12(1,12)AE     &     -4.04 &     80.92 &  162  &   CDMS \\
\tableline
\multicolumn{6}{c}{Acetone (CH$_{3}$COCH$_{3}$ $v_t$=0) } \\
\tableline
216974.46 (0.01) &         19(3,16)-18(4,15)EA &     -3.41 &    115.56 &  156  &    JPL \\
217022.51 (0.01) &         19(3,16)-18(4,15)EE &     -3.41 &    115.50 &  624  &    JPL \\
217070.50 (0.02) &         19(4,16)-18(3,15)AA &     -3.41 &    115.43 &  234  &    JPL \\
229600.40 (0.06) &         29(3,26)-29(2,27)AE &     -4.15 &    245.58 &  354  &    JPL \\
229609.23 (0.06) &         31(6,26)-31(5,27)AA &     -3.96 &    299.63 &  378  &    JPL \\
230412.01 (0.10) &         27(1,26)-27(0,27)EE  &     -4.84 &    194.93 &  880  &    JPL \\
\tableline
\multicolumn{6}{c}{Ethanol (C$_{2}$H$_{5}$OH) } \\
\tableline
213446.98 (0.01) &     5(5,0)-4(4,0), $v_t$=0-1, $gauche$  &     -4.25 &     99.82 &  11  &    JPL \\
213760.08 (0.01) &  13(1,13)-12(1,12), $v_t$=1-1, $gauche$  &     -4.06 &    135.37 &  27  &    JPL \\
213856.26 (0.05) &           13(0,13)-12(1,12), $trans$  &     -4.27 &     74.30 &  27  &    JPL \\
216659.68 (0.01) &     8(4,5)-7(3,5), $v_t$=0-1, $gauche$  &     -4.46 &    106.30 &  17  &    JPL \\
217262.30 (0.01) &  13(0,13)-12(1,12), $v_t$=1-1, $gauche$  &     -6.54 &    135.54 &  27  &    JPL \\
217496.67 (0.01) &  13(1,13)-12(0,12), $v_t$=1-1, $gauche$  &     -6.47 &    135.37 &  27  &    JPL \\
217549.33 (0.01) &  25(3,22)-24(4,20), $v_t$=0-1, $gauche$  &     -4.88 &    342.91 &  51  &    JPL \\
229779.30 (0.05) &           27(4,24)-27(3,25), $trans$  &     -4.13 &    338.83 &  55  &    JPL \\
230672.55 (0.05) &  13(2,11)-12(2,10), $v_t$=0-0, $gauche$  &     -3.97 &    138.62 &  27  &    JPL \\
230696.55 (0.01) &     8(3,5)-8(2,7), $v_t$=1-0, $gauche$  &     -4.40 &    102.57 &  17  &    JPL \\
231558.55 (0.05) &           21(5,17)-21(4,18), $trans$  &     -4.13 &    225.95 &  43  &    JPL \\
231668.73 (0.05) &  14(1,14)-13(1,13), $v_t$=0-0, $gauche$  &     -3.95 &    141.90 &  29  &    JPL \\
231737.61 (0.05) &           19(5,15)-19(4,16), $trans$  &     -4.13 &    191.32 &  39  &    JPL \\
231790.00 (0.05) &           22(5,18)-22(4,19), $trans$  &     -4.13 &    244.54 &  45  &    JPL \\
232034.63 (0.05) &           18(5,14)-18(4,15), $trans$  &     -4.14 &    175.29 &  37  &    JPL \\
232075.83 (0.05) &           15(5,10)-15(4,11), $trans$  &     -4.15 &    132.29 &  31  &    JPL \\
232318.50 (0.05) &           23(5,19)-23(4,20), $trans$  &     -4.12 &    263.99 &  47  &    JPL \\
\tableline
\multicolumn{6}{c}{Glycolaldehyde ($cis$-HCOCH$_{2}$OH $v_t$=0) } \\
\tableline
213181.61 (0.01) &                   18(9,9)-18(8,10)    &     -3.87 &    144.22 &  37  &    JPL \\
213193.52 (0.01) &                  21(1,21)-20(0,20)    &     -3.56 &    114.69 &  43  &    JPL \\
213653.95 (0.01) &                    17(9,8)-17(8,9)    &     -3.88 &    134.19 &  35  &    JPL \\
213689.96 (0.01) &                    12(4,9)-11(3,8)    &     -4.02 &     53.23 &  25  &    JPL \\
217053.75 (0.01) &                 33(10,23)-33(9,24)    &     -3.73 &    374.46 &  67  &    JPL \\
217271.73 (0.02) &                  31(4,27)-31(3,28)    &     -3.92 &    290.41 &  63  &    JPL \\
229746.85 (0.01) &                 29(10,20)-29(9,21)    &     -3.69 &    303.39 &  59  &    JPL \\
230703.71 (0.02) &                  16(4,13)-15(3,12)    &     -3.93 &     85.69 &  33  &    JPL \\
232286.11 (0.01) &                  22(1,21)-21(2,20)    &     -3.51 &    134.52 &  45  &    JPL \\
232335.40 (0.01) &                  22(2,21)-21(1,20)    &     -3.51 &    134.52 &  45  &    JPL \\
\tableline
\multicolumn{6}{c}{Ethylene Glycol (g'Ga-(CH$_2$OH)$_2$) } \\
\tableline
213132.92 (0.01) &          22(0,22)-21(0,21) $v_t$=1-0  &     -3.63 &    115.68 &  405  &   CDMS \\
213646.69 (0.01) &          20(6,14)-19(6,13) $v_t$=1-0  &     -3.67 &    121.65 &  369  &   CDMS \\
213821.90 (0.01) &        23(11,13)-23(10,14) $v_t$=1-1  &     -4.62 &    195.43 &  423  &   CDMS \\
216685.81 (0.01) &          21(3,19)-20(3,18) $v_t$=1-0  &     -3.69 &    117.25 &  387  &   CDMS \\
216826.11 (0.01) &          20(5,15)-19(5,14) $v_t$=1-0  &     -3.75 &    116.78 &  369  &   CDMS \\
217139.72 (0.01) &          21(4,17)-20(4,16) $v_t$=0-1  &     -3.62 &    123.93 &  387  &   CDMS \\
217449.99 (0.01) &          24(1,24)-23(1,23) $v_t$=0-1  &     -3.60 &    136.45 &  441  &   CDMS \\
229817.11 (0.01) &          23(9,14)-22(9,13) $v_t$=0-1  &     -3.60 &    175.60 &  423  &   CDMS \\
230472.53 (0.01) &          21(4,17)-20(4,16) $v_t$=1-0  &     -3.55 &    124.25 &  301  &   CDMS \\
231524.03 (0.01) &          23(6,18)-22(6,17) $v_t$=0-1  &     -3.57 &    154.07 &  329  &   CDMS \\
231564.32 (0.01) &          24(0,24)-23(0,23) $v_t$=1-0  &     -3.52 &    136.79 &  441  &   CDMS \\
232025.85 (0.01) &          22(14,8)-21(14,7) $v_t$=1-0  &     -3.74 &    220.88 &  405  &   CDMS \\
232043.88 (0.01) &          22(13,9)-21(13,8) $v_t$=1-0  &     -3.70 &    207.64 &  405  &   CDMS \\
232065.05 (0.01) &          22(16,6)-21(16,5) $v_t$=1-0  &     -3.84 &    250.32 &  405  &   CDMS \\
232095.74 (0.01) &        22(12,11)-21(12,10) $v_t$=1-0  &     -3.67 &    195.38 &  315  &   CDMS \\
232113.94 (0.01) &          22(17,5)-21(17,4) $v_t$=1-0  &     -3.91 &    266.50 &  405  &   CDMS \\
232350.07 (0.01) &        22(10,12)-21(10,11) $v_t$=1-0  &     -3.62 &    173.84 &  405  &   CDMS \\
\tableline
\multicolumn{6}{c}{Methyl Cyanide ($^{13}$CH$_{3}$CN) } \\
\tableline
232077.27 (0.02) &                13(6)-12(6),F=14-13    &     -3.07 &    335.53 &  116  &    JPL \\
232125.17 (0.01) &                13(5)-12(5),F=14-13    &     -3.04 &    256.88 &  58  &    JPL \\
232164.40 (0.01) &                13(4)-12(4),F=14-13    &     -3.01 &    192.51 &  58  &    JPL \\
232194.91 (0.01) &                13(3)-12(3),F=12-11    &     -2.99 &    142.43 &  100  &    JPL \\
232216.71 (0.01) &                13(2)-12(2),F=13-12    &     -2.98 &    106.65 &  54  &    JPL \\
232229.81 (0.01) &                13(1)-12(1),F=12-11    &     -2.97 &     85.18 &  50  &    JPL \\
232234.19 (0.01) &                13(0)-12(0),F=14-13    &     -2.97 &     78.02 &  58  &    JPL \\
\tableline
\multicolumn{6}{c}{Methyl isocyanate (CH$_{3}$NCO) } \\
\tableline
216781.62 (0.00) &              25(2,24)-24(2,23),m=0    &     -3.32 &    159.14 &  51  &   CDMS \\
217150.93 (0.00) &              25(2,23)-24(2,22),m=0    &     -3.32 &    159.26 &  51  &   CDMS \\
217164.88 (0.00) &              25(-3,0)-24(-3,0),m=1    &     -3.32 &    200.96 &  51  &   CDMS \\
231766.34 (0.00) &                27(0,0)-26(0,0),m=1    &     -3.22 &    169.28 &  55  &   CDMS \\
231793.78 (0.00) &              27(-1,0)-26(-1,0),m=1    &     -3.22 &    175.24 &  55  &   CDMS \\
231815.25 (0.00) &              27(1,27)-26(1,26),m=0    &     -3.23 &    161.79 &  55  &   CDMS \\
232342.23 (0.00) &                27(2,0)-26(2,0),m=2    &     -3.22 &    233.99 &  55  &   CDMS \\
\tableline
\multicolumn{6}{c}{Methylamine (CH$_{3}$NH$_{2}$) } \\
\tableline
213184.17 (0.02) &                    10(1)B1-9(2)B2     &     -4.14 &    122.24 &  5  &  SLAIM \\
217079.40 (0.02) &                   11(2)A1-11(1)A2     &     -4.06 &    156.49 &  13  &  SLAIM \\
229908.12 (0.01) &                     8(2)A2-8(1)A1     &     -4.00 &     92.71 &  9  &  SLAIM \\
\tableline
\multicolumn{6}{c}{Vinyl Cyanide (CH$_{2}$CHCN) } \\
\tableline
213122.22 (0.01) &  22(2,20)-21(2,19),F=21-20 $v_{11}$=1  &     -3.11 &    453.79 &  43  &    JPL \\
213169.01 (0.01) &  23(0,23)-22(0,22),F=23-22 $v_{11}$=1  &     -3.10 &    452.56 &  47  &    JPL \\
216936.71 (0.01) &      23(2,22)-22(2,21),F=23-22 $v$=0  &     -3.08 &    133.93 &  47  &    JPL \\
217272.23 (0.06) &  23(2,22)-22(2,21),F=23-22 $v_{15}$=1  &     -3.08 &    623.81 &  47  &    JPL \\
229647.83 (0.01) &      25(1,25)-24(1,24),F=24-23 $v$=0  &     -3.00 &    145.92 &  49  &    JPL \\
229868.99 (0.01) &  24(3,21)-23(3,20),F=25-24 $v_{11}$=1  &     -3.01 &    485.06 &  51  &    JPL \\
230487.93 (0.01) &      24(1,23)-23(1,22),F=24-23 $v$=0  &     -3.00 &    141.25 &  49  &    JPL \\
230738.56 (0.01) &      25(0,25)-24(0,24),F=26-25 $v$=0  &     -3.00 &    145.54 &  53  &    JPL \\
231952.33 (0.01) &      24(2,22)-23(2,21),F=23-22 $v$=0  &     -2.99 &    146.84 &  47  &    JPL \\
\tableline
\multicolumn{6}{c}{Ethyl Cyanide (CH$_{3}$CH$_{2}$CN $v_t$=0) } \\
\tableline
231854.21 (0.01) &            27(1,27)-26(1,26) &     -2.98 &    157.73 &  55  &    JPL \\
231990.41 (0.01) &            27(0,27)-26(0,26)  &     -2.98 &    157.71 &  55  &    JPL \\
\tableline
\multicolumn{6}{c}{Ethyl Cyanide (CH$_{3}$CH$_{2}$C$^{15}$N) } \\
\tableline
216884.32 (0.01) &                 25(1,24)-24(1,23)     &     -3.07 &    138.37 &  51  &   CDMS \\
217285.95 (0.01) &                 26(1,26)-25(1,25)     &     -3.06 &    142.61 &  53  &   CDMS \\
217474.03 (0.01) &                 26(0,26)-25(0,25)     &     -3.06 &    142.57 &  53  &   CDMS \\
\tableline
\multicolumn{6}{c}{Formamide (NH$_{2}$CHO $v_t$=0) } \\
\tableline
232274.26 (0.41) &             11(2,10)-10(2,9)  &     -3.05 &     78.95 &  69  &    JPL \\
\tableline
\multicolumn{6}{c}{Formamide (NH$_{2}$CDO) } \\
\tableline
229725.95 (0.01) &              11(4,8)-10(4,7) &     -3.12 &    100.23 &  69  &   CDMS \\
229801.61 (0.01) &              11(4,7)-10(4,6) &     -3.12 &    100.24 &  69  &   CDMS \\
229824.75 (0.01) &              11(3,9)-10(3,8) &     -3.09 &     85.32 &  69  &   CDMS \\
\enddata
\end{deluxetable*}

%% file: s5_216GHz_channel_maps_figureset.tex
\begin{figure*}
  \centering
\includegraphics[clip,width=1.3\columnwidth,keepaspectratio]{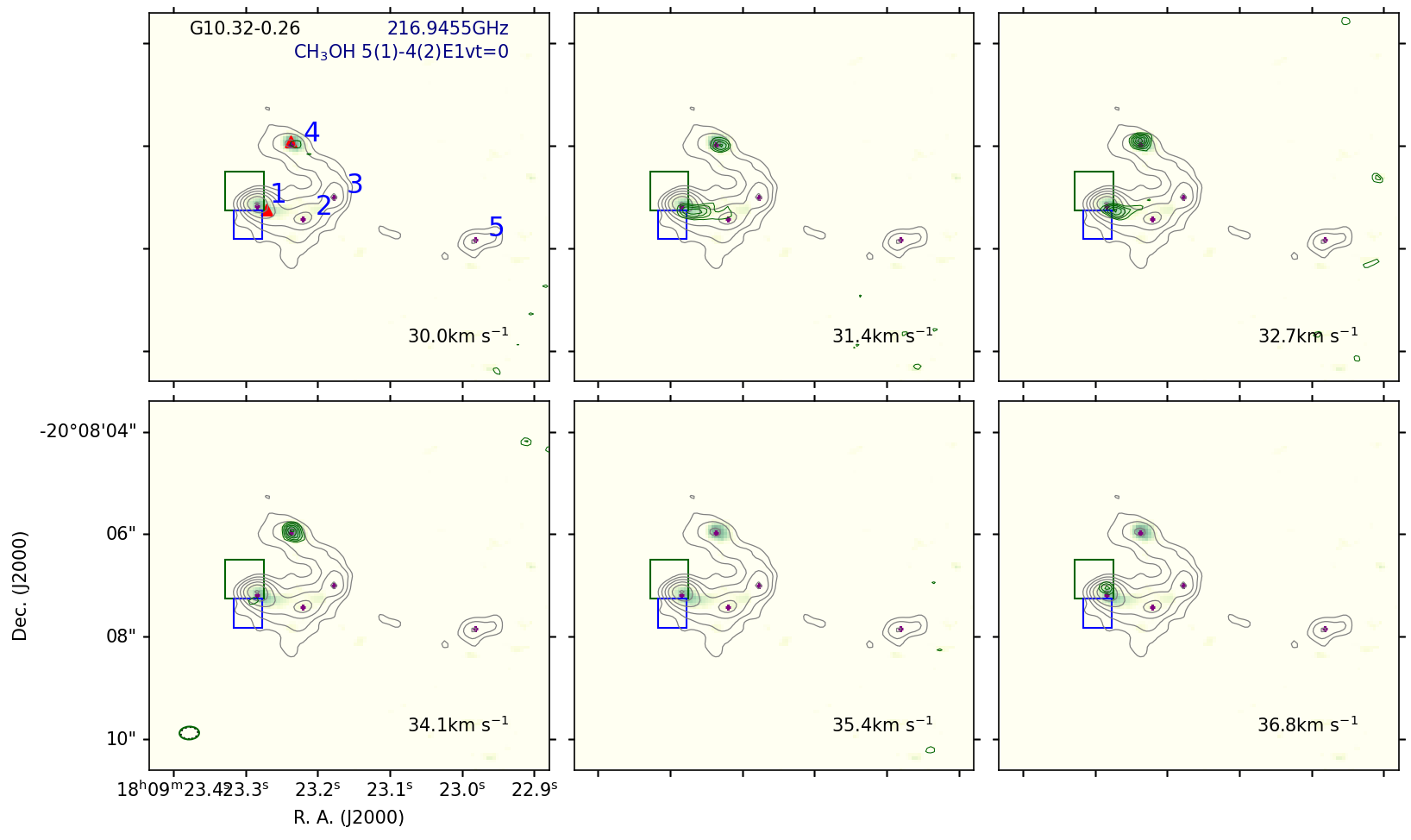} 
\caption{Channel maps of G10.32-0.16. Grey contours depict the dust continuum emission levels. Purple plus symbols indicate the peak positions of continuum images.
Green solid line show the contours of 216.946 GHz methanol emission. Contour levels for 216.946 GHz methanol emission are 0.016, 0.020, 0.023, 0.027, 0.031, 0.035 Jy beam$^{-1}$. Green and blue boxes indicate the previous 6.7 GHz maser detections from Methanol Multibeam Survey (MMB; \citealt{Green2010,Breen2015}) and VLA \citep{Hu2016} observations, with box sizes for the position accuracy of 0.${''}$4 and 0.${''}$3, respectively. Red triangles indicate the CH$_{3}$OH peak position at which the spectrum is extracted. The corresponding coordinates are summarized in Table \ref{tab:216GHz}. \\
The complete figure set (12 images) is available in the online journal.} 
\label{fig:chnmap:G10.32}
\end{figure*}

\begin{figure*}
  \centering
\includegraphics[clip,width=1.3\columnwidth,keepaspectratio]{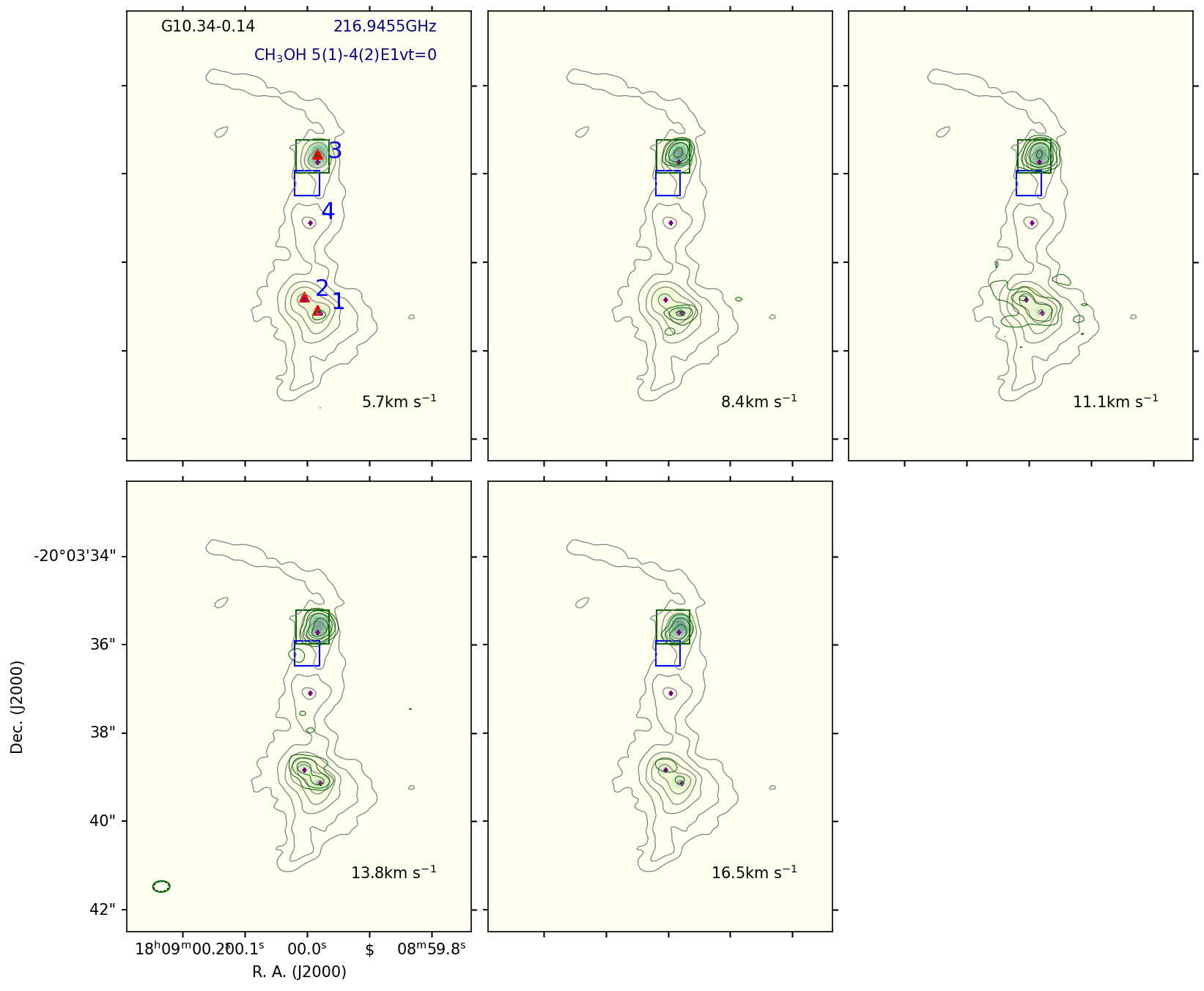}
\caption{The same as Figure \ref{fig:chnmap:G10.32} except for G10.34-0.14. Contour levels for 216.946 GHz methanol emission are 0.009, 0.016, 0.027, 0.048, 0.083, 0.146 Jy beam$^{-1}$.}
\label{fig:chnmap:G10.34}
\end{figure*}

\begin{figure*}
  \centering
\includegraphics[clip,width=1.3\columnwidth,keepaspectratio]{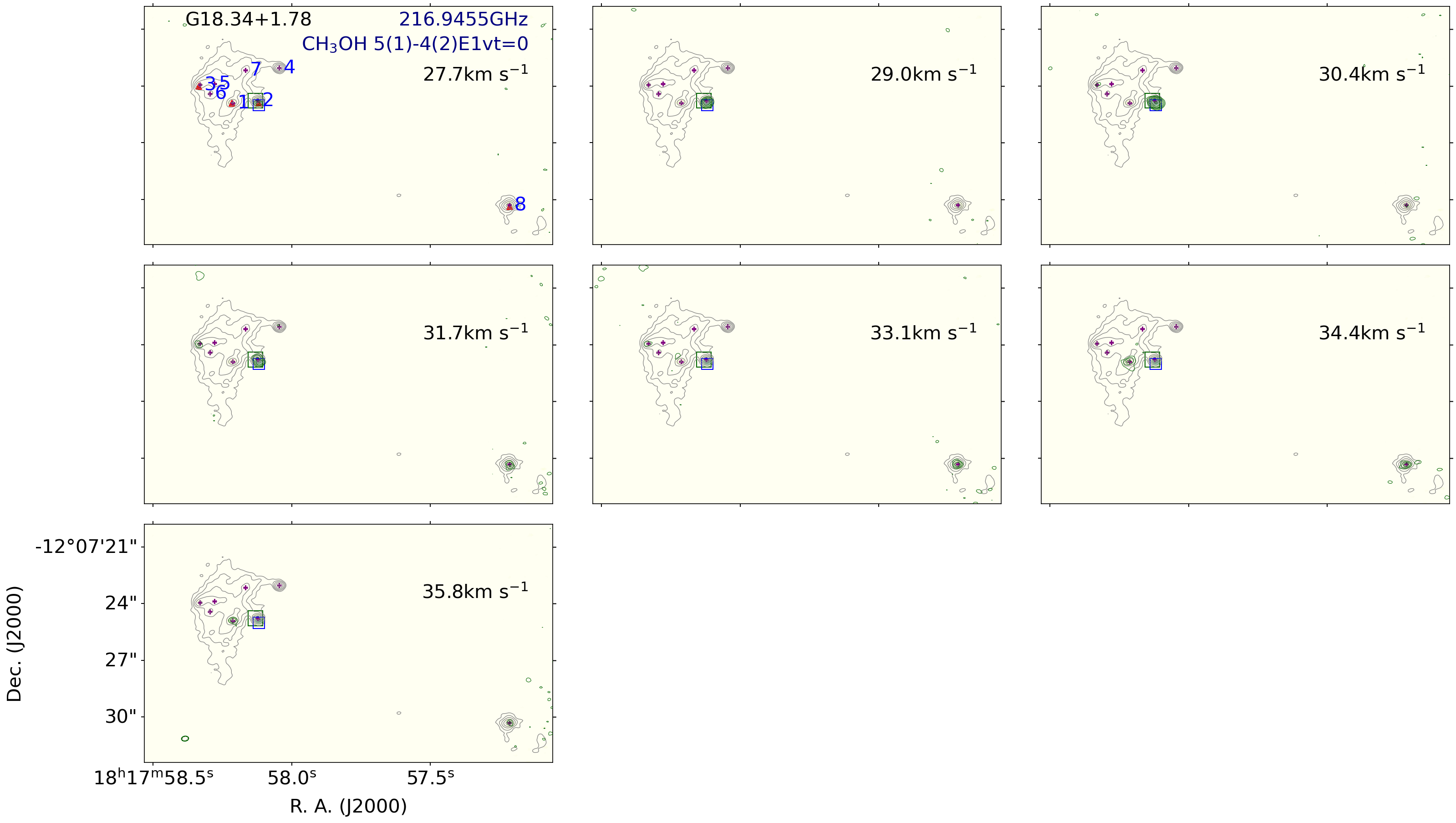}
\caption{The same as Figure \ref{fig:chnmap:G10.32} except for G18.34+1.78. Contour levels for 216.946 GHz methanol emission are 0.009, 0.015, 0.024, 0.038, 0.062, 0.100 Jy beam$^{-1}$.}
\label{fig:chnmap:G18.34}
\end{figure*}

\begin{figure*}
  \centering
\includegraphics[clip,width=1.3\columnwidth,keepaspectratio]{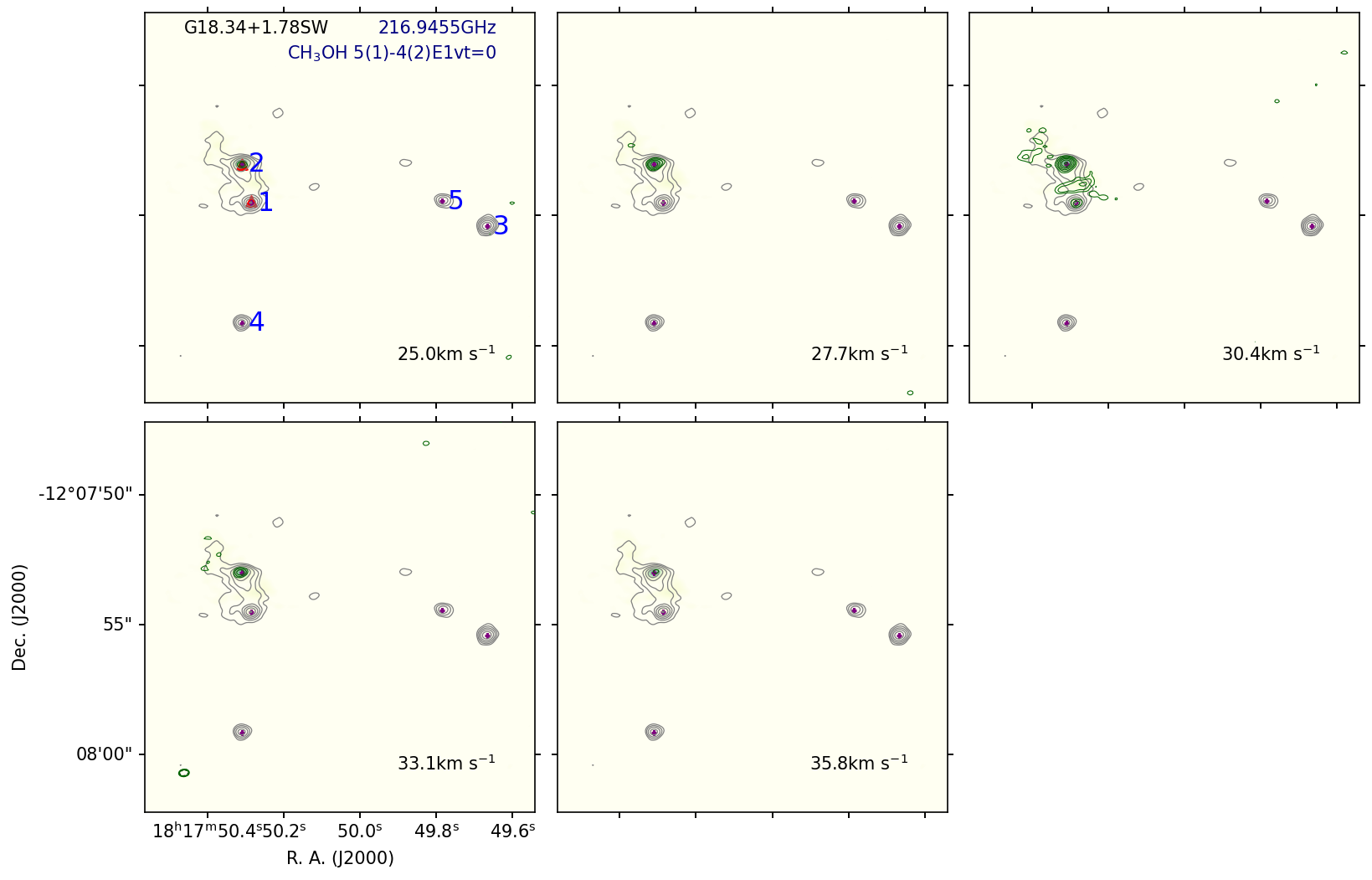}
\caption{The same as Figure \ref{fig:chnmap:G10.32} except for G18.34+1.78SW. Contour levels for 216.946 GHz methanol emission are 0.010, 0.015, 0.022, 0.033, 0.049, 0.073 Jy beam$^{-1}$.}
\label{fig:chnmap:G18.34SW}
\end{figure*}

\begin{figure*}
  \centering
\includegraphics[clip,width=1.3\columnwidth,keepaspectratio]{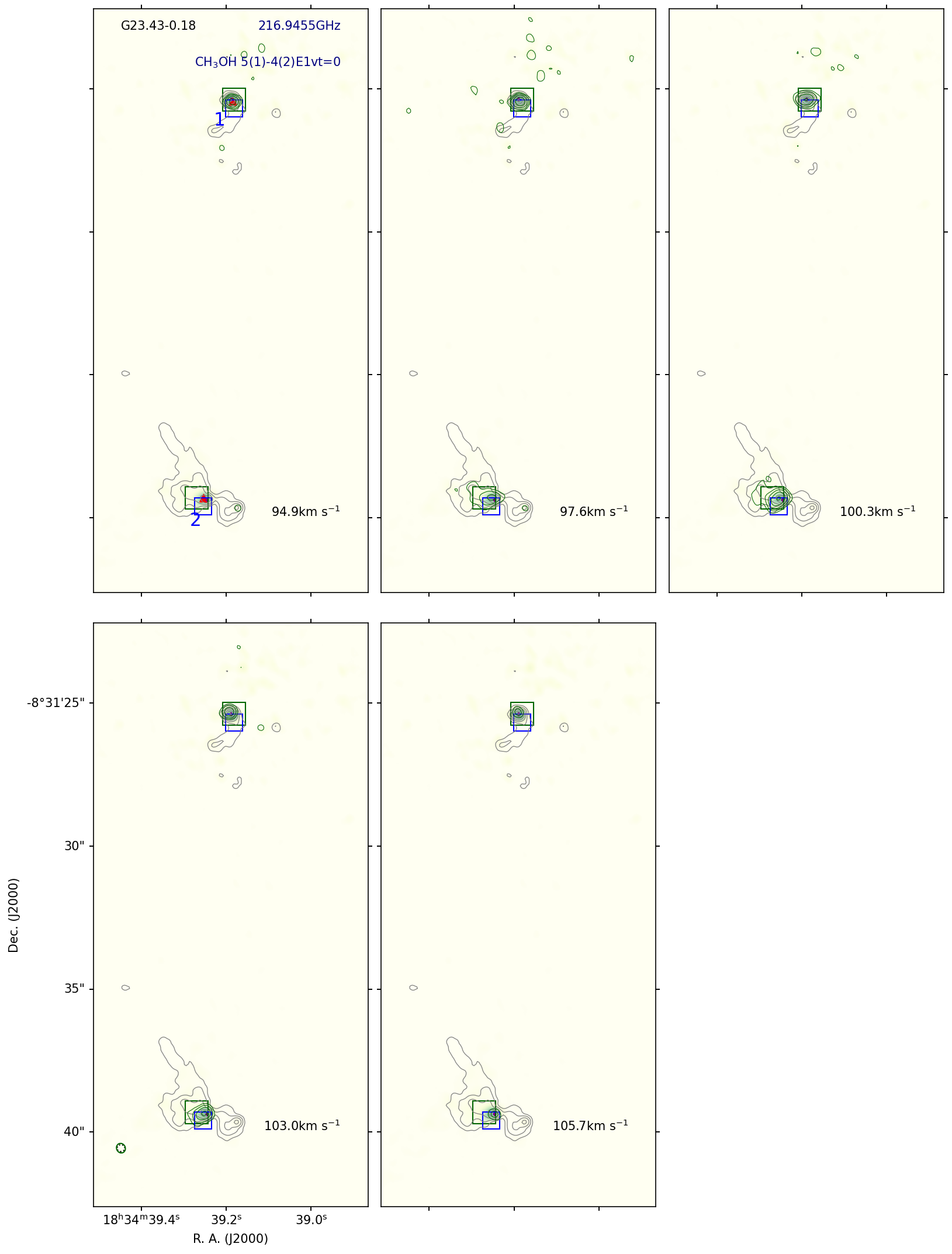}
\caption{The same as Figure \ref{fig:chnmap:G10.32} except for G23.43-0.18. Contour levels for 216.946 GHz methanol emission are 0.009, 0.016, 0.029, 0.052, 0.094, 0.169 Jy beam$^{-1}$.}
\label{fig:chnmap:G23.43}
\end{figure*}

\begin{figure*}
  \centering
\includegraphics[clip,width=1.3\columnwidth,keepaspectratio]{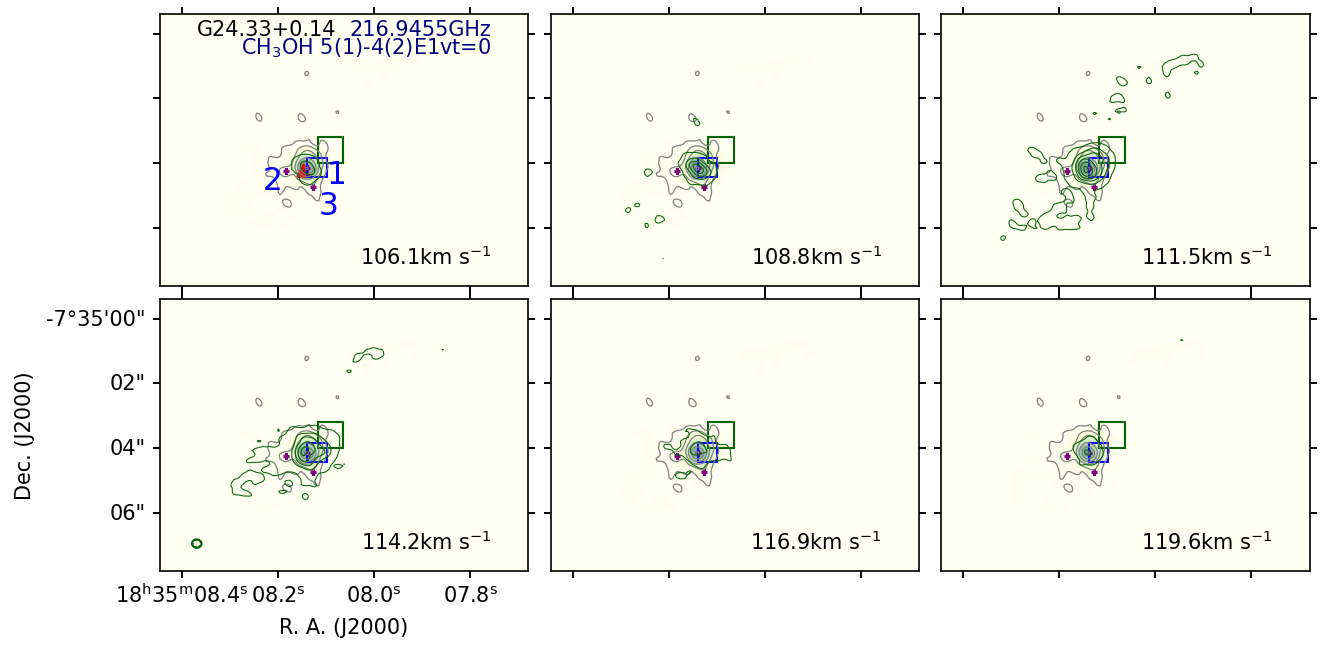}
\caption{The same as Figure \ref{fig:chnmap:G10.32} except for G24.33+0.14. Contour levels for 216.946 GHz methanol emission are 0.010, 0.069, 0.128, 0.186, 0.245, 0.304 Jy beam$^{-1}$.}
\label{fig:chnmap:G24.33}
\end{figure*}

\begin{figure*}
  \centering
\includegraphics[clip,width=1.3\columnwidth,keepaspectratio]{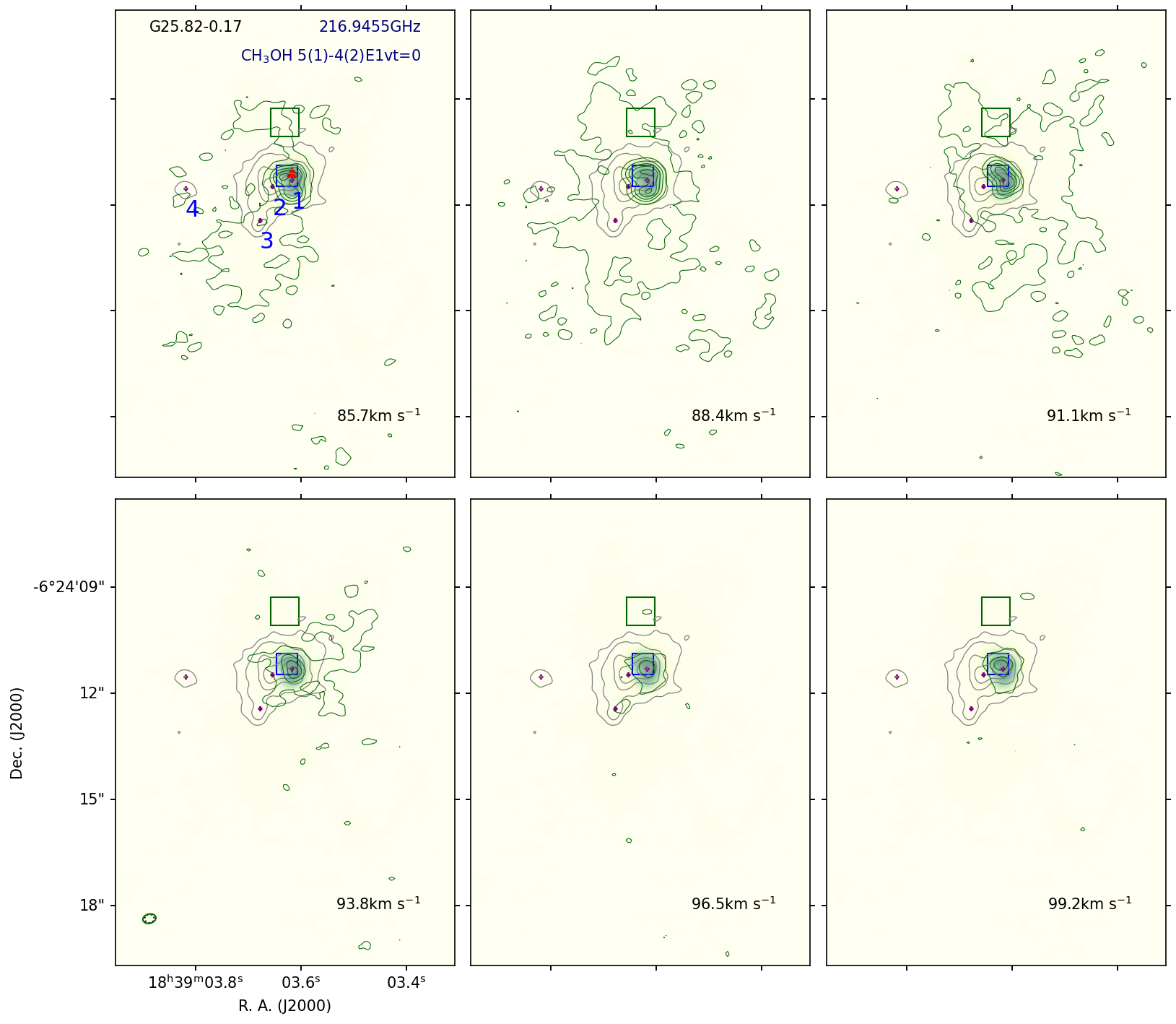}
\caption{The same as Figure \ref{fig:chnmap:G10.32} except for G25.82-0.17. Contour levels for 216.946 GHz methanol emission are 0.008, 0.053, 0.097, 0.142, 0.186, 0.231 Jy beam$^{-1}$.}
\label{fig:chnmap:G25.82}
\end{figure*}

\begin{figure*}
  \centering
\includegraphics[clip,width=1.3\columnwidth,keepaspectratio]{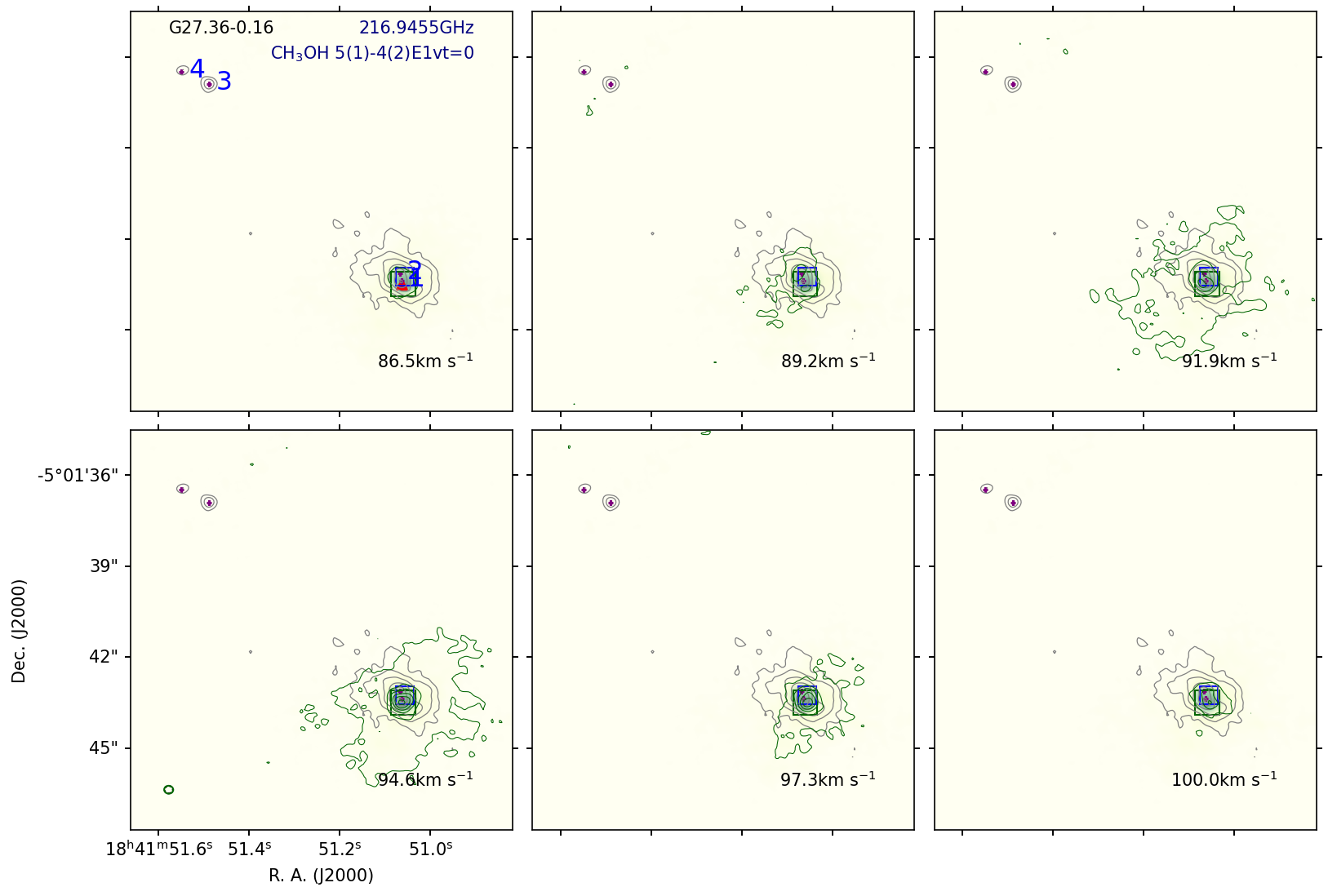}
\caption{The same as Figure \ref{fig:chnmap:G10.32} except for G27.36-0.16. Contour levels for 216.946 GHz methanol emission are 0.010, 0.059, 0.107, 0.156, 0.204, 0.253 Jy beam$^{-1}$.}
\label{fig:chnmap:G27.36}
\end{figure*}

\begin{figure*}
  \centering
\includegraphics[clip,width=1.3\columnwidth,keepaspectratio]{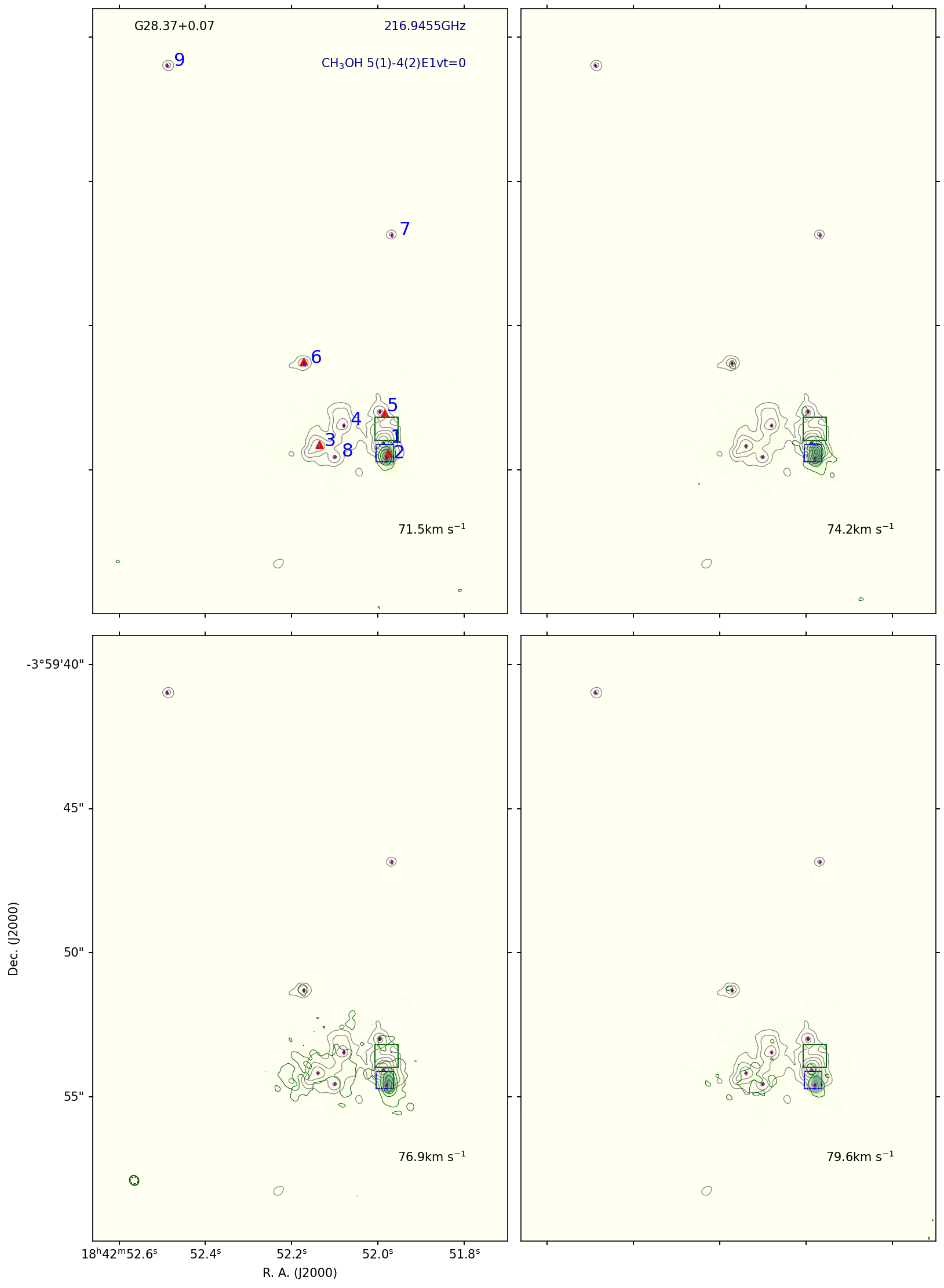}
\caption{The same as Figure \ref{fig:chnmap:G10.32} except for G28.37+0.07. Contour levels for 216.946 GHz methanol emission are 0.012, 0.042, 0.073, 0.103, 0.134, 0.164 Jy beam$^{-1}$.}
\label{fig:chnmap:G28.37}
\end{figure*}

\begin{figure*}
  \centering
\includegraphics[clip,width=1.3\columnwidth,keepaspectratio]{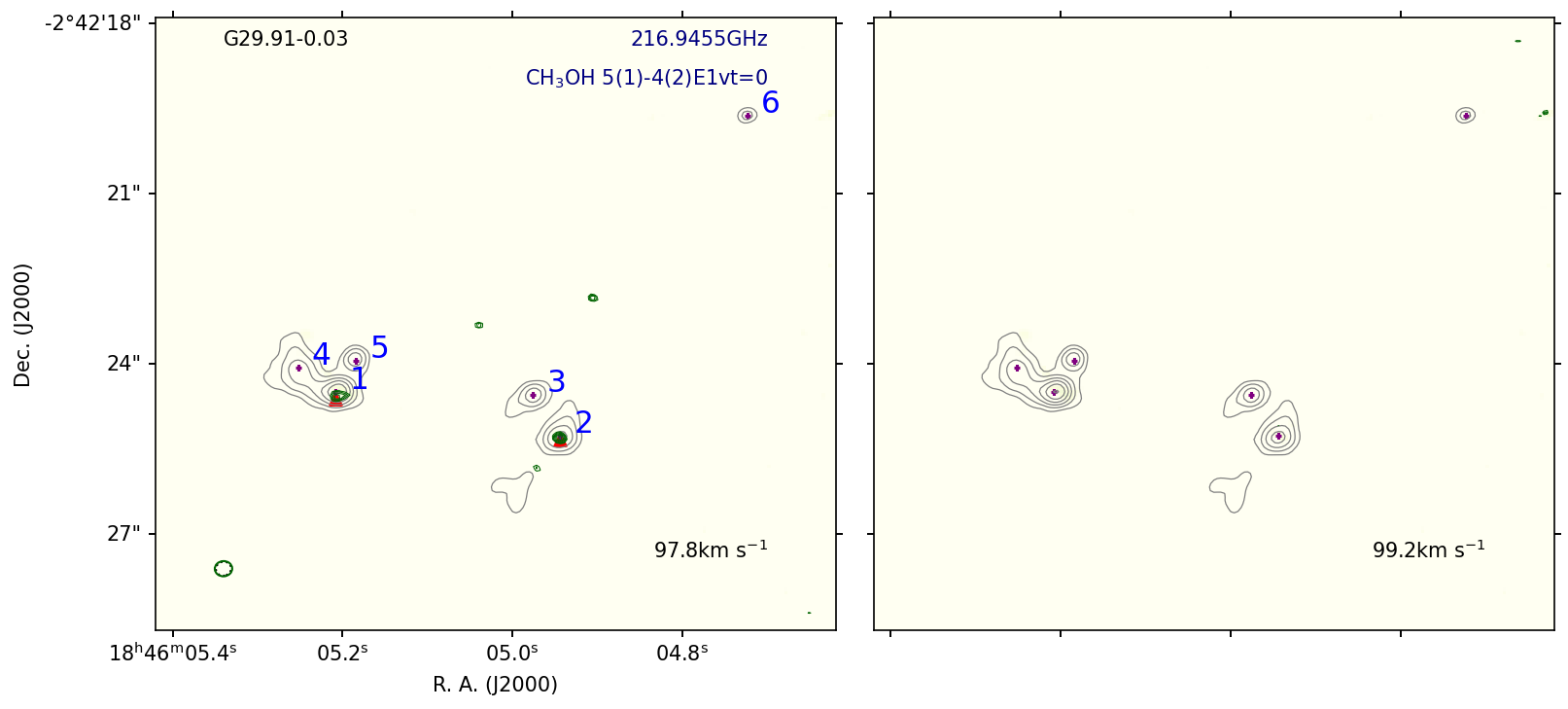}
\caption{The same as Figure \ref{fig:chnmap:G10.32} except for G29.91-0.03. Contour levels for 216.946 GHz methanol emission are 0.009, 0.009, 0.010, 0.010, 0.011, 0.011 Jy beam$^{-1}$.}
\label{fig:chnmap:G29.91}
\end{figure*}

\begin{figure*}
  \centering
\includegraphics[clip,width=1.3\columnwidth,keepaspectratio]{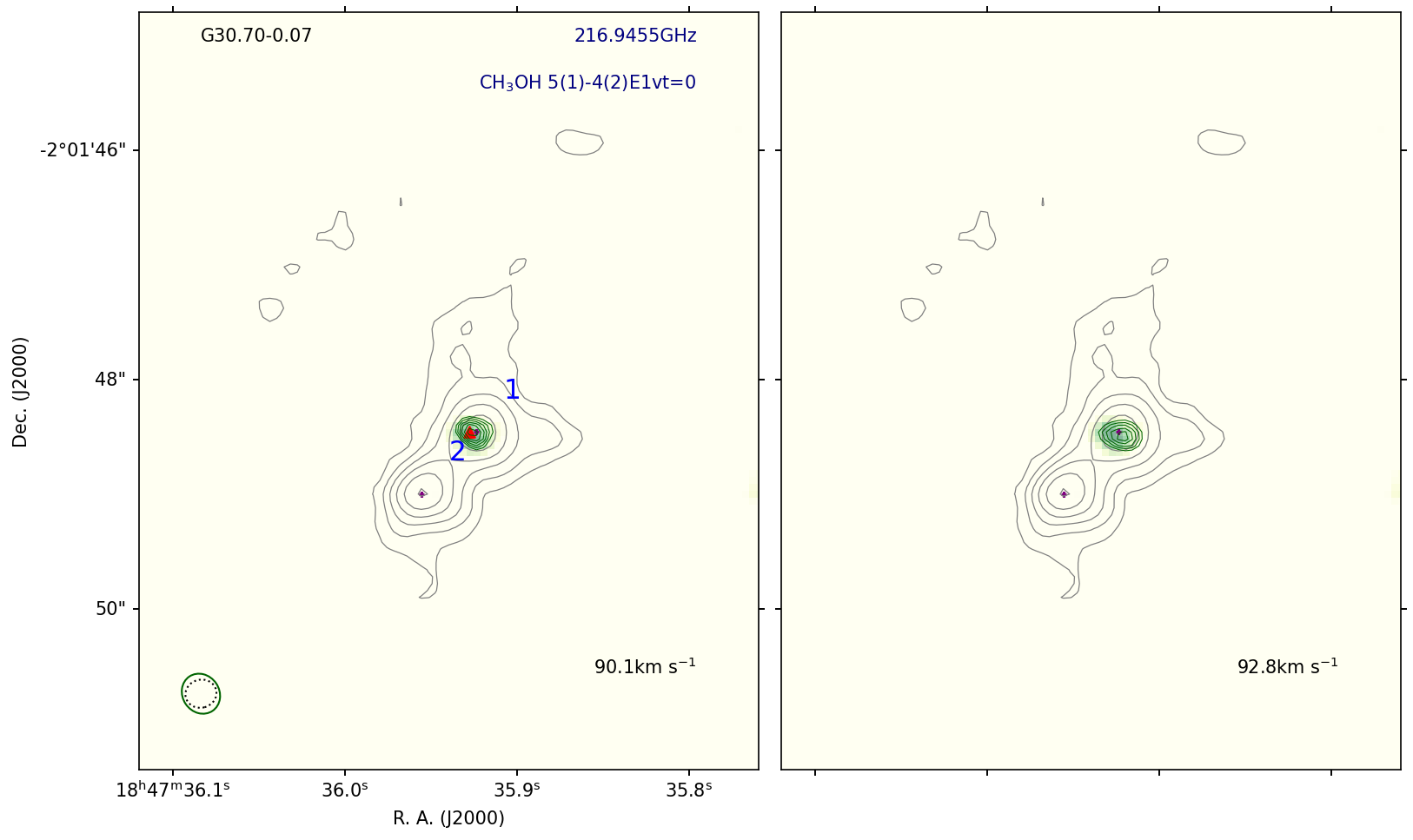}
\caption{The same as Figure \ref{fig:chnmap:G10.32} except for G30.70-0.07. Contour levels for 216.946 GHz methanol emission are 0.009, 0.010, 0.011, 0.013, 0.014, 0.015 Jy beam$^{-1}$.}
\label{fig:chnmap:G30.70}
\end{figure*}

\begin{figure*}
  \centering
\includegraphics[clip,width=1.0\columnwidth,keepaspectratio]{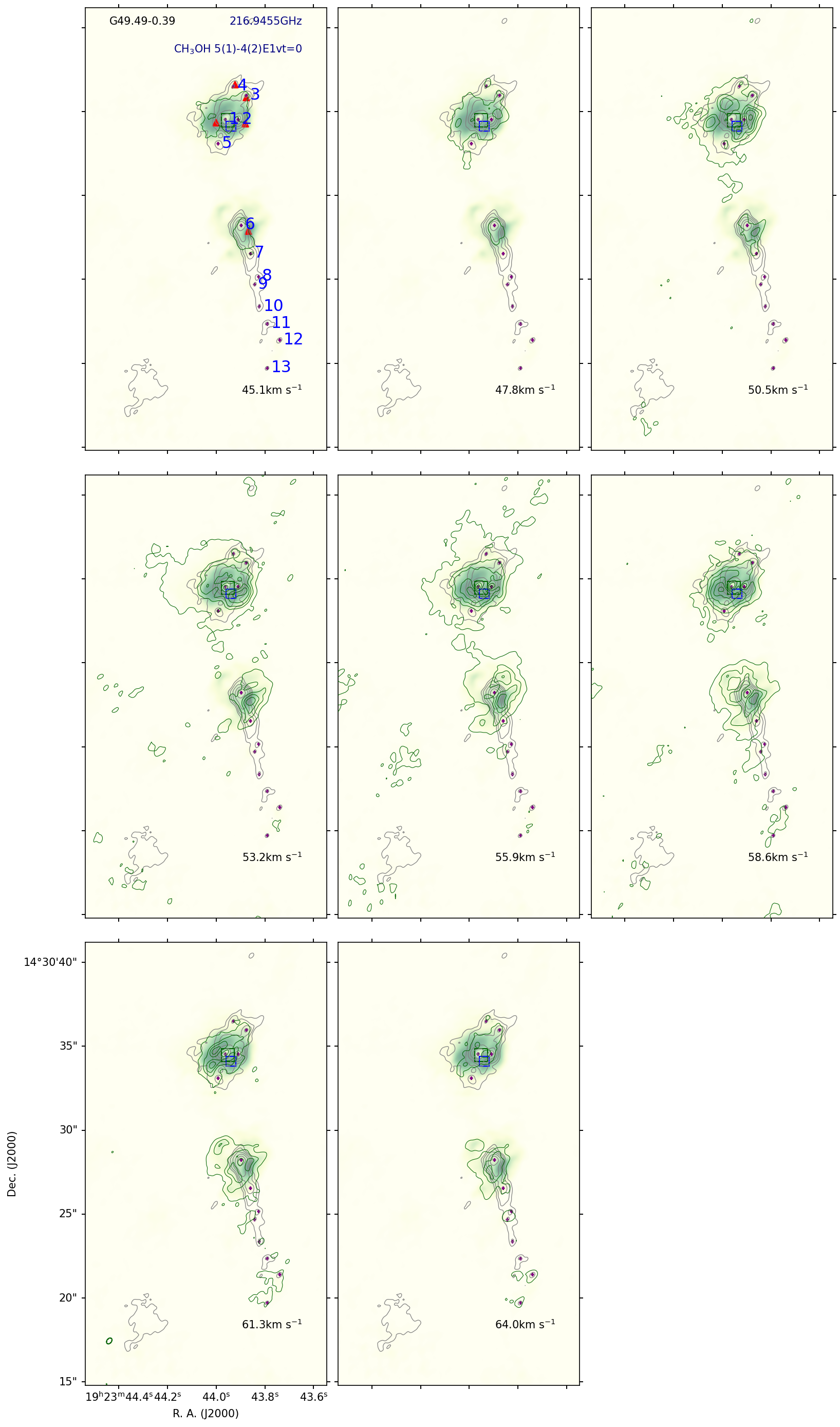}
\caption{The same as Figure \ref{fig:chnmap:G10.32} except for G49.49-0.39. Contour levels for 216.946 GHz methanol emission are 0.020, 0.131, 0.241, 0.352, 0.462, 0.573 Jy beam$^{-1}$.}
\label{fig:chnmap:G49.49}
\end{figure*}

%% file: s3_f4_rotation_diagram.tex
\begin{figure*}
\centering
\includegraphics[clip,width=0.23\textwidth,keepaspectratio]{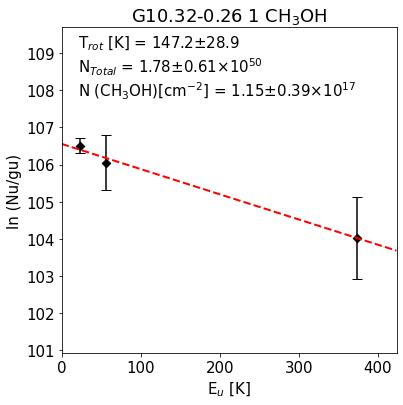}
\includegraphics[clip,width=0.23\textwidth,keepaspectratio]{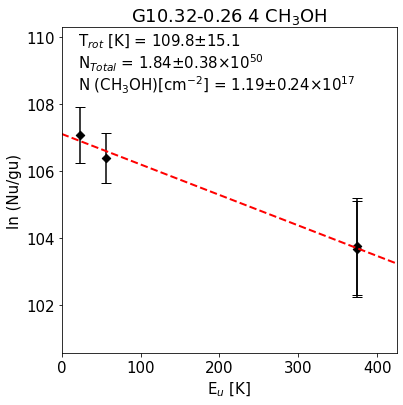}
\includegraphics[clip,width=0.23\textwidth,keepaspectratio]{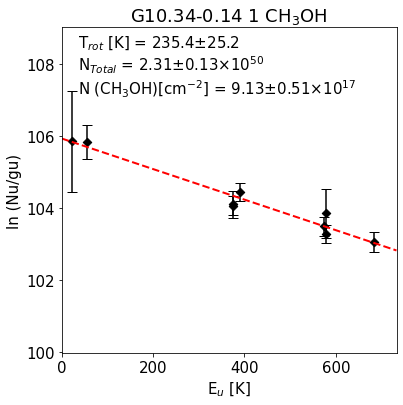}
\includegraphics[clip,width=0.23\textwidth,keepaspectratio]{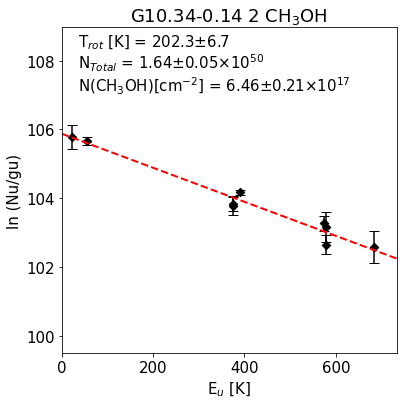}\\
\includegraphics[clip,width=0.23\textwidth,keepaspectratio]{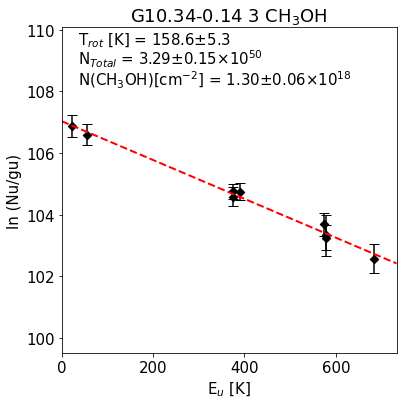}
\includegraphics[clip,width=0.23\textwidth,keepaspectratio]{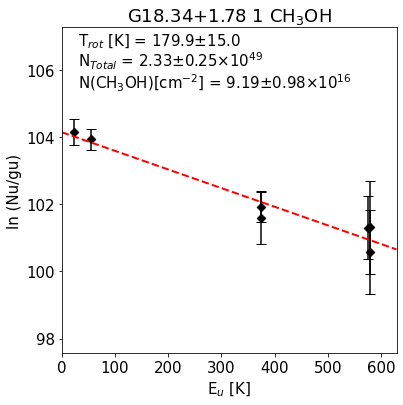}
\includegraphics[clip,width=0.23\textwidth,keepaspectratio]{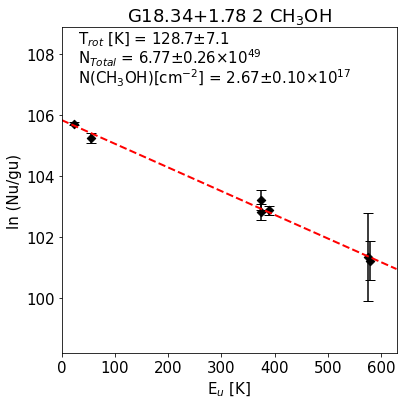}
\includegraphics[clip,width=0.23\textwidth,keepaspectratio]{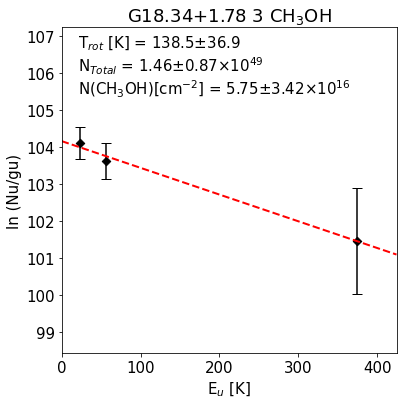}\\
\includegraphics[clip,width=0.23\textwidth,keepaspectratio]{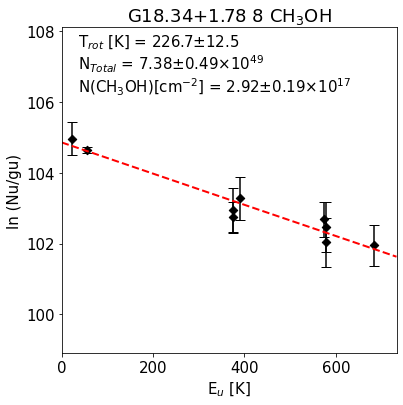}
\includegraphics[clip,width=0.23\textwidth,keepaspectratio]{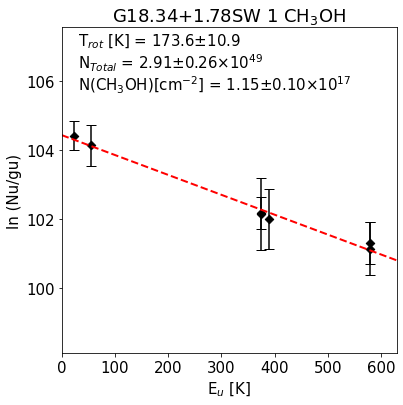}
\includegraphics[clip,width=0.23\textwidth,keepaspectratio]{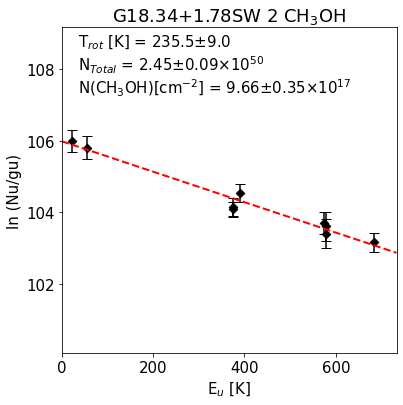}
\includegraphics[clip,width=0.23\textwidth,keepaspectratio]{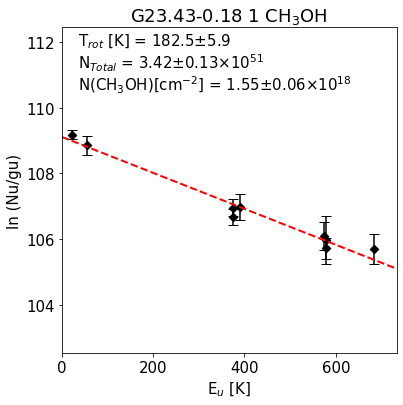}\\
\includegraphics[clip,width=0.23\textwidth,keepaspectratio]{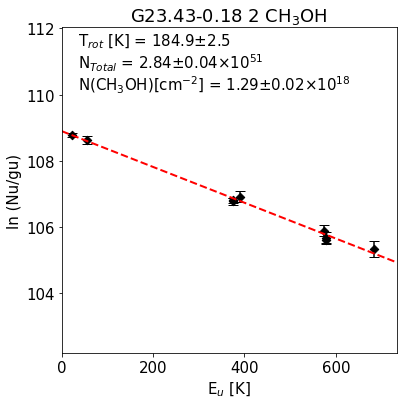}
\includegraphics[clip,width=0.23\textwidth,keepaspectratio]{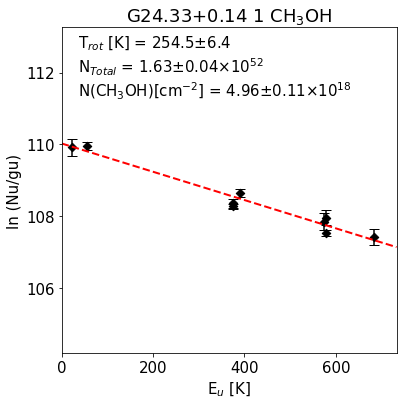}
\includegraphics[clip,width=0.23\textwidth,keepaspectratio]{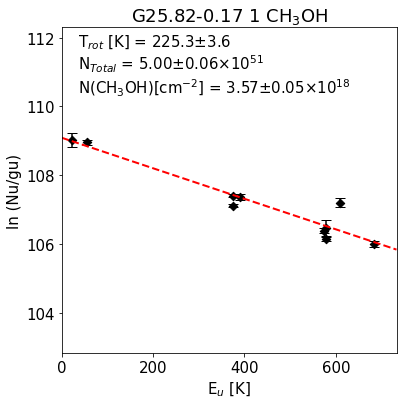}
\includegraphics[clip,width=0.23\textwidth,keepaspectratio]{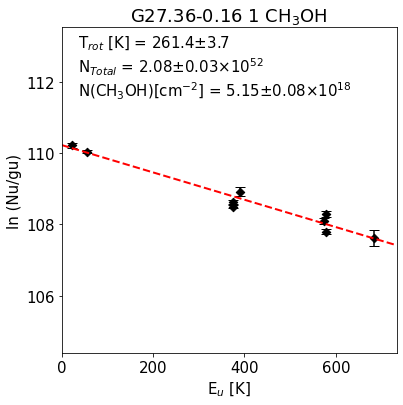}\\
\caption{Rotation diagram for CH$_{3}$OH}
\label{fig:rot_diagram}
\end{figure*}